%% file: sm07.tex
\documentclass[11pt]{cernrepdbg}
\usepackage{graphicx}
\usepackage{epsfig}
\usepackage{color}
\usepackage{bbm}
\usepackage{cite}
\usepackage{url}
\usepackage{amsmath}
\usepackage{amssymb}
\usepackage{here}
\usepackage{xspace}
\usepackage{epsf}
\usepackage{feynmp}

\begin{document}
\bibliographystyle{lesHouchesdbg}

\input{extra/SMHmacro}

\title{\centering{Standard Model Handles and Candles Working Group:\\
          \textbf{Tools and Jets Summary Report}}}

\input{affnumbers.tex}

\newcommand{\nameGlasgow}{Department of Physics and Astronomy, 
  University of Glasgow, G12 8QQ, Glasgow, UK}
\newcommand{\nameVUB}{Vrije Universiteit Brussel, Pleinlaan 2
  B-1050 Brussel, Belgium}
\newcommand{\nameULB}{Universit\'e Libre de Bruxelles (ULB-IIHE), 
  boulevard du Triomphe, 1050 Bruxelles, Belgium}
\newcommand{\nameAachen}{Institut f\"ur Theoretische Physik, RWTH Aachen, 
  D-52056 Aachen, Germany}
\newcommand{\nameLPTHE}{LPTHE, UPMC Univ.\ Paris 6; 
  Univ.\ Paris Diderot (Paris 7); CNRS UMR 7589;
  F-75252 Paris Cedex 05, France}
\newcommand{\nameLPNHE}{LPNHE, UPMC Univ.\ Paris 6; 
  Univ.\ Paris Diderot (Paris 7); IN2P3/CNRS ;
  F-75252 Paris Cedex 05, France}
\newcommand{\nameFNAL}{Fermi National Accelerator Laboratory,
  Batavia, IL-60510, USA}
\newcommand{\namePrinceton}{Department of Physics, Princeton
  University, Princeton, NJ 08544, USA}
\newcommand{\nameDubna}{JINR Dubna, 141980 Russia}
\newcommand{\nameBuffalo}{Department of Physics, University at
  Buffalo, Buffalo, NY 14260, USA}
\newcommand{\nameKarlsruheExp}{Institut f\"ur Experimentelle
  Kernphysik, Universit\"at Karlsruhe, Wolfgang-Gaede-Str.\ 1,
  D-76131~Karlsruhe, Germany}
\newcommand{\nameKarlsruheRech}{Institut f\"ur Wissenschaftliches
  Rechnen, Forschungszentrum Karlsruhe, Hermann-von-Helmholtz-Platz 1,
  D-76344~Eggenstein-Leopoldshafen, Germany}
\newcommand{\nameMSU}{Department of Physics and Astronomy,
Michigan State University, East Lansing, MI 48824, USA}
\newcommand{\nameUCRiverside}{Department of Physics and Astronomy, UC
  Riverside, Riverside, CA 92521, USA}
\newcommand{\nameFrascati}{INFN, Frascati, Italy}
\newcommand{\nameSoton}{School of Physics and Astronomy,
University of Southampton, Southampton SO17 1BJ, UK}
\newcommand{\nameRAL}{Rutherford Appleton Laboratory, Oxon OX11
0QX, UK}
\newcommand{\nameBristol}{H. H. Wills Physics Laboratory,
University of Bristol, UK}
\newcommand{\namePSI}{Paul Scherrer Institut, 5232 Villigen PSI,
Switzerland}
\newcommand{\nameDESY}{Deutsches Elektronen-Synchrotron DESY,
Notkestrasse 85, D-22607 Hamburg, Germany }
\newcommand{\namePavia}{Dipartimento di Fisica Nucleare e Teorica,
Universit\`a di Pavia, Via Bassi 6, Pavia, Italy}
\newcommand{\namePaviaINFN}{INFN, Sezione di Pavia, Via Bassi 6, Pavia, Italy}
\newcommand{\nameAlberta}{Department of Physics, University of
Alberta, Edmonton, AB T6G 2J1, Canada}
\newcommand{\nameINRMoscow}{Institute for Nuclear Research of
Russian Academy of Sciences, 117312 Moscow, Russia}
\newcommand{\nameMPIMunich}{Max-Planck-Institut f\"ur Physik,
F\"ohringer Ring 6, D--80805 Munich, Germany}
\newcommand{\nameKarlsruheTh}{Institut f\"ur Theoretische
Teilchenphysik, Universit\"at Karlsruhe, D-76128 Karlsruhe, Germany}
\newcommand{\nameCavendish}{Cavendish Laboratory, University of
Cambridge, JJ Thomson Avenue, Cambridge, CB3 0HE, UK}
\newcommand{\nameOnLeaveMoscow}{on leave from Moscow State
University, Moscow, Russia}
\newcommand{\nameCernTH}{CERN PH-TH, CH-1211 Geneva 23, Switzerland}
\newcommand{\nameCernEP}{CERN PH-EP, CH-1211 Geneva 23, Switzerland}
\newcommand{\nameFNALTh}{Theoretical Physics, Fermilab MS106, Box
500, Batavia, IL--60510, USA}
\newcommand{\nameUCL}{Department of Physics and Astronomy, University College
  London, WC1E 6BT, UK}
\newcommand{\nameNowUCL}{currently at Department of Physics and Astronomy, University College
  London, WC1E 6BT, UK}
\newcommand{\nameMilan}{Dipartimento di Fisica, Universit\`a di
Milano and INFN, Sezione di Milano, Milano, Italy}
\newcommand{\nameBNL}{Brookhaven National Laboratory, Upton, NY 11973, USA}
\newcommand{\nameArizona}{Department of Physics, University of
  Arizona, Tucson, Arizona 85721, USA}
\newcommand{\nameLouvain}{Universit\'e Catholique de Louvain (UCL), Belgium}
\newcommand{\nameGent}{University of Gent, Belgium}
\newcommand{\nameRockefeller}{Rockefeller University 1230 York Ave New
  York NY 10065}
\newcommand{\nameLouisiana}{Louisiana Tech University, Ruston, Louisiana, USA}
\newcommand{\nameFreiburg}{Physikalisches Institut,
Albert-Ludwigs Universit\"t Freiburg,
Hermann-Herder-Strasse 3a,
790104 Freiburg,
Germany
}

  \author{\underline{Convenors}:
C.~Buttar$^{\affGlasgow}$,
J.~D'Hondt$^{\affVUB}$,
M.~Kr\"amer$^{\affAachen}$, 
G.~Salam$^\affLPTHE$,
M.~Wobisch$^\affLouisiana$,
\\
\underline{Contributing authors}:
N.E.~Adam$^{\affPrinceton}$,
V.~Adler$^{\affULB}$,
A.~Arbuzov$^{\affDubna}$,
D.~Bardin$^{\affDubna}$,
U.~Baur$^{\affBuffalo}$,
A.A.~Bhatti$^{\affRockefeller}$,
S.~Bondarenko$^{\affDubna}$,
V.~B\"uge$^{\affKarlsruheExp,\affKarlsruheRech}$, 
J.M.~Butterworth$^{\affUCL}$,
M.~Cacciari$^{\affLPTHE}$,
M.~Campanelli$^{\affMSU}$,
Q.-H.~Cao$^{\affUCRiverside}$,
C.M.~Carloni~Calame$^{\affFrascati,\affSoton}$,
P.~Christova$^{\affDubna}$,
D.~D'Enterria$^{\affCernEP}$,
J.~D'Hondt$^{\affVUB}$,
S.~Ferrag$^{\affGlasgow}$,
K.~Geerlings$^{\affMSU}$,
V.~Halyo$^{\affPrinceton}$,
M.~Heinrich$^{\affKarlsruheExp}$,
J.~Huston$^{\affMSU}$,
J.~Jackson$^{\affRAL,\affBristol}$,
B.~Jantzen$^{\affPSI}$,
L.~Kalinovskaya$^{\affDubna}$,
D.~Kcira$^{\affLouvain}$, 
B.~Klein$^{\affKarlsruheExp}$, 
A.~Kulesza$^{\affDESY}$, 
P.~Loch$^{\affArizona}$, 
G.~Montagna$^{\affPavia,\affPaviaINFN}$,
S.~Moretti$^{\affSoton}$,
D. Newbold$^{\affRAL}$,
O.~Nicrosini$^{\affPaviaINFN}$,
H.~Nilsen$^{\affFreiburg}$,
A.A.~Penin$^{\affAlberta,\affINRMoscow}$,
F.~Piccinini$^{\affPaviaINFN}$,
S.~Pozzorini$^{\affMPIMunich}$,
K.~Rabbertz$^{\affKarlsruheExp}$,
J.~Rojo Chacon$^{\affLPTHE}$,
R.~Sadykov$^{\affDubna}$,
M.~Schulze$^{\affKarlsruheTh}$,
C.~Shepherd-Themistocleous$^{\affRAL}$,
A.~Sherstnev$^{\affCavendish,\affOnLeaveMoscow}$,
P.Z.~Skands$^{\affCernTH,\affFNALTh}$,
L.~Sonnenschein$^{\affCernEP,\affLPNHE}$,
G.~Soyez$^{\affBNL}$,
R.S.~Thorne$^{\affUCL}$,
M.~Tytgat$^{\affGent}$,
P.~Van~Mulders$^{\affVUB}$,
M.~Vazquez Acosta$^{\affCernEP}$,
A.~Vicini$^{\affMilan}$,
I.~Villella$^{\affVUB}$,
D.~Wackeroth$^{\affBuffalo}$,
C.-P.~Yuan$^{\affMSU}$
\\\mbox{} }
\institute{\centering{\small
  \input{afflist.tex}
}}
 \maketitle
 \begin{center}
   \textit{%
     This report summarises the activity on comparisons of existings
     tools for the standard model and on issues in jet physics by the
     Standard Model Handles and Candles working group during and
     subsequent to the Workshop ``Physics at TeV Colliders'', Les
     Houches, France, 11--29 June, 2007.
     %
   }
\end{center}
\newpage

\setcounter{tocdepth}{1}
\tableofcontents
\setcounter{footnote}{0}

\section[Foreword]{FOREWORD}
\input{extra/LH07foreword.tex}
\part[COMPARISON OF EXISTING TOOLS FOR THE STANDARD MODEL]
{COMPARISON OF EXISTING TOOLS FOR THE STANDARD MODEL}

\section[A tuned comparison of electroweak predictions for $Z$ boson 
observables with HORACE, SANC and ZGRAD2]
{A TUNED COMPARISON OF ELECTROWEAK PREDICTIONS FOR $Z$ BOSON
  OBSERVABLES WITH HORACE, SANC AND ZGRAD2~\protect
  \footnote{Contributed by: A.~Arbuzov, D.~Bardin, U.~Baur,
    S.~Bondarenko, C.M.~Carloni~Calame, P.~Christova, L.~Kalinovskaya,
    G.~Montagna, O.~Nicrosini, R.~Sadykov, A.~Vicini, D.~Wackeroth}}
\input{s_wackeroth/ZcomparisonFeb.tex}

\clearpage

\section[The neutral-current Drell-Yan process in the high invariant-mass
  region]
{THE NEUTRAL-CURRENT DRELL-YAN PROCESS IN THE HIGH INVARIANT-MASS
  REGION~\protect \footnote{Contributed by: U.~Baur, Q.-H.~Cao,
    C.M.~Carloni~Calame, S.~Ferrag, J.~Jackson, B.~Jantzen,
    G.~Montagna, S.~Moretti, D. Newbold, O.~Nicrosini, A.A.~Penin,
    F.~Piccinini, S.~Pozzorini, C.~Shepherd-Themistocleous, A.~Vicini,
    D.~Wackeroth, C.-P.~Yuan}} 
\input{s_jantzen/Dilepton_Ale_final.tex}
\clearpage

\section[Comparison of HORACE and PHOTOS in the $\Zll$ peak region]
{COMPARISON OF HORACE AND PHOTOS IN THE $\Zll$ PEAK REGION~\protect
  \footnote{Contributed by: N.E.~Adam, C.M.~Carloni~Calame, V.~Halyo,
    C.~Shepherd-Themistocleous}}
\input{s_halyo/DiLep_ZHorace.tex}

\clearpage

\section[Electroweak corrections to  $pp\to Wj$]
{ELECTROWEAK CORRECTIONS TO  $pp\to Wj$~\protect
\footnote{Contributed by: A.~Kulesza, S.~Pozzorini, M.~Schulze}}
\input{s_pozzorini/ppwj_main.tex}

\clearpage

\section[Some interesting min-bias distributions for early LHC runs]
{SOME INTERESTING MIN-BIAS DISTRIBUTIONS FOR EARLY LHC RUNS~\protect
\footnote{Contributed by: P.~Z.~Skands}}
\input{s_skands/qcdplots-houches.tex}
\clearpage

\section[Parton distributions for LO generators]
{PARTON DISTRIBUTIONS FOR LO GENERATORS~\protect
\footnote{Contributed by: A.~Sherstnev, R.S.~Thorne}}
\input{s_sherstnev/lomod.tex}

\clearpage

\part[ISSUES IN JET PHYSICS]{ISSUES IN JET PHYSICS}

\section[Jet physics introduction]{JET PHYSICS INTRODUCTION\protect\footnote{%
Convenors: G.P.~Salam and M.~Wobisch;
Contributing authors: 
V.~Adler,            
A.~A.~Bhatti,        
J.~M.~Butterworth,   
V.~B\"uge,           
M.~Cacciari,         
M.~Campanelli,       
D.~D'Enterria,       
J.~D'Hondt,          
J.~Huston,           
D.~Kcira,            
P.~Loch,             
K.~Rabbertz,         
J.~Rojo Chacon,      
L.~Sonnenschein,     
G.~Soyez,            
M.~Tytgat,           
P.~Van Mulders,      
M.~Vazquez Acosta,   
I.~Villella          
}}
\label{sec:jet-intro}

\input{s_jet_intro/les-houches-jet-intro.tex}
\clearpage

\section[Accords related to the hadronic final state]{ACCORDS RELATED
  TO THE HADRONIC FINAL STATE\protect\footnote{%
Convenors: G.P.~Salam and M.~Wobisch;
Contributing authors: 
V.~Adler,            
A.~Bhatti,           
J.~M.~Butterworth,   
V.~B\"uge,           
M.~Cacciari,         
D.~D'Enterria,       
J.~D'Hondt,          
J.~Huston,           
D.~Kcira,            
P.~Loch,             
H.~Nilsen,           
K.~Rabbertz,         
J.~Rojo-Chacon,      
L.~Sonnenschein,     
G.~Soyez,            
M.~Tytgat,           
P.~Van Mulders,      
M.~Vazquez~Acosta,   
I.~Villella          
}}
\label{sec:hadronic-accords}
\input{s_jet_accords/les-houches-jet-accords.tex}
\clearpage 

\section[Quantifying the performance of jet algorithms at the LHC]{%
QUANTIFYING THE PERFORMANCE OF JET ALGORITHMS AT THE LHC%
~\protect
\footnote{Contributed by: 
  M.~Cacciari,      
  J.~Rojo-Chacon,   
  G.~P.~Salam,      
  G.~Soyez          
}}
\label{sec:lhprocs_gavin}
\input{s_cacciari/les-houches-jet-algorithms.tex}
\clearpage

\section[Influence of jet algorithms and jet sizes on the reconstruction
  of the hard process from stable particles at LHC energies]
{INFLUENCE OF JET ALGORITHMS AND JET SIZES ON THE RECONSTRUCTION
  OF THE HARD PROCESS FROM STABLE PARTICLES AT LHC ENERGIES~\protect
\footnote{Contributed by: V.\ B\"uge, M.\ Heinrich, B.\ Klein, K.\
  Rabbertz}}
\label{sec:rabbertz}
\input{s_rabbertz/lhkajets.tex}
\clearpage

\section[A study of jet algorithms using the SpartyJet tool]
{A STUDY OF JET ALGORITHMS USING THE SPARTYJET TOOL~\protect
\footnote{Contributed by: M.~Campanelli, K.~Geerlings, J.~Huston}}
\label{sec:campanelli}
\input{s_campanelli/campanelli.tex}

\clearpage


\bibliography{sm07}
\end{document}

%% file: extra/SMHmacro.tex
{\def\nl{\nonumber\\}
\def\beq{\begin{equation}}
\def\eeq{\end{equation}}
\def\beqar{\begin{eqnarray}}
\def\eeqar{\end{eqnarray}}
\newcommand{\GeV}{\unskip\,\mathrm{GeV}}
\newcommand{\TeV}{\unskip\,\mathrm{TeV}}
\newcommand{\sw}{s_{\scriptscriptstyle{\mathrm{W}}}}
\newcommand{\cw}{c_{\scriptscriptstyle{\mathrm{W}}}}
\newcommand{\cew}{C^{\mathrm{ew}}}
\def\ie{i.e.\ }
\def\wrt{wrt.\ }
\newcommand{\ord}{{\cal O}}
\newcommand{\lsim}
{\;\raisebox{-.3em}{$\stackrel{\displaystyle <}{\sim}$}\;}
\newcommand{\rL}{\mathrm{L}}
\newcommand{\rT}{{\mathrm{T}}}
\newcommand{\rS}{{\mathrm{S}}}
\newcommand{\rd}{{\mathrm{d}}}
\newcommand{\pT}{p_{\mathrm{T}}}
\newcommand{\pTcut}{p_{\mathrm{T}}^{\mathrm{cut}}}
\newcommand{\pTW}{p_{\mathrm{T,\,}W}}
\newcommand{\pTminj}{p_{\mathrm{T},\,j}^{\mathrm{min}}}
\newcommand{\M}{{\cal {M}}}
\newcommand{\NNLLa}{\stackrel{\mathrm{NNLL}}{=}}
\newcommand{\NLLa}{\stackrel{\mathrm{NLL}}{=}}
\newcommand{\shat}{{\hat s}}
\newcommand{\that}{{\hat t}}
\newcommand{\uhat}{{\hat u}}
\newcommand{\MSBAR}{\overline{\mathrm{MS}}}
\newcommand{\Deltamsbar}{\bar \Delta_{\mathrm{UV}}}

\newcommand{\dy}{$ pp \to Z \to l^+ l^- + X $}
\newcommand{\alp}{$\alpha$ }
\newcommand{\alphas}{$\alpha_S$ }
{\newcommand{\MW}{M_\mathrm{W}}
\newcommand{\MZ}{M_\mathrm{Z}}

\newcommand{\TeVx}{\,\mbox{Te\kern-0.2exV}}
\newcommand{\GeVx}{\,\mbox{Ge\kern-0.2exV}}

\newcommand{\plotrow}[2]{
\vspace*{-3mm}\noindent\parbox{0.49\textwidth}{%
\includegraphics[scale=0.425]{#1}\hspace*{-5.2cm}
\raisebox{1.1cm}{\Large\bf\sf Tevatron}}%
\hspace*{2mm}\parbox{0.49\textwidth}{%
\includegraphics[scale=0.425]{#2}\hspace*{-4.85cm}
\raisebox{1.1cm}{\Large\bf\sf LHC}}\\[-5mm]
}

\renewcommand{\l}{\ell}
\newcommand{\Zee}{\mbox{$ Z\to e^{+}  e^{-}$}}
\newcommand{\Zmumu}{\mbox{$ Z\to \mu^{+} \mu^{-}$}}
\newcommand{\Zll}{\mbox{$ Z\to \l^{+} \l^{-}$}}
\newcommand{\Mll}{\mbox{$M_{\l\l}$}}
\newcommand{\gammaZee}{\mbox{$\gamma^{*}/Z\to e^{+} e^{-}$}}
\newcommand{\Zgam}{\mbox{$Z/\gamma^{*}$}}
\newcommand{\BR}{\mathop{\rm BR}\nolimits}
\newcommand{\pTx}{\mbox{$p_{\rm T}$}}
\newcommand{\etal}{{\it et al}}
\def    \beq            {\begin{equation}}
\def    \eeq            {\end{equation}}
\def    \ptel           {\mbox{$p_{\rm T}^{\l}$}}
%

\def\numunue{\nu_\mu\rightarrow\nu_e}
\def\numunutau{\nu_\mu\rightarrow\nu_\tau}
\def\nuebar{\bar\nu_e}
\def\numubar{\bar\nu_\mu}
\def\numubarnuebar{\bar\nu_\mu\rightarrow\bar\nu_e}

\def\dr{\raisebox{2.1ex}{$\scriptsize\lfloor$}\!\raisebox{1ex}{$\rightarrow$}}
\def\bb{b\bar{b}}
\def\ttb{t\bar{t}}
\def\cc{c\bar{c}}
\def\qq{q\bar{q}}
\def\ra{ \rightarrow }
\def\whs{\widehat{\sigma}}
\def\msb{\overline{\rm MS}}
\def\dis{{\rm DIS}}


%% file: affnumbers.tex
%

\newcommand{\affGlasgow}{{1}}
\newcommand{\affVUB}{{2}}
\newcommand{\affAachen}{{3}}
\newcommand{\affLPTHE}{{4}}
\newcommand{\affLouisiana}{{5}}
\newcommand{\affPrinceton}{{6}}
\newcommand{\affULB}{{7}}
\newcommand{\affDubna}{{8}}
\newcommand{\affBuffalo}{{9}}
\newcommand{\affRockefeller}{{10}}
\newcommand{\affKarlsruheExp}{{11}}
\newcommand{\affKarlsruheRech}{{12}}
\newcommand{\affUCL}{{13}}
\newcommand{\affMSU}{{14}}
\newcommand{\affUCRiverside}{{15}}
\newcommand{\affFrascati}{{16}}
\newcommand{\affSoton}{{17}}
\newcommand{\affCernEP}{{18}}
\newcommand{\affRAL}{{19}}
\newcommand{\affBristol}{{20}}
\newcommand{\affPSI}{{21}}
\newcommand{\affLouvain}{{22}}
\newcommand{\affDESY}{{23}}
\newcommand{\affArizona}{{24}}
\newcommand{\affPavia}{{25}}
\newcommand{\affPaviaINFN}{{26}}
\newcommand{\affFreiburg}{{27}}
\newcommand{\affAlberta}{{28}}
\newcommand{\affINRMoscow}{{29}}
\newcommand{\affMPIMunich}{{30}}
\newcommand{\affKarlsruheTh}{{31}}
\newcommand{\affCavendish}{{32}}
\newcommand{\affOnLeaveMoscow}{{33}}
\newcommand{\affCernTH}{{34}}
\newcommand{\affFNALTh}{{35}}
\newcommand{\affLPNHE}{{36}}
\newcommand{\affBNL}{{37}}
\newcommand{\affGent}{{38}}
\newcommand{\affMilan}{{39}}

%% file: afflist.tex
%

$^{\affGlasgow}$ \nameGlasgow \\
$^{\affVUB}$ \nameVUB \\
$^{\affAachen}$ \nameAachen \\
$^{\affLPTHE}$ \nameLPTHE \\
$^{\affLouisiana}$ \nameLouisiana \\
$^{\affPrinceton}$ \namePrinceton \\
$^{\affULB}$ \nameULB \\
$^{\affDubna}$ \nameDubna \\
$^{\affBuffalo}$ \nameBuffalo \\
$^{\affRockefeller}$ \nameRockefeller \\
$^{\affKarlsruheExp}$ \nameKarlsruheExp \\
$^{\affKarlsruheRech}$ \nameKarlsruheRech \\
$^{\affUCL}$ \nameUCL \\
$^{\affMSU}$ \nameMSU \\
$^{\affUCRiverside}$ \nameUCRiverside \\
$^{\affFrascati}$ \nameFrascati \\
$^{\affSoton}$ \nameSoton \\
$^{\affCernEP}$ \nameCernEP \\
$^{\affRAL}$ \nameRAL \\
$^{\affBristol}$ \nameBristol \\
$^{\affPSI}$ \namePSI \\
$^{\affLouvain}$ \nameLouvain \\
$^{\affDESY}$ \nameDESY \\
$^{\affArizona}$ \nameArizona \\
$^{\affPavia}$ \namePavia \\
$^{\affPaviaINFN}$ \namePaviaINFN \\
$^{\affFreiburg}$ \nameFreiburg \\
$^{\affAlberta}$ \nameAlberta \\
$^{\affINRMoscow}$ \nameINRMoscow \\
$^{\affMPIMunich}$ \nameMPIMunich \\
$^{\affKarlsruheTh}$ \nameKarlsruheTh \\
$^{\affCavendish}$ \nameCavendish \\
$^{\affOnLeaveMoscow}$ \nameOnLeaveMoscow \\
$^{\affCernTH}$ \nameCernTH \\
$^{\affFNALTh}$ \nameFNALTh \\
$^{\affLPNHE}$ \nameLPNHE \\
$^{\affBNL}$ \nameBNL \\
$^{\affGent}$ \nameGent \\
$^{\affMilan}$ \nameMilan \\

%% file: extra/LH07foreword.tex
The construction of the LHC and its detectors is nearing completion, and
first collisions are to be expected in 2008. While in essence built to
discover new physics phenomena, the proton collisions at the LHC will
provide a huge number of Standard Model events including jet, W, Z and
top quark processes. These events can be used to further scrutinize the
Standard Model as a theory, but are essential Handles and Candles for the
broad physics commissioning of the experiments. Prior to any discovery of
new phenomena a deep understanding of these background events has to be
obtained. A solid knowledge of the Standard Model is crucial is estimating
the diverse backgrounds in the signal regions and is a pre-requisite for the
correct interpretation of the observed phenomena.

The primary aim of the Standard Model Handles and Candles working group,
which has been set up in the framework of the Les Houches workshop is to
address issues relevant in the programme described above. Several topics
relevant for the Standard Model processes considered as a background or
signal are discussed. Examples are electroweak and QCD processes like Z and
W boson production and the high mass tail of the Drell-Yan spectrum. The
prediction and understanding of the min-bias events and the parton density
distributions are other topics.

The production of jets in the proton collisions at the LHC is abundant.
Therefore a thorough understanding of jet physics is primordial, including
for example a common nomenclature or accord when we speak about a generic
jet of particles. Along this line it becomes relevant to compare the
performance of several jet algorithms. A complete chapter is devoted to this
domain, resulting in a list of recommendations for the physics analyses at
the LHC.

%% file: s_wackeroth/ZcomparisonFeb.tex
\subsection{Introduction}
\label{sec:th_ewkintro}
$W$ and $Z$ bosons will be produced copiously at the LHC and
high-precision measurements of cross sections and their properties
will be used for detector calibration, to understand the background to
many physics analysis, and last but not least, to explore a new
electroweak high-energy regime in tails of $Z$ and $W$ distributions.
In view of the importance of single $W$ and $Z$ production as
'standard candles' and for searches of signals of new physics, it is
crucial to control the theoretical predictions for production cross
section and kinematic distributions. For a review of available
calculations and tools, see Refs.~\cite{Gerber:2007xk}, for instance.
Good theoretical control of the predicitions requires a
good understanding of the residual theoretical uncertainties.  As a
first step, we perform a tuned numerical comparison of the following
publicly available codes that provide precise predictions for $Z$
observables, including electroweak (EW) ${\cal O}(\alpha)$ corrections:
{\sc HORACE}~\cite{Horace-jhep:2005,Calame:2007cd}, {\sc
  SANC}~\cite{Andonov:2004hi,Bardin:2005dp,Arbuzov:2007db}, and {\sc ZGRAD2}~\cite{Baur:2001ze}.
First results of a tuned comparison of $Z$ production cross sections
can be found in Ref.~\cite{Buttar:2006zd}, and predictions for single
$W$ production including QCD and electroweak corrections have been
recently discussed in Ref.~\cite{Gerber:2007xk}. A study of combined
effects of QCD and electroweak corrections to the neutral-current
process in the high invariant-mass region can be found in these
procceedings.

\subsection{Results of a tuned comparison of {\sc HORACE}, {\sc SANC} and {\sc ZGRAD2}}
\label{sec:th_ewkcomp}

\begin{center}
{\it Setup for the tuned comparison} 
\end{center}
\noindent
For the numerical evaluation of the cross sections 
at the LHC ($\sqrt{s}=14$ TeV) we chose the
following set of Standard Model input parameters:
\begin{eqnarray}\label{eq:parsx}
G_{\mu} = 1.16637\times 10^{-5} \; {\rm GeV}^{-2}, 
& \qquad & \alpha= 1/137.03599911, \quad \alpha_s\equiv\alpha_s(M_Z^2)=0.1176 
\nonumber \\ 
M_Z = 91.1876 \; {\rm GeV}, & \quad & \Gamma_Z =  2.4924  \; {\rm GeV}
\nonumber  \\
M_W = 80.37399 \; {\rm GeV}, & \quad & \Gamma_W = 2.0836 \; {\rm GeV}
\nonumber  \\
M_H = 115 \; {\rm GeV}, & \quad & 
\nonumber  \\
m_e  = 0.51099892 \; {\rm keV}, &\quad &m_{\mu}=0.105658369 \; {\rm GeV},  
\quad m_{\tau}=1.77699 \; {\rm GeV}
\nonumber  \\
m_u=0.06983 \; {\rm GeV}, & \quad & m_c=1.2 \; {\rm GeV}, 
\quad m_t=174 \; {\rm GeV}
\nonumber  \\
m_d=0.06984 \; {\rm GeV}, & \quad & m_s=0.15 \; {\rm GeV}, \quad m_b=4.6 \; {\rm GeV} 
\nonumber \\
|V_{ud}| = 0.975, & \quad & |V_{us}| = 0.222 
\nonumber \\
|V_{cd}| = 0.222, & \quad & |V_{cs}| = 0.975 
\nonumber \\
|V_{cb}|=|V_{ts}|=|V_{ub}|& =& |V_{td}|= |V_{tb}|=0  
\end{eqnarray}
The $W$ and Higgs boson masses, $M_W$ and $M_H$, are related via loop
corrections. To determine $M_W$ we use a parametrization which, for
$100~{\rm GeV}<M_H<1$~TeV, deviates by at most 0.2~MeV from the
theoretical value including the full two-loop
contributions~\cite{Awramik:2003rn} (using Eqs.~(6,7,9)).  Additional
parametrizations can also be found
in~\cite{Degrassi:1997iy,Ferroglia:2002rg}.

We work in the constant width scheme and fix the weak mixing angle by
$c_w=M_W/M_Z$, $s_w^2=1-c_w^2$.  The $Z$ and $W$-boson decay widths
given above are calculated including QCD and electroweak corrections,
and are used in both the LO and NLO evaluations of the cross sections.
The fermion masses only enter through loop contributions to the vector
boson self energies and as regulators of the collinear singularities
which arise in the calculation of the QED contribution. The light
quark masses are chosen in such a way, that the value for the hadronic
five-flavor contribution to the photon vacuum polarization, $\Delta
\alpha_{had}^{(5)}(M_Z^2)=0.027572$~\cite{Jegerlehner:2001wq}, is
recovered, which is derived from low-energy $e^+ e^-$ data with the
help of dispersion relations.  The finestructure constant,
$\alpha(0)$, is used throughout in both the LO and NLO calculations of
the $Z$ production cross sections.

In the course of the calculation of $Z$ observables the
Kobayashi-Maskawa-mixing has been neglected.

To compute the hadronic cross section we use the MRST2004QED set of
parton distribution functions~\cite{Martin:2004dh}, and take the
renormalization scale, $\mu_r$, and the QED and QCD factorization
scales, $\mu_{\rm QED}$ and $\mu_{\rm QCD}$, to be $\mu_r^2=\mu_{\rm
  QED}^2=\mu_{\rm QCD}^2=M_Z^2$. In the
MRST2004QED structure functions, the factorization of the photonic
initial state quark mass singularities is done in the QED DIS scheme
which we therefore use in all calculations reported here. It
is defined analogously to the usual DIS~\cite{Owens:1992hd} schemes
used in QCD calculations, i.e.~by requiring the same expression for
the leading and next-to-leading order structure function $F_2$ in deep
inelastic scattering, which is given by the sum of the quark
distributions. Since $F_2$ data are an important ingredient in
extracting PDFs, the effect of the ${\cal O}(\alpha)$ QED corrections
on the PDFs should be reduced in the QED DIS scheme.

The detector acceptance is simulated by imposing the following
transverse momentum ($p_T$) and pseudo-rapidity ($\eta$) cuts:
\begin{equation}
p_T^\ell>20~{\rm GeV,}\qquad\qquad |\eta_\ell|<2.5, \qquad\qquad
\ell=e,\,\mu ,
\label{eq:lepcut}
\end{equation}
These cuts approximately model the acceptance of the ATLAS and CMS detectors
at the LHC. Uncertainties in the energy measurements of the charged
leptons in the detector are simulated in the calculation by Gaussian
smearing of the particle four-momentum vector with standard deviation
$\sigma$ which depends on the particle type and the detector. The
numerical results presented here were calculated using $\sigma$ values
based on the ATLAS specifications. 
In addition to the separation cuts of
Eq.~\ref{eq:lepcut}, we apply a cut on the invariant mass of the
final-state lepton pair of $M_{ll} >$ 50 GeV.

The granularity of the detectors and the size of the electromagnetic
showers in the calorimeter make it difficult to discriminate between
electrons and photons with a small opening angle. In such cases we
recombine the four-momentum vectors of the electron and photon to an
effective electron four-momentum vector.  
We require that the electron and photon momentum four-vectors are
combined into an effective electron momentum four-vector if their
separation in the pseudorapidity -- azimuthal angle plane,
\begin{equation}
\Delta R(e,\gamma)=\sqrt{(\Delta\eta(e,\gamma))^2+(\Delta\phi(e,
\gamma))^2}, 
\end{equation}
is $\Delta R(e,\gamma)<0.1$. 
For
$0.1<\Delta R(e,\gamma)<0.4$ events are rejected if
$E_{\gamma}>0.1 \; E_e$. Here $E_\gamma$ ($E_e$) is the energy of the
photon (electron) in the laboratory frame. 

Muons are identified by hits in the muon chambers and the requirement
that the associated track is consistent with a minimum ionizing
particle. This limits the photon energy for small muon -- photon
opening angles. For muons, we require that
the energy of the photon is $E_{\gamma}<2$~GeV for $\Delta
R(\mu,\gamma)<0.1$, and $E_{\gamma}<0.1 E_{\mu}$~GeV for $0.1<\Delta
R(\mu,\gamma)<0.4$.  We summarize the lepton identification
requirements in Table~\ref{tab:th_ewk_c}.
\begin{table}
\begin{center}
\begin{tabular}{|c|c|} \hline
\multicolumn{1}{|c|}{electrons} & \multicolumn{1}{|c|}{muons} \\
\hline
combine $e$ and $\gamma$ momentum four vectors, & reject events with 
$E_\gamma>2$~GeV \\ 
if $\Delta R(e,\gamma)<0.1$ & for $\Delta R(\mu,\gamma)<0.1$ \\
\hline
reject events with 
$E_\gamma>0.1~E_e$ & reject events with 
$E_\gamma>0.1~E_\mu$ \\  
for $0.1<\Delta R(e,\gamma)<0.4$ & for $0.1<\Delta R(\mu,\gamma)<0.4$ \\ \hline  
\end{tabular}
\caption{Summary of lepton identification requirements. } 
\label{tab:th_ewk_c}
\end{center}
\end{table}
For each observable we will provide ``bare'' results, i.e.~without
smearing and recombination (only lepton separation cuts are applied)
and ``calo'' results, i.e.~including smearing and recombination.  We
will show results for kinematic distributions and total cross sections,
at LO and NLO, and the corresponding relative
corrections, $\delta(\%)=d\sigma_{NLO}/d\sigma_{LO}-1$, at the LHC.  
We consider the
following neutral current processes: $pp \to Z,\gamma \to l^-
l^+$ with $l=e,\mu$.

\begin{figure}
\begin{center}
  \includegraphics[width=7.1cm,
  keepaspectratio=true]{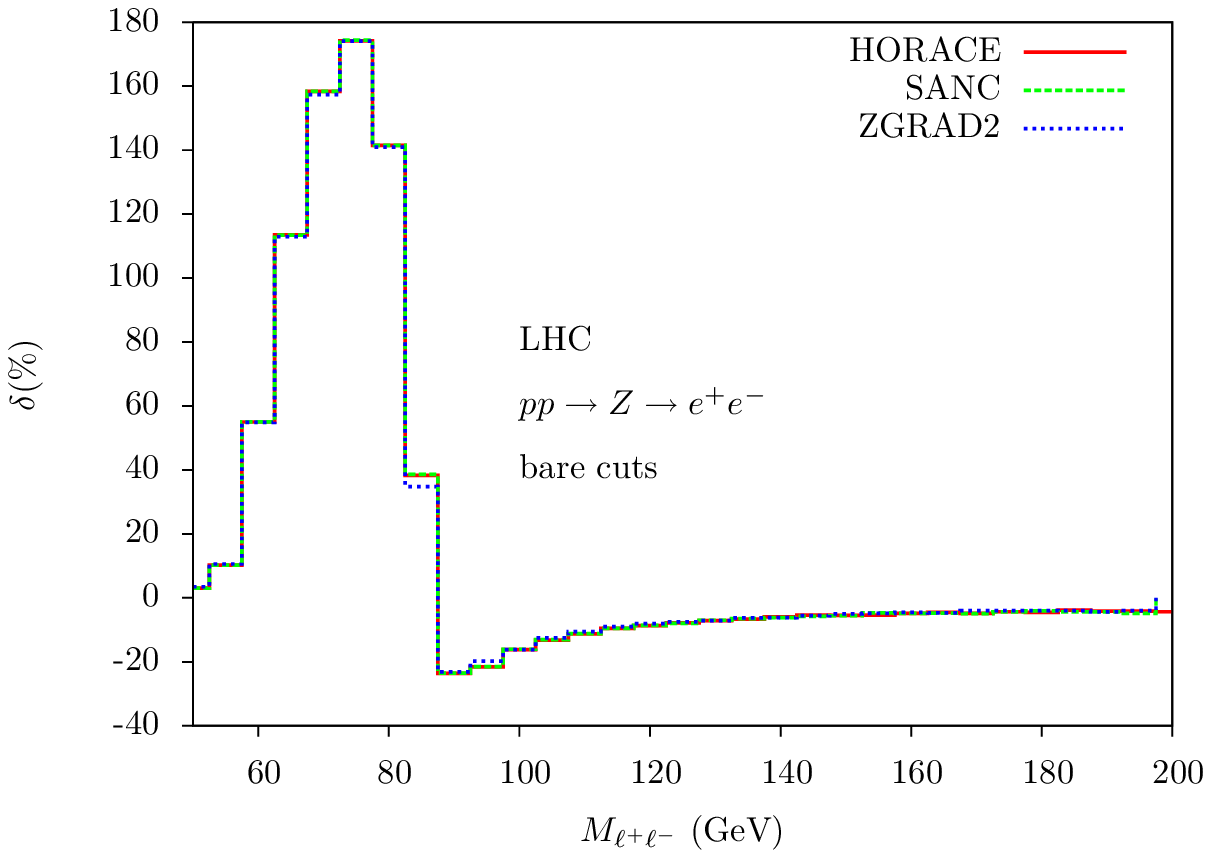}
\hspace*{-.0cm}
  \includegraphics[width=7.1cm,
  keepaspectratio=true]{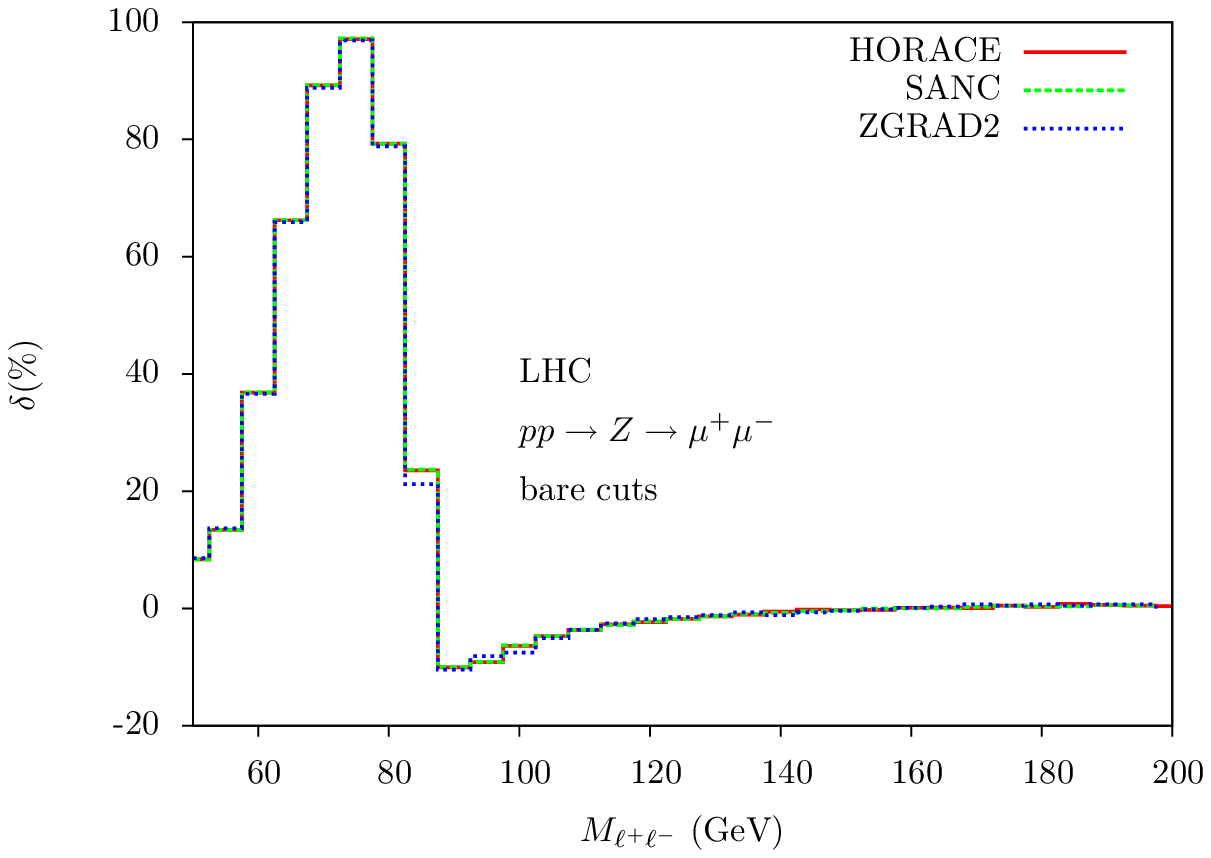}
\hspace*{-.0cm}
  \includegraphics[width=7.1cm,
  keepaspectratio=true]{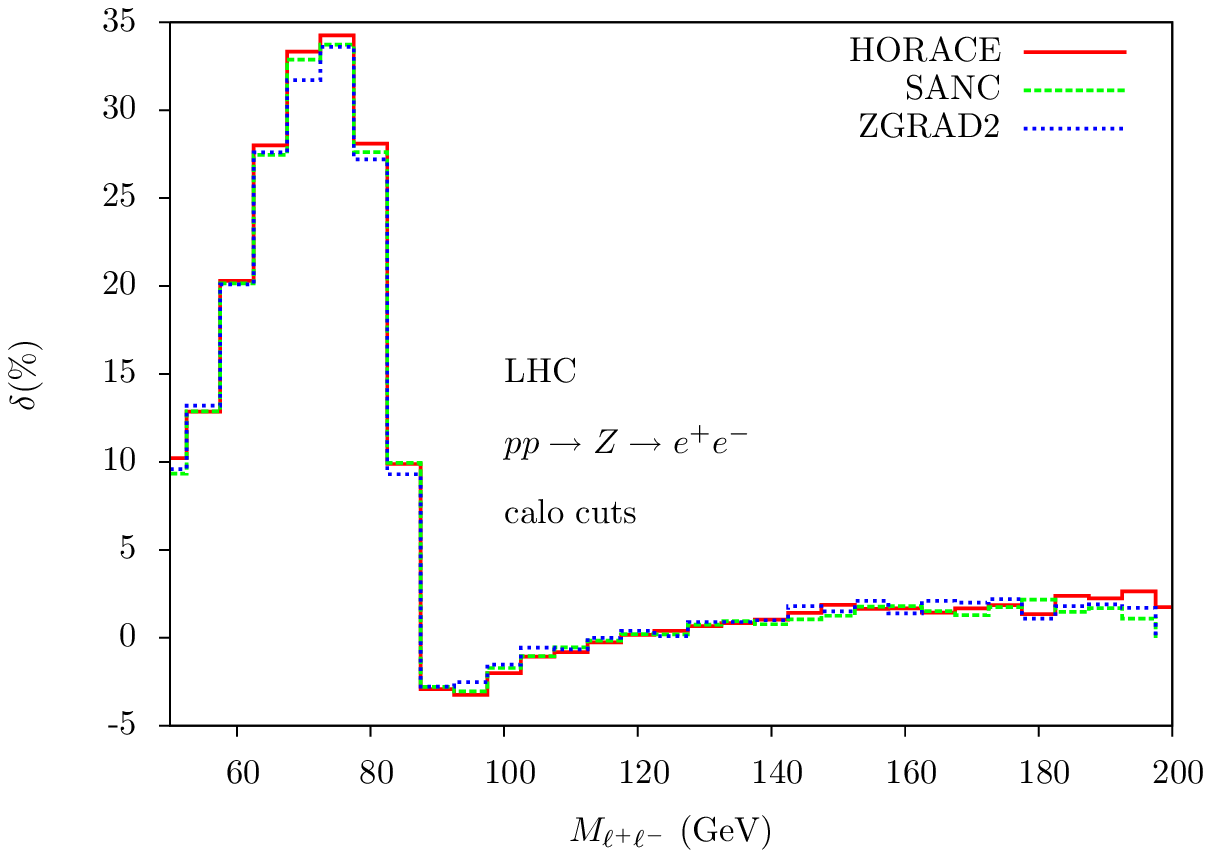}
\hspace*{-.0cm}
  \includegraphics[width=7.1cm,
  keepaspectratio=true]{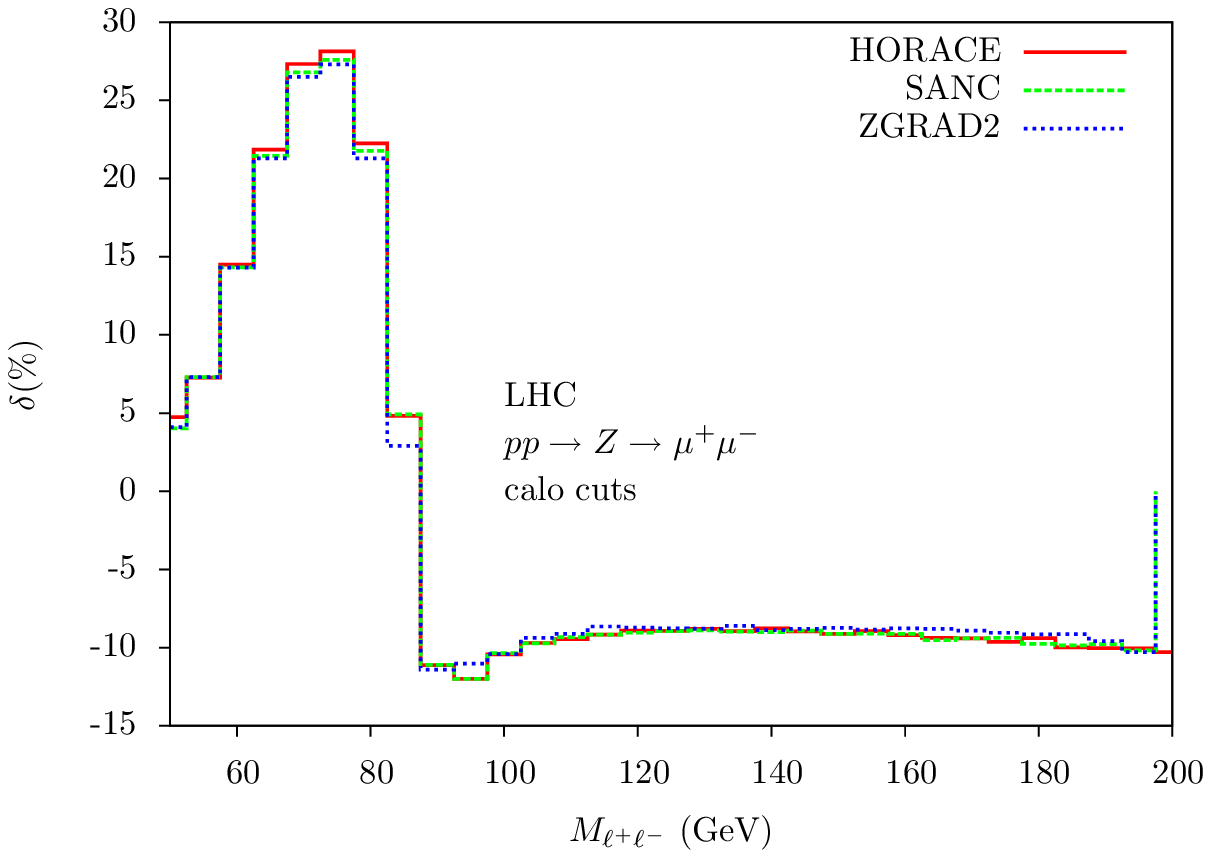}
\end{center}
\caption{The relative correction $\delta$ due to electroweak ${\cal O}(\alpha)$ corrections to the $M(l^+l^-)$ distribution
for $Z$ production with bare and calo cuts at the LHC.}\label{fig:th_ewk_mll1}
\end{figure}

\begin{figure}
\begin{center}
  \includegraphics[width=7.1cm,
  keepaspectratio=true]{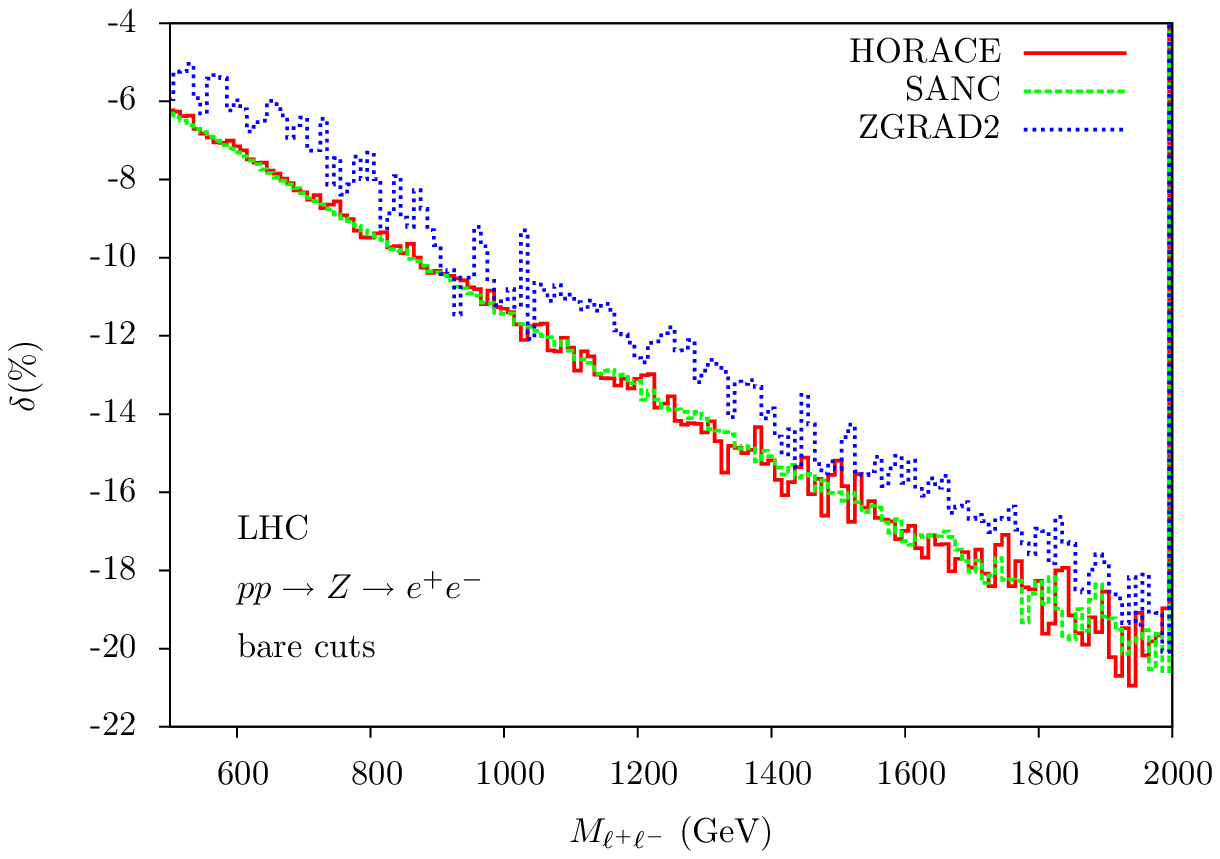}
\hspace*{-.0cm}
  \includegraphics[width=7.1cm,
  keepaspectratio=true]{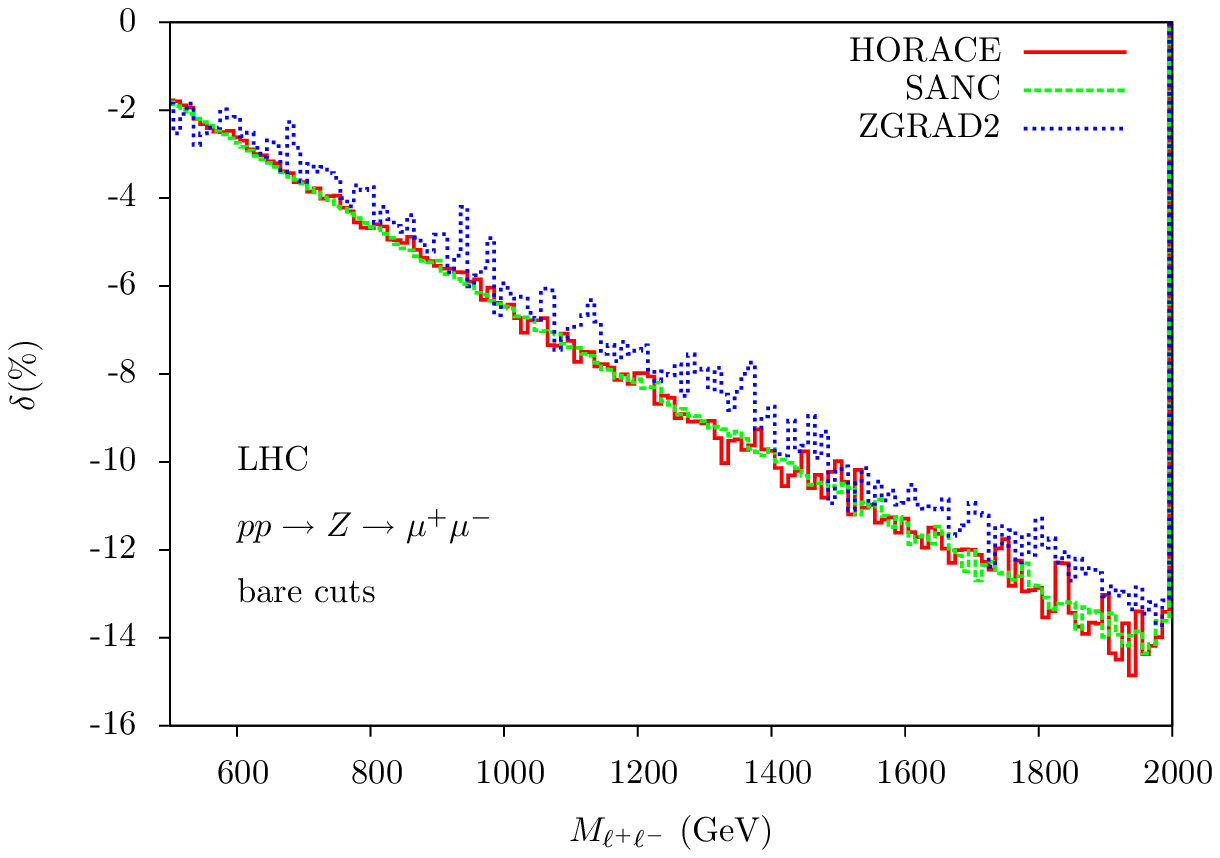}
\hspace*{-.0cm}
  \includegraphics[width=7.1cm,
  keepaspectratio=true]{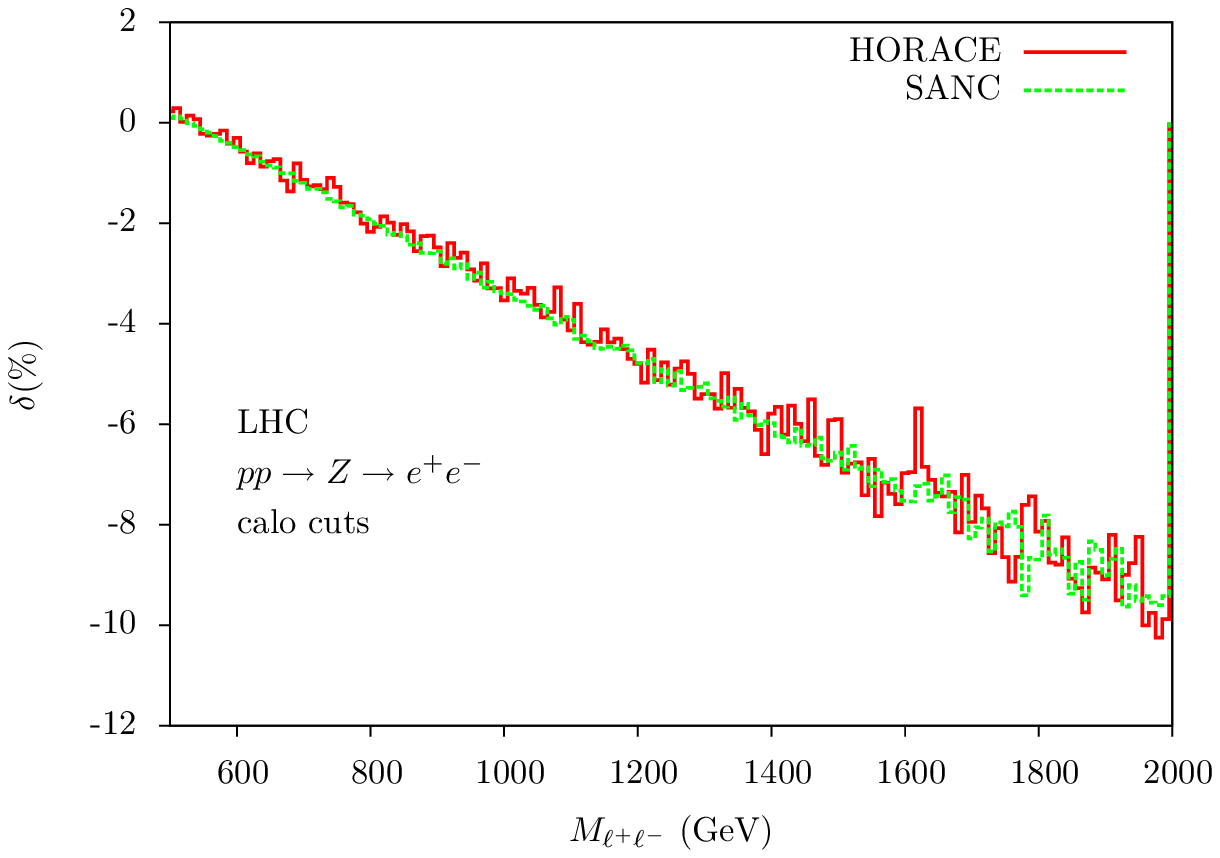}
\hspace*{-.0cm}
  \includegraphics[width=7.1cm,
  keepaspectratio=true]{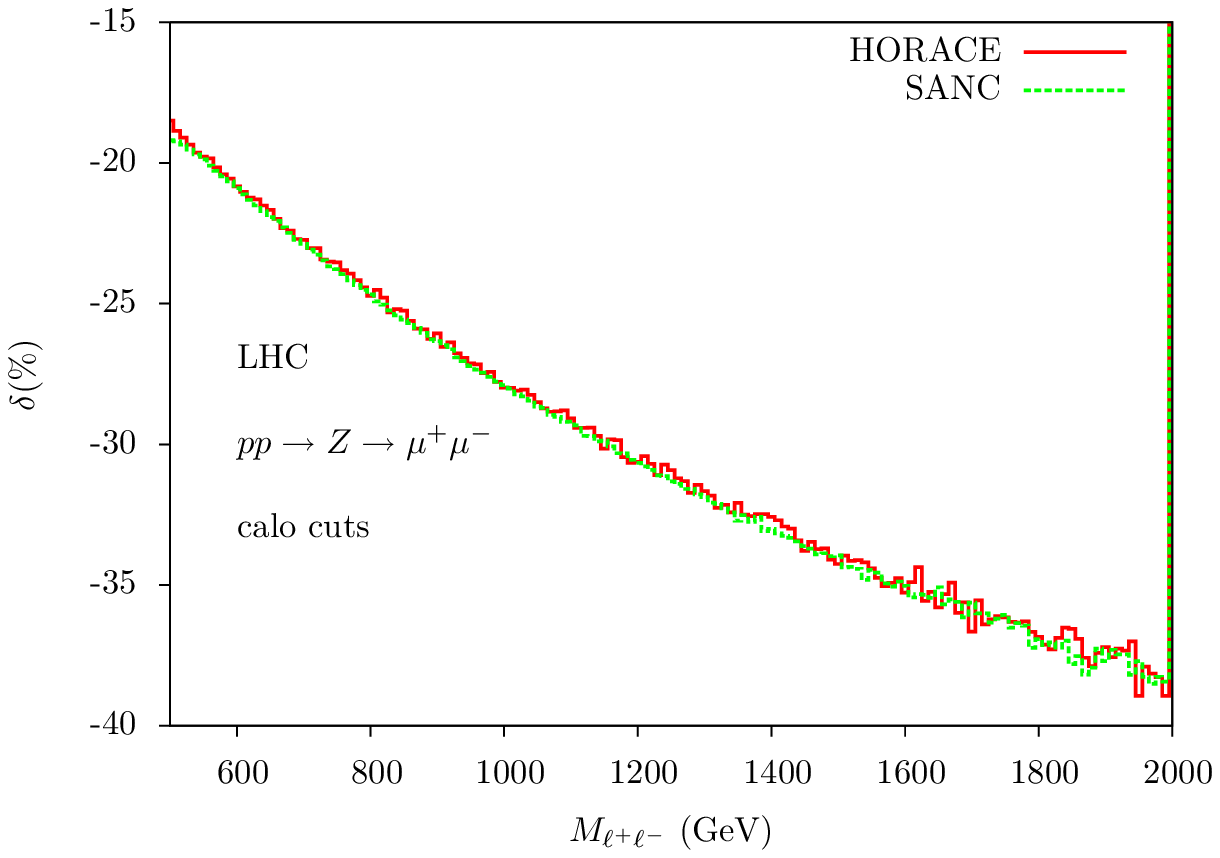}
\end{center}
\caption{The relative correction $\delta$ due to electroweak ${\cal O}(\alpha)$ corrections to the $M(l^+l^-)$ distribution
for $Z$ production with bare and calo cuts at the LHC.}\label{fig:th_ewk_mll2}
\end{figure}

\begin{figure}
\begin{center}
  \includegraphics[width=7.1cm,
  keepaspectratio=true]{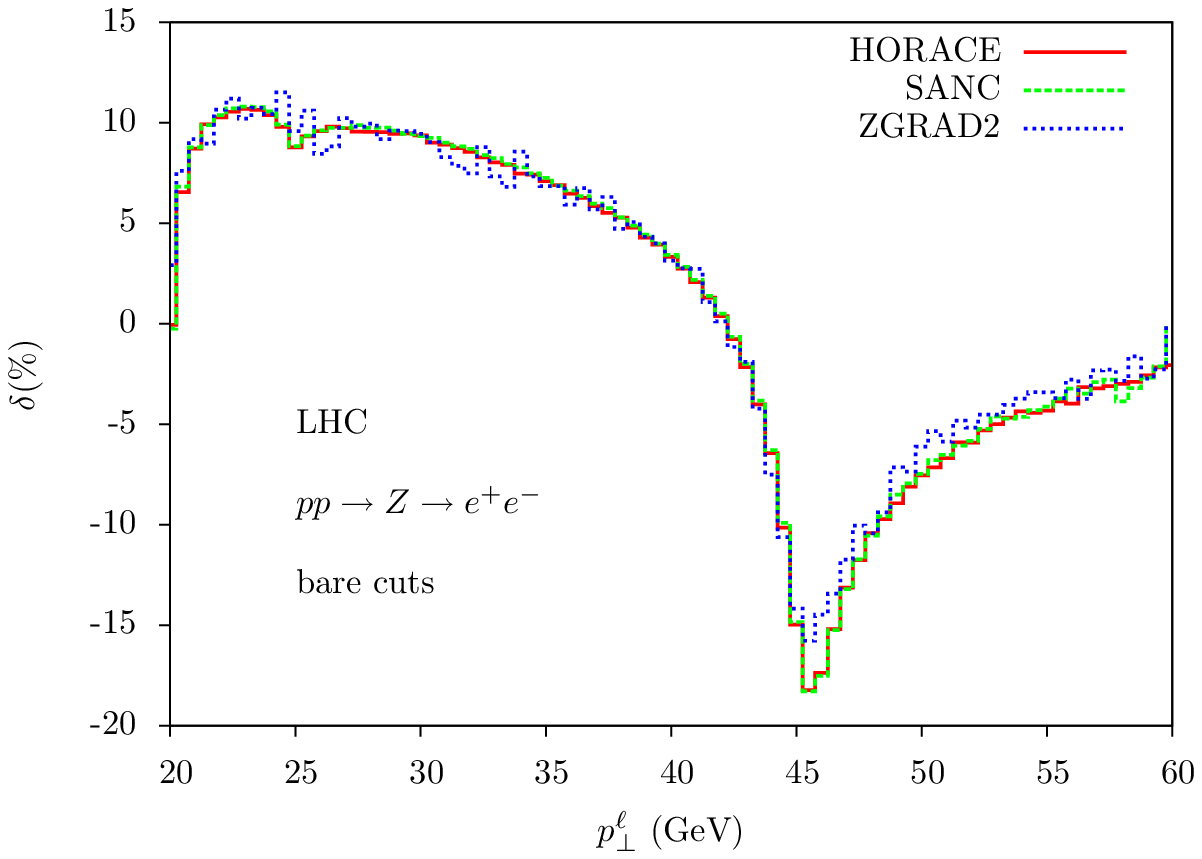}
\hspace*{-.0cm}
  \includegraphics[width=7.1cm,
  keepaspectratio=true]{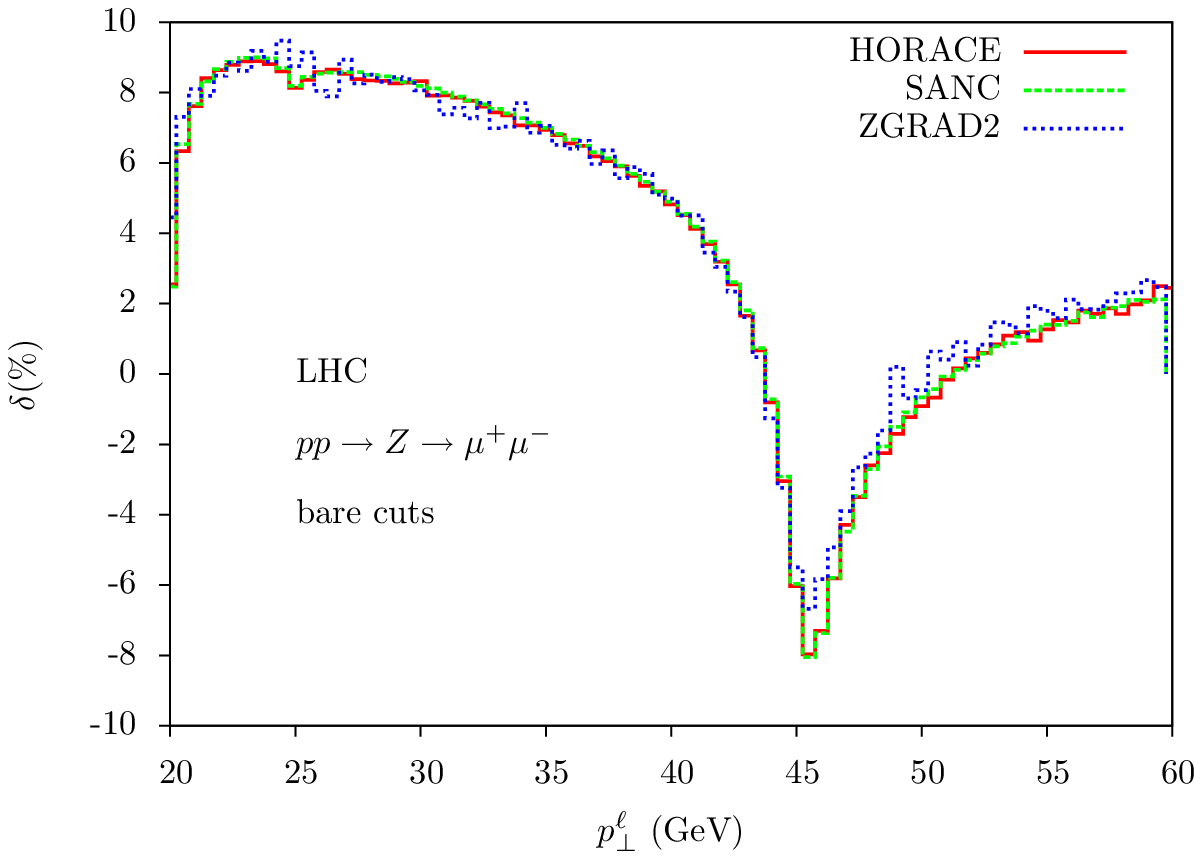}
\hspace*{-.0cm}
  \includegraphics[width=7.1cm,
  keepaspectratio=true]{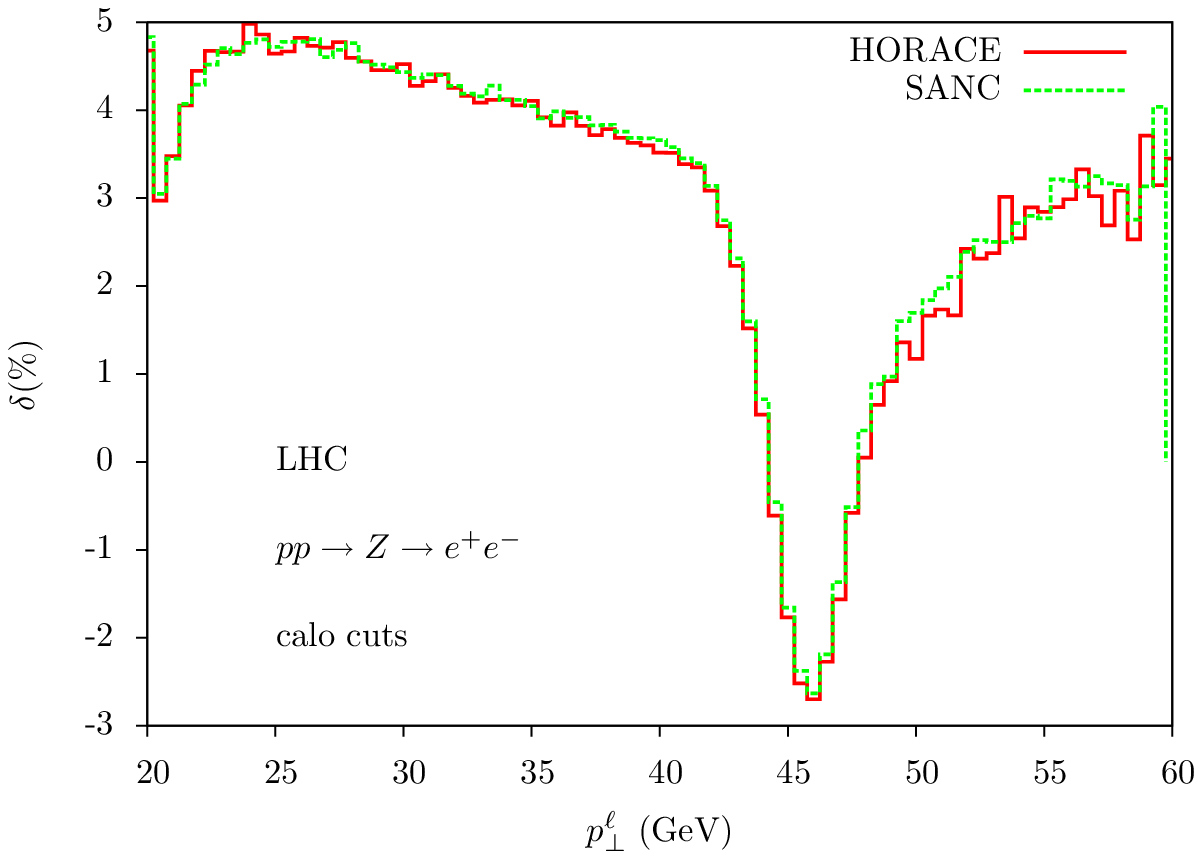}
\hspace*{-.0cm}
  \includegraphics[width=7.1cm,
  keepaspectratio=true]{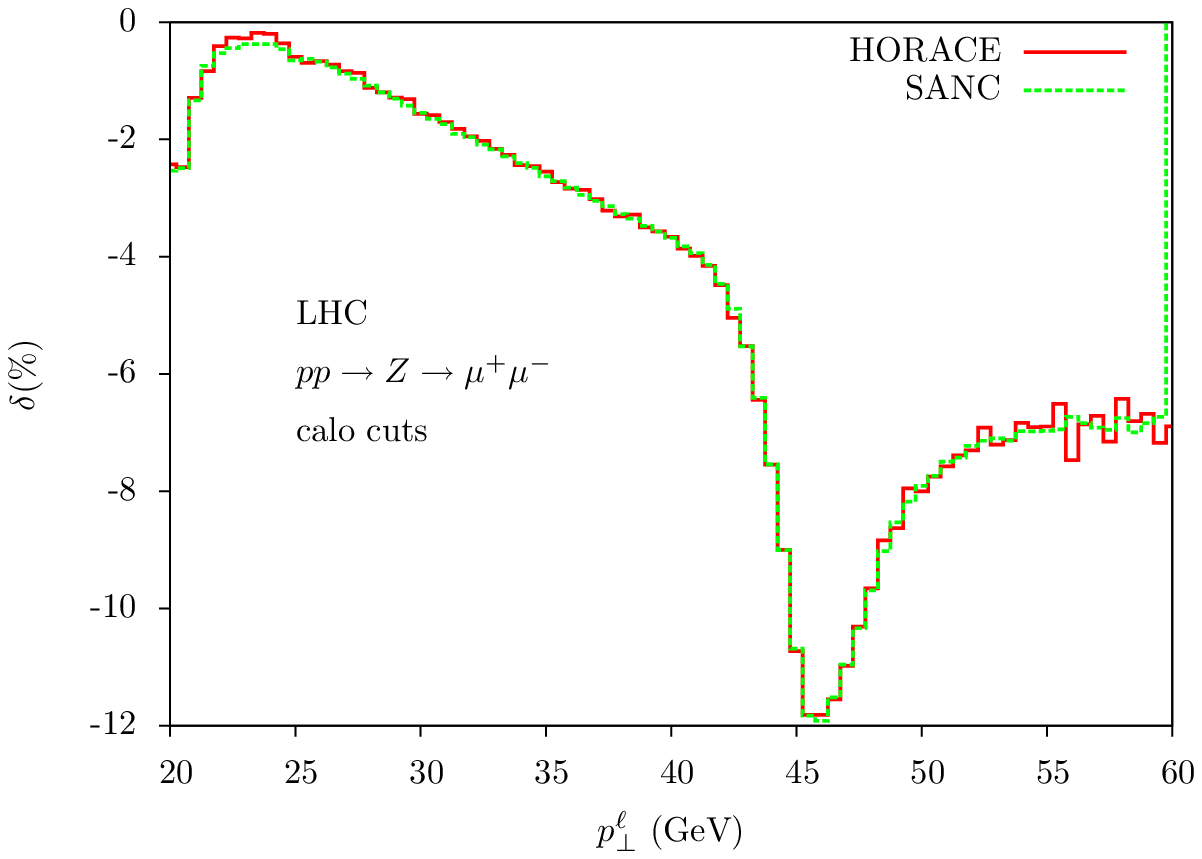}
\end{center}
\caption{The relative correction $\delta$ due to electroweak ${\cal O}(\alpha)$ corrections to the $p_T^l$ distribution
for $Z$ production with bare and calo cuts at the LHC.}\label{fig:th_ewk_pt}
\end{figure}

\begin{figure}
\begin{center}
  \includegraphics[width=7.1cm,
  keepaspectratio=true]{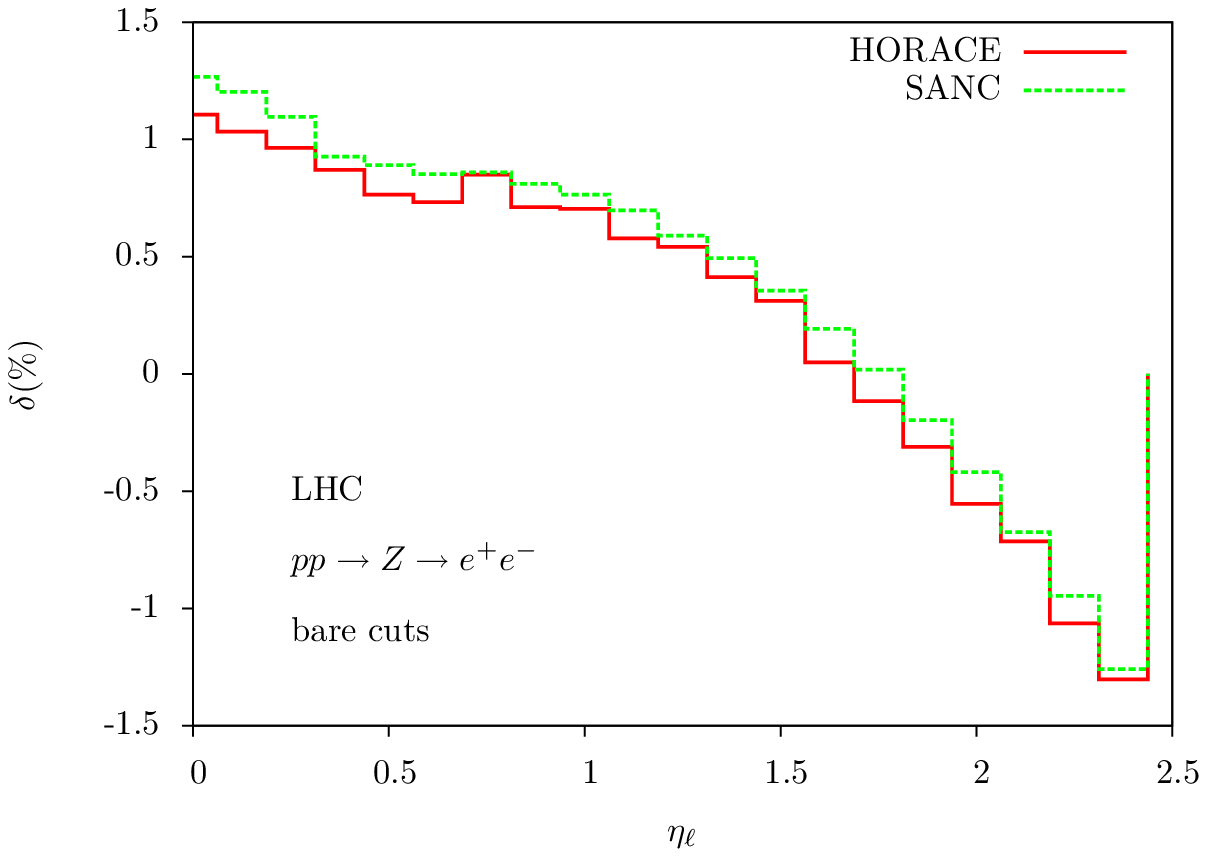}
\hspace*{-.0cm}
  \includegraphics[width=7.1cm,
  keepaspectratio=true]{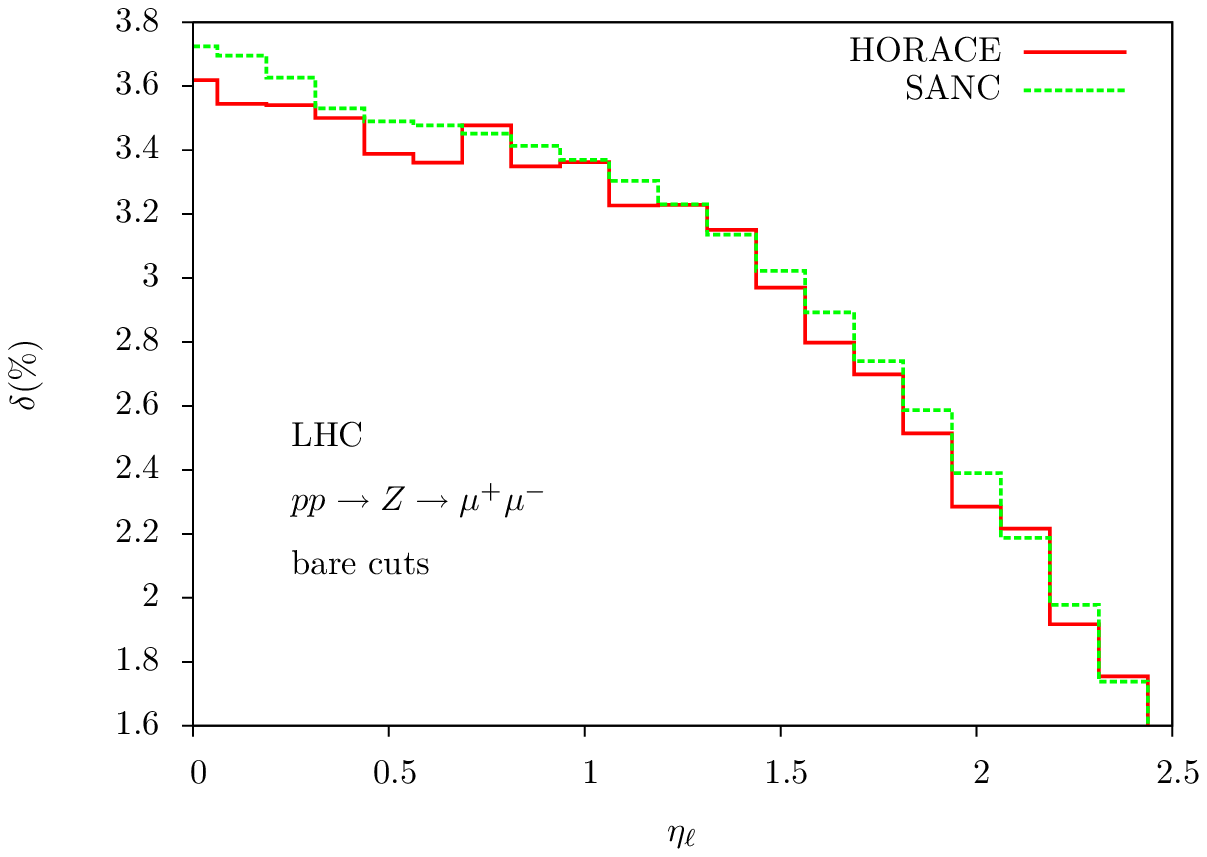}
\hspace*{-.0cm}
  \includegraphics[width=7.1cm,
  keepaspectratio=true]{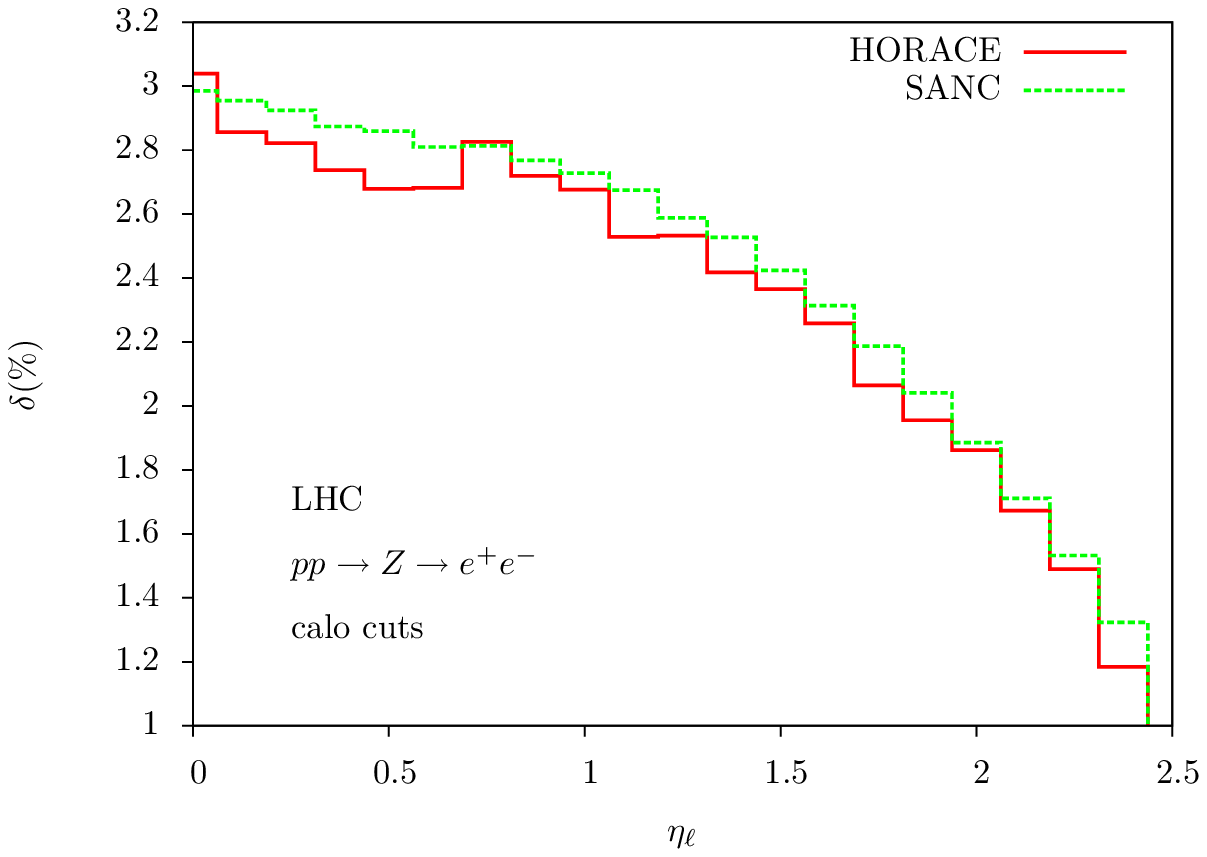}
\hspace*{-.0cm}
  \includegraphics[width=7.1cm,
  keepaspectratio=true]{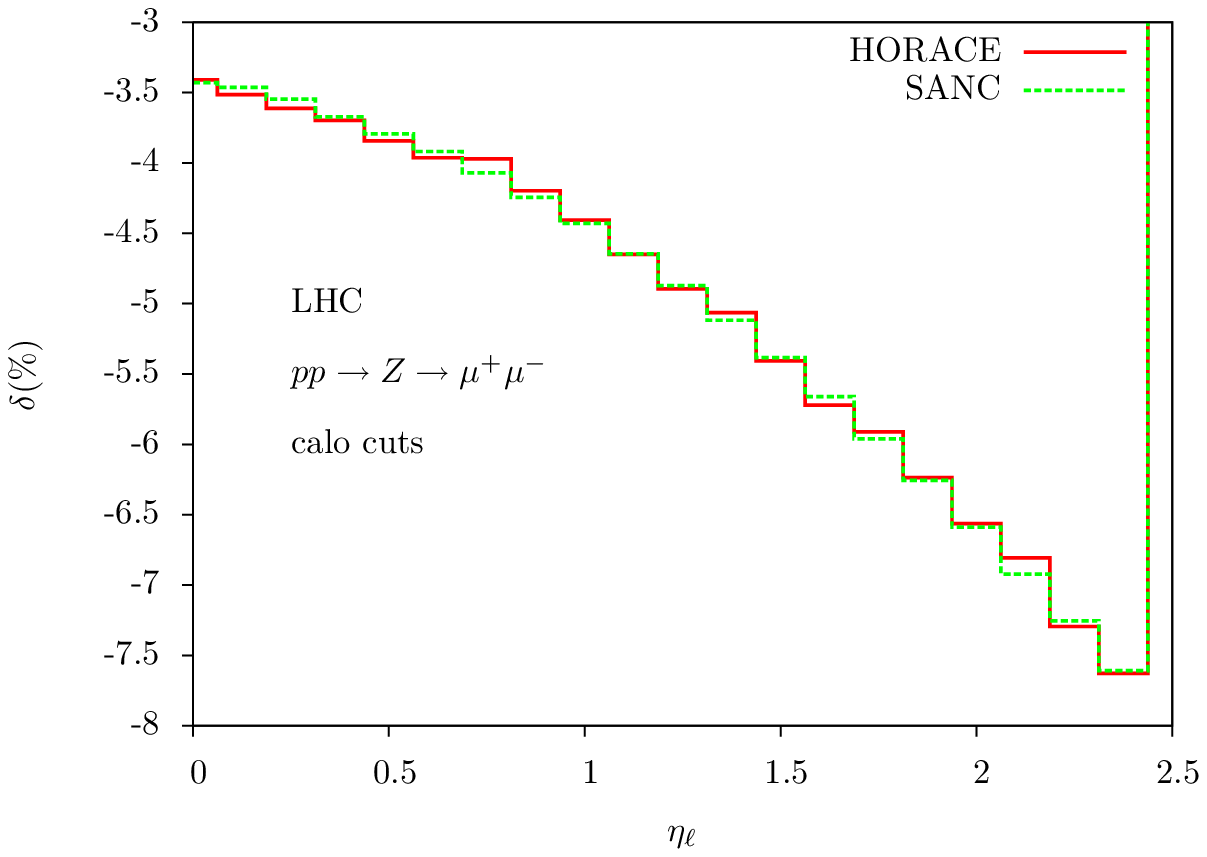}
\end{center}
\caption{The relative correction $\delta$ due to electroweak ${\cal O}(\alpha)$ corrections to the $\eta_l$ distribution
for $Z$ production with bare and calo cuts at the LHC.}\label{fig:th_ewk_eta}
\end{figure}

\begin{figure}
\begin{center}
  \includegraphics[width=7.1cm,
  keepaspectratio=true]{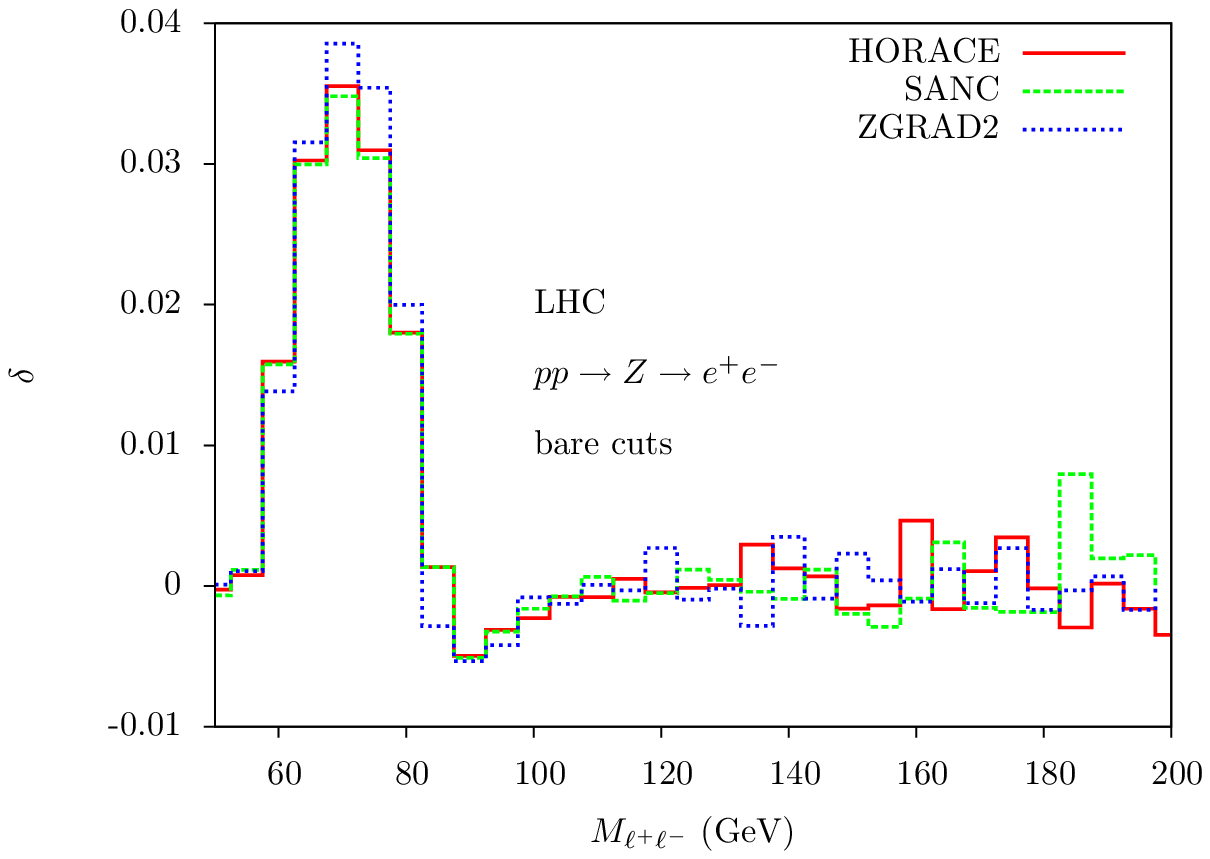}
\hspace*{-.0cm}
  \includegraphics[width=7.1cm,
  keepaspectratio=true]{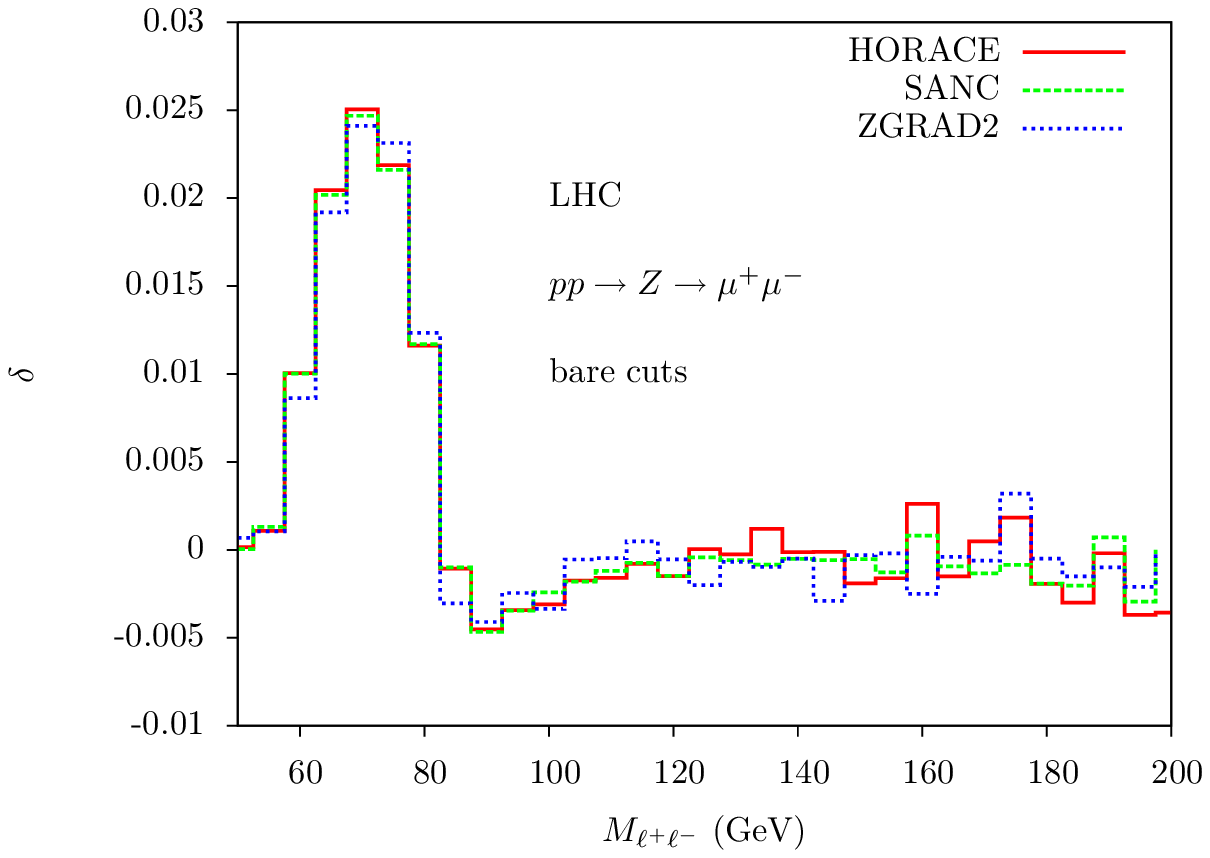}
\hspace*{-.0cm}
  \includegraphics[width=7.1cm,
  keepaspectratio=true]{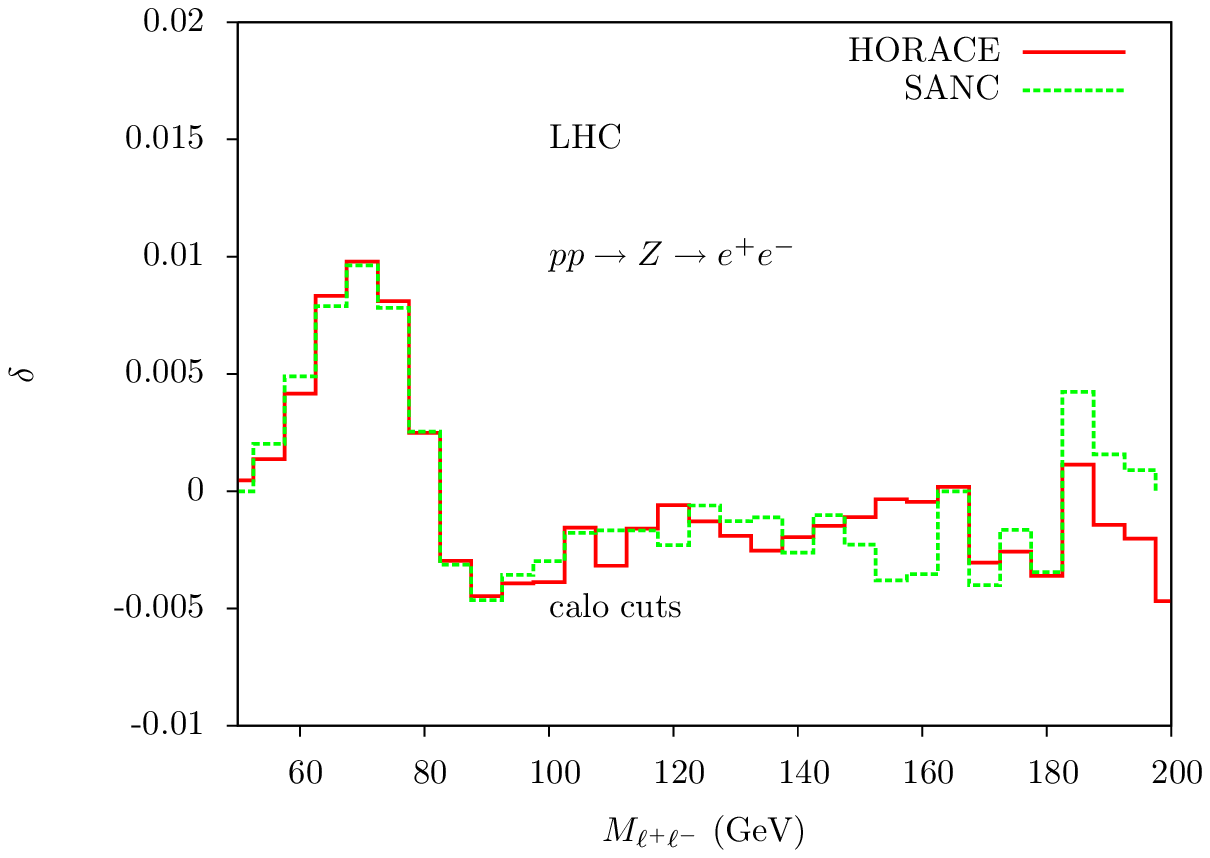}
\hspace*{-.0cm}
  \includegraphics[width=7.1cm,
  keepaspectratio=true]{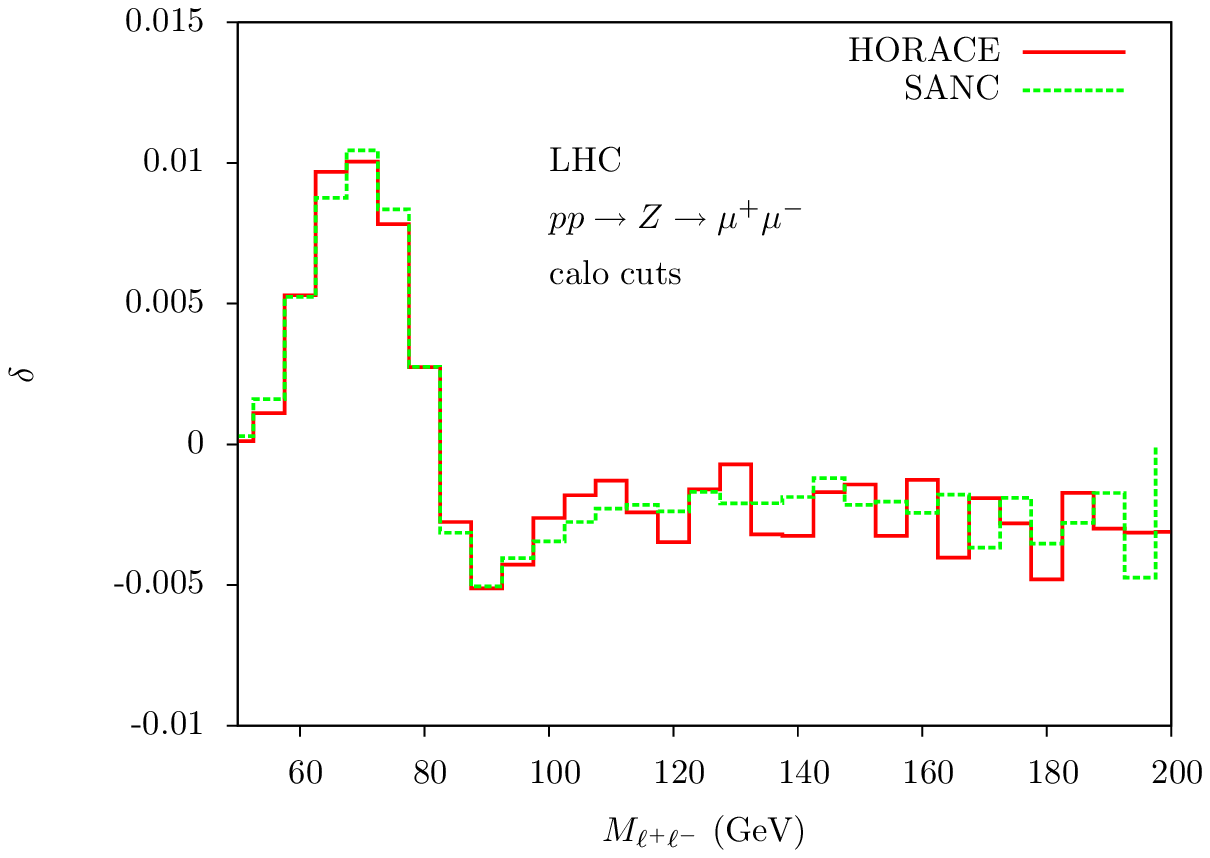}
\end{center}
\caption{The difference between the NLO and LO predictions for $A_{FB}$ due to electroweak ${\cal O}(\alpha)$ corrections
for $Z$ production with bare and calo cuts at the LHC.}\label{fig:th_ewk_afb1}
\end{figure}

\begin{figure}
\begin{center}
  \includegraphics[width=7.1cm,
  keepaspectratio=true]{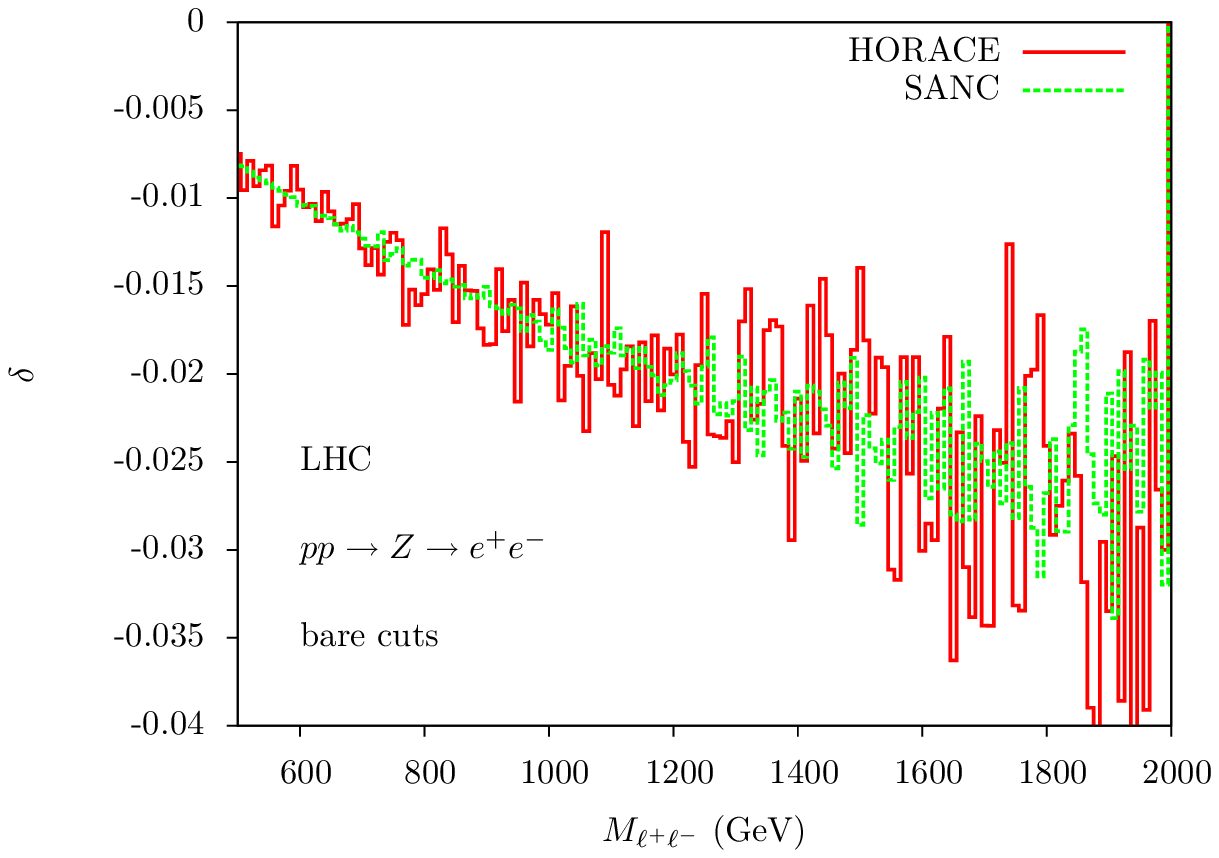}
\hspace*{-.0cm}
  \includegraphics[width=7.1cm,
  keepaspectratio=true]{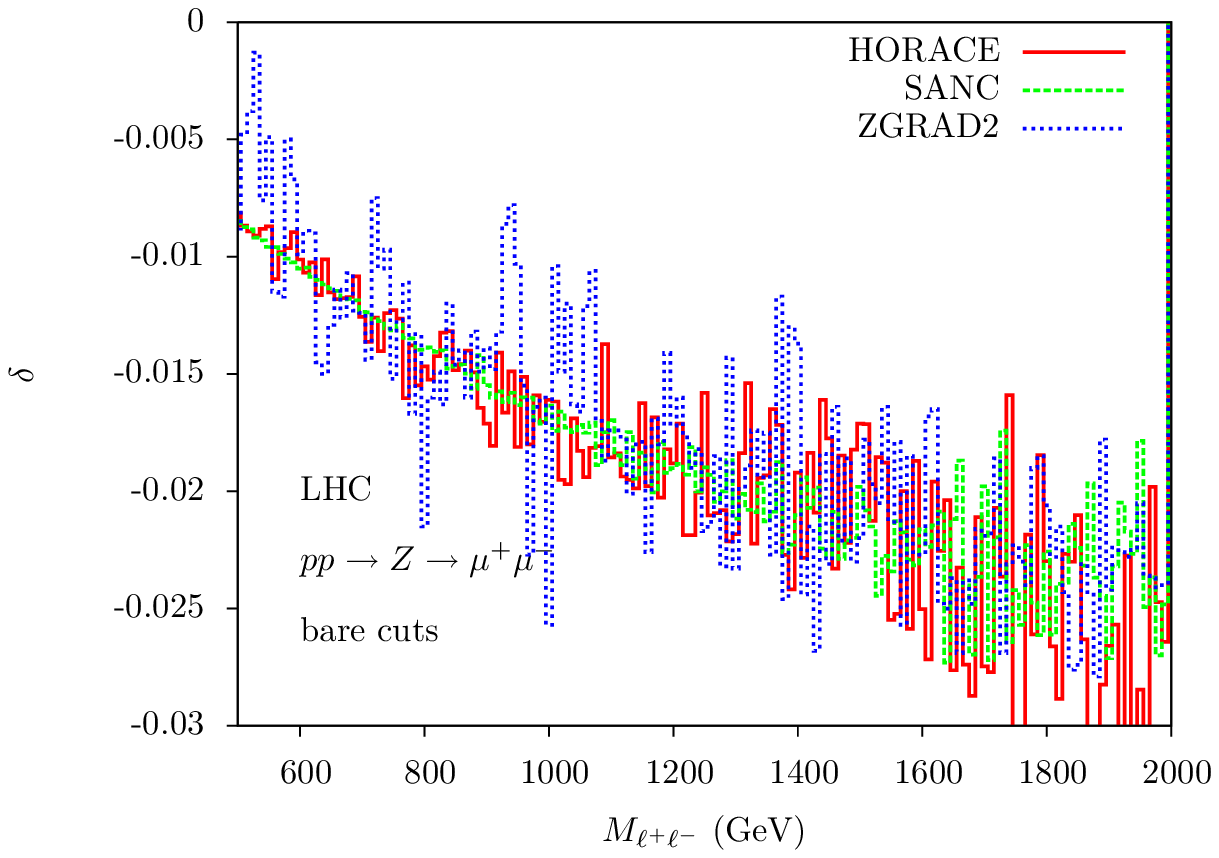}
\hspace*{-.0cm}
  \includegraphics[width=7.1cm,
  keepaspectratio=true]{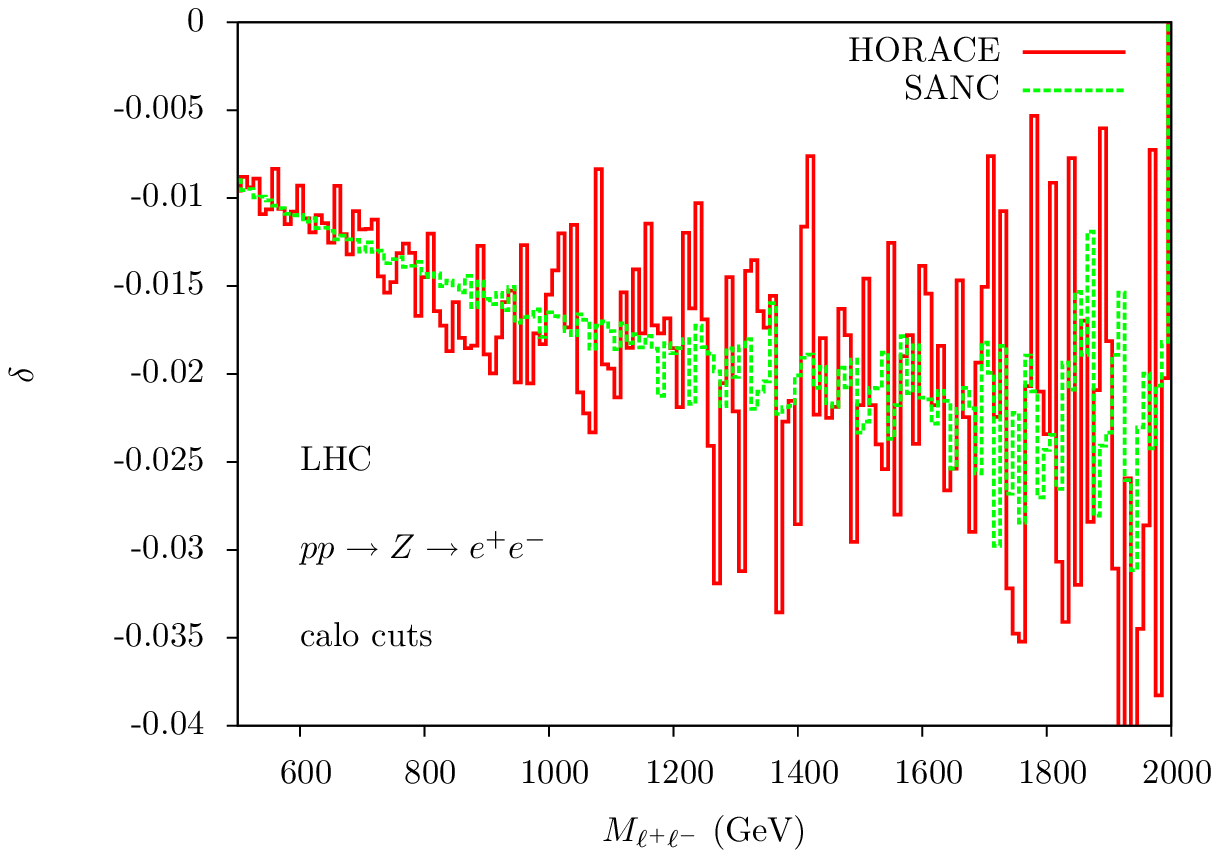}
\hspace*{-.0cm}
  \includegraphics[width=7.1cm,
  keepaspectratio=true]{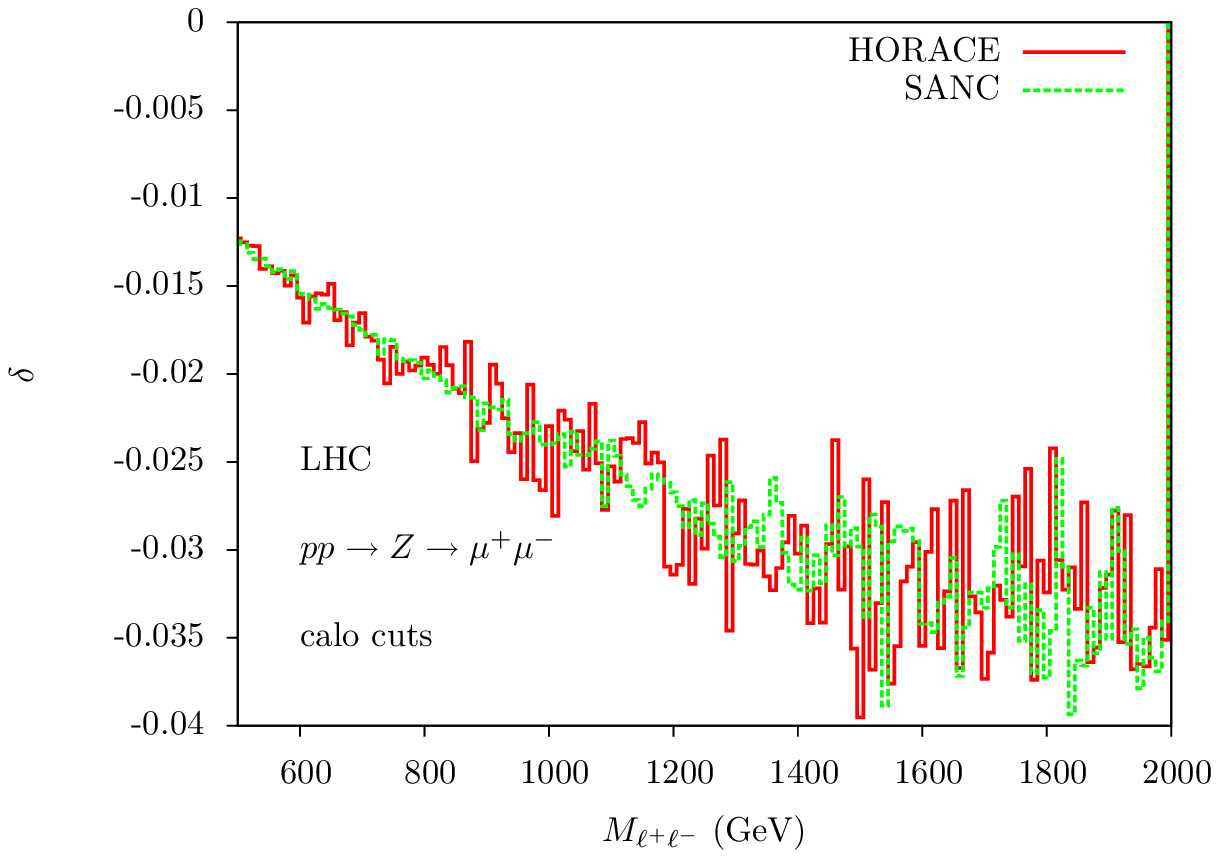}
\end{center}
\caption{The difference between the NLO and LO predictions for $A_{FB}$ due to electroweak ${\cal O}(\alpha)$ corrections
for $Z$ production with bare and calo cuts at the LHC.}\label{fig:th_ewk_afb2}
\end{figure}

\begin{center} {\it $Z$ boson observables} \end{center}
\begin{itemize}
\item
$\sigma_Z$: total inclusive cross section of $Z$ boson production.\\
\noindent
The results for $\sigma_Z$ at LO and EW NLO and the corresponding
relative corrections $\delta$ are provided in Table~\ref{tab:th_ewk_d}.
\item
$\frac{d\sigma}{dM(l^+l^-)}$: invariant mass distribution of the final-state lepton-pair.\\
\noindent
The relative corrections $\delta$ for different $M(l^+l^-)$ ranges 
are shown for bare and calo cuts in Figs.~\ref{fig:th_ewk_mll1},\ref{fig:th_ewk_mll2}.
\item
$\frac{d\sigma}{dp_T^l}$: transverse lepton momentum distribution. \\
\noindent
The relative corrections $\delta$ are shown in Fig.~\ref{fig:th_ewk_pt}
for bare and calo cuts. 
\item
$\frac{d\sigma}{d\eta_l}$: pseudo rapidity distribution of the lepton. \\
\noindent
The relative corrections $\delta$ are shown in Fig.~\ref{fig:th_ewk_eta}
for bare and calo cuts. 
\item
$A_{FB}$: forward-backward asymmetries (as a function of $M_{l^+l^-}$).\\
\noindent
For $p\bar p$ collisions at Tevatron energies, $A_{\rm FB}$ usually is 
defined by~\cite{Baur:2001ze} 
\begin{equation}
A_{\rm FB}=\frac{F-B}{F+B}~,
\end{equation}
where
\begin{equation}
F=\int_0^1{d\sigma\over d\cos\theta^*}\,d\cos\theta^*, \qquad
B=\int_{-1}^0{d\sigma\over d\cos\theta^*}\,d\cos\theta^*.
\end{equation}
Here, $\cos\theta^*$ is given by
\begin{equation}
\cos\theta^*={2\over m(l^+l^-)\sqrt{m^2(l^+l^-)+p_T^2(l^+l^-)}}\left 
[p^+(l^-)p^-(l^+)-p^-(l^-)p^+(l^+)\right ]
\label{EQ:CSTAR}
\end{equation}
with
\begin{equation}
p^\pm={1\over\sqrt{2}}\left (E\pm p_z\right ),
\end{equation}
where $E$ is the energy and $p_z$ is the longitudinal component of the
momentum vector. In this definition of $\cos\theta^*$, the polar axis 
is taken to be the bisector of the
proton beam momentum and the negative of the anti-proton beam momentum
when they are boosted into the $l^+l^-$ rest frame. In $p\bar p$ 
collisions at Tevatron energies, the flight direction of the incoming 
quark coincides with the proton beam direction for a large fraction of
the events. The definition of $\cos\theta^*$ in Eq.~(\ref{EQ:CSTAR})
has the advantage of minimizing the effects of the QCD corrections (see
below). In the limit of vanishing di-lepton $p_T$,
$\theta^*$ coincides with the angle between the lepton and the incoming 
proton in the $l^+l^-$ rest frame. 

For the definition of $\cos\theta^*$ given in Eq.~(\ref{EQ:CSTAR}),
$A_{\rm FB}=0$ for $pp$ collisions. The easiest way to obtain a non-zero
forward-backward asymmetry at the LHC is to extract the 
quark direction in the initial state from the boost direction of the 
di-lepton system with respect 
to the beam axis. The cosine of the angle between the 
lepton and the quark in the $l^+l^-$ rest frame is then
approximated by~\cite{Baur:2001ze}
\begin{equation}
\cos\theta^*={|p_z(l^+l^-)|\over p_z(l^+l^-)}~{2\over 
m(l^+l^-)\sqrt{m^2(l^+l^-)
+p_T^2(l^+l^-)}}\left [p^+(l^-)p^-(l^+)-p^-(l^-)p^+(l^+)\right ].
\label{EQ:CSTAR1}
\end{equation}
In Fig.~\ref{fig:th_ewk_afb1} (resonance region) and Fig.~\ref{fig:th_ewk_afb2} (tail region) we show the difference $\delta A_{FB}$ between the NLO EW and LO predictions for the forward-backward asymmetries for bare and calo cuts at the LHC. 
\end{itemize}
\begin{table}
\begin{center}
\begin{tabular}{|c|l|l|l|l|l|l|} \hline
\multicolumn{7}{|c|}{\bf LHC, $p p \to Z,\gamma \to e^+ e^-$} \\ \hline
& \multicolumn{3}{|c|}{bare cuts} & \multicolumn{3}{|c|}{calo cuts} \\ \hline
           & LO [pb]& NLO [pb]& $\delta$ [\%] & LO [pb]& NLO [pb]& $\delta$ [\%] \\ \hline
{\sc HORACE} & 739.34(3) & 742.29(4) & 0.40(1) & 737.51(3) & 755.67(6)  & 2.46(1) \\
{\sc SANC}   & 739.3408(3)  &  743.072(7) & 0.504(1)  & 737.857(2)  &  756.54(1) & 2.532(2) \\ 
{\sc ZGRAD2} & 737.8(7)  &  743.0(7) & 0.71(9)  & 737.8(7)   &  756.9(7) & 2.59(9) \\
\hline
\multicolumn{7}{|c|}{\bf LHC, $p p \to Z,\gamma \to \mu^+ \mu^-$} \\ \hline
& \multicolumn{3}{|c|}{bare cuts} & \multicolumn{3}{|c|}{calo cuts} \\ \hline
           & LO [pb]& NLO [pb]& $\delta$ [\%] & LO [pb]& NLO [pb] & $\delta$ [\%]  \\ \hline
{\sc HORACE} &  739.33(3) & 762.20(3) & 3.09(1) & 738.28(3) & 702.87(5) & -4.79(1) \\
{\sc SANC}   &  739.3355(3) & 762.645(3) & 3.1527(4) & 738.5331(3)  &  703.078(3) & -4.8006(3) \\ 
{\sc ZGRAD2} &  740(1) &  764(1) & 3.2(2) & 740(1)   &  705(1) & -4.7(2) \\
\hline
\end{tabular}
\caption{Tuned comparison of LO and EW NLO predictions for $\sigma_Z$ from {\sc HORACE}, {\sc SANC}, and {\sc ZGRAD2}. The statistical error of the Monte Carlo integration is given in parentheses.} 
\label{tab:th_ewk_d}
\end{center}
\end{table}

\noindent
The predictions of HORACE, SANC and ZGRAD2 show a satisfactory
level of agreement.
The effect of the EW NLO corrections,
calculated for the total cross sections within the specified cuts, 
agrees within the statistical uncertainties of the MC integration,
differs for the three codes at most by two per mille and in general by
few tenth of per mille.
Some discrepancies are present in specific observables. 
This requires further investigation, which is left to a future publication.

\subsection*{Conclusions}
\label{sec:th_ewkconcl}

In this report we performed a tuned comparison of the Monte Carlo
programs {\sc HORACE, SANC} and {\sc ZGRAD2}, taking into account
realistic lepton identification requirements.  We found good 
numerical agreement of the predictions for
the total $Z$ production cross section, the $M(ll)$, $p_T^l$ and
$\eta_l$ distributions and the forward-backward asymmetry at the LHC.
To find agreement between the available electroweak tools is only a
first, albeit important step towards controlling the predictions for the
neutral-current Drell-Yan process at the required precision level.  More
detailed studies of the residual uncertainties of predictions obtained
with the available tools are needed, in particular of the impact of
multiple photon radiation, higher-order electroweak Sudakov logarithms and
combined QCD and EW effects (see contribution to these proceedings).
Moreover, such a study should include PDF uncertainties, EW input
scheme and QED/QCD scale uncertainties.

%% file: s_jantzen/Dilepton_Ale_final.tex
\subsection{Introduction}
The Neutral-Current (NC) Drell-Yan (DY) process, which can give rise to a
high invariant-mass lepton pair, is a background to searches for new
phenomena. Examples of these are new heavy resonances Z' and G*
or possible excess resulting from the exchange of new
particles such as the leptoquarks.  These searches are an
important part of the LHC physics program and require a precise
knowledge of the Standard Model (SM) background in order to enable the
observation of new physics signatures, which may only give rise to
small deviations from the SM cross section. 

The DY process has been studied in great detail
(cf. \cite{Nadolsky:2004vt,Baur:2005rx} for a review), but
independently in the strong (QCD) and electroweak (EW) sectors.  In the high
invariant-mass region QCD effects are known to be large and
positive. These must be studied including both fixed order results and,
for some classes of results, resummation to all orders of the
contributions. The EW corrections tend to increase in size with
energy, because of the virtual Sudakov EW logarithms. In the high
invariant-mass region, these can be of the same order of magnitude as
the QCD corrections, but have opposite sign.  In addition, multiple
photon radiation plays a non-negligible role in the determination of
the invariant-mass distribution and induces negative corrections of
the order of a few percent.  In the light of this, it is a worthwhile
and non-trivial exercise to combine all of these different sets of
corrections, with the ultimate objective of determining the DY
NC cross section, in the high invariant-mass region, to a precision of
a few percent.  The results presented in this contribution represent
the first stage of a longer term project, with the objective of
systematically investigating all of the various sources of theoretical
uncertainty, which can induce effects of the order of a few percent.

\subsection{Available calculations and codes}

QCD corrections have been very well studied and a variety of
calculations and Monte Carlo (MC) generators exist. These include,
next-to-leading-order (NLO) and next-to-next-to-leading-order (NNLO)
corrections to the $W/Z$ total production rate
\cite{Altarelli:1979ub,Hamberg:1990np}, NLO calculations for $W, Z +
1, 2 \, \, {\rm jets}$ signatures \cite{Giele:1993dj,Campbell:2002tg}
(available in the codes {\tt DYRAD} and {\tt MCFM}), resummation of
leading and next-to-leading logarithms due to soft gluon radiation
\cite{Balazs:1997xd,Landry:2002ix} (implemented in the MC
{\tt ResBos}), NLO corrections merged with QCD Parton Shower (PS)
evolution (for instance in the event generators {\tt MC@NLO}
\cite{Frixione:2002ik} and {\tt POWHEG} \cite{Frixione:2007vw}), NNLO
corrections to neutral- and charged-current DY in fully
differential
form~\cite{Anastasiou:2003yy,Anastasiou:2003ds,Melnikov:2006di,
Melnikov:2006kv} (available in the MC program {\tt FEWZ}), as
well as leading-order multi-parton matrix element generators matched
with PS, such as, for instance, {\tt ALPGEN}
\cite{Mangano:2002ea}, {\tt MADEVENT} \cite{Stelzer:1994ta,
Maltoni:2002qb}, {\tt SHERPA} \cite{Gleisberg:2003xi} and {\tt HELAC}
\cite{Kanaki:2000ey,Papadopoulos:2005ky,Papadopoulos:2006mh}.

Complete ${\cal O}(\alpha)$ EW corrections to DY processes
have been computed independently by various authors in
\cite{Baur:2001ze,Calame:2007cd,Arbuzov:2007db,Zykunov:2005tc} for NC
production.  The EW tools which implement exact NLO corrections to NC
production are {\tt ZGRAD2}~\cite{Baur:2001ze}, {\tt HORACE}
\cite{Calame:2007cd} and {\tt SANC} \cite{Arbuzov:2007db}.  In {\tt
HORACE} the effect of multiple photon radiation to all orders via
PS is matched with the exact NLO-EW calculation.

\subsection{Electroweak Sudakov logarithms}
At high invariant masses $Q^2\gg \MW^2$, the EW corrections
are enhanced by Sudakov logarithms of the form $\ln(Q^2/\MW^2)$, which
originate from the exchange of soft and collinear virtual EW
gauge bosons as well as from the running of the EW couplings.
At the LHC, these corrections can reach tens of percent at the 
one-loop level and several percent at the two-loop level
\cite{Maina:2004rb,Kuhn:2005az,Kuhn:2007cv}.
The EW Sudakov corrections to the
NC four-fermion  scattering amplitude
 \cite{Kuhn:2000hx,Kuhn:2001hz,Jantzen:2005xi,Jantzen:2005az}
can schematically be written as
\begin{eqnarray}
\mathcal{A}&=&
\mathcal{A}_\mathrm{B}(Q^2)
\left[
1+\sum_{n\ge1}\left(\frac{\alpha}{4\pi}\right)^n
\sum_{k=0}^{2n}
C_{n,k}\ln^k\left(\frac{Q^2}{\MW^2}\right)
\right],
\end{eqnarray}
where $\mathcal{A}_\mathrm{B}(Q^2)$ is the Born amplitude with running
EW couplings at the scale $Q^2$.  The logarithmic corrections
are known to next-to-next-to-next-to-leading-logarithmic
(NNNLL) accuracy at the two-loop level
\cite{Jantzen:2005xi,Jantzen:2005az}, i.e.  $C_{2,k}$ with $4 \ge k
\ge 1$ are known.
Due to very strong cancellations between 
dominant and subdominant logarithmic terms, the two-loop
corrections to the $\mathrm{e}^+\mathrm{e}^- \to \mu^+\mu^-$ and
$\mathrm{e}^+\mathrm{e}^- \to q \bar q$ total cross sections are much
smaller than what might naively be expected and do not exceed a few per mil in the
TeV region.

\begin{figure}[htb] 
   \centering
   \includegraphics[width=0.40\linewidth]{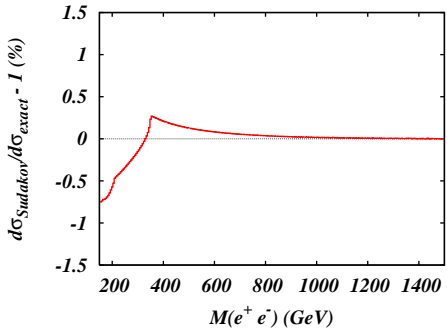} 
\caption{ Relative precision (in percent) of the Sudakov
approximation: the one-loop predictions for the
$\mathrm{e}^+\mathrm{e}^-$ invariant mass at the LHC are compared with
{\tt ZGRAD2}. The results have been obtained with the following
separation cuts: 
$ p_T(l) > 20$ GeV and 
$|\eta(l)| < 2.5$.
} 
   \label{fig:1loopComp}
\end{figure}

%
%
Nevertheless, for the DY process, kinematic cuts and
differential distributions
might partially destroy the cancellations and thus lead to much bigger
corrections.
It is therefore important to investigate higher-order Sudakov EW
corrections to differential DY distributions at the LHC.
To this end we have written a {\tt FORTRAN} code that implements the
results of Ref.~\cite{Jantzen:2005az} in fully differential form and
permits the interfacing of these to the programs {\tt
ZGRAD2}~\cite{Baur:2001ze} and {\tt HORACE}~\cite{Calame:2007cd}.  The
one-loop Sudakov expansion has been validated and agrees with the weak
corrections of {\tt ZGRAD2} with a precision at the few
per mil level or better for $Q\ge 200 ~\mathrm{GeV}$ (see
Fig.~\ref{fig:1loopComp}). 
The small deviations, at low invariant mass, are of the order of the
    mass-suppressed terms neglected in the Sudakov approximation.
\begin{figure}[htb] 
   \centering
   \includegraphics[width=0.7\linewidth]{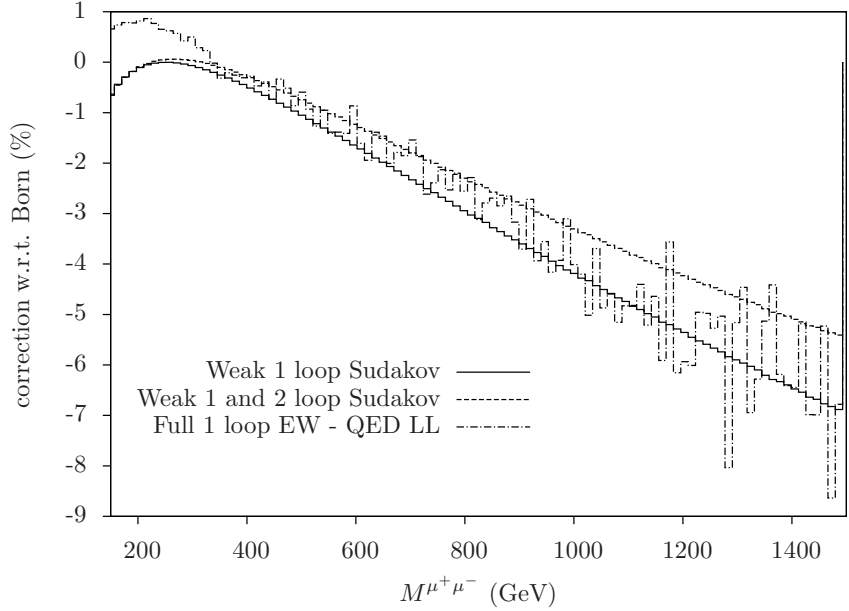} 
\caption{ EW corrections to the $\mu^+\mu^-$
  invariant mass at the LHC: one-loop predictions of {\tt HORACE}
  (dashed-dotted, see text);
  one-loop (solid) and two-loop (dashed) Sudakov approximation.}
   \label{fig:EWcorr}
\end{figure}
Fig.~\ref{fig:EWcorr} shows 
the Sudakov expansion up to two loops, wherein
virtual photonic contributions 
are subtracted as in Ref.~\cite{Jantzen:2005az}
and real photon emission is not included.
%
%
At the one-loop level, the Sudakov approximation (solid curve)
is in good agreement with the  {\tt HORACE} prediction (dashed-dotted curve),
which was obtained by using the set of input parameters appearing in Section
\ref{subsec:mcatnlohorace},  from the full EW
correction by subtracting 
${\cal O}(\alpha)$ 
photon emission in the leading-logarithmic (LL) approximation.\footnote{
Electromagnetic matching
corrections will be addressed in a forthcoming publication,
but the good agreement suggests that they should be quite small.}
The subtraction of the QED-LL correction
makes the results presented in Fig.~\ref{fig:EWcorr} independent, up to
terms of order ${\cal O}(m_l^2/M_{\scriptscriptstyle W}^2)$, of the
final state lepton flavour.
The one-loop Sudakov correction yields a negative contribution
that reaches $-7\%$ at 1.5~TeV.
The combination of one- and two-loop Sudakov corrections
is shown by the dashed line.
The two-loop effects are positive,
reach 1--2\%  in the plotted invariant-mass range 
and tend to reduce the one-loop contributions.

\subsection{Combining QCD and EW corrections}
In the high invariant-mass region both QCD and EW effects are large
and therefore, in view of the high accuracy needed by new physics
searches, it is important to combine both corrections consistently, at
the event generator level, to perform a realistic simulation of this
process.  A first attempt to combine QED and QCD corrections can be
found in \cite{Cao:2004yy} and results for the 
high invariant-mass distribution of charged lepton pairs 
are shown in Section~\ref{subsec:resbos}. The combination of
QCD and EW effects presented in Section~\ref{subsec:mcatnlohorace}
follows the approach first devised in
\cite{BCMMNPTV,Balossini:2007zz,Balossini:2007zza}.

\subsubsection{Combined QCD and EW effects with {\tt MC@NLO} and {\tt HORACE}}
\label{subsec:mcatnlohorace}
The formula for the combination of QCD and EW effects is given by
\cite{BCMMNPTV,Balossini:2007zz,Balossini:2007zza}:
\begin{equation}
\left\{\frac{d\sigma}{d{\cal O}}\right\}_{\rm QCD\oplus EW}= \left\{
\frac{d\sigma}{d{\cal O}}\right\}_{\rm best~QCD}+\left(
\left\{\frac{d\sigma}{d{\cal O}}\right\}_{\rm best~EW} -  
\left\{\frac{d\sigma}{d{\cal O}}\right\}_{\rm born}\right)_{\rm HERWIG PS}
\end{equation}   
where the differential cross-section, with respect to any observable
${\cal O}$, is given by two terms:
i) the results of a code which describes at best
the effect of QCD corrections;
ii) the effects due to NLO-EW corrections and to higher-order QED
effects of multiple photon  radiation computed with {\tt HORACE}. 
In the EW calculation, the effect of the Born distribution is subtracted
to avoid  double counting since this is included in the QCD generator.
In addition, the EW corrections are convoluted with a QCD PS and
include, in the collinear approximation, the bulk of the 
${\cal  O}(\alpha\alpha_s) $ corrections.

\begin{figure}[htb] 
   \centering
   \includegraphics[width=0.49\linewidth]{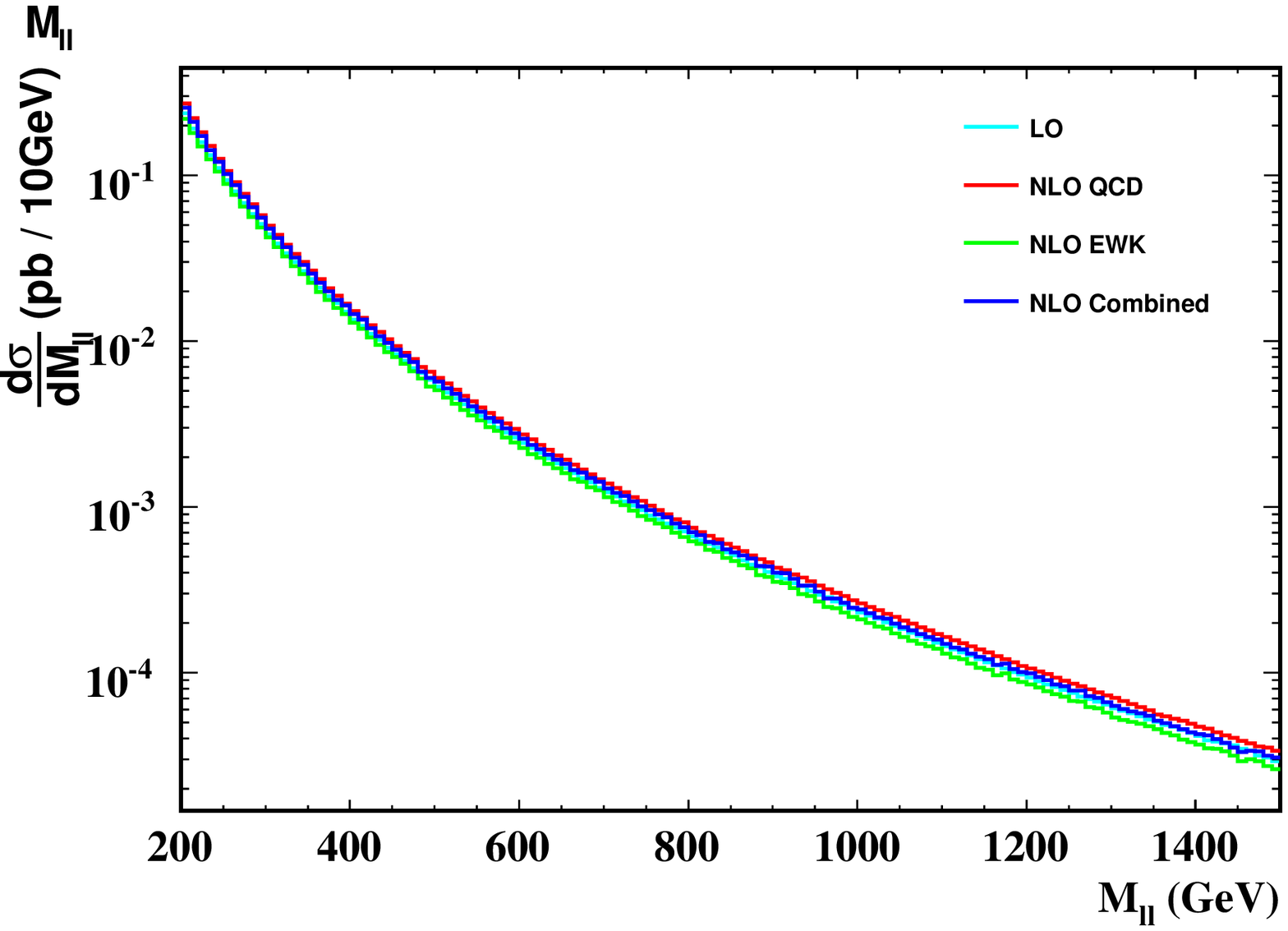}
   \includegraphics[width=0.49\linewidth]{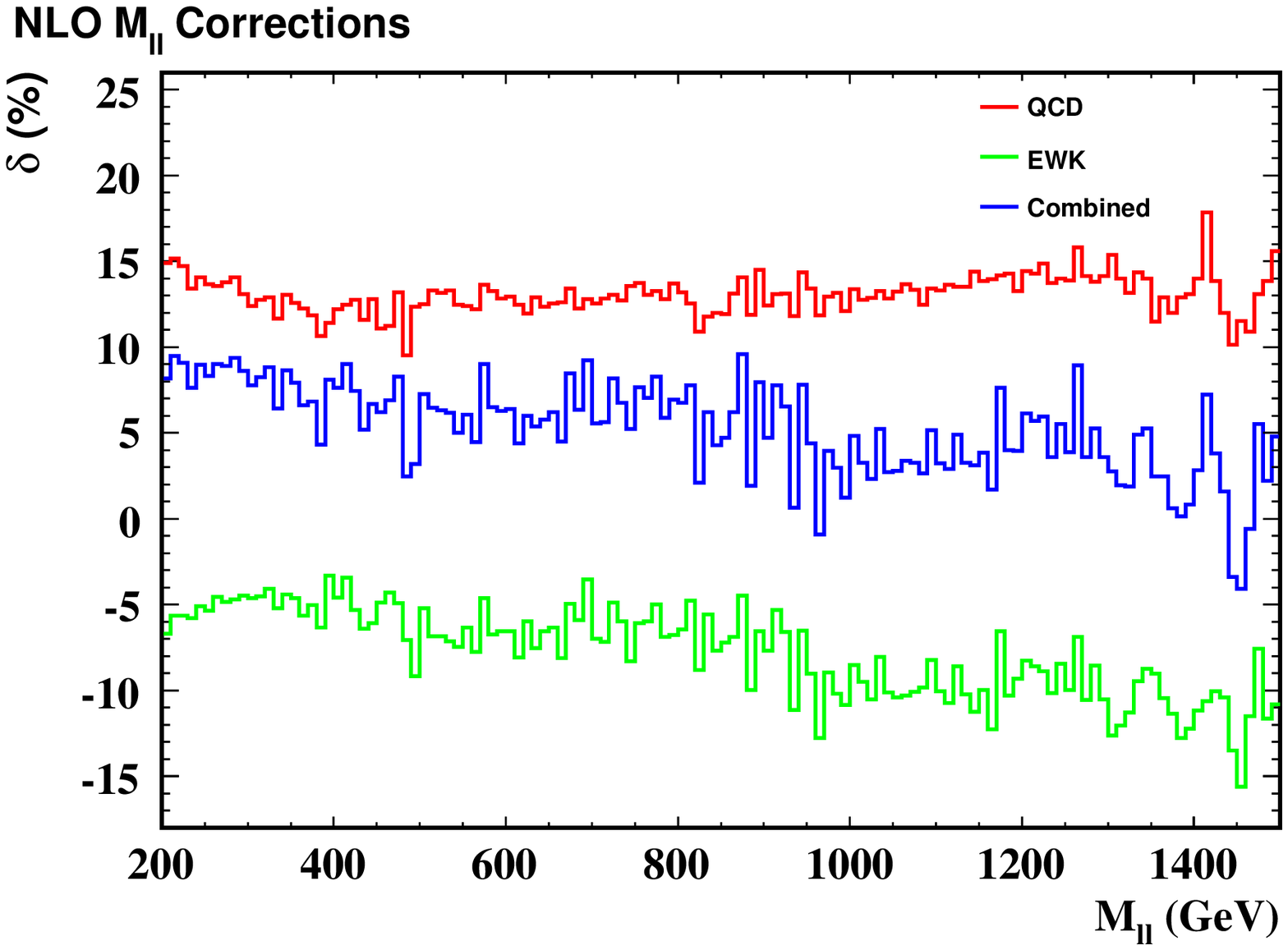}
   \caption{QCD and EW corrections to the di-electron
   invariant mass.} 
   \label{fig:QCDEWmass}
\end{figure}

\begin{figure}[htb] 
   \centering
   \includegraphics[width=0.49\linewidth]{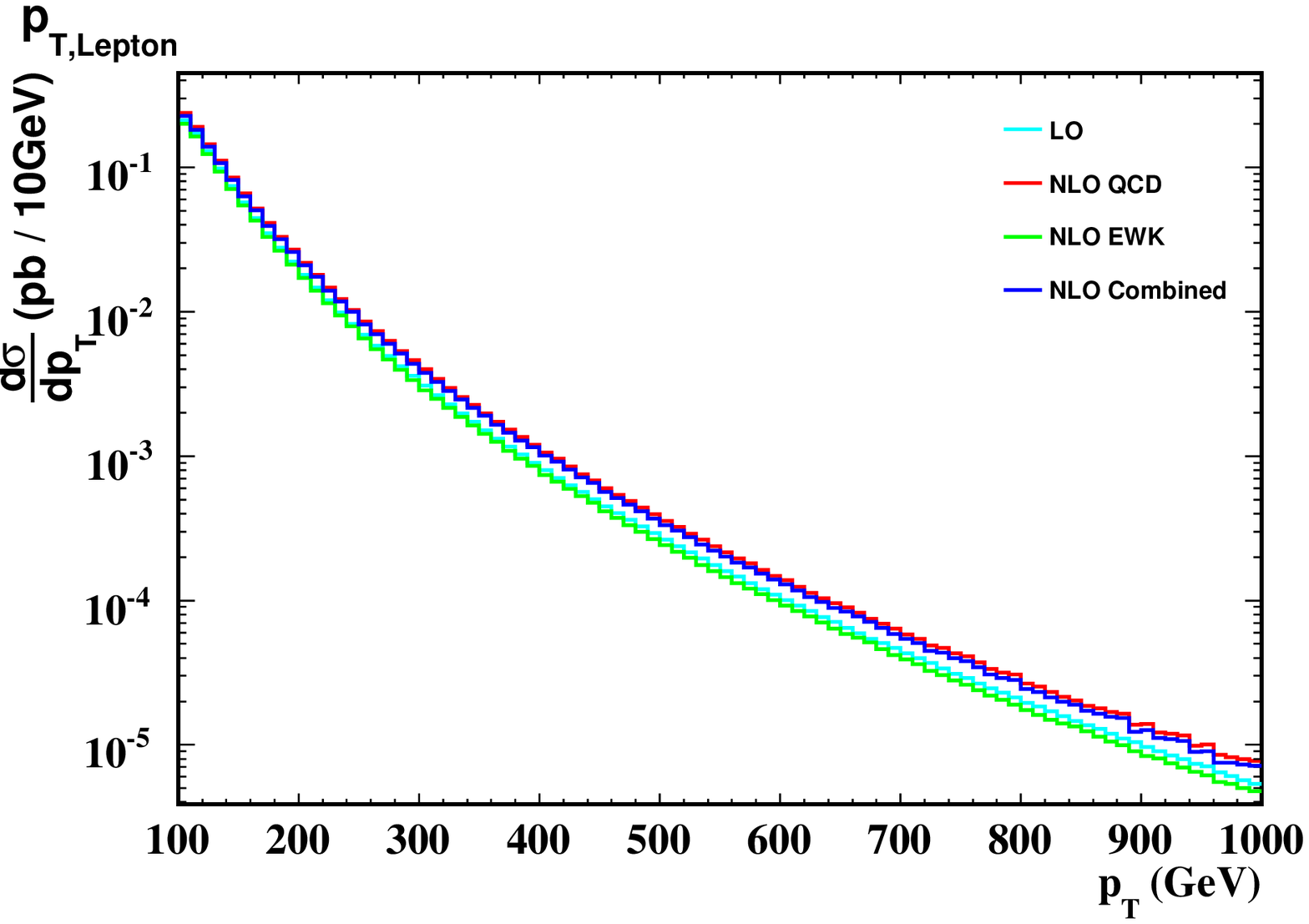}
   \includegraphics[width=0.49\linewidth]{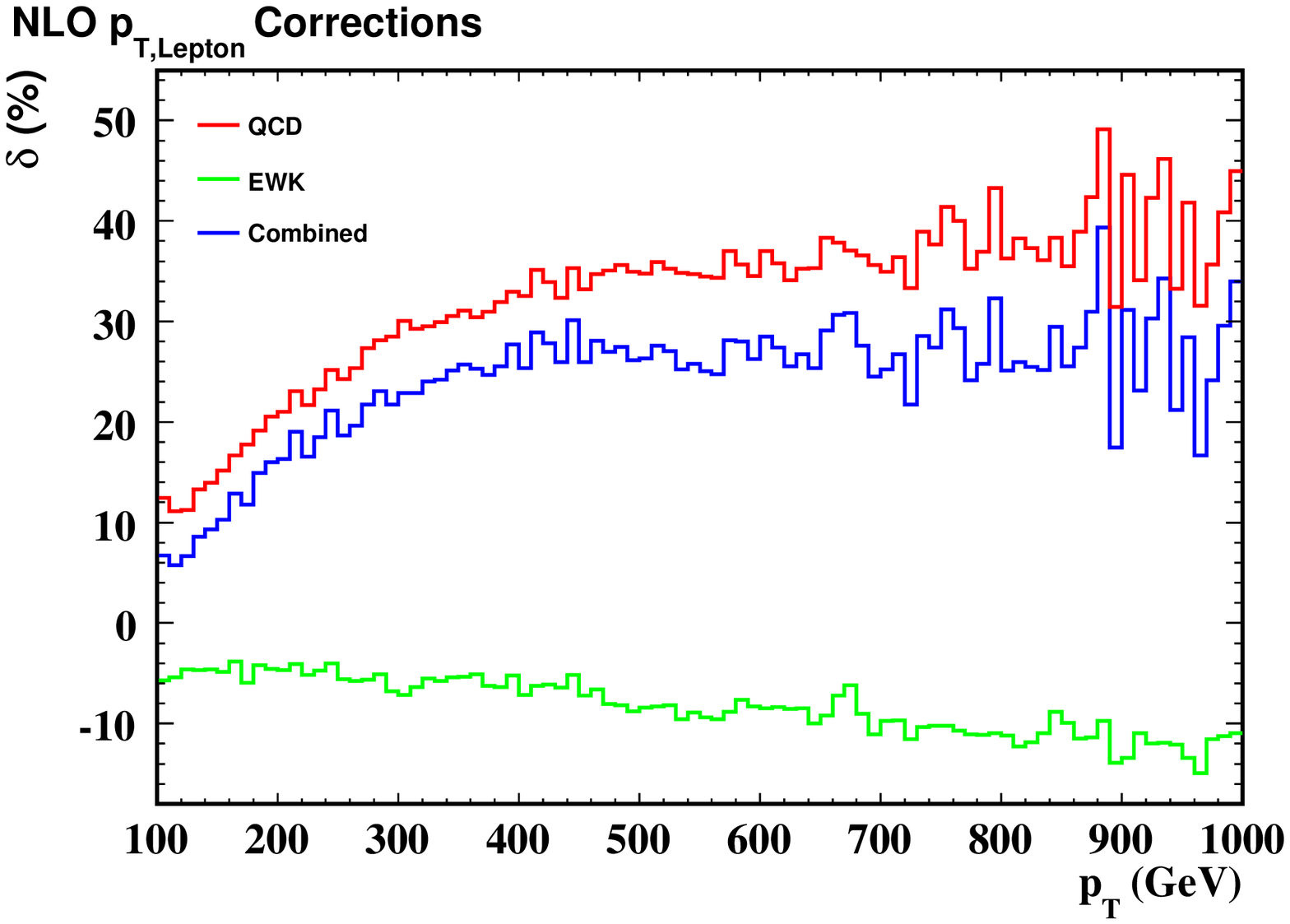}
  \caption{QCD and EW corrections to the electron
  transverse momentum.}
   \label{fig:QCDEWptlep}
\end{figure}

Preliminary numerical results have been obtained, for an $e^+e^-$ final state, 
with the following set of input parameters:
\begin{eqnarray}\label{eq:pars}
G_{\mu} = 1.16639\times 10^{-5} \; {\rm GeV}^{-2}, 
& \qquad & \alpha= 1/137.03599911, \quad \alpha_s\equiv\alpha_s(M_Z^2)=0.118,
\nonumber \\ 
M_W = 80.419 \; {\rm GeV}, 
& \quad & M_Z = 91.188 \; {\rm GeV},~~~~ 
\Gamma_Z =  2.4952  \; {\rm GeV},\nonumber  \\
m_e  = 0.51099892 \; {\rm MeV}, &\quad &m_{\mu}=0.105658369 \; {\rm GeV},  
\quad m_{t}=174.3 \; {\rm GeV}.
\nonumber
\end{eqnarray}
The parton distribution function (PDF) set {\tt MRST2004QED} \cite{Martin:2004dh}
has been used to describe the proton partonic content.
The PDF factorization scale has been set equal to
$\mu_F=\sqrt{\left(p_{\perp}^{\scriptscriptstyle Z}\right)^2+
M^2_{e^+e^-}}$, where $M_{e^+e^-}$
is the invariant mass of the lepton pair.
The following cuts have been imposed to select the events:
\begin{equation}
p_\perp^{e^\pm} > 25~ {\rm GeV},~~~~|\eta^{e^\pm}| < 2.5,~~~~
M_{e^+e^-}>200~{\rm GeV}.
\label{eq:cuts}
\end{equation}
The percentage corrections shown in the right panels of
Figs.~\ref{fig:QCDEWmass}  and \ref{fig:QCDEWptlep}
have been defined as
$\delta = \linebreak
  \left(\sigma_{NLO}-\sigma_{Born+PS}\right)/\sigma_{Born+PS}   $. 
The granularity of the detectors and the size of the electromagnetic
showers in the calorimeter  
make it difficult to discriminate between electrons and photons with a
small opening angle. We adopt the following procedure to select the event:
we recombine the four-momentum vectors of the electron and
photon into an effective electron four-momentum vector if, defining
\begin{equation}
\Delta R(e,\gamma) = \sqrt{
  \Delta\eta(e,\gamma)^2+\Delta\phi(e,\gamma)^2 },
\end{equation}
$\Delta R(e,\gamma)<0.1$ (with $\Delta\eta,\Delta\phi$ the distances
of electrons and photons along the longitudinal and azimuthal directions).
We do not recombine electrons and photons if $\eta_{\gamma}>2.5$ (with
$\eta_\gamma$ the photon pseudo-rapidity).
We apply the event selection cuts only after the recombination procedure.

We have used {\tt MC@NLO} as the best QCD generator
and have tuned it with {\tt MCFM/FEWZ} at NLO. With the same
settings, the two codes, when run at LO, give the same results as {\tt HORACE}.
The tuning procedure validates the interpretation of
the various relative effects as due to the radiative corrections and
not to a mismatch in the setups of the two codes.
The results presented have been obtained using {\tt HORACE}
where the exact NLO-EW corrections are included, but no higher-order
effects due to QED multiple emissions.
Fig.~\ref{fig:QCDEWmass} shows the interplay between the QCD and 
EW  corrections for the di-lepton invariant mass.
The QCD corrections are quite flat and positive 
with a value of about $15\%$ over the mass range 200--1500~GeV. The
EW corrections are negative and vary from about $-5\%$ to
$-10\%$ and thus partially cancel the NLO-QCD effect.
The 2-loop Sudakov logarithms (absent in this plot) would give an
additional positive contribution to the cross-section.
In Fig.~\ref{fig:QCDEWptlep} the lepton transverse-momentum
distribution is shown.
The NLO-QCD corrections 
rise from
10\% to 35\% in the interval considered (100--1000 GeV). 
The NLO-EW corrections are negative and fall from $-5\%$ to
$-10\%$ over the same range. 

\subsubsection{Combined QCD and EW effects with {\tt ResBos}}
\label{subsec:resbos}
In this work we also examine the effects of the initial-state multiple soft-gluon emission and the
dominant final-state EW correction (via box diagrams)
on the high invariant-mass distribution of the charged lepton pairs produced at
the LHC. We shall focus on the region of $200\,{\rm GeV}<m_{\ell\ell}<1500\,{\rm GeV}$,
where $m_{\ell\ell}$ denotes the invariant mass of the two final-state charged
leptons.
The fully differential cross section including the contributions from the
initial-state multiple soft-gluon emission
is given by the resummation formula presented
in Refs.~\cite{Balazs:1997xd,Cao:2004pt,Cao:2004yy,Dobbs:2004bu}.
Furthermore, it has been shown that, above the $Z$ pole region, the
EW correction contributed from the box diagrams involving $Z$ and $W$
exchange is no longer negligible~\cite{Baur:2001ze}. It
increases strongly with energy and contributes significantly at
high invariant mass of the lepton pair. Hence, we will also include
the dominant EW correction via box diagrams in this study.
\begin{figure}[htb]

\includegraphics*[scale=0.5]{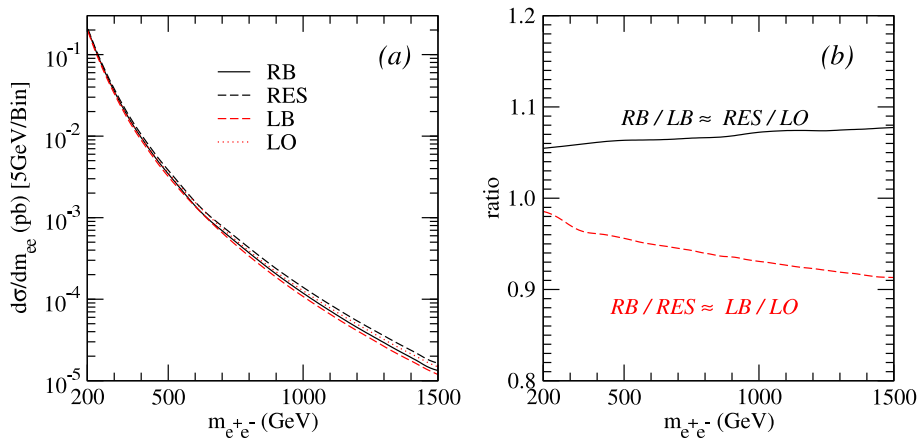}
\centering
\caption{(a) Invariant-mass distributions of the charged lepton pair; (b) ratios
of various contributions.\label{fig:resbos}}
\end{figure}

For clarity, we introduce below the four shorthand notations:
\begin{itemize}
\item LO: leading-order initial state,
\item LO+BOX (LB): leading-order initial state plus the $ZZ/WW$ box diagram contribution,
\item RES: initial-state QCD resummation effects,
\item RES+BOX (RB): initial-state QCD resummation effects plus the $ZZ/WW$
box-diagram contribution.
\end{itemize}
For this exercise, we consider the electron lepton pairs only and adopt the
CTEQ6.1M PDFs~\cite{Pumplin:2002vw}.
Fig.~\ref{fig:resbos}(a) shows the distributions of the invariant
mass $m_{e^{+}e^{-}}$ for
RES+BOX (RB) (black solid line),
RES only (black dashed line),
LO+BOX (LB) (red dashed line)
and LO only (red dotted line).
It is instructive to also examine the ratios of various contributions, as shown in
in Fig.~\ref{fig:resbos}(b). We note that the initial-state QCD resummation effect
and the EW correction via box diagrams are almost factorized in the
high invariant-mass region, e.g. \begin{eqnarray}
{\displaystyle \frac{d\sigma_{RB}}{dm_{\ell\ell}}}/{\displaystyle \frac{d\sigma_{LB}}{dm_{\ell\ell}}} & \simeq & {\displaystyle \frac{d\sigma_{RES}}{dm_{\ell\ell}}}/{\displaystyle \frac{d\sigma_{LO}}{dm_{\ell\ell}}},\\
{\displaystyle \frac{d\sigma_{RB}}{dm_{\ell\ell}}}/{\displaystyle \frac{d\sigma_{RES}}{dm_{\ell\ell}}} & \simeq & {\displaystyle \frac{d\sigma_{LB}}{dm_{\ell\ell}}}/{\displaystyle \frac{d\sigma_{LO}}{dm_{\ell\ell}}}.\end{eqnarray}
The EW correction from the box diagrams reduces the invariant-mass distribution
slightly around
\linebreak\mbox{$m_{e^{+}e^{-}}\sim200\,{\rm GeV}$} and largely ($\sim9\%$)
around $m_{e^{+}e^{-}}\sim1500\,{\rm GeV}$. On the other hand, the
initial state soft-gluon resummation effect increases the invariant-mass distribution
by an amount of $5\%$ at 200\,GeV and $8\%$ at 1500\,GeV.
Therefore, the QCD resummation effect dominates
over the EW correction induced by the $ZZ/WW$ box diagrams in the relatively
low invariant-mass
region, and they become comparable in the high invariant-mass region.
The cancellation between both contributions in the high invariant-mass region
causes the net contribution to be close to the leading order prediction.
Finally, we note that the final state QED correction should also be included
for predicting precision measurements. A detailed study including the soft-gluon resummation
effect and the full EW correction will be presented elsewhere.

\subsection{Outlook and conclusions}
The preliminary results of this contribution show the 
non-trivial interplay between EW and QCD corrections in the high
invariant-mass region of the NC DY process.  For most of the
observables, the NLO EW corrections are negative and
partially cancel the QCD ones. \\ The NC DY process has been
studied in great detail in the literature.  This contribution is a
first step towards collecting these different results and augmenting
them with further studies to obtain an accurate prediction of this
process.  We have shown a preliminary investigation which includes,
separately, results on the EW 2-loop Sudakov logarithms, QCD
resummation, and combination of QCD and EW NLO corrections.
The ongoing investigation aims to combine the effects above in the
simulation and complete them with multiple photon emission and
photon-induced partonic subprocesses.  All these effects induce
corrections of the order of a few percent.  In addition, the
di-electron and di-muon final states will be studied separately in
more detail.  We also aim to include the effect of real $W$ and $Z$ boson
emission.  This could result in the partial cancellation of virtual
EW corrections, but it is dependent upon the definition of the
observables and the experimental analysis.  For completeness, we will
include the systematic uncertainties from the PDFs, energy scale,
choice of calculation scheme, higher-order contributions, showering
model and the EW-QCD combination.

\subsection*{Acknowledgment}
The authors would like to thank the Scottish Universities Physics 
Alliance (SUPA) and the
Department of Physics and Astronomy of the University of Glasgow for 
their financial support.
SM thanks the Royal Society (London, UK) for financial support in the
form of a conference grant to attend this workshop.


%% file: s_halyo/DiLep_ZHorace.tex
%
\subsection{Introduction}

Precise measurement of gauge boson production cross-sections for 
$pp$ scattering will be crucial at the LHC. $W/Z$ bosons will be 
produced copiously, and a careful measurement of their production
cross-sections will be important in testing the Standard Model (SM) more
rigorously than ever before to potentially uncover signs of new physics.

Currently, no Monte Carlo (MC) event generators exist that include {\it both} higher 
order QCD and electroweak corrections. In what follows therefore,  
we evaluate whether it is possible to accurately describe the $Z$ production
cross-section under the $Z$ peak with an event-level generator that
includes only Final State QED Radiation (FSR)
corrections (in the leading-log approximation) instead of the complete
electroweak corrections included in the HORACE generator. In addition,
we estimate the error that results if one chooses to use this MC event generator scheme.

\subsection{Impact of Electroweak Corrections on $Z$ Production Cross-Section.}

The lack of a MC event generator that incorporates beyond leading order
corrections in both the electroweak and QCD calculations, leads us to
study which of the corrections contribute dominantly under the $Z$
peak. By far the largest correction comes from inclusion of NLO QCD
calculations. These produce a change in the cross-section of 20\% or
more~\cite{dixon2003}, depending on the $Z$ kinematic region
considered.
What we wish to determine then is the error imposed through including only the
leading-log FSR contributions instead of the exact ${\cal O}(\alpha)$
corrections matched with
higher-order QED radiation that exist in HORACE.
(since these are currently all that can be incorporated in addition to the NLO
QCD corrections). 

In order to study this error we used HORACE~\cite{HORACE1,HORACE2,HORACE3,HORACE4}, a MC event
generator that includes exact ${\cal O}(\alpha)$ electroweak radiative
corrections matched to a leading-log QED parton shower, and compared it
to a Born-level calculation with final-state QED corrections added. The
latter QED corrections were calculated by the program PHOTOS~\cite{PHOTOS1,PHOTOS2,PHOTOS3}, a
process-independent module for adding multi-photon emission to events
created by a host generator.

\begin{figure}[htb]
\begin{center}
\resizebox{!}{8cm}{\includegraphics{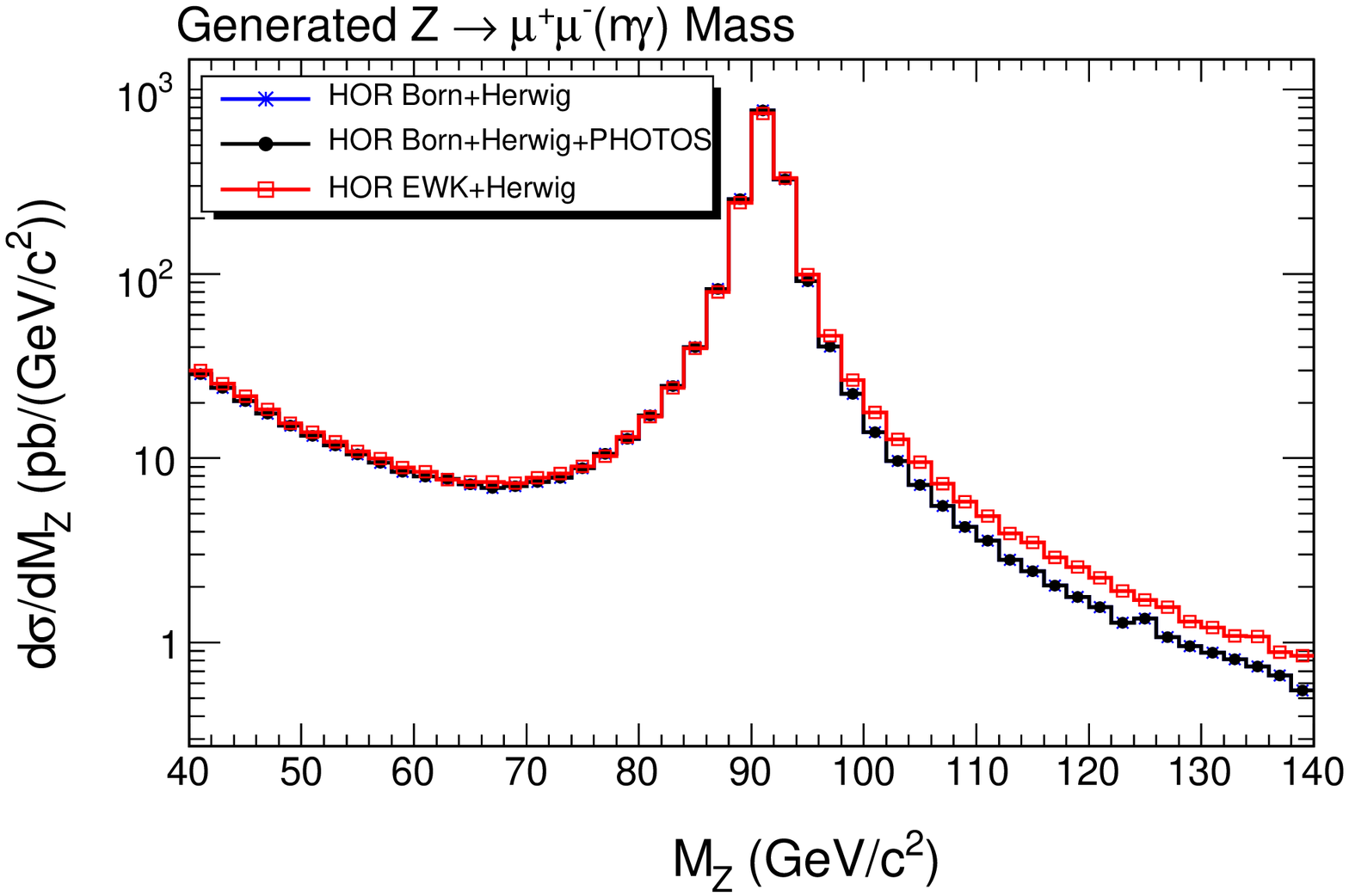}}
\caption{ Comparison of $Z$ boson invariant mass distributions for the
 process $Z/\gamma^* \to \ell^{+}\ell^{-}(n\gamma)$ in 
 HORACE 3.1 including electroweak and QED corrections showered with
 HERWIG (open red squares), HORACE Born-level showered with HERWIG plus
 PHOTOS (black circles), and HORACE Born-level (blue stars).}
\label{fig:horace_zmass}
\end{center}
\end{figure}

\begin{figure}[htb]
\begin{center}
\resizebox{!}{8cm}{\includegraphics{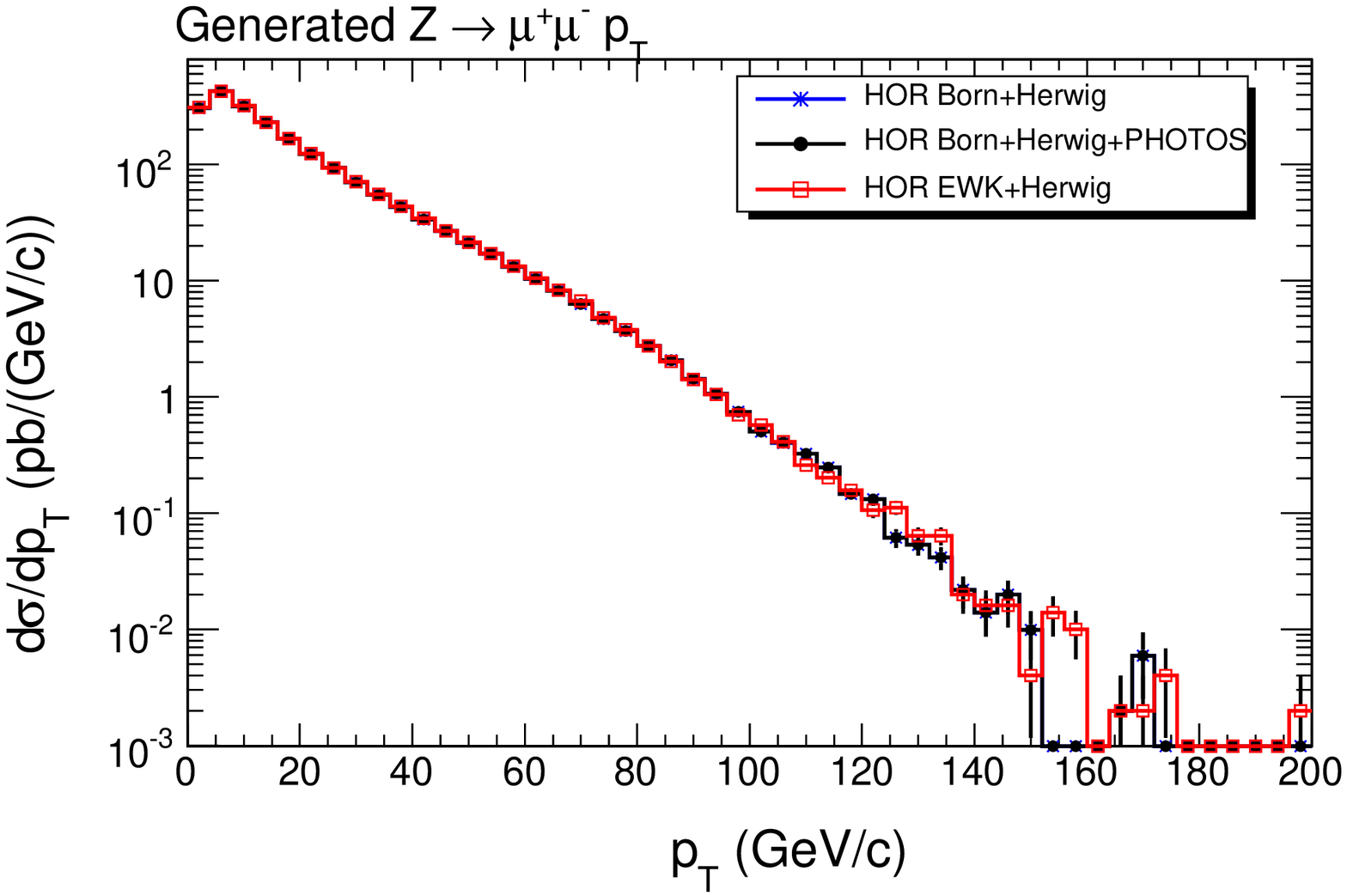}}
\caption{ Comparison of $Z$ boson transverse momentum distributions for the
 process $Z/\gamma^* \to \ell^{+}\ell^{-}(n\gamma)$ in 
 HORACE 3.1 including electroweak and QED corrections showered with
 HERWIG (open red squares), HORACE Born-level showered with HERWIG plus
 PHOTOS (black circles), and HORACE Born-level (blue stars).}
\label{fig:horace_zpt}
\end{center}
\end{figure}

\begin{figure}[htb]
\begin{center}
\resizebox{!}{8cm}{\includegraphics{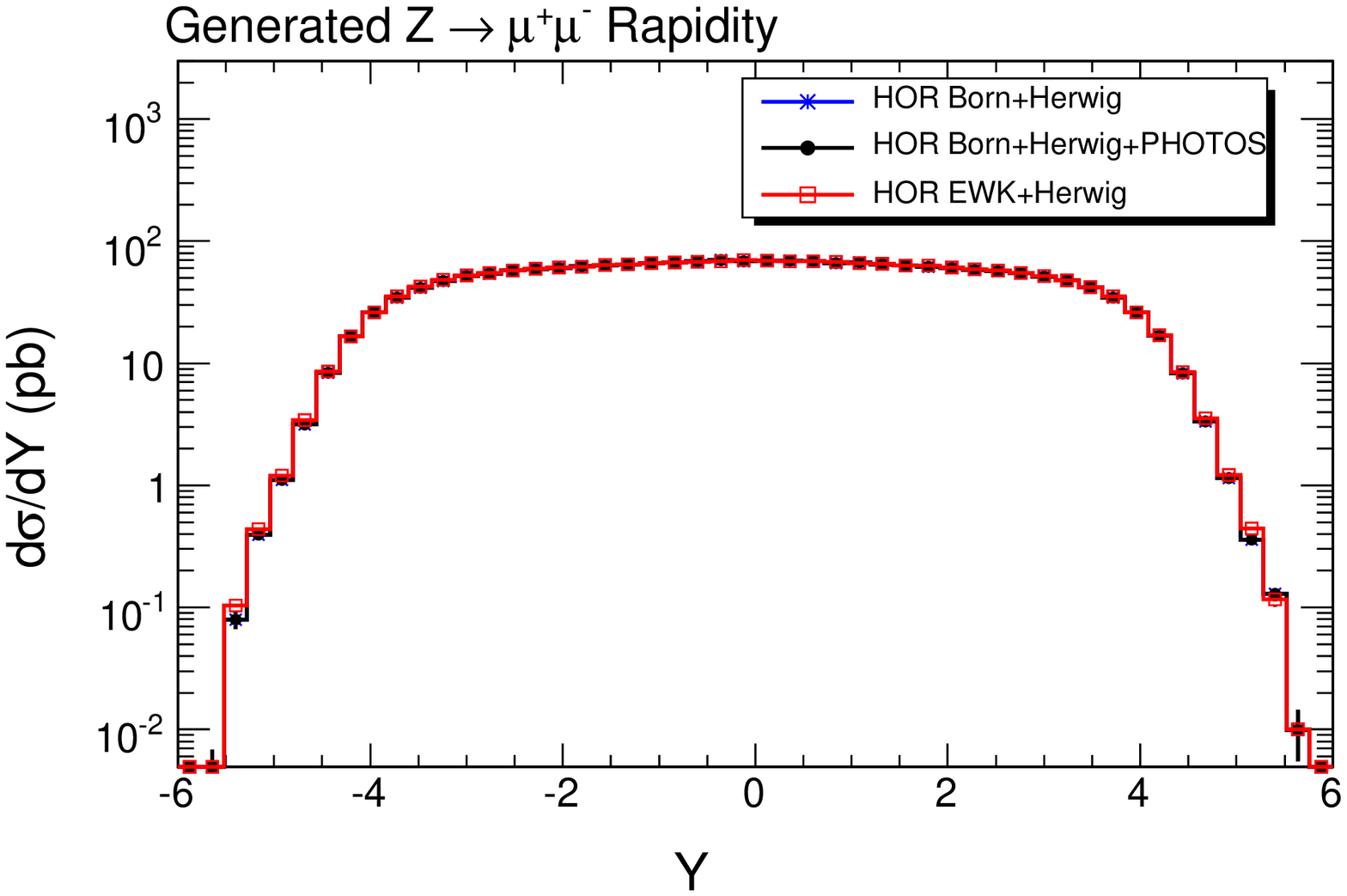}}
\caption{ Comparison of $Z$ boson rapidity distributions for the
 process $Z/\gamma^* \to \ell^{+}\ell^{-}(n\gamma)$ in 
 HORACE 3.1 including electroweak and QED corrections showered with
 HERWIG (open red squares), HORACE Born-level showered with HERWIG plus
 PHOTOS (black circles), and HORACE Born-level (blue stars).}
\label{fig:horace_zy}
\end{center}
\end{figure}

\begin{figure}[htb]
\begin{center}
\resizebox{!}{8cm}{\includegraphics{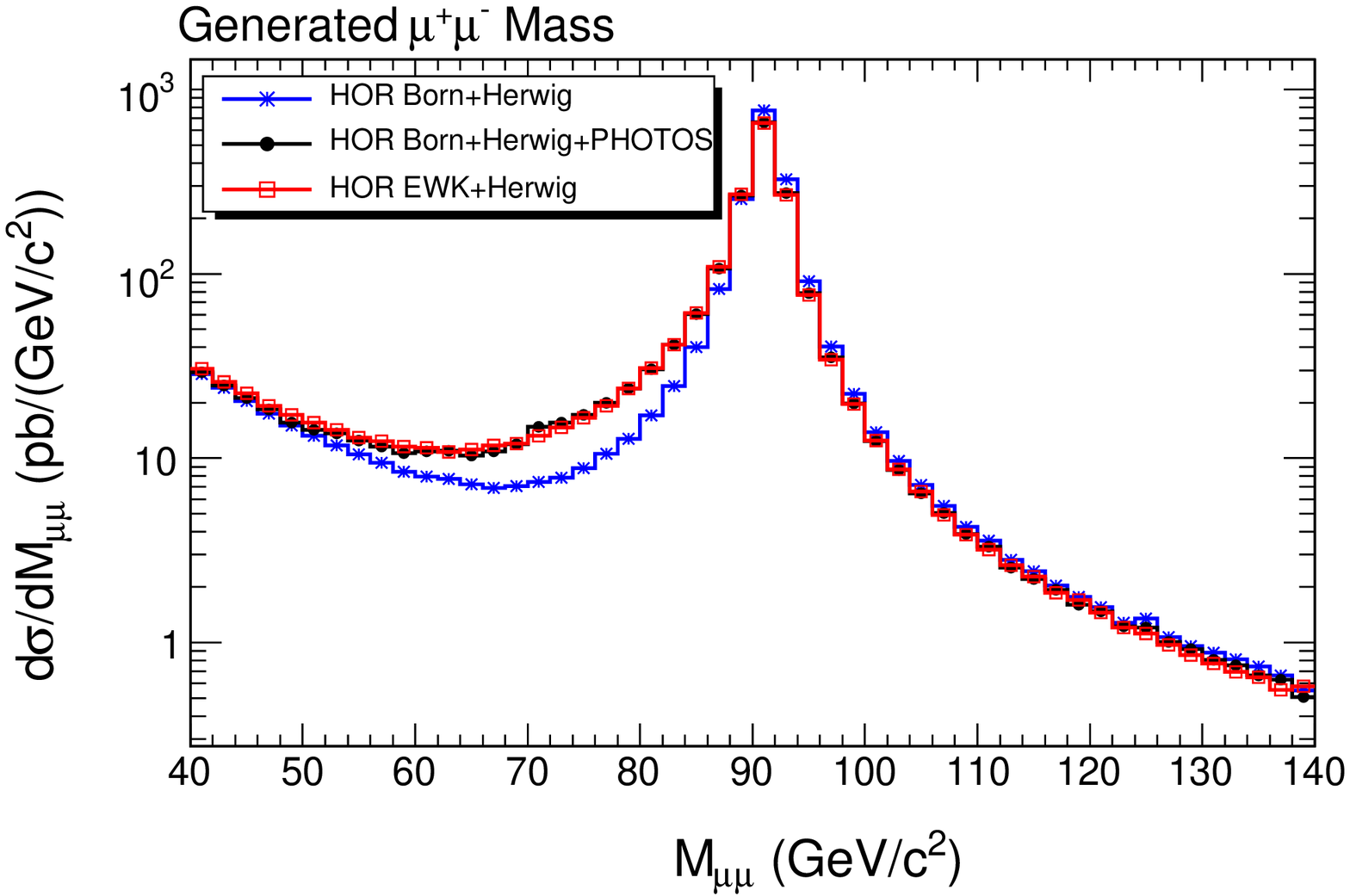}}
\caption{ Comparison of $\ell^+\ell^-$ invariant mass distributions for the
 process $Z/\gamma^* \to \ell^{+}\ell^{-}(n\gamma)$ in 
 HORACE 3.1 including electroweak and QED corrections showered with
 HERWIG (open red squares), HORACE Born-level showered with HERWIG plus
 PHOTOS (black circles), and HORACE Born-level (blue stars).}
\label{fig:horace_mass}
\end{center}
\end{figure}

\begin{figure}[htb]
\begin{center}
\resizebox{!}{8cm}{\includegraphics{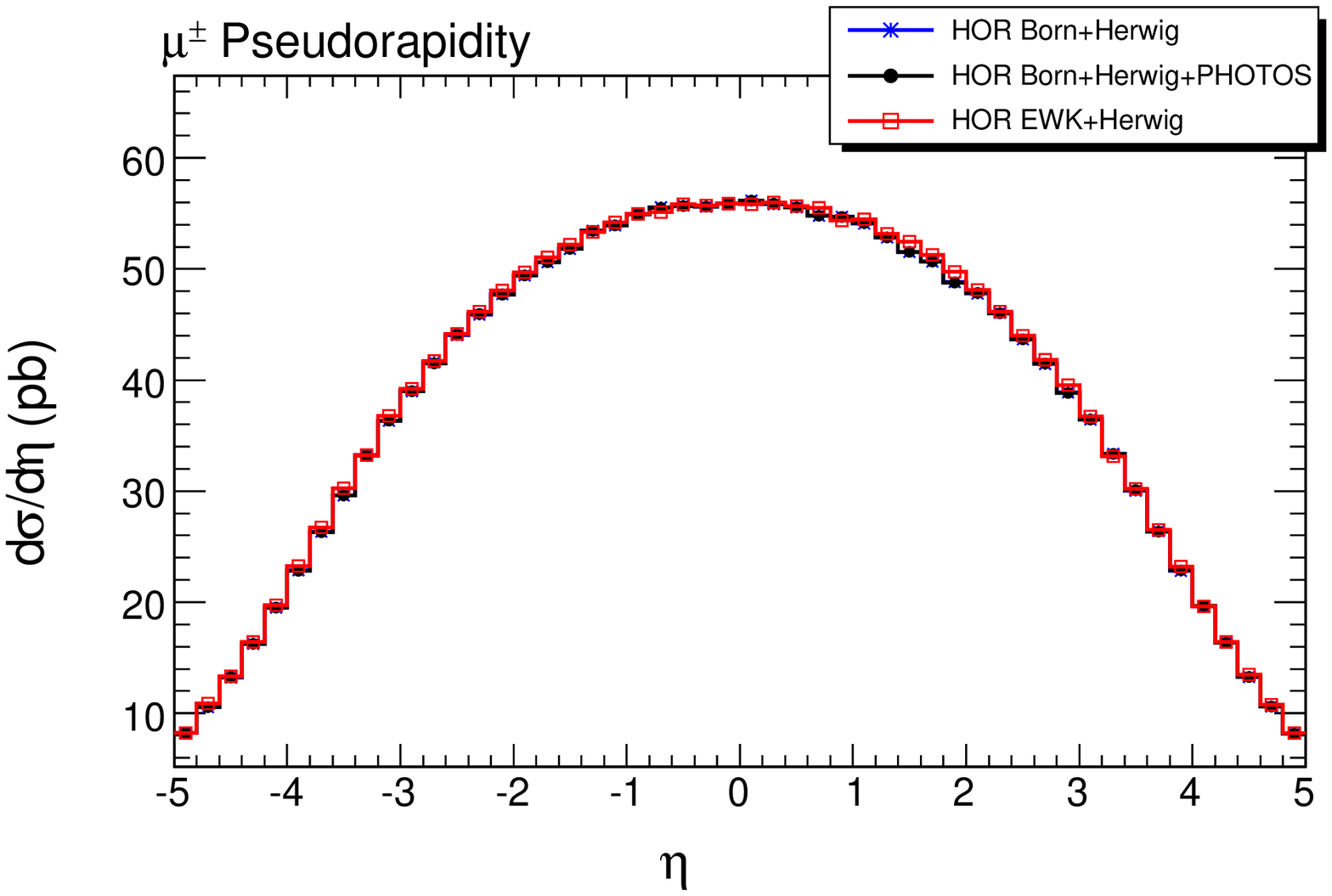}}
\caption{ Comparison of $\ell^+\ell^-$ lepton pseudo-rapidity distributions for the
 process $Z/\gamma^* \to \ell^{+}\ell^{-}(n\gamma)$ in 
 HORACE 3.1 including electroweak and QED corrections showered with
 HERWIG (open red squares), HORACE Born-level showered with HERWIG plus
 PHOTOS (black circles), and HORACE Born-level (blue stars).}
\label{fig:horace_mu_eta}
\end{center}
\end{figure}

\begin{figure}[htb]
\begin{center}
\resizebox{!}{8cm}{\includegraphics{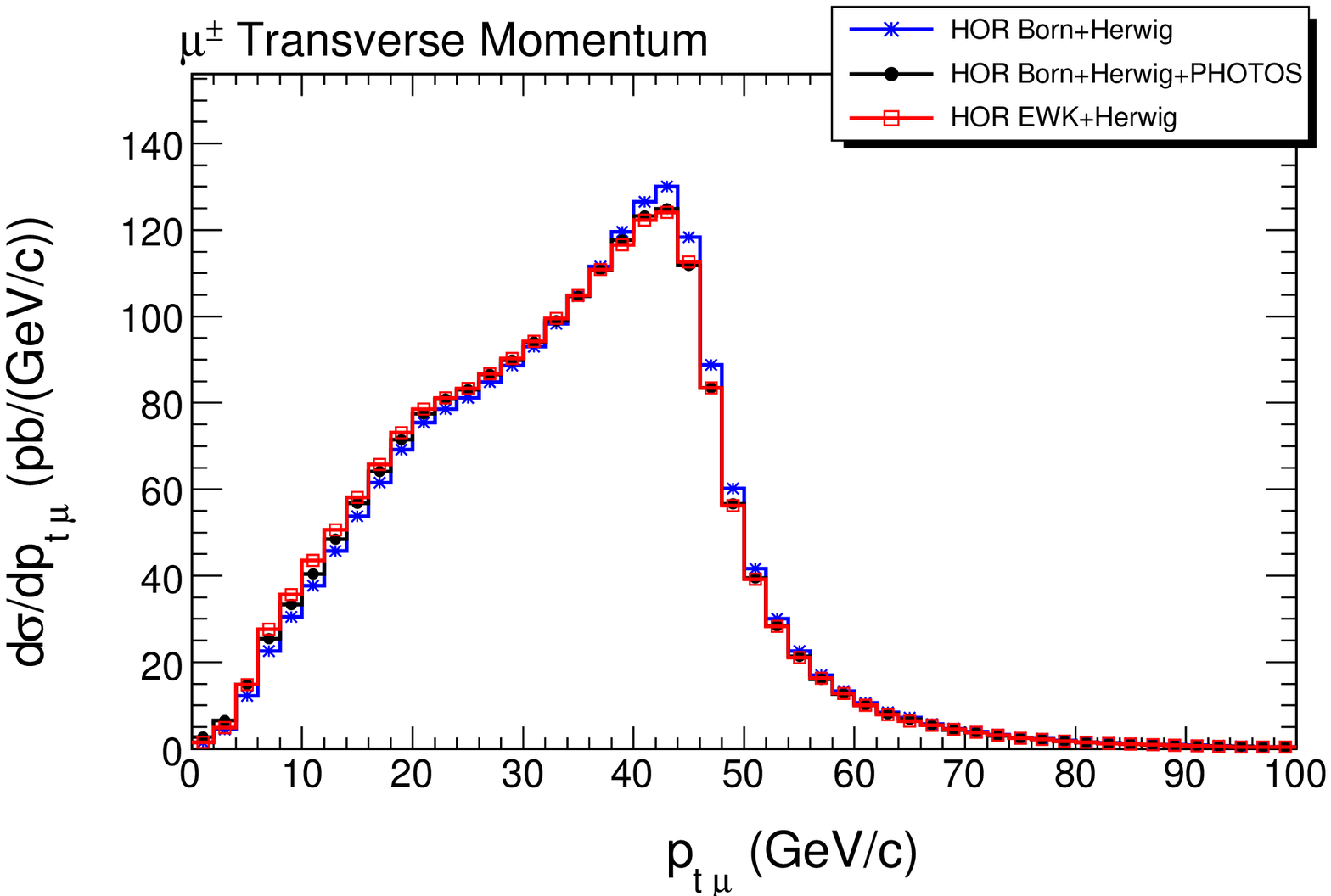}}
\caption{ Comparison of $\ell^+\ell^-$ lepton transverse momentum distributions for the
 process $Z/\gamma^* \to \ell^{+}\ell^{-}(n\gamma)$ in 
 HORACE 3.1 including electroweak and QED corrections showered with
 HERWIG (open red squares), HORACE Born-level showered with HERWIG plus
 PHOTOS (black circles), and HORACE Born-level (blue stars).}
\label{fig:horace_mu_pt}
\end{center}
\end{figure}

\begin{figure}[htb]
\begin{center}
\resizebox{!}{8cm}{\includegraphics{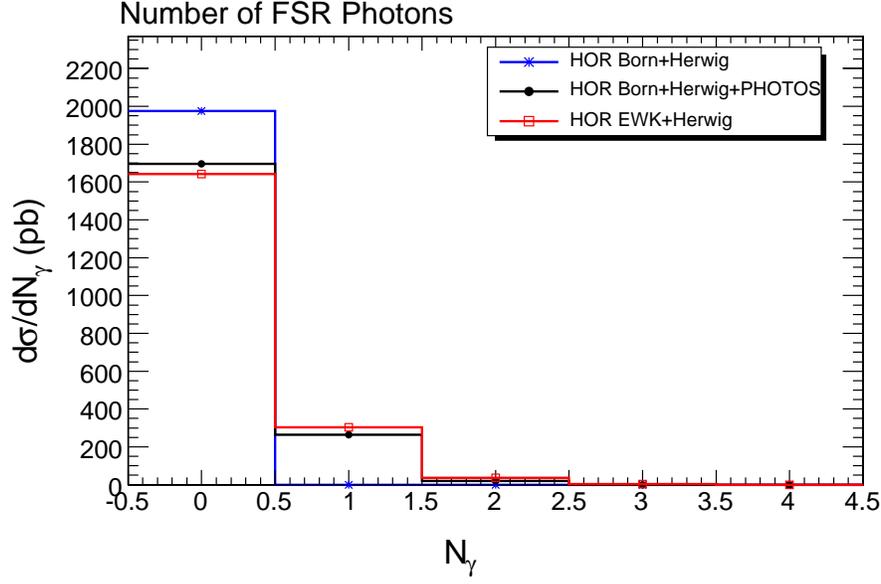}}
\caption{ Comparison of the number $n$ of final state radiation (FSR) photons 
  in $Z/\gamma^* \to \ell^{+}\ell^{-}(n\gamma)$ for
 HORACE 3.1 including electroweak and QED corrections showered with
 HERWIG (open red squares), HORACE Born-level showered with HERWIG plus
 PHOTOS (black circles), and HORACE Born-level (blue stars).}
\label{fig:horace_ngamma}
\end{center}
\end{figure}

\begin{figure}[htb]
\begin{center}
\resizebox{!}{8cm}{\includegraphics{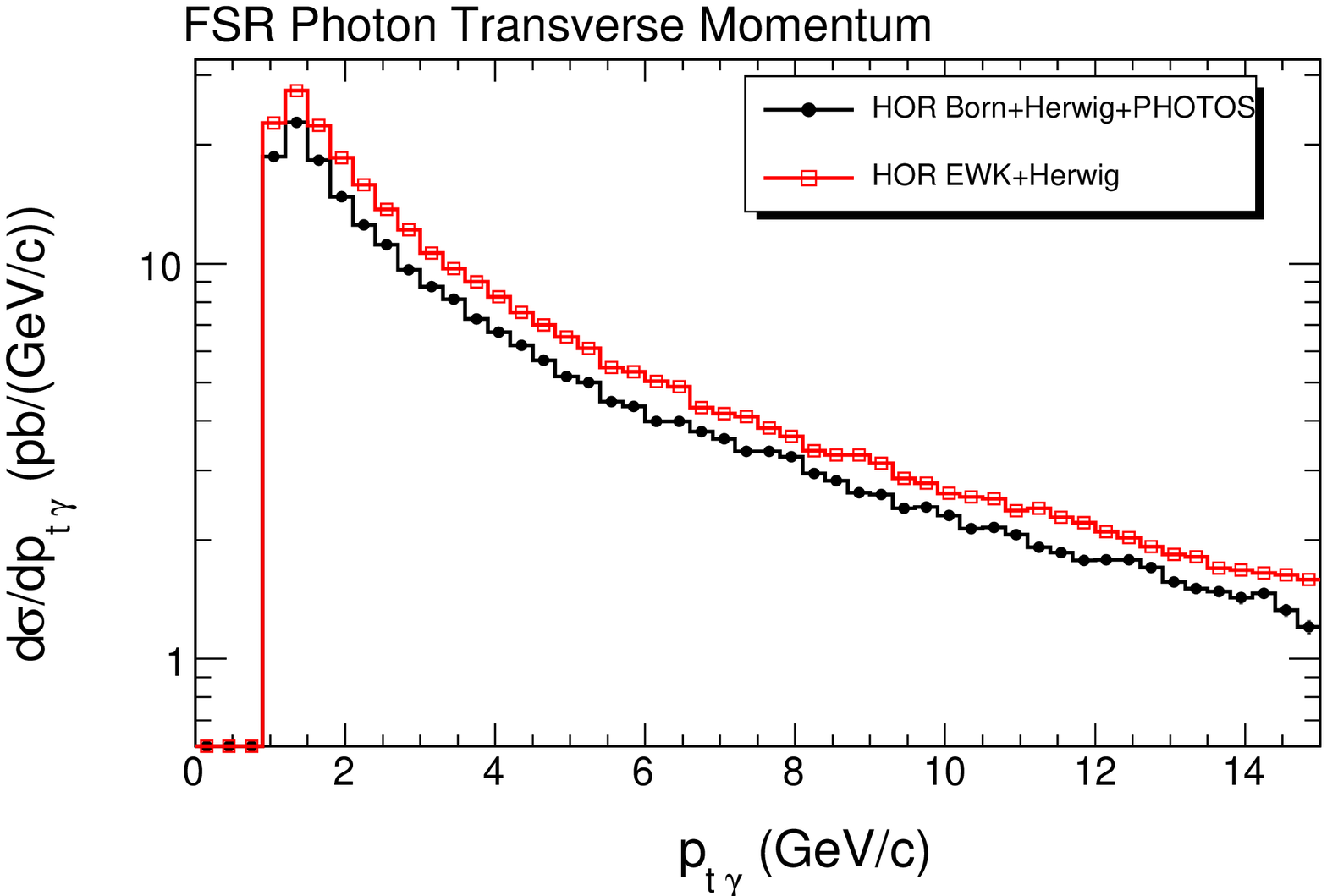}}
\caption{ Comparison of $Z/\gamma^* \to \ell^{+}\ell^{-}(n\gamma)$ final state
  radiation (FSR) transverse momentum distributions for
  HORACE 3.1 including electroweak and QED corrections showered with
  HERWIG (open red squares) and HORACE Born-level showered with HERWIG plus
  PHOTOS (black circles).}
\label{fig:horace_gamma_pt}
\end{center}
\end{figure}

\begin{figure}[htb]
\begin{center}
\resizebox{!}{8cm}{\includegraphics{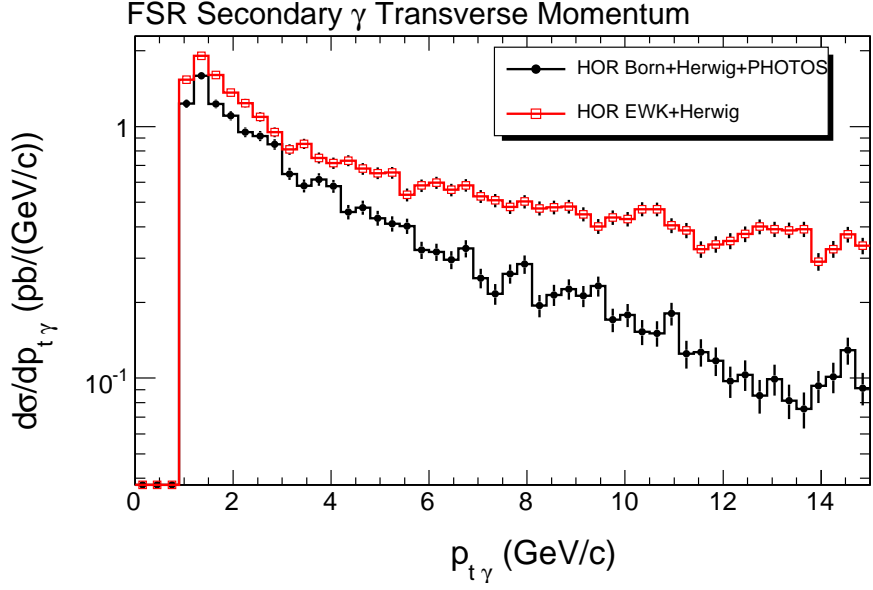}}
\caption{ Comparison of $Z/\gamma^* \to \ell^{+}\ell^{-}(n\gamma)$
  secondary final state radiation (FSR) transverse momentum distributions for
  HORACE 3.1, including electroweak and QED corrections showered with
  HERWIG (open red squares), and HORACE Born-level showered with HERWIG plus
  PHOTOS (black circles). Secondary FSR includes any FSR photons other
  than the first hard photon. }
\label{fig:horace_gammaII_pt}
\end{center}
\end{figure}

In the following we compared $pp \to Z/\gamma^* \to l^{+}l^{-}$ events
generated by HORACE with the full 1-loop corrections (as described above) and 
parton-showered with HERWIG, to these events generated again by 
HORACE, but with only the Born-level calculation, and showered with HERWIG+PHOTOS. 
The results are shown in Figs.~\ref{fig:horace_zmass}--\ref{fig:horace_gammaII_pt}. In addition, the total production cross-sections of
 $Z \to \ell^{+}\ell^{-}$ with and without a mass cut around the $Z$ peak and kinematic
acceptance cuts are provided in Table~\ref{table:xs_comp}. 

\begin{figure}[htb]
\begin{center}
\resizebox{!}{8cm}{\includegraphics{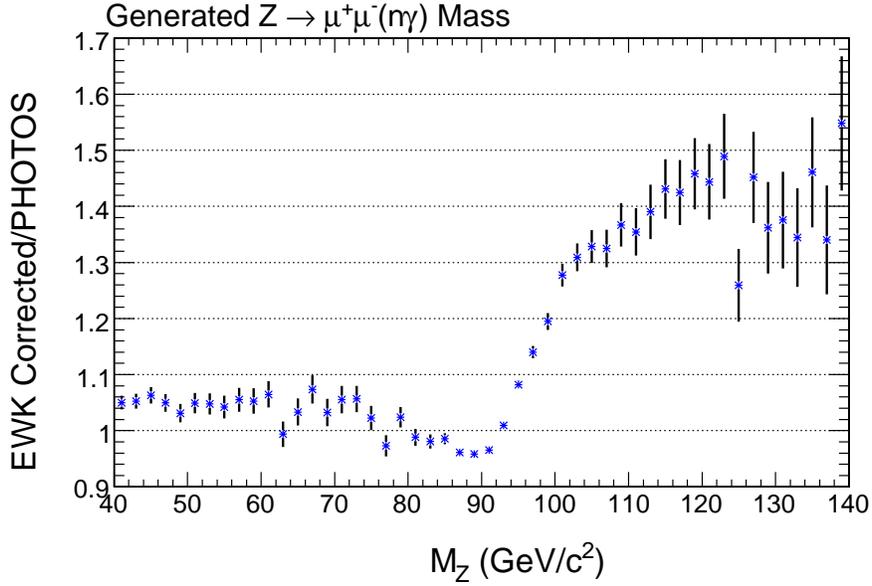}}
\caption{ Ratio of HORACE $Z/\gamma^* \to \ell^{+}\ell^{-}(n\gamma)$
  differential cross-section with full EWK corrections, to HORACE with
  PHOTOS corrections, for the generated $Z$ mass. In this case PHOTOS
  corrections do not contribute. }
\label{fig:horace_ratio_zmass}
\end{center}
\end{figure}

\begin{figure}[htb]
\begin{center}
\resizebox{!}{8cm}{\includegraphics{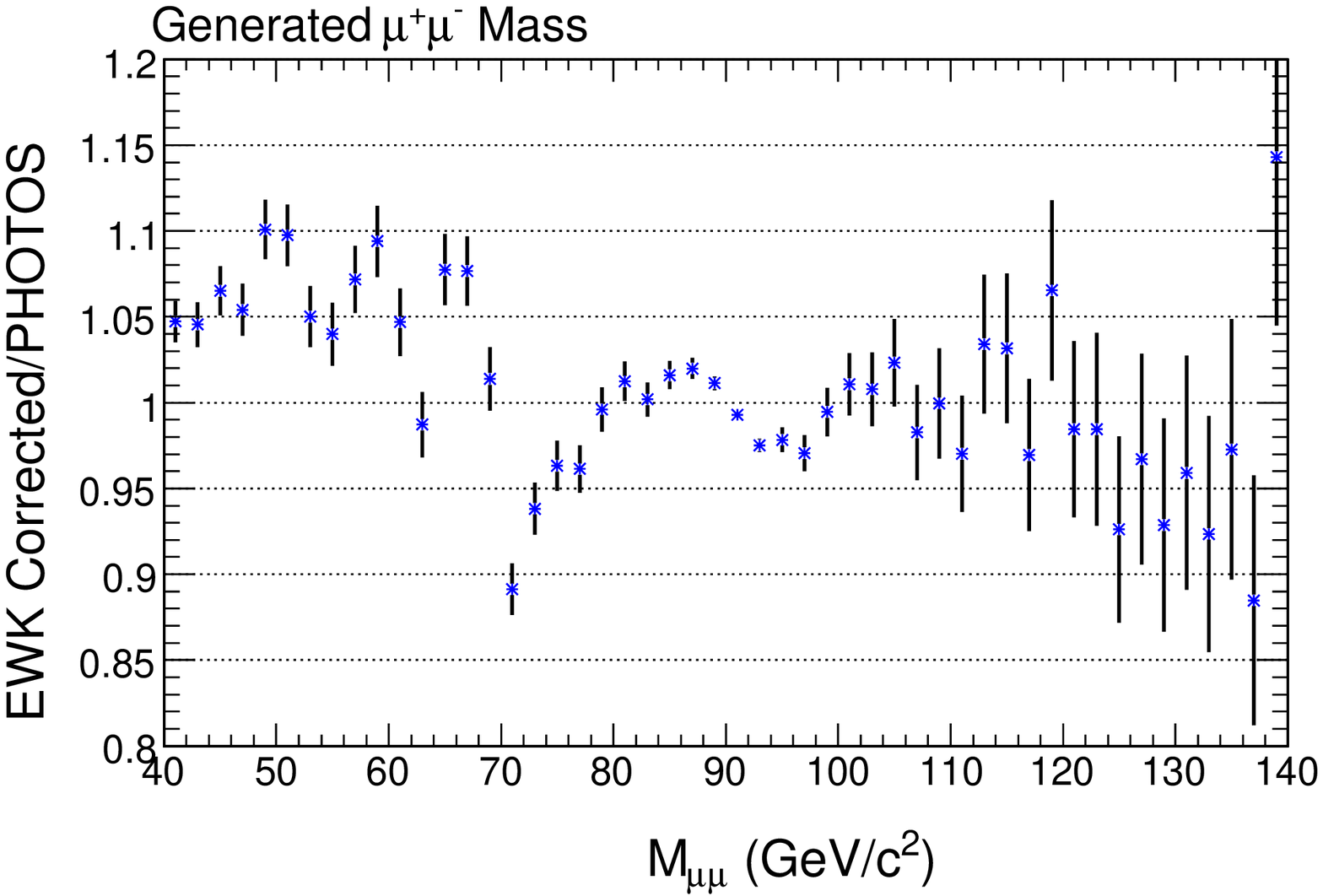}}
\caption{ Ratio of HORACE $Z/\gamma^* \to \ell^{+}\ell^{-}(n\gamma)$
  differential cross-section with full EWK corrections, to HORACE with
  PHOTOS corrections, for the generated $\mu^+\mu^-$ invariant mass after parton and
  QED showering.}
\label{fig:horace_ratio_mumumass}
\end{center}
\end{figure}

The histograms of the $Z$ boson distributions
(Figs.~\ref{fig:horace_zmass}--\ref{fig:horace_zy}) show that the HORACE Born-level calculation and Born-level with PHOTOS
FSR are the same. This is expected, since PHOTOS does not modify the
properties of the parent $Z$. The higher order calculation gives a visible difference
in cross-section for $M_Z > 100$ GeV/$c^2$, as is shown in
Fig.~\ref{fig:horace_ratio_zmass}. For the invariant mass of the
lepton pair (in Fig.~\ref{fig:horace_mass} we show this for muons),
however, the two calculations agree nicely. The much better agreement
(from the PHOTOS corrections) is highlighted in Fig.~\ref{fig:horace_ratio_mumumass}.
Similarly there is good
shape agreement for the other
lepton kinematic quantities shown in Figs.~\ref{fig:horace_mu_eta}
and~\ref{fig:horace_mu_pt}. In terms of the acceptance, this agreement is
quantitatively demonstrated to be better than 1\%, as shown in
Table~\ref{table:xs_comp}. A reasonable agreement in the number of FSR
photons emitted, and their transverse momentum spectra, between PHOTOS
and HORACE is also shown in Figs.~\ref{fig:horace_ngamma}--\ref{fig:horace_gammaII_pt}.    

We conclude that the errors due to not including the complete electroweak
one-loop corrections are below the $1\%$ in the region of the $Z$
peak as far as integrated cross sections are considered.


\begin{table}[ht]
\begin{center}
\begin{tabular}{|l|c|c|c|}
\multicolumn{3}{c}{$\Zll$ Production Cross-Section}\\
\hline
 & $\sigma$(No PS) & $\sigma$(Cuts Loose)  & $\sigma$(Cuts Tight) \\
\hline
HORACE Born & $1984.2 \pm 2.0 $& $1984.2 \pm 2.0$ & $612.5 \pm 1.1$  \\
HORACE Born+PHOTOS & $1984.2 \pm 2.0$ & $1964.6 \pm 2.0$ & $597.6 \pm 1.1$  \\
HORACE EWK Corr. & $1995.7 \pm 2.0$ & $1961.4 \pm 2.0$ & $595.3 \pm 1.1$  \\
\hline
Error & 0.58 $\pm$ 0.14\% & 0.16 $\pm$ 0.14\% & 0.38 $\pm$ 0.26 \%  \\
\hline
\end{tabular}
\end{center}
\caption{Calculation of the $Z/\gamma^* \to \ell^{+}\ell^{-}$
  cross-section at various orders of electroweak corrections using
  HORACE 3.1~\cite{HORACE1,HORACE2,HORACE3,HORACE4}. The first column
  gives the generator level cross-section with no QCD parton showering
  (No PS). This cross-section is the same for the Born calculation, and the Born
  calculation with PHOTOS corrections, since PHOTOS does not modify the
  inital cross-section.  
  The PDF calculations are from CTEQ6.5M and the loose cut region is defined as
  $M_{\ell\ell} > 40$ GeV/$c^2$, $p_{T}^{\ell} > 5$ GeV/$c$, and
  $|\eta_{\ell}| < 50.0$, while the tight cut region is defined as
  $40 < M_{\ell\ell} < 140$ GeV/$c^2$, $p_{T}^{\ell} > 20$ GeV/$c$, and
  $|\eta_{\ell}| < 2.0$. In the first column we show the total
  generator-level cross-section before parton showering. The events are
  generated in the kinematic region defined by   $M_{Z} > 40$
  GeV/$c^2$, $p_{T}^{\ell} > 5$ GeV/$c$, and $|\eta_{\ell}| < 50.0$. }
\label{table:xs_comp}
\end{table}

\subsection*{Acknowledgments}
We thank the organizers and conveners of the Les Houches workshop where this work originated. We also thank Fulvio Piccinni for useful discussions. 
This work was also supported in part by USDOE grant DE-FG02-91ER40671

%% file: s_pozzorini/ppwj_main.tex
\subsection{Introduction}
\label{ppwj_sect1}
At the LHC, electroweak gauge bosons can recoil against hard jets
reaching very high transverse momenta, up to \mbox{2 TeV} or even beyond.  
These reactions represent an important background for new-physics searches.
Moreover they can be used to determine 
the parton distribution functions or to measure $\alpha_\rS$
at the TeV scale.
In this kinematic region,
the electroweak corrections are strongly
enhanced by  Sudakov logarithms of the form $\ln(\shat/M_W^2)$
and may amount to tens of percent at one loop and several percent at two
loops.\footnote{For a recent survey of the literature on electroweak
  Sudakov logarithms and their impact at the LHC 
see \mbox{Refs.~\cite{Denner:2006jr,Kuhn:2007cv}}.}
The electroweak corrections to  
$p p \to Zj $ and $p p \to \gamma j$  were studied in 
\mbox{Refs.~\cite{Maina:2004rb,Kuhn:2004em,Kuhn:2005az,Kuhn:2005gv}}.
The electroweak corrections to $p p \to Wj $ have been recently
completed  by two groups \cite{Kuhn:2007qc,Kuhn:2007cv,Hollik:2007sq}.
Besides the full set of quark- and gluon-induced $\ord(\alpha)$
reactions, these two calculations include different additional
contributions that turn out to be important at high transverse momenta: two-loop
Sudakov logarithms \cite{Kuhn:2007qc,Kuhn:2007cv} and photon-induced
processes \cite{Hollik:2007sq}.
We also observe that, while the calculation of  \mbox{Ref.~\cite{Hollik:2007sq}} is
completely inclusive with respect to photon emission, the definition of
the $Wj$ cross section adopted in \mbox{Refs.~\cite{Kuhn:2007qc,Kuhn:2007cv}} is
more exclusive: $W\gamma$ final states are rejected 
requiring that the final-state jet has a minimum transverse 
momentum.
However, the numerical results indicate that
this difference in the definition of the observable 
has a quite small impact on the size of the
corrections.
In the following we present the results of
\mbox{Refs.~\cite{Kuhn:2007qc,Kuhn:2007cv}}.  In Sect.~\ref{ppwj:sect2} we define
the exclusive $pp\to Wj$ cross section and discuss the treatment of
final-state collinear singularities using quark fragmentation
functions.
Compact analytic formulae for 
the high-energy behaviour of the one- and two-loop 
virtual corrections 
are presented in Sect.~\ref{ppwj:sect3}.
Real-photon bremsstrahlung is briefly discussed in 
Sect.~\ref{ppwj:sect4} and the numerical results are 
given in Sect.~\ref{ppwj:sect5}.
For a discussion of QCD corrections we refer to \mbox{Refs.~\cite{Ellis:1981hk,Arnold:1988dp,Gonsalves:1989ar,Giele:1993dj,Campbell:2003hd}}.

\subsection{Observable definition}
\label{ppwj:sect2}

The hadronic reaction $pp\to W^\pm j(\gamma)$ receives contributions from 
various partonic subprocesses of the type
$\bar q q'\to W^\pm g(\gamma)$, $g q\to W^\pm q'(\gamma)$, and $\bar q g\to W^\pm \bar q'(\gamma)$.
Details concerning the implementation of PDFs and quark-mixing effects
can be found in \mbox{Ref.~\cite{Kuhn:2007cv}}. In the following 
we focus on the transverse momentum ($\pT$) distribution\footnote{Summing and averaging over colour and polarization is implicitly understood.}
for a generic partonic subprocess  $a b \to W^\pm k (\gamma)$,
\beqar
\frac{\rd \hat{\sigma}^{a b \to W^\pm k (\gamma)} 
}{\rd \pT}
&=& 
\frac{1}{2 \shat}
\left[
\int \rd  \Phi_2\, 
|\M^{a b \to W^\pm k} |^2 \, F_{\mathrm{O}, 2}(\Phi_2)
+
\int \rd  \Phi_3\, 
|\M^{a b \to W^\pm k \gamma} |^2 \, F_{\mathrm{O}, 3}(\Phi_3)
\right]
.
\eeqar
Here  $\rd \Phi_N$ and $F_{\mathrm{O}, N}(\Phi_N)$ denote the phase-space measure 
and the observable function in the $N$-particle final-state phase space.
The soft and collinear divergences arizing from virtual and real photons
need to be extracted in analytic form and,
after factorization of initial-state collinear singularities,
the singular parts of virtual and real corrections must 
cancel. Since we are interested in 
$W$-boson production in association with a hard jet, 
we define
\beq
F_{\mathrm{O}, N}(\Phi_N)= \delta (\pT - \pTW) \theta (p_{\rT,k} -\pTminj),
\eeq
requiring  a minimum transverse momentum
$\pTminj$ for the final-state parton  $k=g,q,\bar q$.
This observable is free from singularities associated with 
soft and collinear QCD partons. 
However, for partonic channels involving final-state quarks
 (or anti-quarks),
the cut on $p_{\rT,q}$ restricts the emission of collinear photons
off quarks and gives rise to  collinear singularities.
These singularities can be factorized into 
quark fragmentation functions
\cite{Glover:1993xc,Buskulic:1995au}.
Let us consider the  quark-photon collinear region,
\beq\label{ppwj_eq:Rsep}
R_{q\gamma}=\sqrt{(\eta_q-\eta_\gamma)^2+(\phi_q-\phi_\gamma)^2}< R_{\mathrm{sep}},
\eeq  
where the rapidity and azimuthal-angle
separation  
between photon and quark becomes small.
In practice one can split the 3-particle phase space according to
$F_{\mathrm{O}, 3}(\Phi_3)
=
F^{\mathrm{rec}}_{\mathrm{O}, 3}(\Phi_3)-
\Delta F_{\mathrm{O}, 3}(\Phi_3)
$, where in
\beqar
F^{\mathrm{rec}}_{\mathrm{O}, 3}(\Phi_3)= \delta (\pT - \pTW) 
\left[
\theta (R_{q\gamma}-R_{\mathrm{sep}}) \theta (p_{\rT,q}-\pTminj)
+\theta (R_{\mathrm{sep}}-R_{q\gamma})\right]
\eeqar 
the $p_{\rT,q}$-cut is imposed only outside 
the collinear region.
This contribution  is collinear safe and
corresponds to the case where collinear photon-quark pairs 
with $R_{q\gamma}<R_{\mathrm{sep}}$ are recombined.
The remainder,
\beqar
\Delta F_{\mathrm{O}, 3}(\Phi_3)= \delta (\pT - \pTW) \theta (R_{\mathrm{sep}}-R_{q\gamma}) \theta (\pTminj-p_{\rT,q}),
\eeqar 
describes the effect of the $p_{\rT,q}$-cut inside  the collinear region.
This contribution can
be described by means of 
quark fragmentation functions ${\mathcal{D}_{q \gamma}}(z)$
as\footnote{For a detailed discussion we refer to App.~A of \mbox{Ref.~\cite{Kuhn:2007cv}}.} 
\beqar\label{ppwj_eq:fragm}
\frac{1}{2 \shat}\int \rd  \Phi_3\, 
|\M^{q' g \to W^\pm q \gamma} |^2 \, \Delta F_{\mathrm{O}, 3}(\Phi_3)
=
\frac{\rd \hat{\sigma}^{q'g \to W^\pm   q} 
}{\rd \pT}\,
\int_{z_\mathrm{min}}^1 \rd z\, 
{\mathcal{D}_{q \gamma}}(z)
,
\eeqar
where 
$z=
p_{\rT,\gamma}/p_{\rT,W}
$
and
$z_{\mathrm{min}}=1-\pTminj/p_{\rT,W}$.
The collinear singularities were factorized into 
the fragmentation function, and 
using a parametrization 
derived from measurements 
of isolated hard photons in hadronic $Z$ decays
\cite{Buskulic:1995au}
we obtained
${\mathcal D}_{q \gamma}(z)
=
\frac{\alpha Q_q^2}{2\pi}[
P_{q\gamma}(z)
\ln\left({z R_\mathrm{sep}p_{\rT,W}}/{0.14\,\GeV}\right)^2
+z-13.26]
$.
For  $R_\mathrm{sep}\lsim\ord(1)$ and a wide range of transverse momenta, $2 \pTminj\le p_{\rT,W} \le 2\TeV $, we found that 
the
$\Delta F_{\mathrm{O}, 3}$-contribution 
(\ref{ppwj_eq:fragm})
does not exceed two permille 
of the cross section.
Therefore 
we could safely neglect this contribution and 
perform the calculation
using 
$F_{\mathrm{O}, 3}(\Phi_3)\simeq F^{\mathrm{rec}}_{\mathrm{O}, 3}(\Phi_3)$
for final-state (anti-) quarks.
We also checked that this approximation is very stable against variations
of $R_\mathrm{sep}$ \cite{Kuhn:2007cv}.

\subsection{Virtual corrections}
\label{ppwj:sect3}
The electroweak couplings were renormalized in the
$G_\mu$-scheme, where $\alpha=\sqrt{2}\,G_\mu M_W^2\sw^2/\pi$ and
$\sw^2=1-\cw^2=1-M_W^2/M_Z^2$.
For transverse momenta of $\ord(100 \GeV)$ or beyond, the virtual
corrections are dominated by logarithms of
the type $\ln(\shat/M_W^2)$.  In addition, the virtual corrections
involve divergent logarithms of electromagnetic origin.
The logarithms resulting from
photons with virtuality smaller than $M_W$ have been subtracted 
from the virtual corrections and combined with real-photon emission.
As a result,
the (subtracted) virtual and real corrections 
are free from large logarithms involving light-fermion masses,
and the bulk of the electroweak effects
is isolated in the virtual part (see Sect.~\ref{ppwj:sect5}).
At one loop, the double and single electroweak logarithms (NLL approximation)
can be derived from the general results of \mbox{Ref.~\cite{Denner:2000jv}}.
For the $u \bar d\to W^+ g$ subprocess,
\beqar
|\M_1^{u \bar d\to W^+ g}|^2
&\NLLa& 
|\M_{0}^{u \bar d\to W^+ g}|^2
\Biggl\{
1
+\left(\frac{\alpha}{2\pi}\right)
\Biggl\{
-
\cew_{q_\rL}\left[
\ln^2\left(\frac{|\shat|}{M_W^2}\right)
-
3\ln\left(\frac{|\shat|}{M_W^2}\right)
\right]
\nl&&{}-
\frac{C_{\mathrm{A}}}{2 \sw^2}
\left[
\ln^2\left(\frac{|\that|}{M_W^2}\right)
+\ln^2\left(\frac{|\uhat|}{M_W^2}\right)
-\ln^2\left(\frac{|\shat|}{M_W^2}\right)
\right]
\Biggr\}
\Biggr\}
,
\eeqar
where 
$\shat=(p_{u}+ p_{\bar d})^2$,
$\that=(p_{u}- p_W)^2$, 
$\uhat=(p_{\bar d}-p_W)^2$,
$\cew_{q_\rL}=C_{\mathrm{F}}/\sw^2+1/(36\cw^2)$,
$C_{\mathrm{F}}=3/4$, 
$C_{\mathrm{A}}=2$ and 
$|\M_{0}^{u \bar d\to W^+ g}|^2
=
32 \pi^2  \alpha_\rS (\alpha/{\sw^2})
(\that^2+\uhat^2+2 M_W^2 \shat)/(\that\uhat)
$.
This result is easily extended 
to all relevant partonic reactions by means of CP and crossing symmetries.

The exact one-loop expression for the (subtracted) virtual corrections has the general form
\beqar
&&|\M^{u \bar d\to W^+ g}_{1}|^2 
=
\left[1+2\, \mathrm{Re}\left(\delta C^{\mathrm{A}}+\delta C^{\mathrm{N}}\right)\right]
|\M^{u \bar d\to W^+ g}_{0}|^2 
+
\frac{ 16 \pi \alpha^2 \alpha_\rS}{\sw^2} 
\mathrm{Re}\,
\Biggl\{\frac{1}{2\sw^2}
H_1^{\mathrm{X}}(M_W^2)
\\
&&{}\hspace{-4mm}+
\sum_{V=\mathrm{A,Z}}\Biggl[
\left(\frac{3\delta^{\mathrm{SU}(2)}_{VV}}{4\sw^2}+\frac{\delta^{\mathrm{U}(1)}_{VV}}{36\cw^2} \right)
H_1^{\mathrm{A}}(M^2_V)
+\frac{\delta^{\mathrm{SU}(2)}_{VV}}{2\sw^2}
\left(
2 H_1^{\mathrm{N}}(M^2_V)
-H_1^{\mathrm{X}}(M^2_V)
\right)
-\frac{X_V}{6}
H_1^{\mathrm{Y}}(M^2_V)
\Biggr]
\Biggr\},
\nonumber
\eeqar 
where $\delta^{\mathrm{SU}(2)}_{AA}=\sw^2$, $X_A=-1$,
$\delta^{\mathrm{U}(1)}_{AA}=\cw^2$, 
$\delta^{\mathrm{SU}(2)}_{ZZ}=\cw^2$, $X_Z=1$ and
$\delta^{\mathrm{U}(1)}_{ZZ}=\sw^2$.
Explicit expressions for the functions
$H_1^{\mathrm{I}}(M^2_V)$
and the counterterms  $\delta C^{\mathrm{A}}$, $\delta
C^{\mathrm{N}}$ 
can be found in  \mbox{Ref.~\cite{Kuhn:2007cv}}. 
Here we present compact NNLL 
expressions in the high-energy limit.
This approximation 
includes all terms that are not suppressed by powers of $M_W^2/\shat$.
The NNLL expansion of the loop diagrams involving massive gauge
bosons ($M_V=M_Z,M_W$) yields
\beqar
H^{\mathrm{A}}_1 (M_V^2)&=&
\frac{ \that^2+\uhat^2}{\that\uhat}\,
\Biggl\{
\Deltamsbar +\ln\left(\frac{M_Z^2}{M_V^2}\right)
-\ln^2\left(\frac{-\shat}{M_V^2}\right)
+3\ln\left(\frac{-\shat}{M_V^2}\right)
+\frac{3}{2}\Biggl[
\ln^2\left(\frac{\that}{\shat}\right)
\nl&&\hspace{-20mm}{}
+\ln^2\left(\frac{\uhat}{\shat}\right)
+\ln\left(\frac{\that}{\shat}\right)
+\ln\left(\frac{\uhat}{\shat}\right)
\Biggr]
+\frac{7\pi^2}{3}-3
\Biggr\}
+
\frac{ \that^2-\uhat^2}{2\that\uhat}
\,\Biggl\{
\ln^2\left(\frac{\that}{\shat}\right)
-
\ln^2\left(\frac{\uhat}{\shat}\right)
+3\ln\left(\frac{\uhat}{\shat}\right)
\nl&&\hspace{-20mm}{}
-3\ln\left(\frac{\that}{\shat}\right)
\Biggr\}
+2\Biggl[
\ln^2\left(\frac{\that}{\shat}\right)
+\ln^2\left(\frac{\uhat}{\shat}\right)
+\ln\left(\frac{\that}{\shat}\right)
+\ln\left(\frac{\uhat}{\shat}\right)
\Biggr]
-2\ln\left(\frac{M_V^2}{M_W^2}\right)+4{\pi^2}
,\nl
H^{\mathrm{N}}_1 (M_V^2)
&=&
\frac{ \that^2+\uhat^2}{\that\uhat}\,
\Biggl\{
2\left[\Deltamsbar +\ln\left(\frac{M_Z^2}{M_W^2}\right)+\ln\left(\frac{M_V^2}{M_W^2}\right)\right]
+\ln^2\left(\frac{-\shat}{M_V^2}\right)
-\frac{1}{2}\Biggl[
\ln^2\left(\frac{-\that}{M_V^2}\right)
\nl&&\hspace{-20mm}{}
+
\ln^2\left(\frac{-\that}{M_W^2}\right)
+
\ln^2\left(\frac{-\uhat}{M_V^2}\right)
+
\ln^2\left(\frac{-\uhat}{M_W^2}\right)
\Biggr]
+\ln^2\left(\frac{\that}{\uhat}\right)
-\frac{3}{2}\Biggl[
\ln^2\left(\frac{\that}{\shat}\right)
+\ln^2\left(\frac{\uhat}{\shat}\right)
\Biggr]
\nl&&\hspace{-20mm}{}
-\frac{20\pi^2}{9}
-\frac{2\pi}{\sqrt{3}}
+4 
\Biggr\}
+
\frac{ \that^2-\uhat^2}{2\that\uhat}
\,\Biggl\{
\ln^2\left(\frac{\uhat}{\shat}\right)
-
\ln^2\left(\frac{\that}{\shat}\right)
\Biggr\}
-2\Biggl[
\ln^2\left(\frac{\that}{\shat}\right)
+\ln^2\left(\frac{\uhat}{\shat}\right)
+\ln\left(\frac{\that}{\shat}\right)
\nl&&\hspace{-20mm}{}
+\ln\left(\frac{\uhat}{\shat}\right)
\Biggr]
+2\ln\left(\frac{M_V^2}{M_W^2}\right)-4{\pi^2}
,\nl
H^{\mathrm{Y}}_1 (M_V^2)
&=&
\frac{ \that^2+\uhat^2}{\that\uhat}\,
\Biggl\{
\ln^2\left(\frac{-\that}{M_W^2}\right)
-\ln^2\left(\frac{-\that}{M_V^2}\right)
-\ln^2\left(\frac{-\uhat}{M_W^2}\right)
+\ln^2\left(\frac{-\uhat}{M_V^2}\right)
\Biggr\}
+2\ln\left(\frac{\that}{\uhat}\right)
,\nl
H^{\mathrm{X}}_1 (M_V^2)
&=&
-2\left[
2\ln\left(\frac{-\shat}{M_V^2}\right)
+\ln\left(\frac{\that}{\shat}\right)
+\ln\left(\frac{\uhat}{\shat}\right)
-3
\right]
,
\eeqar
where 
$
\Deltamsbar = 
1/\varepsilon -\gamma_{\mathrm{E}} +\ln(4\pi)+\ln\left({\mu^2}/{M_Z^2}\right)
$.
For the loop functions associated with photons
we obtain
$H^{\mathrm{I}}_1 (M_A^2)
=
H^{\mathrm{I}}_1 (M_W^2)
+
\frac{ \that^2+\uhat^2}{\that\uhat}
K^{\mathrm{I}}$
with
$K^{\mathrm{A}}=\pi^2$ ,
$K^{\mathrm{N}}=2\pi/\sqrt{3}-7\pi^2/9$,
and
$K^{\mathrm{X}}=K^{\mathrm{Y}}=0$.
The functions describing the photonic and the $W$-boson contributions 
differ only by non-logarithmic terms,
since the logarithms from photons with virtuality smaller than $M_W$
have been subtracted.

At two loops , using the general results 
for  leading- and next-to-leading electroweak logarithms
in \mbox{Refs.~\cite{Denner:2003wi, Melles:2001ye}}
and subtracting logarithms from photons with virtuality smaller than $M_W$,
we obtain
$
|\M_2^{u \bar d\to W^+ g}|^2=
|\M_1^{u \bar d\to W^+ g}|^2+(\frac\alpha {2\pi})^2A^{(2)}
|\M_0^{u \bar d\to W^+ g}|^2
$ with
\beqar
A^{(2)}&=&
\frac{1}{2}
\left(\cew_{q_\rL}+\frac{C_{\mathrm{A}}}{2\sw^2}\right)
\Biggl[\cew_{q_\rL}\left[
\ln^4\left(\frac{|\shat|}{M_W^2}\right)
-6\ln^3\left(\frac{|\shat|}{M_W^2}\right)
\right]
+
\frac{C_{\mathrm{A}}}{2\sw^2}
\Bigl[
\ln^4\left(\frac{|\that|}{M_W^2}\right)
\nl&&\hspace{-14mm}{}
+\ln^4\left(\frac{|\uhat|}{M_W^2}\right)
-\ln^4\left(\frac{|\shat|}{M_W^2}\right)
\Bigr]
\Biggr]
+\frac{1}{6}
\Biggl[
\frac{b_1}{\cw^2}\left(\frac{Y_{q_\rL}}{2}\right)^2
+\frac{b_2}{\sw^2} \left(
C_{\mathrm{F}}
+\frac{C_{\mathrm{A}}}{2}\right)
\Biggr]
\ln^3\left(\frac{|\shat|}{M_W^2}\right)
,
\eeqar
where $b_1=-41/(6\cw^2)$ and $b_2=19/(6\sw^2)$.

\subsection{Real photon radiation}
\label{ppwj:sect4}
We  performed two independent calculations of real photon bremsstrahlung using the dipole subtraction method
\cite{Dittmaier:1999mb,Catani:1996vz,Catani:2002hc}.
In the first calculation, we used the subtraction method for massive
fermions \cite{Dittmaier:1999mb} regularizing soft and collinear
singularities by means of small photon and fermion masses.
In the second calculation we used massless fermions 
and we subtracted the singularities in the framework of
dimensional regularization  \cite{Catani:1996vz,Catani:2002hc}.
%
The initial-state collinear singularities were factorized in the
$\MSBAR$ scheme.  This procedure introduces a logarithmic dependence
on the QED factorization scale $\mu_{\rm QED}$, which must be
compensated by the QED evolution of the PDFs.  Since our calculation
is of LO in $\alpha_\rS$, for consistency we should use LO QCD parton
distributions including NLO QED effects.
However, such a PDF set is not available.\footnote{ The currently
  available PDFs incorporating NLO QED corrections (MRST2004QED)
  include QCD effects at the NLO.}  Thus we used a LO QCD PDF set
without QED corrections \cite{Martin:2002dr}, and we chose the value
of $\mu_{\mathrm{QED}}$ in such a way that the neglected QED effects
are small.  In \mbox{Ref.~\cite{Roth:2004ti}} it was shown that the QED
corrections to the quark distribution functions grow with
$\mu_\mathrm{QED}$ but do not exceed one percent for
$\mu_\mathrm{QED}\lsim 100\GeV$.  Thus we set
$\mu_{\mathrm{QED}}=M_W$.
Photon-induced processes were not included in our calculation.  These
contributions are parametrically suppressed by a factor $\alpha /
\alpha_{\mathrm{S}}$.  However in \mbox{Ref.~\cite{Hollik:2007sq}} it was found
that, at very large $\pT$, these 
photon-induced effects can amount to several
percent.

\subsection{Numerical results}
\label{ppwj:sect5}
The hadronic cross
section was obtained using LO MRST2001 PDFs~\cite{Martin:2002dr} at the
factorization and renormalization scale $\mu^2_{\rm QCD}=p^2_{{\rm T}}$.  For the jet we required a minimum
transverse momentum $\pTminj=100 \GeV$, and the value of the
separation parameter in (\ref{ppwj_eq:Rsep}) was set to $R_{\mathrm{sep}}=0.4$.
The input parameters are specified in \mbox{Ref.~\cite{Kuhn:2007cv}}.  
Here we present the electroweak corrections to $pp\to W^+
j$ at $\sqrt{s}=14\TeV$.  The corrections to $W^-$ production are
almost identical \cite{Kuhn:2007cv}.  
\begin{figure}
\begin{center}
\begin{minipage}{0.48\textwidth}
\includegraphics[width=1.0\textwidth]{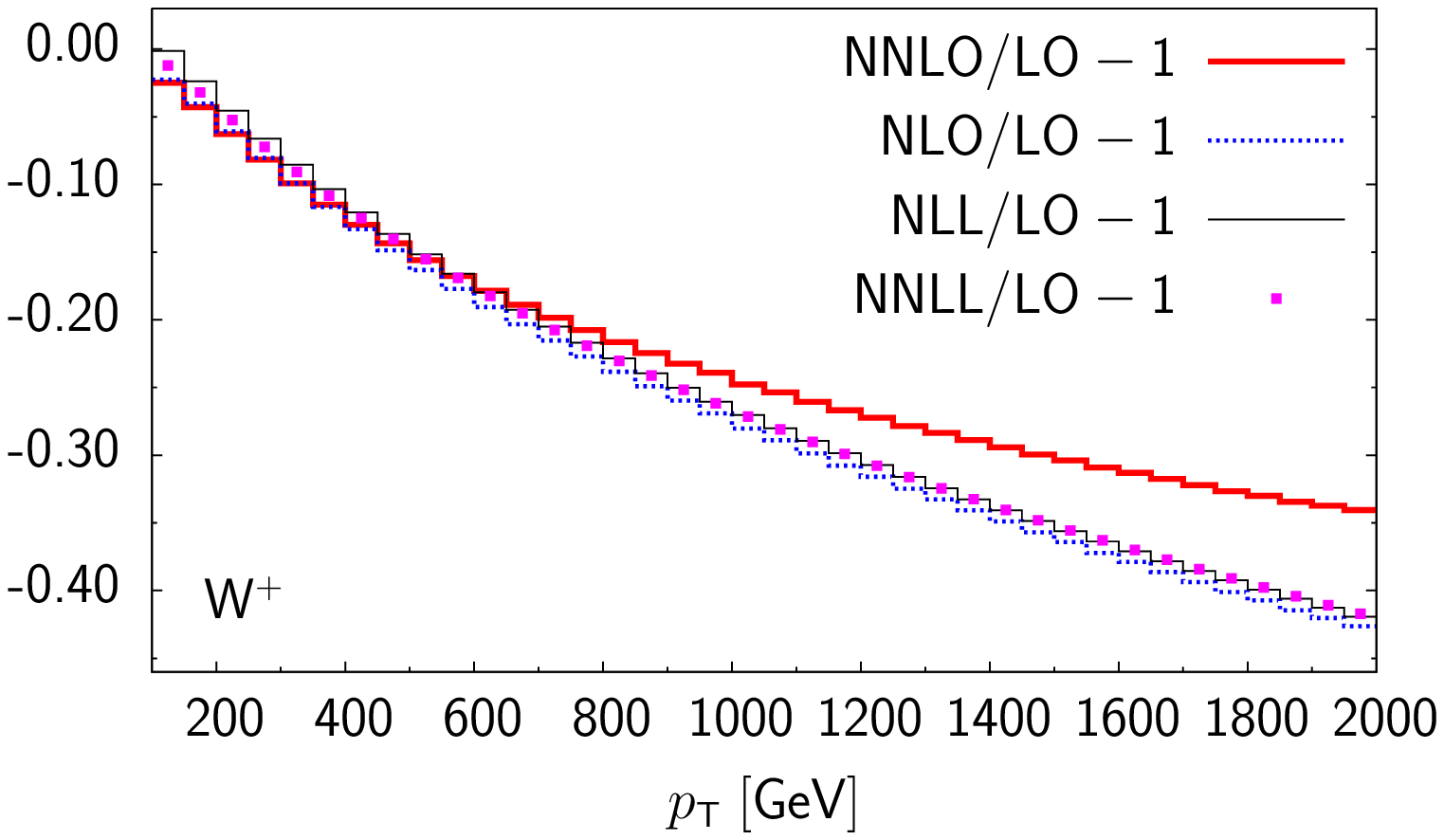}
\end{minipage}
\hfill
\begin{minipage}{0.48\textwidth}
\includegraphics[width=1.0\textwidth]{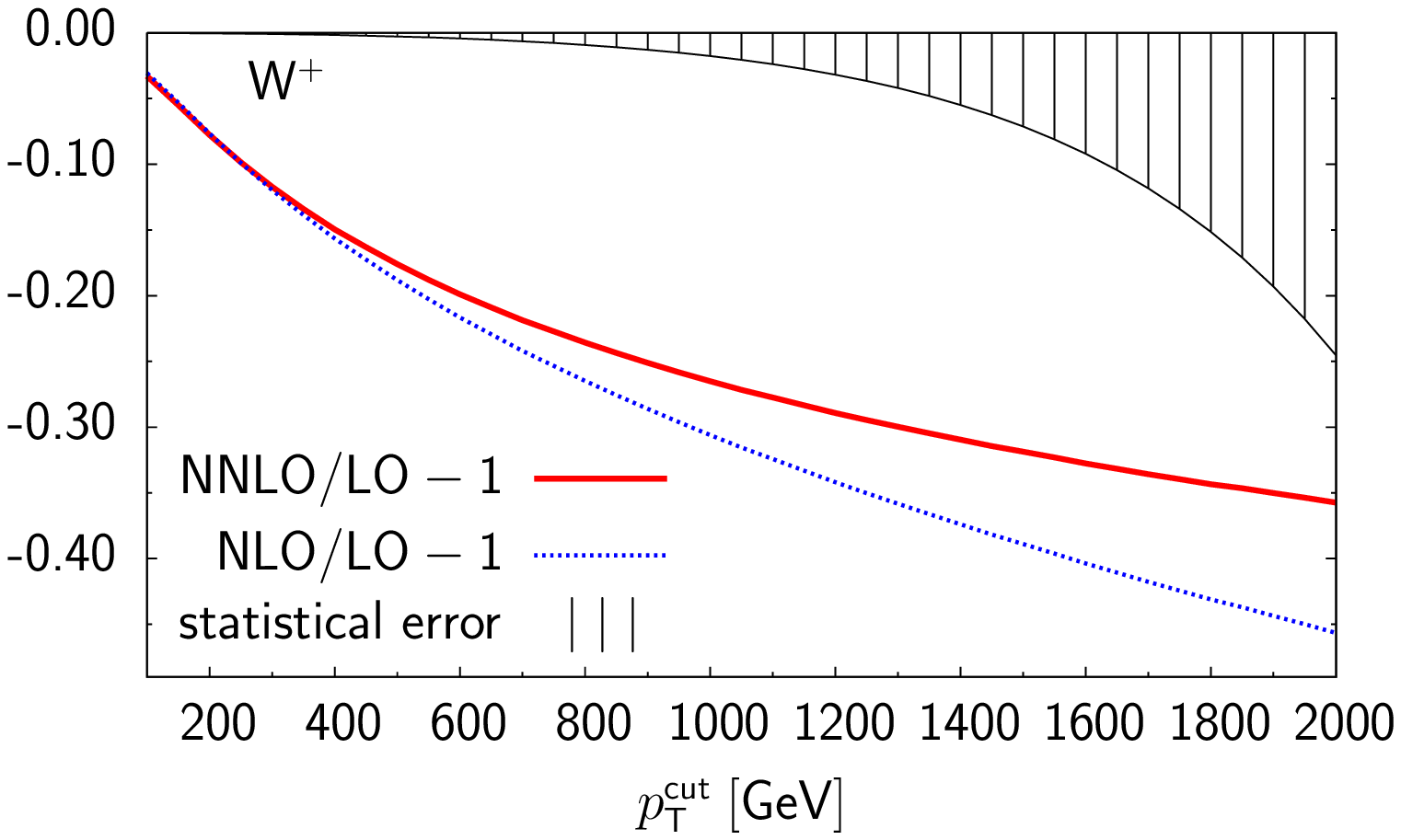}
\end{minipage}
\caption{
Electroweak correction to $pp\to W^+ j$ at
$\sqrt{s}=14\TeV$:
(a)
relative NLO (dotted), NLL (thin solid), NNLL (squares) and NNLO (thick solid)
correction \wrt the LO $\pT$-distribution;
(b)  NLO (dotted) and NNLO (solid) corrections
to the integrated cross section and estimated statistical error (shaded area).}
\label{ppwj_fig1}
\end{center}
\end{figure}
In Fig.~\ref{ppwj_fig1}a we plot the relative size of the
electroweak corrections \wrt the LO $W$-boson $\pT$-distribution.
The exact $\ord(\alpha)$ correction (NLO curve) increases
significantly with $\pT$ and ranges from  $-15\%$ at \mbox{$\pT=500\GeV$} to $-43\%$ at $\pT=2\TeV$.
This enhancement is clearly due to the Sudakov logarithms that are
present in the virtual corrections.  Indeed the one-loop NLL and NNLL
approximations, which describe the virtual part of the corrections in
the Sudakov regime, are in very good agreement with the full NLO
result.
The difference between the NLO and NNLO curves corresponds to the
two-loop Sudakov logarithms.  Their contribution is
positive and becomes significant at high $\pT$.  It amounts to $+3\%$
at $\pT=1\TeV$ and $+9\%$ at $\pT=2\TeV$. 
In  Fig.~\ref{ppwj_fig1}b we consider 
the integrated cross
section for $\pT > \pTcut$ and,
to underline the relevance of the large electroweak corrections, 
we compare 
the relative NLO and NNLO corrections 
with the statistical
accuracy at the LHC. This latter is estimated using the integrated luminosity $\mathcal{L}=300
\mathrm{fb}^{-1}$ and the branching ratio BR($W\to
e\nu_e+\mu\nu_\mu)=2/9$.  
The size of the NLO corrections is clearly
much bigger than the statistical error.  
Also the two-loop logarithmic effects are significant. 
In terms of the estimated statistical error they 
 amount to 1--3 standard deviations for $\pT$ of
$\ord(1\TeV)$.
\begin{figure}
\begin{center}
\begin{minipage}{0.48\textwidth}
\includegraphics[width=1.0\textwidth]{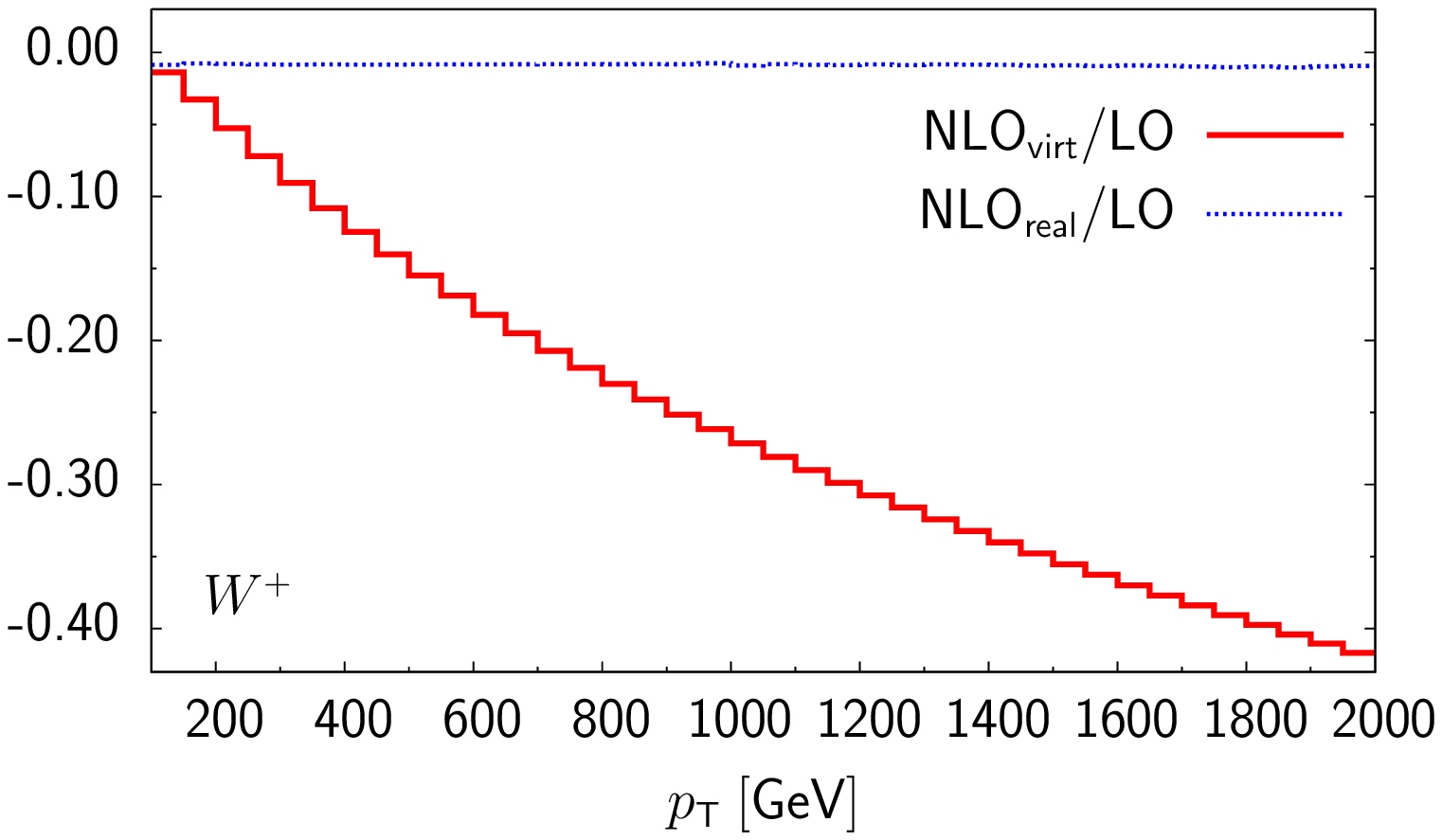}
\end{minipage}
\hfill
\begin{minipage}{0.48\textwidth}
\includegraphics[width=1.0\textwidth]{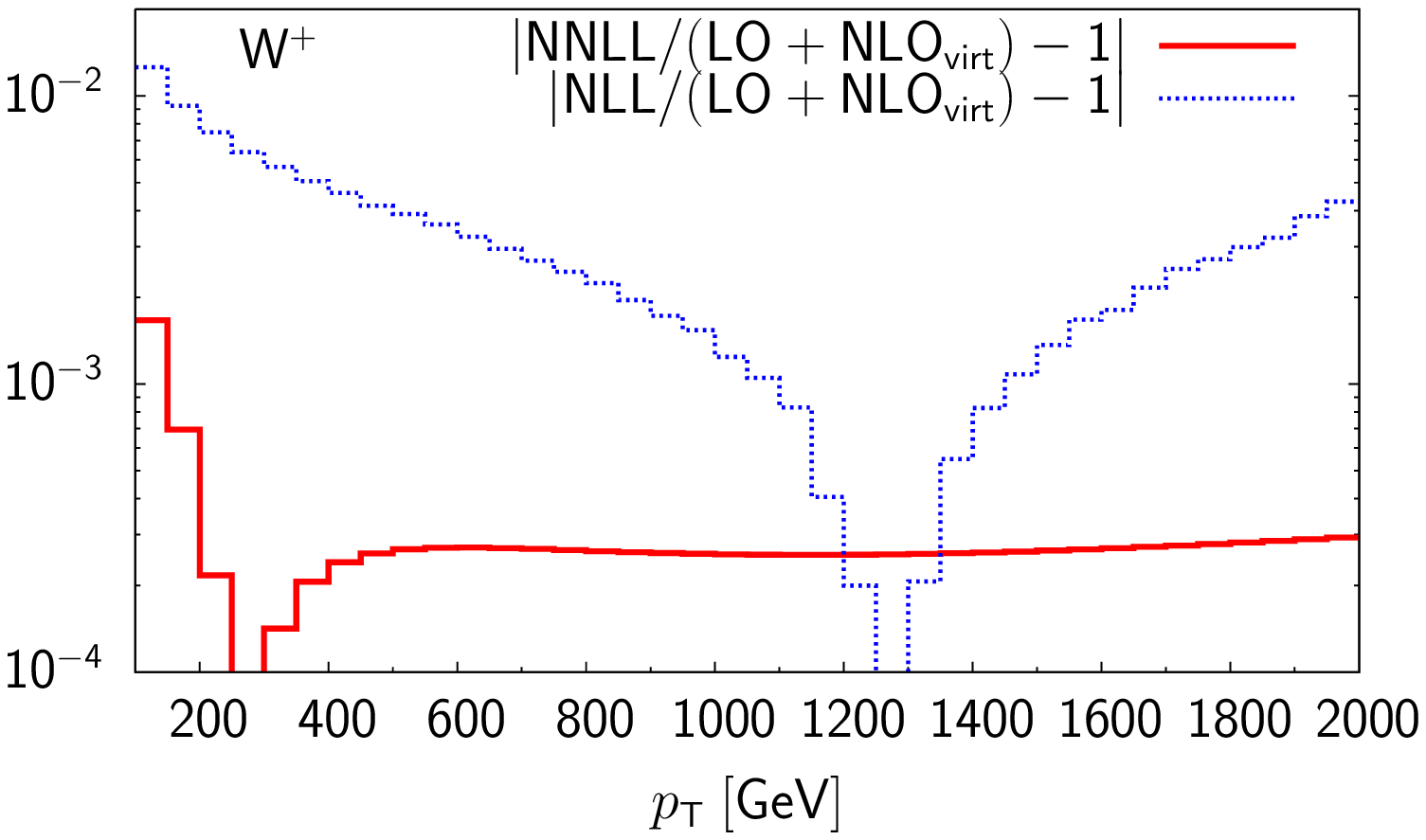}
\end{minipage}
\caption{
$\pT$-distribution of $W$ bosons in the process $pp\to W^+
  j$ at $\sqrt{s}= 14\TeV$:
(a)
relative importance of the 
virtual ($\mathrm{NLO}_{\mathrm{virt}}$)
and real ($\mathrm{NLO}_{\mathrm{real}}$) corrections;
(b)
precision of the 
 NNLL (solid) and NLL (dashed)
one-loop approximations.
} 
\label{ppwj_fig2}
\end{center}
\end{figure}
The relative importance of the virtual
($\mathrm{NLO}_{\mathrm{virt}}$) and real
($\mathrm{NLO}_{\mathrm{real}}$) contributions is shown in
Fig.~\ref{ppwj_fig2}a.  The electromagnetic logarithms have been
subtracted from the virtual part and added to the real one as
explained in Sect.~\ref{ppwj:sect3}
As a consequence, the bulk of the corrections is isolated in the
virtual part, which grows with $\pT$ and amounts up to $-42\%$ at
$\pT=2\TeV$.  In contrast, the real part represents a small and nearly
constant corrections of about $-1\%$.
In presence of additional cuts on hard photons,
$\mathrm{NLO}_{\mathrm{real}}$ becomes more negative and can amount up
to $-5\%$ for $\pT \simeq 1\TeV$ \cite{Kuhn:2007cv}.
As illustrated in Fig.~\ref{ppwj_fig2}b, the NLL and NNLL
one-loop approximations provide a very precise
description of the high-energy behaviour of the
$\mathrm{NLO}_{\mathrm{virt}}$ part.  For $\pT \ge 200 \GeV$, the
precision of the NLL and NNLL approximations is better than $10^{-2}$
and $10^{-3}$, respectively.

\subsection*{Conclusions}

We evaluated the electroweak corrections to 
large transverse momentum production of $W$ bosons at the LHC,
including the contributions from virtual and real photons.
The singularities resulting from photons with virtuality smaller than
$M_W$ have been subtracted from the virtual contributions and combined
with real-photon bremsstrahlung.  As a result, the bulk of the
electroweak effects is isolated in the virtual contributions, which 
are enhanced by Sudakov logarithms and give rise to corrections of
tens of percent at high $\pT$.
We presented compact analytic approximations
that  describe these virtual effects with high precision.
The complete $\ord(\alpha)$ corrections range between -15\% and -40\%
for $500\GeV \le \pT \le 2\TeV$.  Considering the large event rate at
the LHC, leading to a fairly good statistical precision even at
transverse momenta up to 2 TeV, we evaluated also the dominant
two-loop Sudakov logarithms.  In the high-$p_\rT$ region, these
two-loop effects increase the cross section by 5-10\% and thus become
of importance in precision studies.

\subsection*{Acknowledgements}
We would like to thank S.~Dittmaier, B.~J\"ager and P.~Uwer for
helpful discussions.


%% file: s_skands/qcdplots-houches.tex
\subsection{Introduction}
At first glance, the confined nature of both the initial and final
state implies that there are no perturbatively calculable observables
in inelastic hadron-hadron collisions. Under ordinary circumstances,
however, two powerful tools are used to circumvent this problem,
factorisation and infrared safety. The trouble with minimum-bias and
underlying-event (MB/UE) physics is that the applicability of both of
these tools is, at best, questionable for a wide range of interesting
observables.

To understand why the main perturbative tools are ineffective, let us
begin with factorisation. When applicable, factorisation allows us to
subdivide the calculation of an observable (regardless of whether it
is infrared safe or not) into a perturbatively calculable
short-distance part and a universal long-distance part, the latter of
which may be modeled and constrained by fits to data. However, in the
context of hadron collisions the oft
made separation into ``hard scattering'' and ``underlying event''
components is not necessarily equivalent to a clean separation in
terms of formation/fluctuation time, since the underlying event may
contain short-distance physics of its own. Regardless of which
definition is more correct, any breakdown of the assumed factorisation
could introduce a process-dependence of the long-distance part,
leading to an unknown systematic uncertainty in the procedure of
measuring the corrections in one process and applying them to another.

The second tool, infrared safety, provides us
with a class of observables which are insensitive to the details of
the long-distance physics. This works up to corrections of order the
long-distance scale divided by the short-distance scale,
$Q_{\mathrm{IR}}^n/Q_{\mathrm{UV}}^n$, where the power $n$ depends on the
observable in question and $Q_\mathrm{IR,UV}$ denote
generic infrared and ultraviolet scales in the problem.  Since
$Q_{\mathrm{IR}}/Q_{\mathrm{UV}}\to 0$ for large $Q_{\mathrm{UV}}$, such
observables ``decouple'' from the infrared physics as long as all
relevant scales are $\gg Q_{\mathrm{IR}}$. Infrared sensitive 
quantities, on the other hand, 
contain logarithms $\log^n(Q_{\mathrm{UV}}^2/Q_{\mathrm{IR}}^2)$ which grow
increasingly large as $Q_{\mathrm{IR}}/Q_{\mathrm{UV}}\to 0$.  
In MB/UE studies, many  of the important measured distributions are
not infrared safe in the perturbative sense. Take particle
multiplicities, for instance; in the absence of non-trivial infrared
effects, the number of partons that would be
mapped to hadrons in a na\"ive local-parton-hadron-duality
\cite{Azimov:1984np} picture depends logarithmically on the infrared
cutoff. 

We may thus classify collider observables in four categories: 
least intimidating are the factorisable infrared
safe quantities, such as the $R$ ratio in $e^+e^-$ annihilation, 
which are only problematic at low scales (where the
above-mentioned power corrections can  be large).  
Then come the factorisable infrared sensitive quantities, with the
long-distance part parametrised by process-independent non-perturbative
functions, such as parton distributions. Somewhat nastier are
non-factorised infrared safe observables. An example could here be the
energy flow into one of Rick Field's ``transverse regions''
\cite{Field:2005sa}. The
energy flow is nominally infrared safe, but in these regions where
bremsstrahlung is suppressed there can be large contributions from
pairwise balancing minijets which are correlated to the hard
scattering and hence do not factorise according to at
least one of the definitions outlined above (see also
\cite{Korotkikh:2004bz,Treleani:2007gi}).  
The nastiest beasts by all accounts are non-factorised infrared
sensitive quantities, such as the particle multiplicity in the
transverse region. 

The trouble, then, is that MB/UE physics is full of 
distributions of the very nastiest kinds
imaginable. Phenomenologically, the implication is that the
theoretical treatment of non-factorised and non-perturbative effects
becomes more important and the interpretation of experimental
distributions correspondingly more involved. The problem may also be
turned around, noting that MB/UE offers an ideal lab for studying
these theoretically poorly understood phenomena; the most
interesting observables and cuts, then, are those which minimise the 
``backgrounds'' from better-known physics.

As part of the effort to spur more interplay between theorists and
experimentalists in this field, we here present a collection of simple
min-bias distributions that carry interesting and complementary information
about the underlying physics, both perturbative and non-perturbative. The main
point is that, while each plot represents a complicated cocktail of physics
effects, such that most models could probably be tuned to give an acceptable
description observable by observable, it is very difficult to simultaneously
describe the entire set. It should therefore be possible to carry out
systematic physics studies beyond simple tunings. For brevity, this text only
includes a representative selection, with more results available on the web
\cite{lhplots}. Note also that we have here left out several important
ingredients which are touched on elsewhere in these proceedings, such as
observables involving explicit jet reconstruction and observables in
leading-jet, dijet, jet + photon, and Drell-Yan events. See also the
underlying-event sections in the HERA-and-the-LHC \cite{Alekhin:2005dx} and
Tevatron-for-LHC \cite{Albrow:2006rt} writeups.

\subsection{Models}

We have chosen to consider a set of 
six different tunes of the \textsc{Pythia} event generator
\cite{Sjostrand:2006za}, called A,
DW, and DWT \cite{Field:2005sa,Albrow:2006rt}, 
S0 and S0A \cite{Skands:2007zg}, and ATLAS-DC2 / Rome 
\cite{atlas}. For min-bias, all of these start from
leading order QCD $2\to2$ matrix elements, augmented by initial- and
final-state showers (ISR and FSR, respectively) 
and perturbative multiple parton interactions (MPI)
\cite{Sjostrand:1987su,Sjostrand:2004ef},
folded with CTEQ5L parton distributions \cite{Lai:1999wy} 
on the initial-state side and
the Lund string fragmentation model \cite{Andersson:1998tv} 
on the final-state side. In addition, the initial state is 
characterised by a transverse mass distribution roughly representing
the degree of lumpiness in the proton\footnote{Note that the
  impact-parameter dependence is still assumed factorised from the $x$
  dependence in these models, $f(x,b)=f(x)g(b)$, where $b$ denotes
  impact parameter, a simplifying 
assumption that by no means should be treated
  as inviolate, see e.g.\ \cite{Renner:2005sm,Hagler:2007xi,Treleani:2007gi}.} 
and by correlated
multi-parton densities 
derived from the standard ones by imposing elementary sum rules such
as momentum conservation \cite{Sjostrand:1987su} and flavour conservation
\cite{Sjostrand:2004pf}. 
The final state, likewise, is subject to several 
effects unique to hadronic collisions, such as the treatment of beam
remnants (e.g., affecting the flow of baryon number)
and colour (re-)connection effects between the
MPI final states
\cite{Sjostrand:1987su,Sandhoff:2005jh,Skands:2007zg}. 

\begin{table}
\begin{center}
\begin{tabular}{lcccccc}
     & Showers & MPI $\ensuremath{p_\perp}$ cutoff
at  & FS Colour   & Shower   & Proton &
  Tevatron\\
Model& off MPI &$1.96\to14\TeVx$& Correlations& Ordering &
  Lumpiness& Constraints\\
\hline
A  &      No & $2.04\stackrel{\mathrm{fast}}{\to}3.34$ & Strong& $Q^2$  & More & MB, UE\\
DW &      No & $1.94\stackrel{\mathrm{fast}}{\to}3.17$ & Strong& $Q^2$ & More & MB, UE, DY\\
DWT &     No & $1.94\stackrel{\mathrm{slow}}{\to}2.66$ & Strong& $Q^2$ & More & MB, UE, DY\\
S0 &      Yes& $1.88\stackrel{\mathrm{slow}}{\to}2.57$ & Strong& $\ensuremath{p_\perp}^2$& Less & MB, DY\\
S0A &     Yes& $1.89\stackrel{\mathrm{fast}}{\to}3.09$ & Strong& $\ensuremath{p_\perp}^2$& Less & MB, DY\\
ATLAS&No & $2.00\stackrel{\mathrm{slow}}{\to}2.75$ & Weak  & $Q^2$  & More & UE\\
\hline
\end{tabular}
\caption{Brief overview of models. Note that the IR cutoff in these
  models is not imposed as a step function, but rather as a smooth
  dampening, see \cite{Sjostrand:1987su,Sjostrand:2004ef}. The labels
  $\stackrel{\mathrm{fast}}{\to}$ and $\stackrel{\mathrm{slow}}{\to}$ refer
  to the pace of the scaling of the cutoff with collider energy.
  \label{tab:models}}
\end{center}\vspace*{-3mm}
\end{table}
Although not perfectly orthogonal in ``model space'',
these tunes are still reasonably complementary on a number of important
points, as illustrated in tab.~\ref{tab:models}. 
Column by column in tab.~\ref{tab:models}, 
these differences are as follows: 1) showers off the
  MPI are only included in S0(A). 2)
  the MPI infrared cutoff scale evolves faster 
  with collision energy in tunes A, DW, and S0A than in S0 and
  DWT. 3) all models except the ATLAS tune have very strong
  final-state colour correlations.
  4) tunes A, DW(T), and ATLAS use $Q^2$-ordered showers and the old
  MPI framework, whereas tunes S0(A) use the new
  interleaved \ensuremath{p_\perp}-ordered model. 5) tunes A and DW(T) have 
  transverse mass distributions which are significantly more peaked
  than Gaussians, with ATLAS following close behind, and S0(A) having
  the smoothest distribution. 6) the models were
  tuned to describe one or more of min-bias (MB), underlying-event (UE), and/or
  Drell-Yan (DY) data at the Tevatron.

Tunes DW and DWT only differ in the energy extrapolation away from the
Tevatron and hence are only shown separately at the LHC. Likewise for
S0 and S0A.  We regret not including a comparison to other MB/UE Monte
Carlo generators, but note that the S0(A) models are very similar to
\textsc{Pythia} 8 
\cite{Sjostrand:2007gs}, apart from the colour (re-)connection
model and some subtleties connected with the parton shower, and
that the \textsc{Sherpa} \cite{Gleisberg:2003xi} model closely resembles the
$Q^2$-ordered models considered here, with the addition of showers off
the MPI. The \textsc{Jimmy} add-on to \textsc{Herwig}
\cite{Butterworth:1996zw,Corcella:2000bw} is 
currently only applicable to underlying-event and not to min-bias.

\subsection{Results}
\begin{figure}[t!]
\plotrow{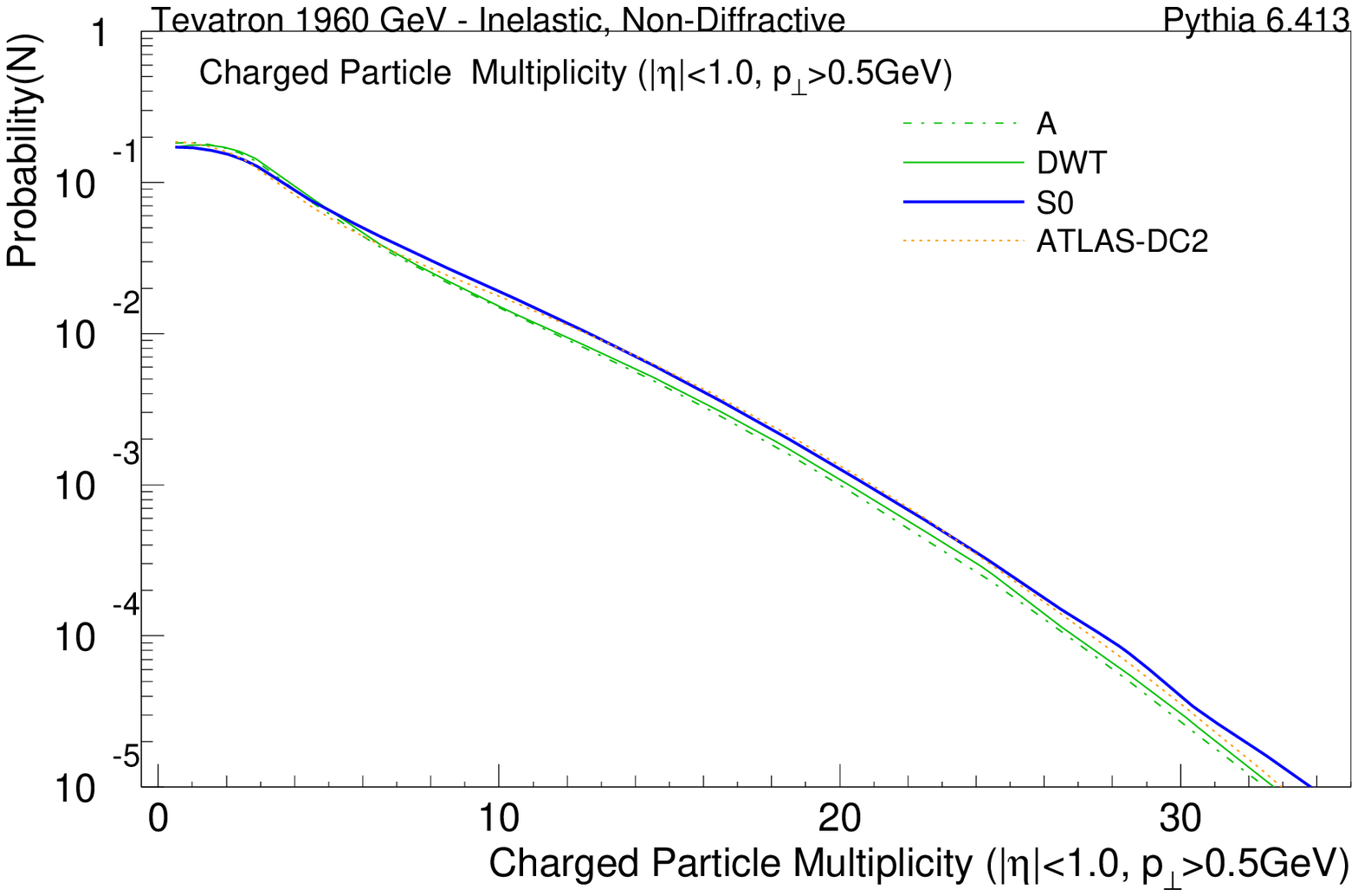}{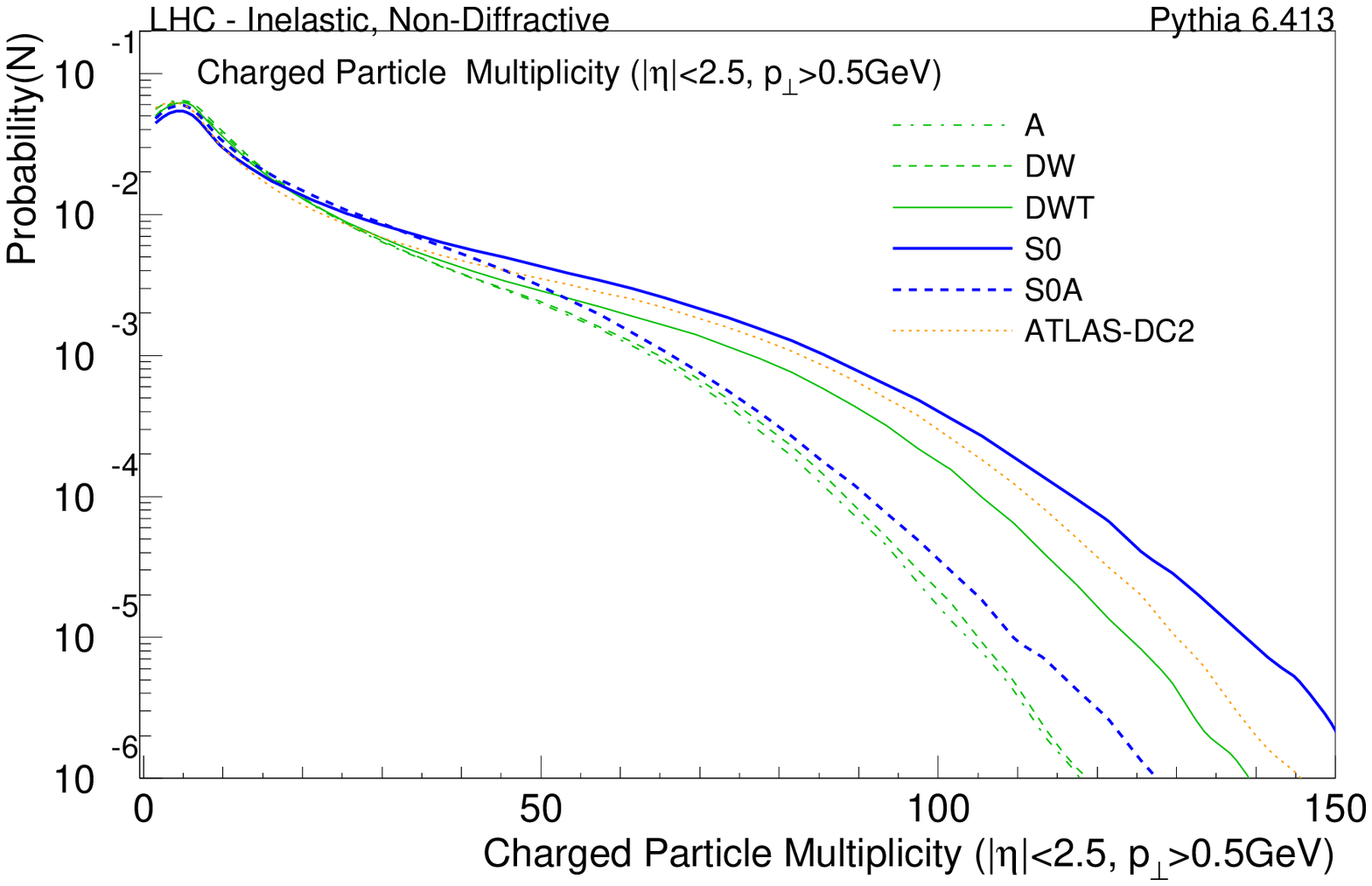}\vspace*{-3mm}
\plotrow{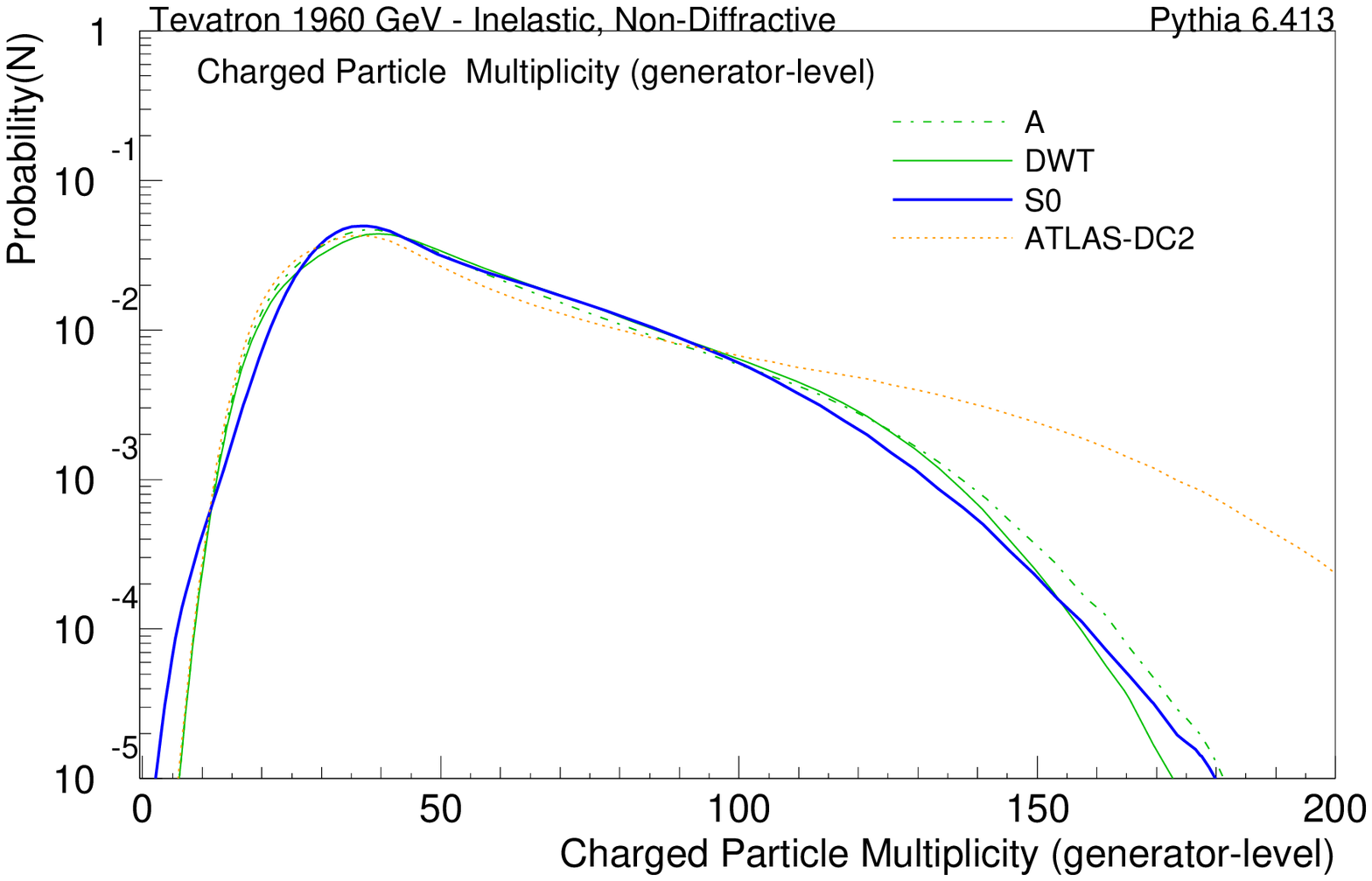}{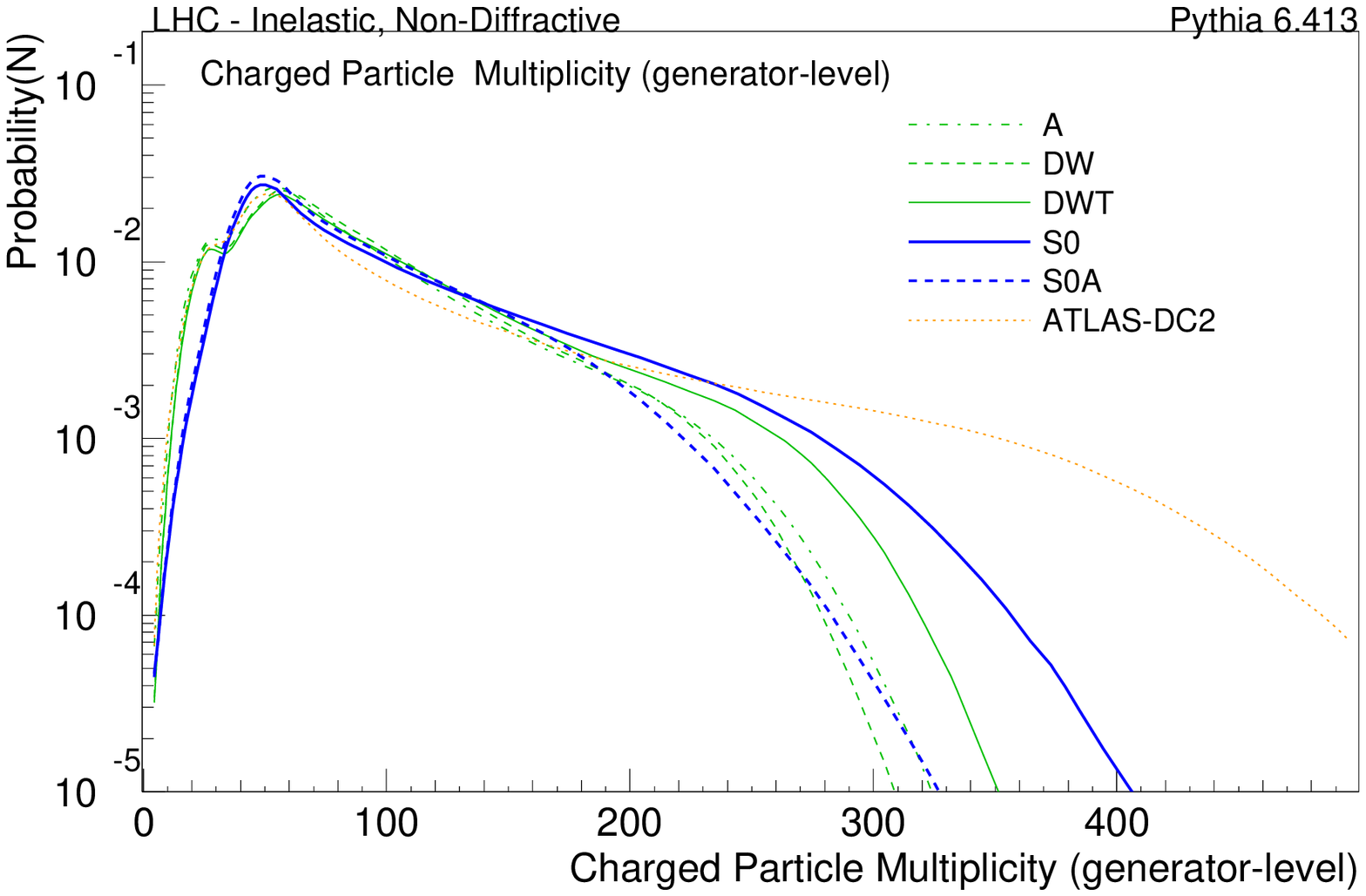}\vspace*{5mm}
\caption{Charged particle multiplicity distributions, at fiducial
  (top) and generator (bottom) levels, for the Tevatron (left) and LHC
  (right). The fiducial averages range from $3.3 <
  \left<N_{\mathrm{ch}}\right> < 3.6$ at the Tevatron to $13.0 <
  \left<N_{\mathrm{ch}}\right> < 19.3$ at the LHC.\label{fig:nch}}
\end{figure}

In this section we focus on the following distributions 
for inelastic non-diffractive events at the 
Tevatron and LHC:
charged particle multiplicity $P(N_{\mathrm{ch}})$, $dN_\mathrm{ch}/d\ensuremath{p_\perp}$,
$dN_{\mathrm{ch}}/d\eta$, the average $\ensuremath{p_\perp}$ vs.\ $N_\mathrm{ch}$
correlation, the forward-backward $N_\mathrm{ch}$ and $E_\perp$ correlations 
vs.\ $\eta$, as well as a few plots of theoretical interest showing the
multiplicity distribution of multiple interactions $P(N_\mathrm{int})$. 
On most of the plots we include the effects of fiducial
cuts, which are represented by the cuts $\ensuremath{p_\perp}>0.5\GeVx$ and $|\eta|<1.0$
($|\eta|<2.5$) at the Tevatron (LHC). 

The charged particle 
multiplicity is shown in fig.~\ref{fig:nch}, both including
fiducial cuts (top row) and at generator-level (bottom row). 
Tevatron results are shown to the
left and LHC ones to the right. Given the amount of tuning that went
into all of these models, it is not surprising that there 
is general agreement on the charged 
track multiplicity in the fiducial region at the Tevatron (top left
plot). In the top right plot, however, it is clear that this near-degeneracy
is broken at the LHC, due to the different energy extrapolations, 
and hence even a small amount of data on the charged track multiplicity will
yield important constraints. The bottom row of plots shows how things
look at the generator-level, i.e., without fiducial cuts. 
An important difference between the ATLAS tune and the other models
emerges. The ATLAS tune has a significantly higher component of
unobserved charged multiplicity. This highlights the fact that 
extrapolations from the measured distribution to the generator-level
one are model-dependent. 

\begin{figure}[t!]
\plotrow{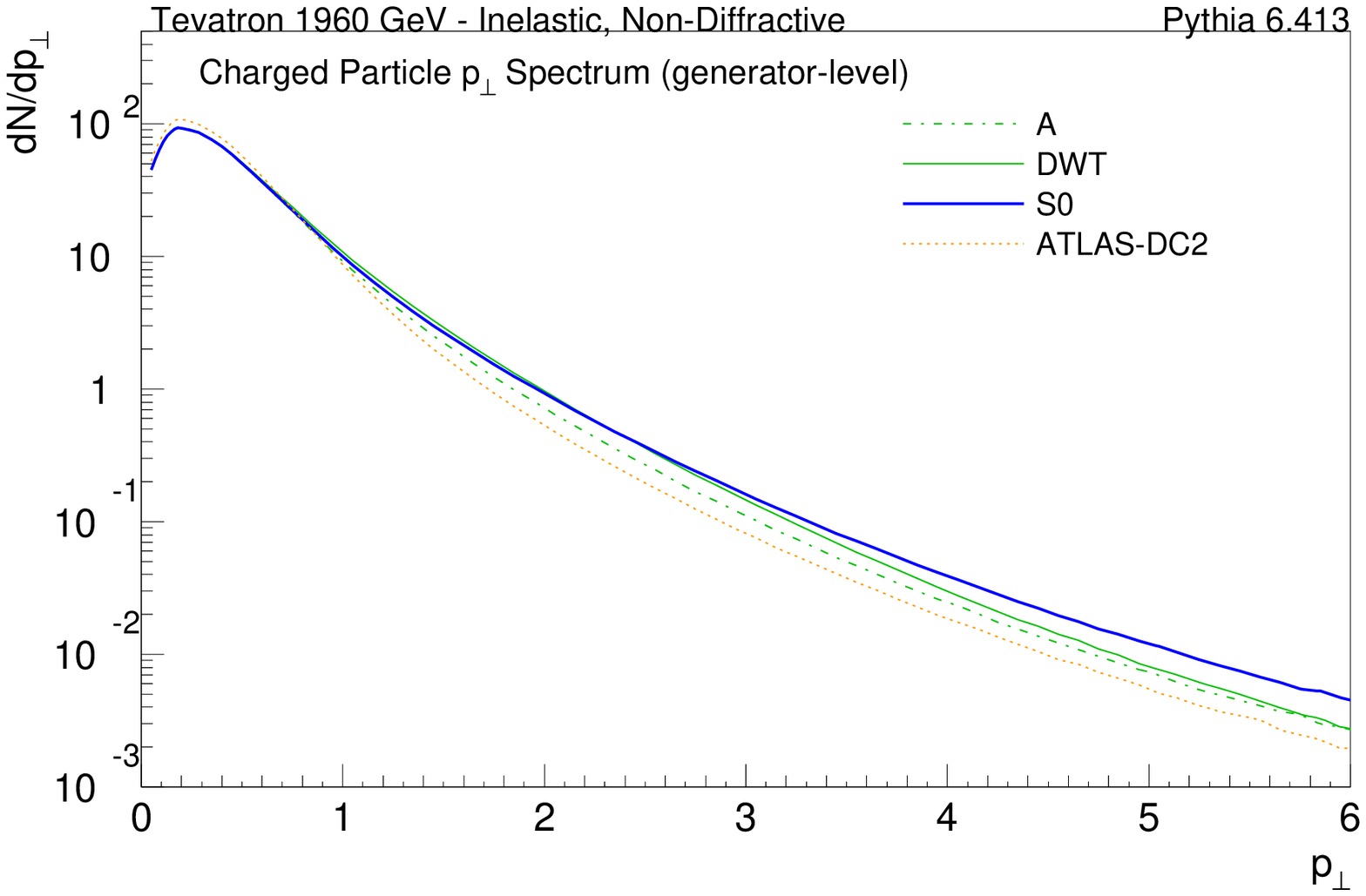}{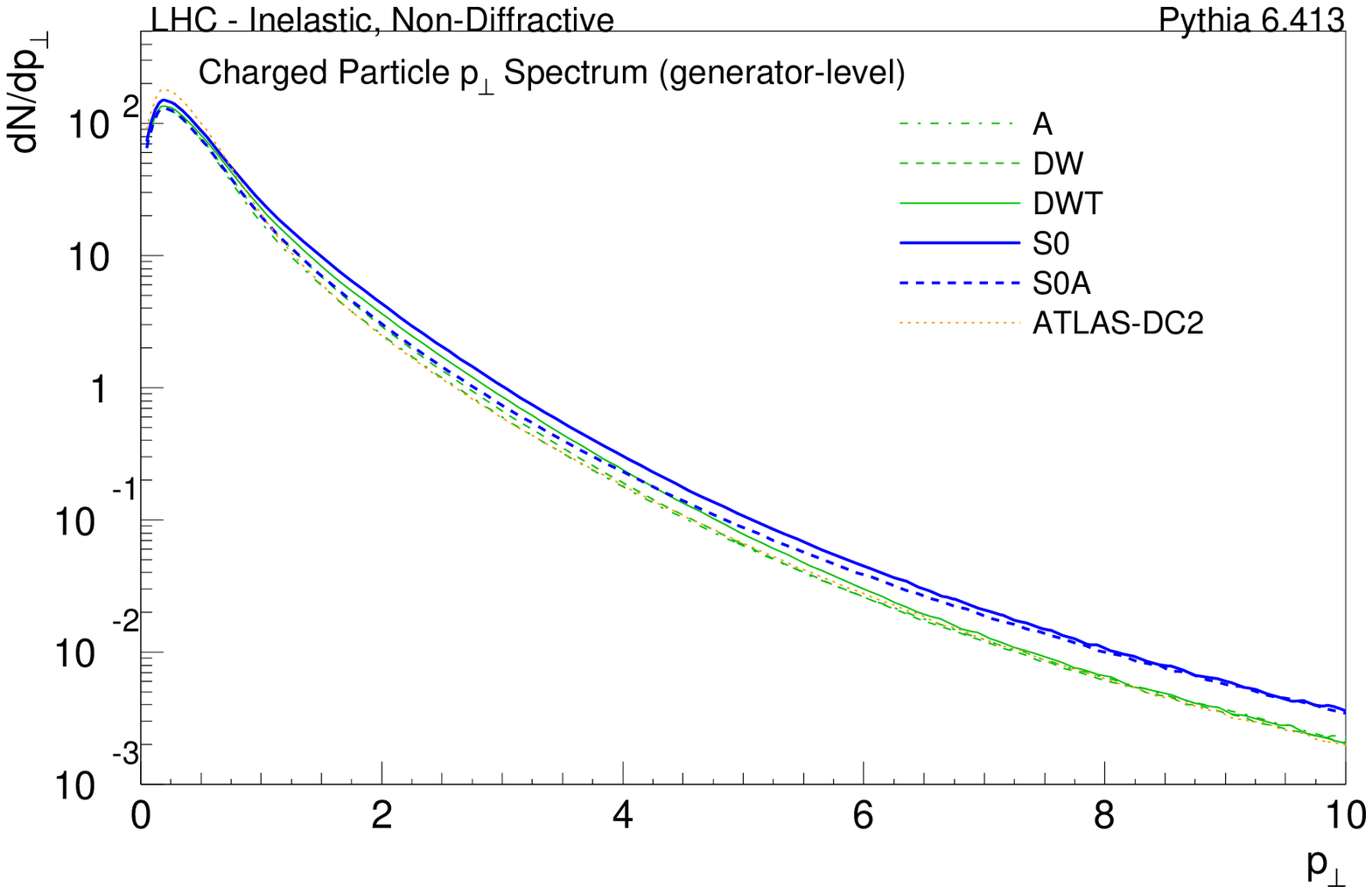}\\[-3cm]
\raisebox{0cm}{\hspace*{9mm}\includegraphics*[scale=0.25]{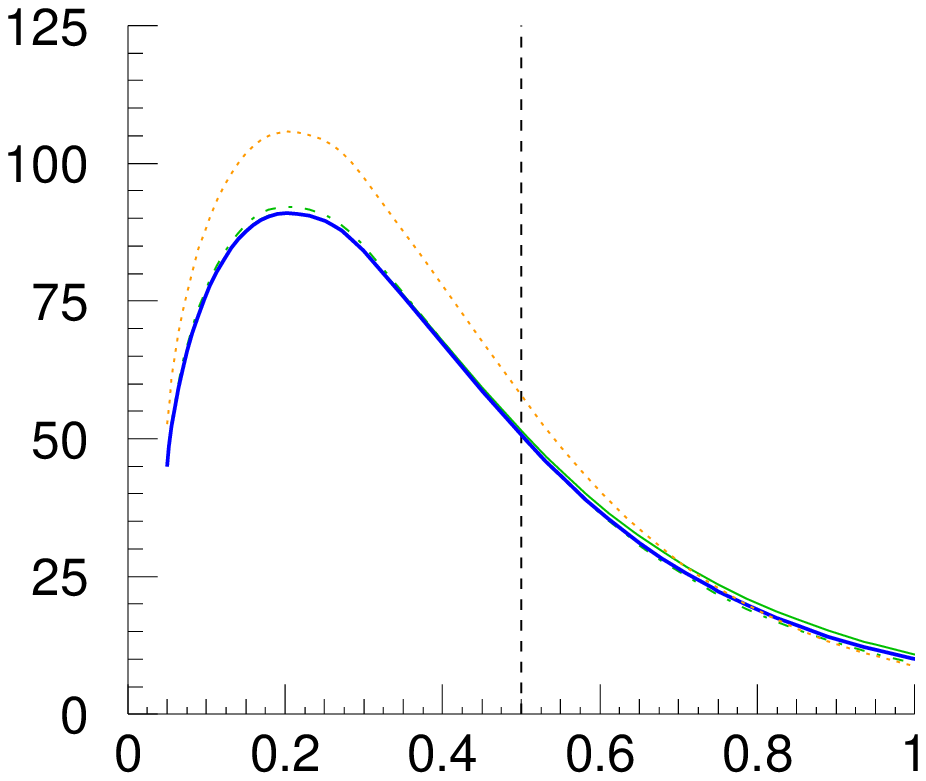}\hspace*{4.85cm}\includegraphics*[scale=0.25]{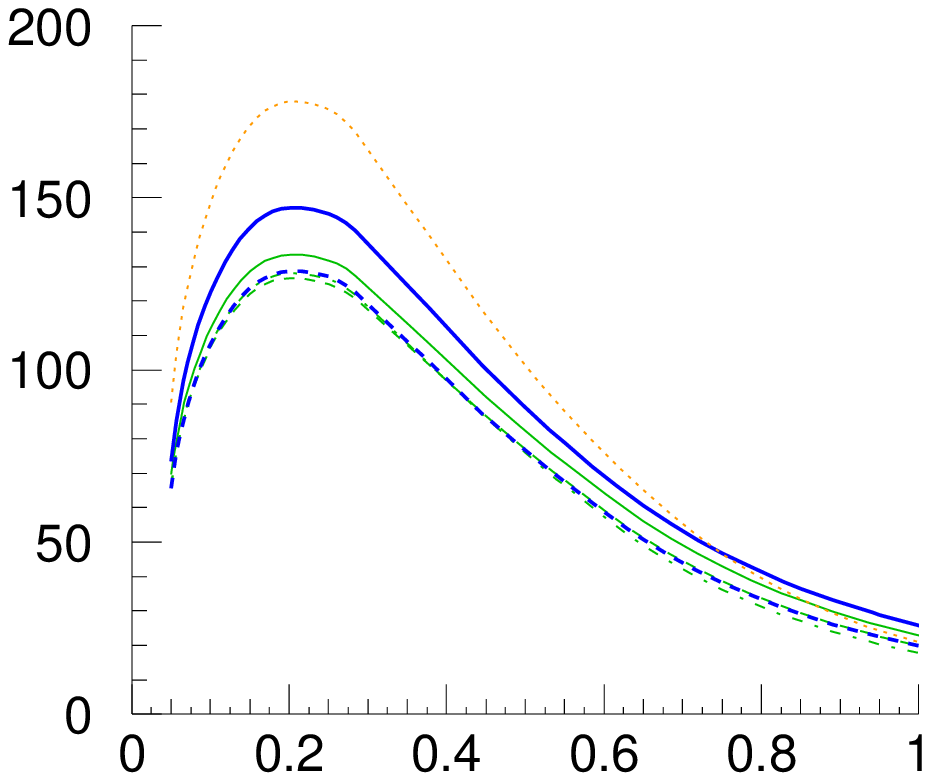}}\\[-3mm]
\caption{Charged particle \ensuremath{p_\perp}\ spectrum, generator-level only. Insets
  show the region  below 1 GeV on a linear scale. The
  fiducial distributions \cite{lhplots} are very similar, apart from an overall
  normalisation and the cut at $\ensuremath{p_\perp}=0.5\GeVx$. \label{fig:pt}}
\end{figure}
\begin{figure}
\plotrow{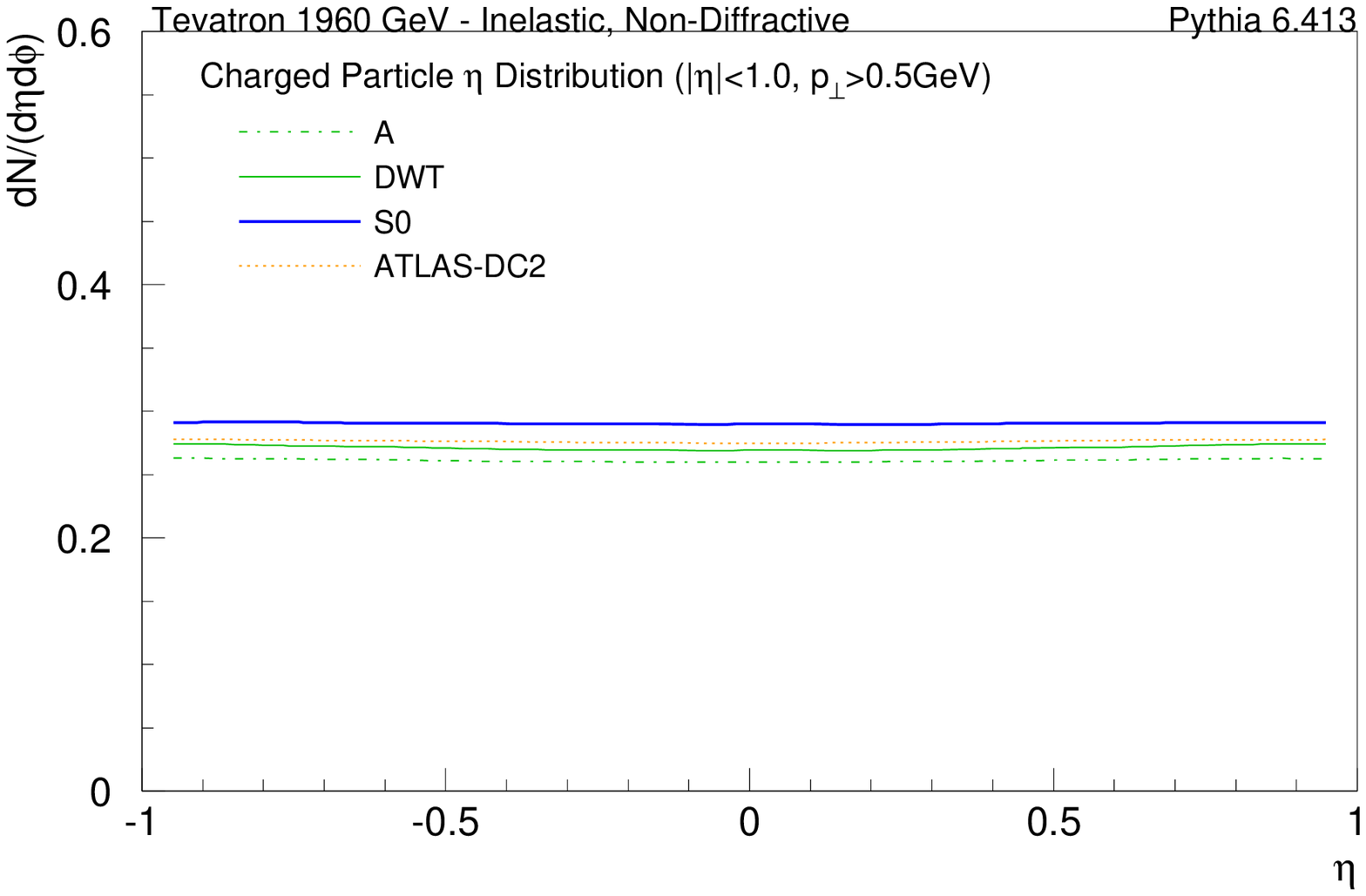}{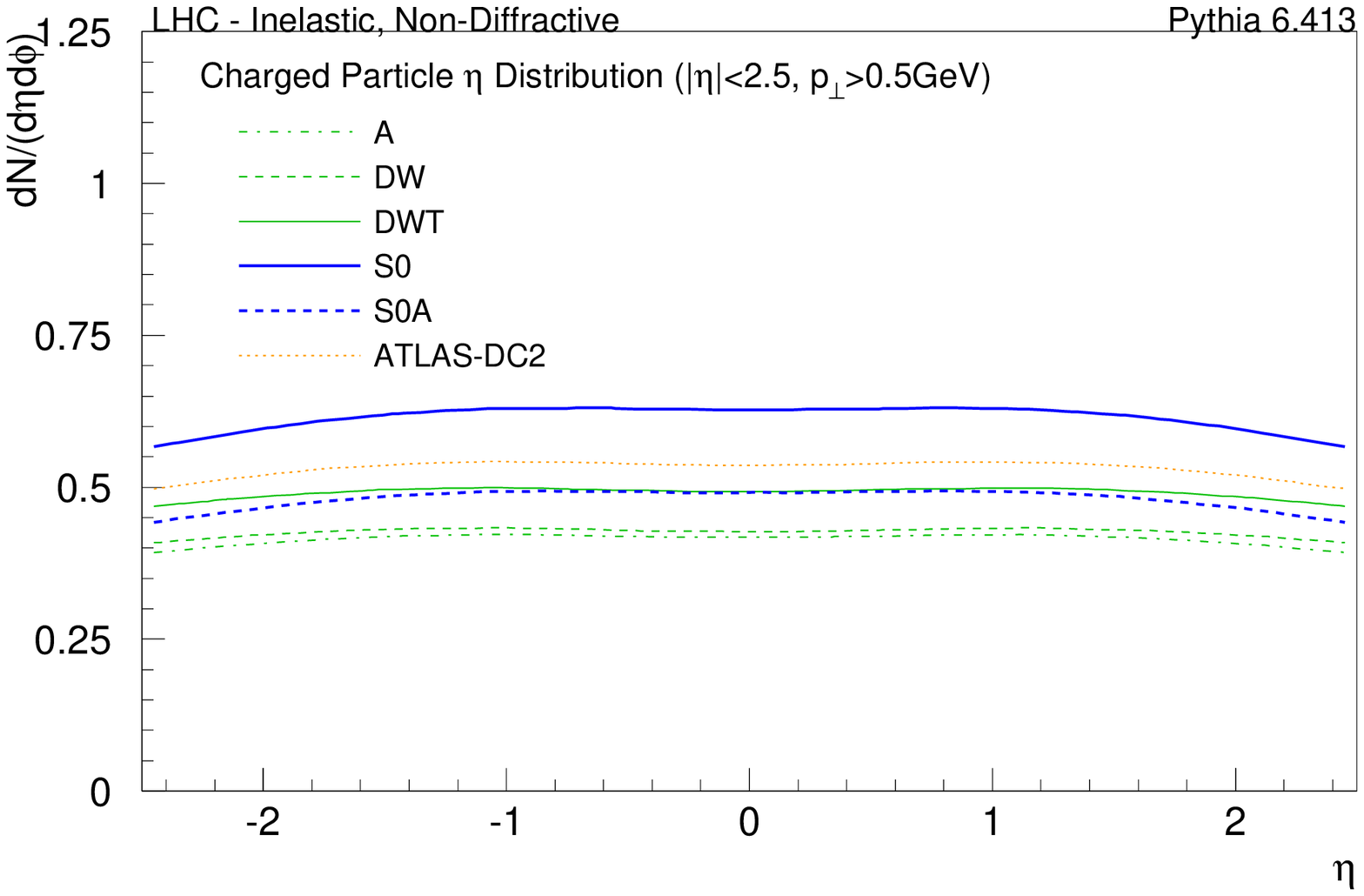}
\caption{Charged particle density vs.\ pseudorapidity, fiducial
  distribution only. The generator-level ones can be found at
  \cite{lhplots}. \label{fig:eta}}
\end{figure}

The cause for the difference in unobserved multiplicity
can be readily identified by considering the generator-level \ensuremath{p_\perp}\ spectra of
charged particles, fig.~\ref{fig:pt}. 
The small insets show the region below 1 GeV on a
linear scale, with the cut at $\ensuremath{p_\perp}=0.5\GeVx$ shown as a dashed
line. Below the fiducial cut, the 
ATLAS tune has a significantly larger soft peak than
the other models. The S0 model, on the other hand,  
has a harder distribution in the tail, which also 
causes S0 to have a slightly larger overall
multiplicity in the central region, as illustrated in the
fiducial pseudorapidity distributions, fig.~\ref{fig:eta}. 
Apart from the overall normalisation, however, the pseudorapidity 
distribution is almost featureless except for
the tapering off towards large $|\eta|$ at the LHC. Nonetheless, we note
that to study possible non-perturbative fragmentation differences between 
LEP and hadron colliders, 
quantities that would be interesting to plot vs.\ this axis would be 
strangeness and baryon fractions, such as $N_{K^0_S}/N_\mathrm{ch}$ and
$N_{\Lambda^0}/(N_{\Lambda^0}+N_{\bar{\Lambda}^0})$, as well as the
the $\ensuremath{p_\perp}$ spectra of these particles. With good statistics, also 
multi-strange baryons would carry interesting information, as has 
been studied in $pp$ collisions in particular by the STAR experiment 
\cite{Abelev:2006cs,Heinz:2007hg}. 

\begin{figure}
\plotrow{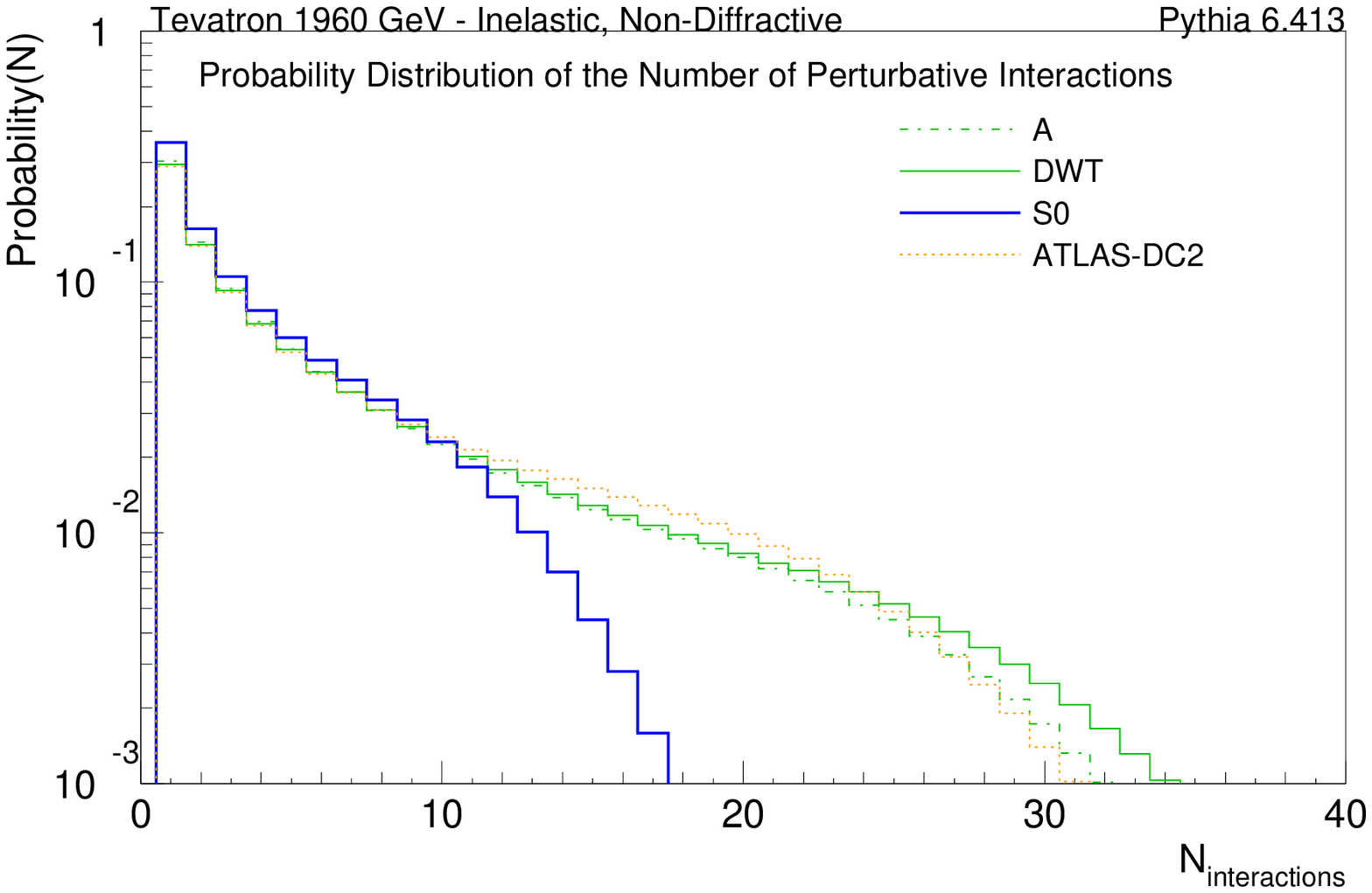}{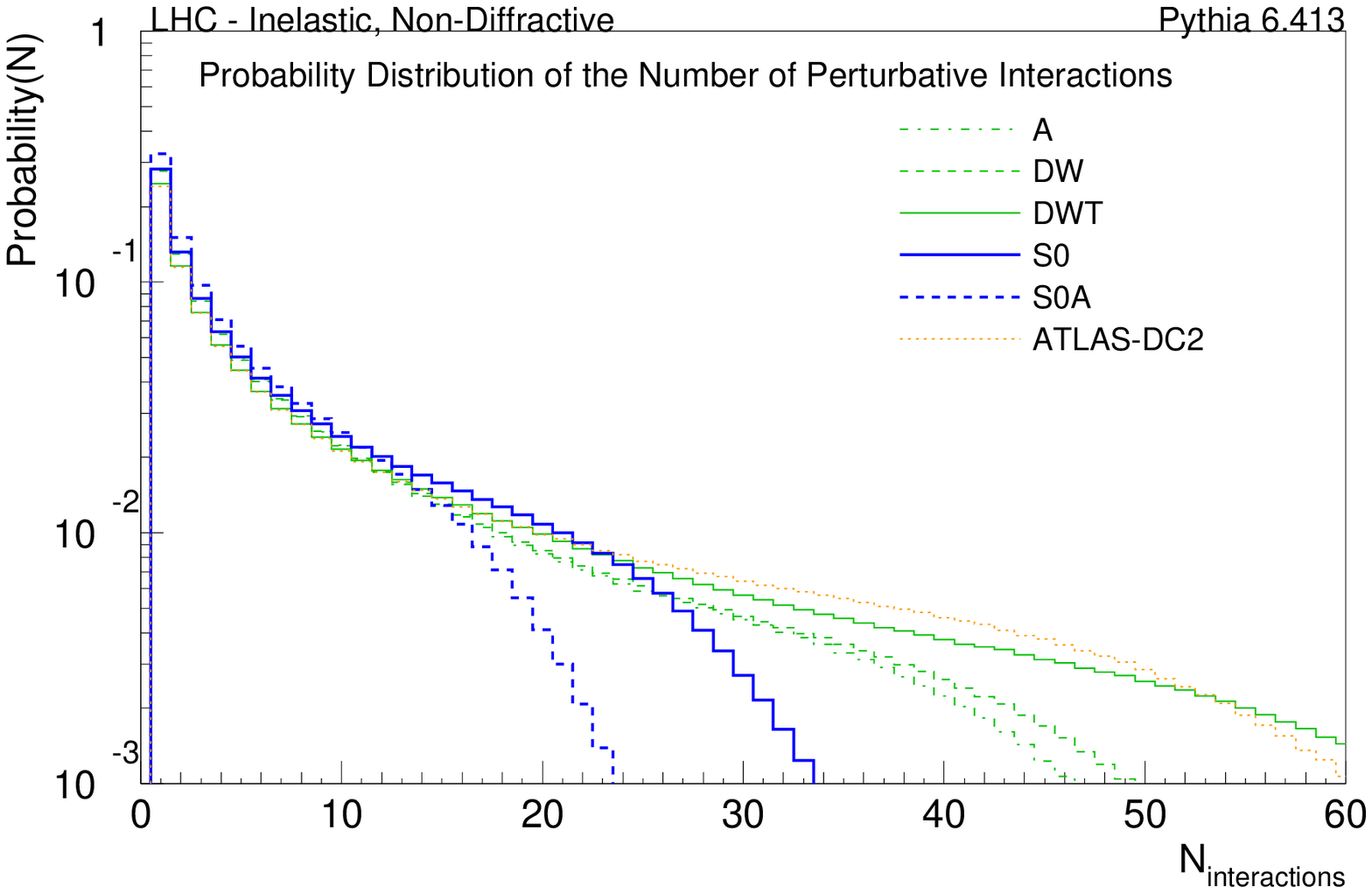}
\caption{
Probability distribution of the number of multiple interactions. The
averages range from $3.7 < \left<N_{\mathrm{int}}\right> < 6.1$ at the
Tevatron to $4.7 < \left<N_{\mathrm{int}}\right> < 11.2$ at the LHC.
\label{fig:nint}}
\end{figure}

Before going on to correlations, let us briefly consider how the
multiplicity is built up in the various models. 
Fig.~\ref{fig:nint} shows the probability distribution of the number
of multiple interactions. This distribution essentially represents a
folding of the multiple-interactions cross section above the infrared
cutoff with the assumed transverse matter distribution. Firstly, the
ATLAS and Rick Field tunes have almost identical infrared cutoffs and 
transverse mass profiles and hence look very similar. 
(Since ATLAS and DWT have the same energy extrapolation,
these are the most similar at LHC.) On the other hand, the S0(A)
models exhibit a significantly smaller tail towards large numbers of
interactions caused by a combination of the smoother mass profile and
the fact that the MPI are associated with ISR showers of their own,
hence each takes a bigger $x$ fraction. 

\begin{figure}
\plotrow{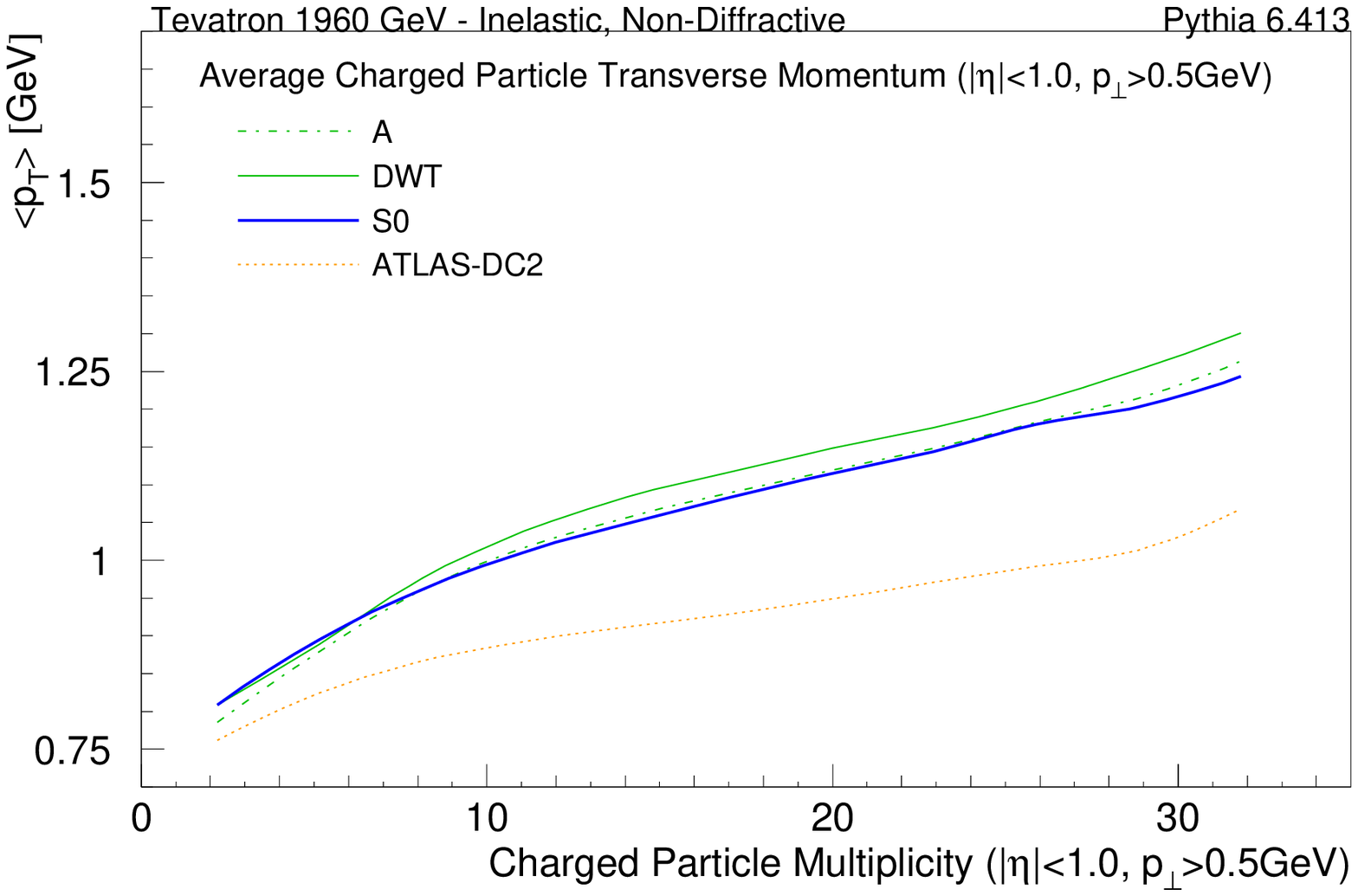}{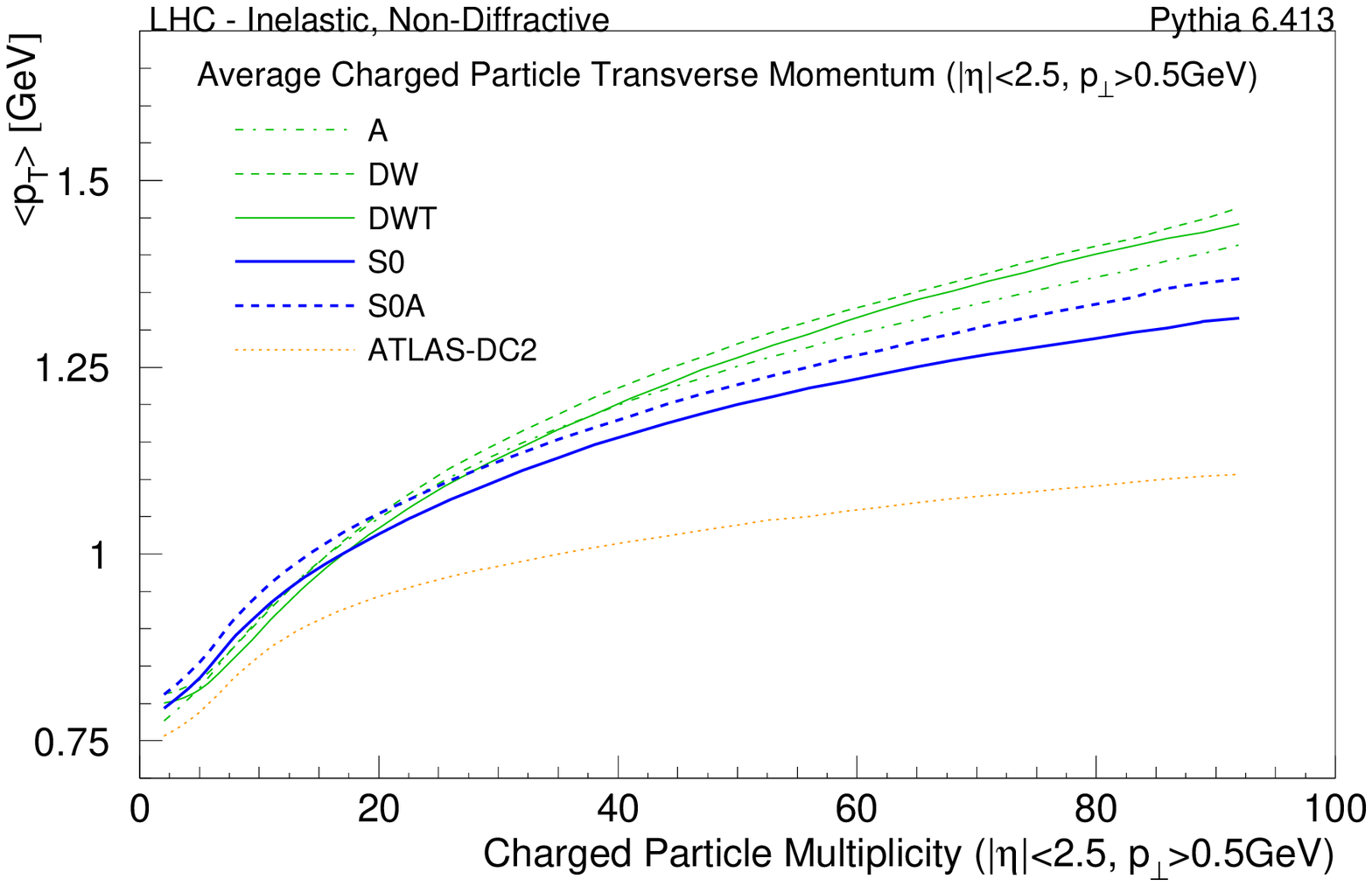}
\caption{The average track transverse momentum vs.~the number of
  tracks, counting fiducial tracks only, for events with at least one
  fiducial track.\label{fig:ptofnch}}
\end{figure}

Fig.~\ref{fig:ptofnch} shows the first non-trivial correlation, the 
average track momentum (counting fiducial tracks only) vs.\
multiplicity for events with at least one charged particle passing the 
fiducial cuts. 
The general trend is that the tracks in high-multiplicity events are
harder on average than in low-multiplicity ones. This agrees with
collider data and is an interesting observation in itself. 
We also see that the tunes roughly agree for low-multiplicity events, while 
the ATLAS tune falls below at high multiplicities. In the models here
considered, this is tightly linked to the weak final-state 
colour correlations in the ATLAS tune; 
the naive expectation from an uncorrelated system of strings decaying
to hadrons would be that $\ensuremath{\left<\ensuremath{p_\perp}\right>}$ should be independent of 
$\ensuremath{N_{\mathrm{ch}}}$.
To make the average \ensuremath{p_\perp}\ rise sufficiently to agree with Tevatron
data, tunes A, DW(T), and S0(A) incorporate strong 
colour correlations between final-state partons 
from different interactions, chosen in such a way as to 
minimise the resulting string length. An alternative possible
explanation could be Cronin-effect-type rescatterings of the outgoing
partons, a preliminary study of which is in progress
\cite{sjostrandrescattering}.  

An additional important correlation, which carries information on local vs.\
long-distance fluctuations, is the forward-backward
correlation strength, $b$, defined as
\cite{Sjostrand:1987su,Ansorge:1988fg,Capella:1992yb} 
\begin{equation}
b = \frac{\left<n_Fn_B\right> - \left<n_F\right>^2}{\left<n_F^2\right>-\left<n_F\right>^2}~,
\end{equation} 
where $n_F$ ($n_B$) is the number of charged particles in a forward (backward)
pseudorapidity bin of fixed size, separated by a central 
interval $\Delta\eta$ centred at zero. The UA5 study
\cite{Ansorge:1988fg} used 
pseudorapidity bins one unit wide and plotted the correlation vs.\ the
rapidity difference, $\Delta\eta$. For comparison, STAR, which has 
a much smaller coverage, uses 0.2-unit wide bins
\cite{Srivastava:2007ei}. However, as shown in a recent study
\cite{Yan:2007ia}, small bins increase the relative importance of
statistical fluctuations, washing out the genuine correlations. For
the Tevatron and LHC detectors, which also have small coverages, we
therefore settle on a compromise of
0.5-unit wide bins. We also choose to plot the result vs.\ the
pseudorapidity of the forward bin, 
$\eta_F\sim\Delta\eta/2$,  
such that the $x$ axis corresponds directly to a pseudorapidity in the
detector (the backward bin is then situated symmetrically on
the other side of zero). 
\begin{figure}
\plotrow{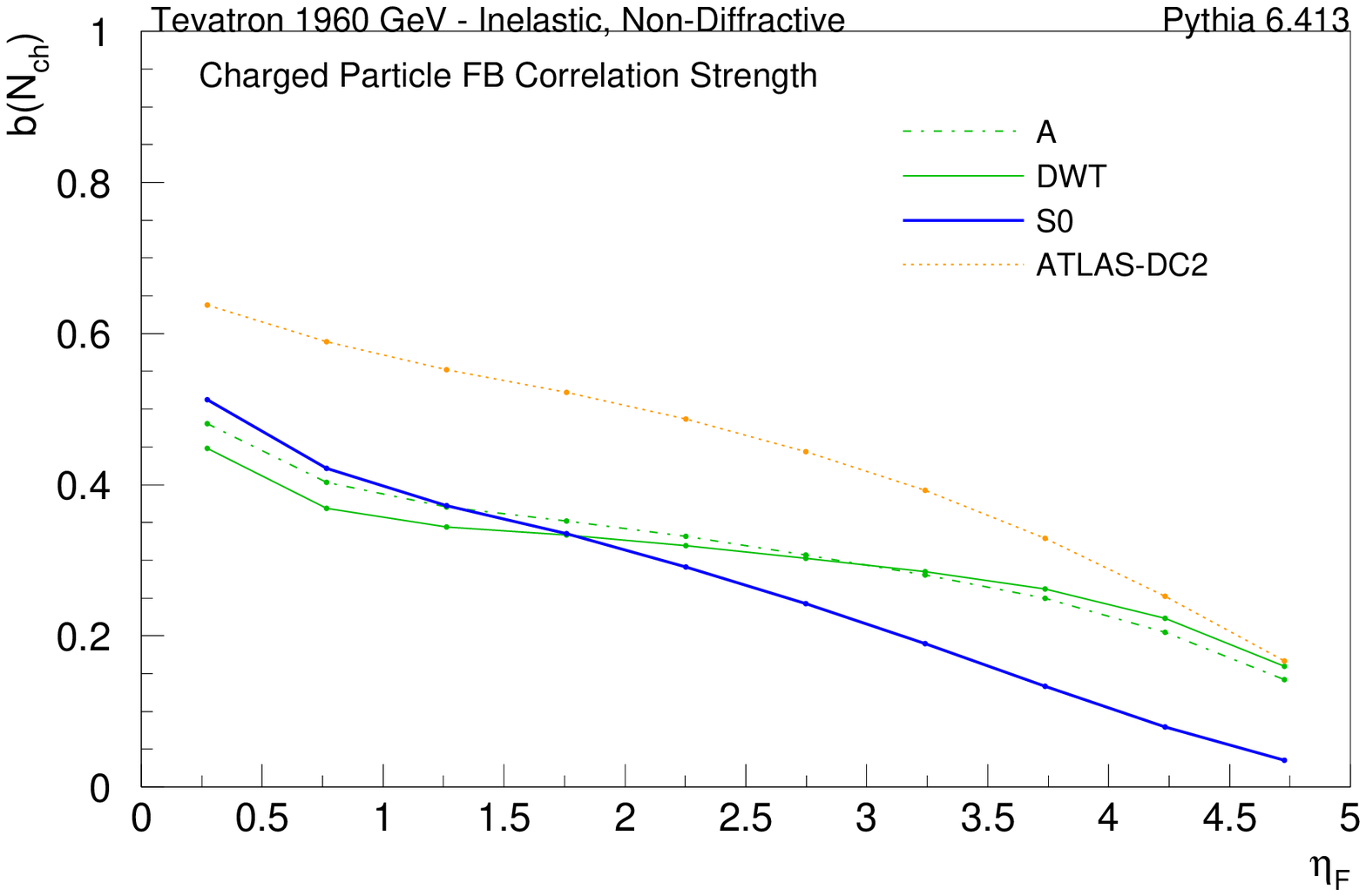}{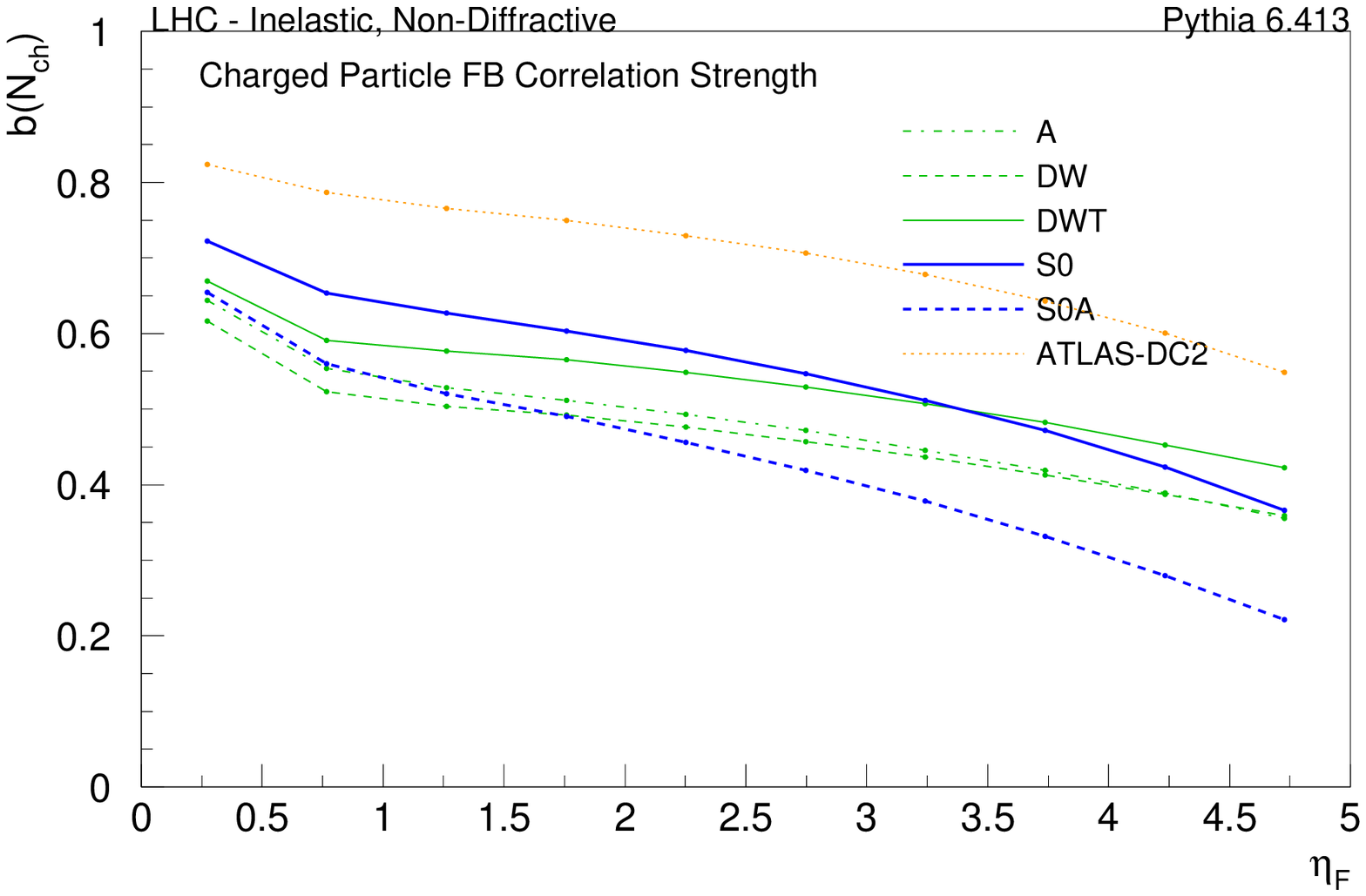}\vspace*{-1mm}
\plotrow{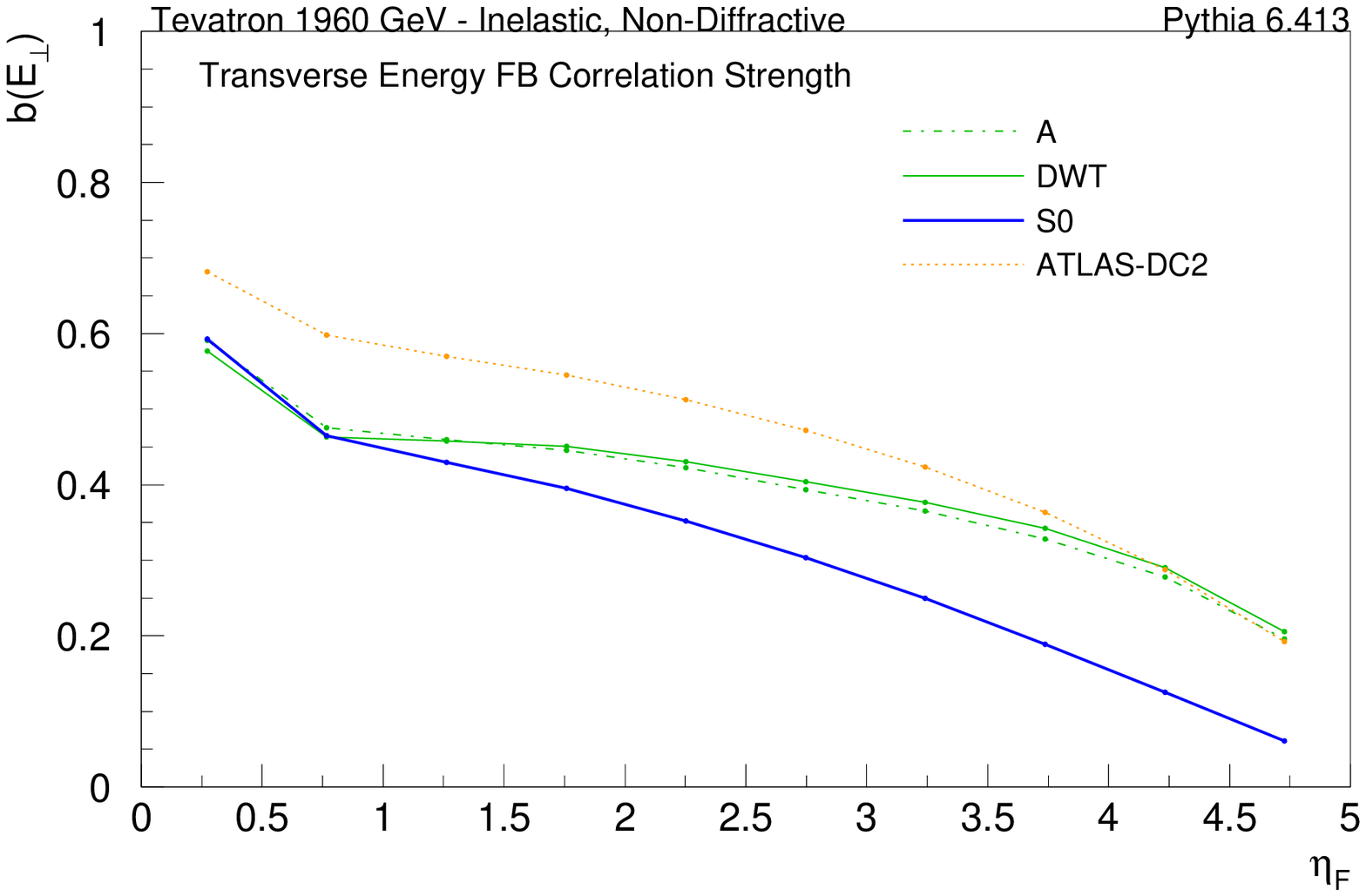}{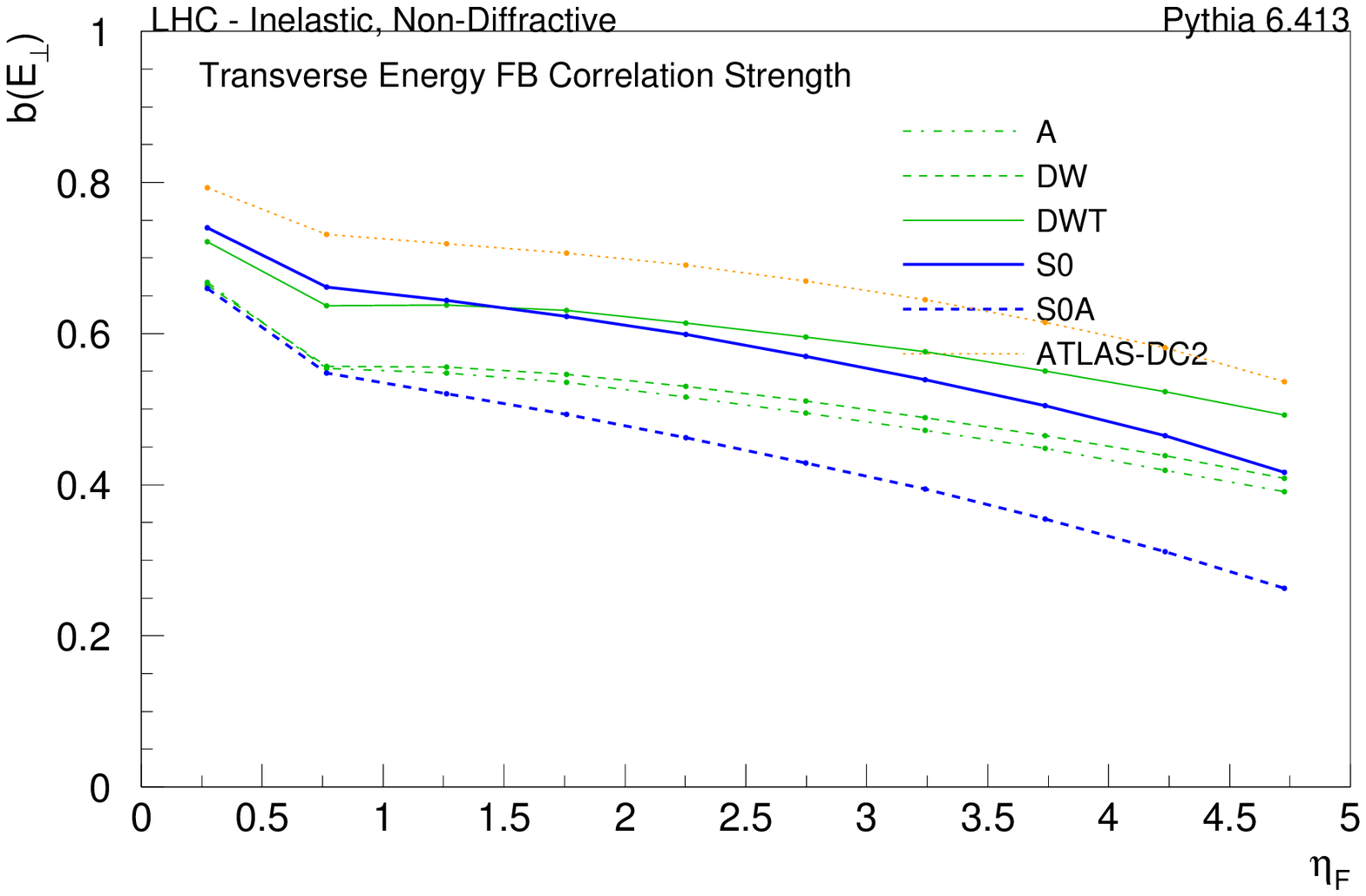}\vspace*{3mm}
\caption{Generator-level forward-backward correlation strength, $b$, 
for charged particles (top) and transverse energy (bottom). 
\label{fig:fbcor}}
\end{figure}
Fig.~\ref{fig:fbcor} shows the generator-level
correlations, both for charged particles (top row) and for 
a measure of transverse energy (bottom row), here defined as 
the \ensuremath{p_\perp}\ sum of all neutral and charged particles inside the
relevant rapidity bins. Note that we let the
$x$ axis extend to pseudorapidities of 5, outside the measurable region, in
order to get a more comprehensive view of the behaviour of the
distribution. The fact that the ATLAS and S0(A) distributions have a more
steeply falling tail than A and DW(T) again reflects the qualitatively
different physics cocktails represented by these models. Our tentative
conclusions are as follows: Rick Field's tunes A, DW, and DWT 
have a large number of multiple interactions, cf.~fig.~\ref{fig:nint},
but due to the strong final-state colour correlations in these tunes,
the main effect of each additional interaction 
is to add ``wrinkles'' and energy to already existing
string topologies. Their effects on short-distance
correlations are therefore suppressed relative to the ATLAS tune,
which exhibits similar long-distance correlations but stronger
short-distance ones. S0(A) has a smaller total number of MPI,
cf.~fig.~\ref{fig:nint}, which leads to smaller long-distance
correlations, but it still has strong short-distance ones. In summary,
the $b$  
distributions are clearly sensitive to the relative mix of MPI
and shower activity. They also depend on the detailed shape of
fig.~\ref{fig:nint}, which in turn is partly controlled by the
transverse matter density profile. 
Measurements of these distributions, both at present and
future colliders, would therefore add 
another highly interesting and complementary piece of information on
the physics cocktail. 


\subsection{Conclusion and outlook}

We have illustrated some elementary distributions in inelastic,
non-diffractive events at the Tevatron and LHC, as they look with
various tunes of the two underlying-event models in the
\textsc{Pythia} event generator. In particular, taking the charged particle
multiplicity distribution to set the overall level 
of the MB/UE physics, 
the $\ensuremath{p_\perp}$ spectrum of charged particles and the
\ensuremath{\ensuremath{\left<\ensuremath{p_\perp}\right>}(N_\mathrm{ch})}\ 
correlations then add important information on aspects such as
final-state colour correlations. Identified-particle spectra
would yield further insight on beam remnants and 
hadronization in a hadron-collider environment. Finally,
correlations in multiplicity and energy vs.\
pseudorapidity can be used to extract information on
the importance of short-distance vs.\ long-distance correlations,
which (very) roughly correspond to the type of fluctuations produced by
shower- and multiple-interaction-activity, respectively.  

By comparing the multiplicity distributions with and without fiducial
cuts, we note that the extrapolation from observed to generator-level
distributions can be highly model-dependent. It is therefore important
to extend the measured region as far as possible in both $\eta$ and
$\ensuremath{p_\perp}$. 

On the phenomenological side, several remaining issues could still be
addressed without requiring a more formal footing (see below). These
include parton rescattering effects (Cronin effect)
\cite{sjostrandrescattering},  
correlations between $x$- and impact-parameter-dependence in the
multi-parton PDFs \cite{Korotkikh:2004bz,Renner:2005sm,Hagler:2007xi},
saturation and small-$x$ effects \cite{Avsar:2006jy}, improved
modeling of baryon production
\cite{Sjostrand:2002ip,Sjostrand:2004pf,DuranDelgado:2007tg},   
possible breakdowns of jet universality between LEP, HERA, and hadron
colliders, and closer studies of the correspondence between coherent
phenomena, such as diffraction and elastic scattering, and inelastic
non-diffractive processes  
\cite{Treleani:2007gi,Gustafson:2007sb}.  

Further progress would seem to require a systematic way of improving
on the phenomenological models, both on the perturbative and
non-perturbative sides, which necessitates some degree of formal
developments in addition to more advanced model building. 
The correspondence with fixed-order QCD is already being elucidated by
parton-shower / matrix-element matching methods, already a
well-developed field. Though these methods are currently applied
mostly to $X$+jet-type topologies, there is no reason they should not
be brought to bear on MB/UE physics as well. Systematic inclusion of
higher-order effects in showers (beyond that offered by ``clever
choices'' of ordering, renormalisation, and kinematic variables) would
also provide a more solid foundation for the perturbative side of the
calculation, though this is a field still in its infancy
\cite{Kato:1990as,Nagy:2007ty}. To go further, however, 
factorisation in the context of hadron collisions 
needs to be better understood, probably including by now 
well-established short-distance phenomena such as 
multiple perturbative interactions on the ``short-distance'' side and,
correspondingly, correlated multi-parton PDFs on the ``long-distance''
side. It is also interesting to note that current
multiple-interactions models effectively amount to a resummation of
scattering cross sections, in much the same way as parton showers
represent a resummation of emission cross sections. 
However, whereas a wealth of 
higher-order analytical results exist for emission-type corrections, which
can be used as useful cross-checks and tuning benchmarks for
parton showers, corresponding results
for multiple-interactions corrections are almost entirely absent. This
is intimately linked to the absence of a satisfactory formulation of
factorisation. 

On the experimental
side, it should be emphasised that there is much more than Monte Carlo
tuning to be done in MB/UE studies, and that data is vital to guide us
in both the phenomenological and formal directions discussed above. 
Dedicated Tevatron studies have
already had a large impact on our understanding of hadron collisions,
but much remains uncertain. Results of future measurements are likely
to keep challenging that understanding and could provide for a very
fruitful interplay between experiment and theory.


%% file: s_sherstnev/lomod.tex
\subsection{Introduction}
It has long been known that for certain regions of $x$ there can be large 
differences between PDFs extracted at different orders of perturbative QCD. 
It happens due to missing higher order corrections both in the parton 
evolution and in the MEs, which govern their extraction by comparison to 
experimental data. In particular, use of PDFs of the wrong order can lead 
to wrong conclusions for the small-$x$ gluon. Traditionally, LO PDFs are 
usually thought to be the best choice for use with LO ME, usually available 
in Monte-Carlo programs, though it has been recognised that all such results 
should be treated with care. However, recently another viewpoint has appeared, 
namely it has been suggested that NLO PDFs may be more appropriate~\cite{Campbell:2006wx}.
The argument is that NLO corrections to MEs are often small, and the 
main change in the total cross-section in going from LO to NLO is due to the 
PDFs. 

In this paper we present another approach, which is based on advantages 
of both the LO and NLO PDF approximations, and compare all three predications 
for several processes with the {\it truth} -- NLO PDFs combined with NLO 
MEs\footnote{Since NLO matrix elements are most readily available in $\msb$ 
scheme, we will take this as the default, and henceforth NLO is intended to 
mean NLO in $\msb$ scheme.}. We interpret the features of the results noting 
that there are significant faults if one uses exclusively either LO or NLO 
PDFs. We hence attempt to minimise this problem, and investigate how a best 
set of PDFs for use with LO matrix elements may be obtained. 

\subsection{Parton Distributions at Different Orders}
Let us briefly explain the reasons for the origins of the differences between 
the PDFs at different perturbative orders. The LO gluon is much larger at small 
$x$ than any NLO gluon at low $Q^2$. The evolution of the gluon at LO and 
NLO is quite similar, so at larger $Q^2$ the relative difference is smaller, 
but always remains significant. This difference in the gluon PDF is 
a consequence of quark evolution, rather than gluon evolution. The small-$x$ 
gluon is determined by $dF_2/d\ln Q^2$, which is directly related to the 
$Q^2$ evolution of the quark distributions. The quark-gluon splitting function 
$P_{qg}$ is finite at small $x$ at LO, but develops a small-$x$ divergence 
at NLO (and further $\ln(1/x)$ enhancements at higher orders), so the small $x$ 
gluon needs to be much bigger at LO in order to fit structure function evolution. 
There are also significant differences between the LO and NLO quark distributions. 
Most particularly the quark coefficient functions for structure functions 
in $\msb$ scheme have $\ln(1-x)$ enhancements at higher perturbative order, 
and the high-$x$ quarks are smaller as the order increases. Hence, the LO gluon is 
much bigger at small $x$, and the LO valence quarks are much bigger at high-$x$. 
This is then accompanied by a significant depletion of the quark distribution for 
$x\sim 0.01$, despite the fact this leads to a poor fit to data. 

\begin{figure}
\begin{center}
\centerline{
\includegraphics[width=0.35\textwidth]{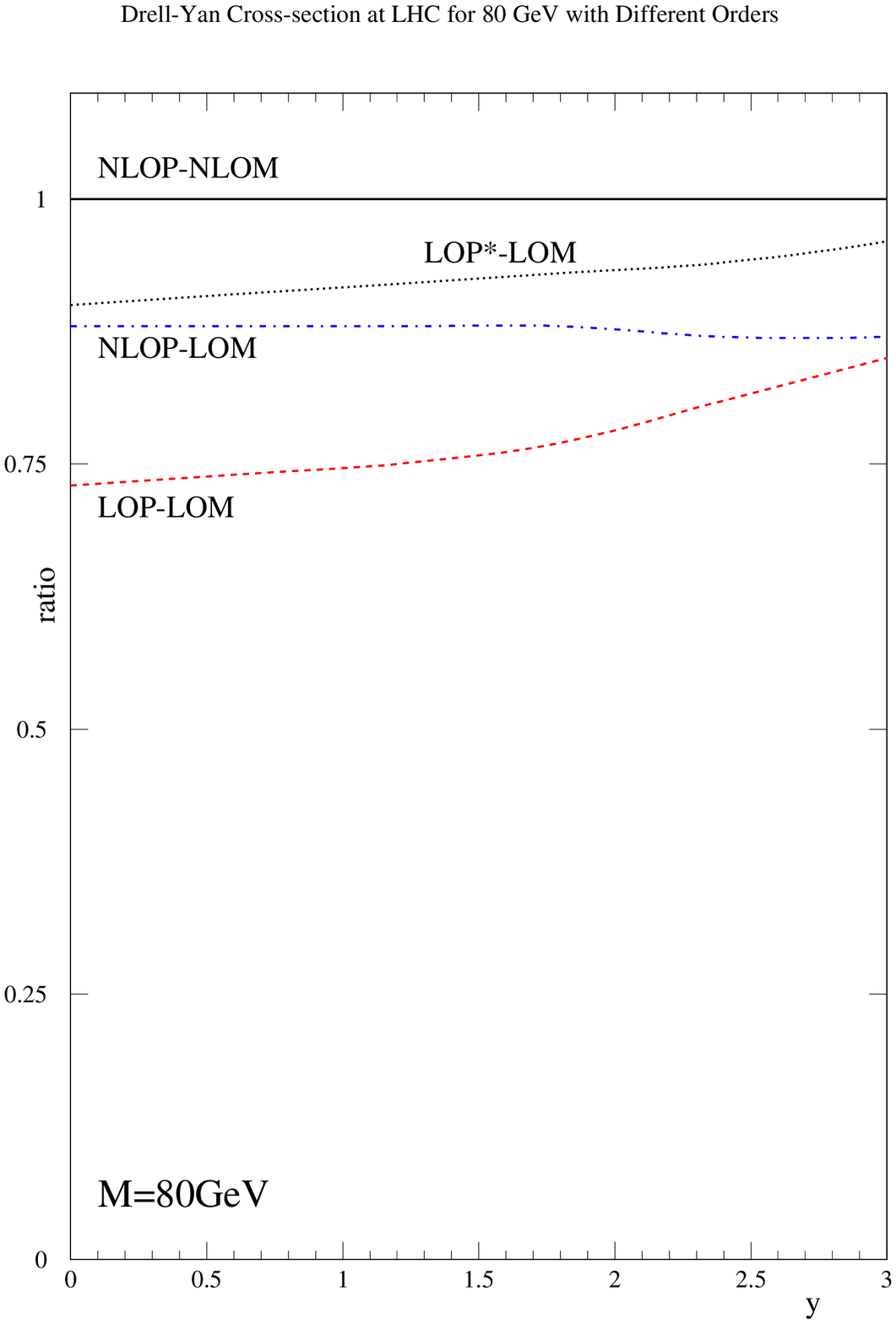}
\hspace{0.7cm}
\includegraphics[width=0.35\textwidth]{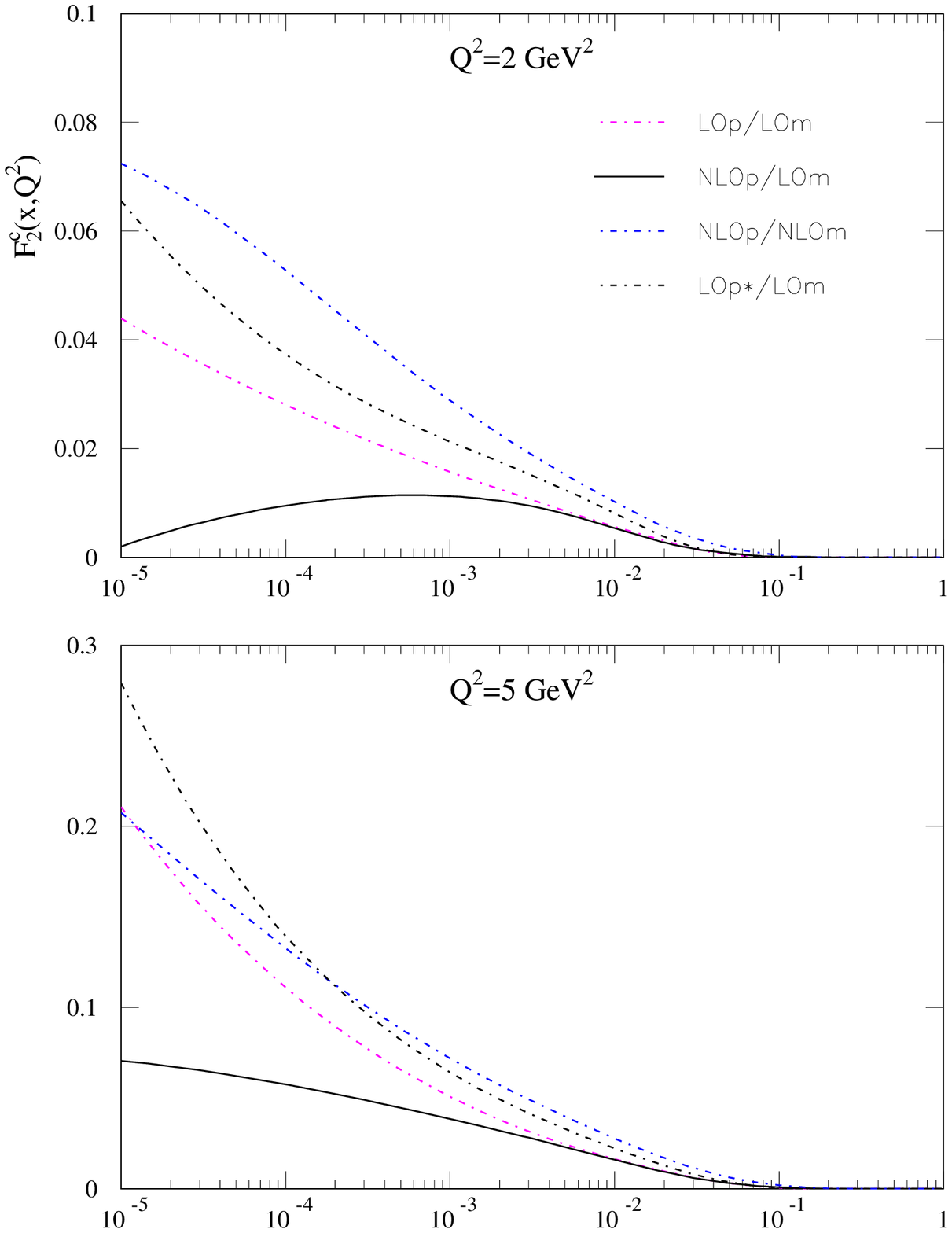}
}
\caption{
Comparison of boson production at the LHC and charm production at HERA using 
combinations of different orders of ME and PDF. 
}
\vspace{-1.0cm}
\label{w_c}
\end{center}
\end{figure}

Let us examine these differences using concrete examples. In the right of 
Fig.~\ref{w_c} we show the ratio of rapidity distributions for $W$-boson 
production at the LHC for several combination of PDF and ME to {\em the truth}. 
In this case the quark distributions are probed. Clearly we are generally 
nearer to the {\it truth} with the LO ME and NLO PDF~\cite{Martin:2004ir} 
than with the LO ME and LO PDF~\cite{Martin:2002dr}. However, this is always 
too small, since the NLO correction to the ME is large and positive. 
The depletion of the LO quark distributions for $x\sim 0.006$ (corresponding to
the central $y$) leads to the extra suppression in the PDF[LO]-ME[LO] calculation. 
However, when probing the high $x$ quarks the increase in the LO parton 
compensates for the increase in NLO matrix element, and for $y>2$ this gives 
the more accurate result. However, overall the shape as a function of $y$ 
is much worse using the LO parton distributions than the NLO distributions. 
The general conclusion is the NLO PDFs provide a better normalization and a 
better shape. 

This example suggests that the opinion in~\cite{Campbell:2006wx} is correct. 
However, let us consider a counter-example, the production of charm in DIS, 
i.e. $F^{c\bar c}_{2}(x,Q^2)$. In this case the NLO coefficient function, 
$C^{c\bar c,(2)}_{2,g}(x,Q^2,m_c^2)$ has a divergence at small $x$ not presented 
at LO, in the same way that the quark-gluon splitting function does, the latter 
being responsible for the large difference between the LO and NLO gluons at small 
$x$. In the right of Fig.\ref{w_c} we see the large effect of the NLO coefficient functions. When 
using NLO partons the LO ME result is well below the {\it truth} at low scales. 
In this case the distribution is suppressed due to a lack of the divergence in 
both the NLO gluon and the LO coefficient function. While the LO PDFs combined 
with LO coefficient functions is not a perfect match to the {\it truth}, 
after all the small-$x$ divergences are not exactly the same in matrix element 
and splitting function, it is better. In particular, in this case the NLO PDFs 
together with the LO matrix elements fail badly. 

Hence, from these two simple examples alone we can conclude that both the NLO 
partons and the LO partons can give incorrect results in some processes. 
Let us try to find some {\it optimal} set of PDFs for use with LO matrix 
elements. Due to missing terms in $\ln(1-x)$ and $\ln(1/x)$ in coefficient 
functions and/or evolution the LO gluon is much 
bigger as $x \to 0$ and valence quarks are much larger as $x \to 1$. From 
the momentum sum rule there are then not enough partons to go around, hence 
the depletion in the quark distributions at moderate to small $x$.
This depletion leads to a bad global fit at LO, particularly for HERA 
structure function data, which is very sensitive to quark distributions at 
moderate $x$. In practice the lack of partons at LO is partially compensated 
by a LO extraction of much larger $\alpha_S(M_Z^2) \sim 0.130$. So, the 
first obvious modification is to use $\alpha_S$ at NLO in a LO fit to parton 
distributions. Indeed the NLO coupling with $\alpha_S(M_Z^2)=0.120$ does 
a better job of fitting the low-$Q^2$ structure function data. 

However, even with this modification the LO fit is still poor compared with 
NLO. The problems caused due to the depletion of partons has led to a 
suggestion by T.~Sj\"ostrand\footnote{private comments at ATLAS Generators 
  meeting, CERN, December 2006.} that relaxing the momentum sum rule for the 
input parton distributions could make LO partons rather more like NLO partons 
where they are normally too small, while allowing the resulting partons still 
to be bigger than NLO where necessary, i.e the small-$x$ gluon and high-$x$ 
quarks. Relaxing the momentum sum rule at input and using the NLO definition 
of the strong coupling does improve the quality of the LO global 
fit. The $\chi^2=3066/2235$ for the standard LO fit, and becomes $\chi^2=
2691/2235$ for the modified fit with the same data set as in~\cite{Martin:2004ir} 
and using $\alpha_S(M_Z^2)=0.120$ at NLO. The momentum carried by input partons 
goes up to $113\%$. We denote the partons resulting from 
this fit as the LO* parton distribution functions. 

We can make a simple test of the potential of these LO* partons by repeating 
the previous comparisons. For the W-boson production we are indeed nearer 
to the {\it truth} with the LO ME and LO* PDF than with either LO or 
NLO PDF. Moreover, the shape using the LO* PDF is of similar quality to that 
using the NLO partons with the LO ME. So in this case LO* PDF and NLO PDF 
are comparably successful. The exercise is also repeated for the charm 
structure function at HERA. When using the LO coefficient function the LO* 
PDF result is indeed nearest to the {\it truth} at low scales, being generally 
a slight improvement on the result using LO PDF, and clearly much better 
than that using NLO PDF. 

These simple examples suggest that the LO* PDFs may well be a useful tool 
for use with Monte Carlo generators at LO, combining much of the advantage 
of using the NLO PDF while avoiding the major pitfalls. However, the examples 
so far are rather unsophisticated. In order to determine the best set of 
PDFs to use it is necessary to work a little harder. We need to examine a 
wide variety of contributing parton distributions, both in type of 
distribution and range of $x$. Also, the above examples are both fully 
inclusive, they have not taken into account cuts on the data. Nor have they 
taken account of any of the possible effects of parton showering, which is 
one of the most important features of Monte Carlo generators. Hence, before 
drawing any conclusions we will make a wide variety of comparisons for 
different processes at the LHC, using Monte Carlo generators to produce 
the details of the final state. 

\subsection{More examples at the LHC.} 
We consider a variety of final states for $pp$ collisions at LHC energies. 
In each case we compare the total $\sigma$ with LO MEs and full parton 
showering for the three cases of LO, LO* and NLO parton distributions. 
As the {\it truth} we use the results obtained with MC@NLO~\cite{Frixione:2002ik}, 
which combines NLO QCD corrections and parton showers. As the main LO 
generator we use CompHEP~\cite{Boos:2004kh}, interfaced to HERWIG~\cite{Corcella:2000bw}, 
but $pp\to\bb$ was calculated by HERWIG only. 

\begin{table}
\centerline{\begin{tabular}{|c|c|c|c||c|c||c|c||c|c|}
  \hline
    PDF & ME &$\sigma(pp\to Z/\gamma)$ &$K$  &$\sigma(pp\to tq)$ &$K$  &$\sigma (pp\to\bb)$ &$K$  &$\sigma(pp\to\ttb)$ &$K$  \\
  \hline
    NLO &NLO &    2.40 pb              &     &     259.4 pb      &     &     2.76 $\mu$b    &     &    812.8 pb        &     \\
    LO  &LO  &    1.85 pb              &1.30 &     238.1 pb      &1.09 &     1.85 $\mu$b    &1.49 &    561.4 pb        &1.45 \\
    NLO &LO  &    1.98 pb              &1.26 &     270.0 pb      &0.96 &     1.56 $\mu$b    &1.77 &    531.0 pb        &1.53 \\
    LO* &LO  &    2.19 pb              &1.09 &     297.5 pb      &0.87 &     2.63 $\mu$b    &1.05 &    699.4 pb        &1.16 \\
  \hline
\end{tabular}}
\caption{
The total cross sections for $pp\to tq$, $pp\to b\bar b$, $pp\to t\bar t$, and 
$\sigma(pp\to Z/\gamma\to \mu\mu)$ at the LHC. 
Applied cuts: for $\bb$ ($p_T>20$ GeV, $|\eta(b)|<5.0$, $\Delta R(b,\bar b)>0.5$); 
for $Z/\gamma$ ($p_T(\mu)>10$ GeV, $|\eta{\mu}|<5.0$); no cuts for $\ttb$ and single t. 
K-factor is defined according to $K=\sigma_{NLO}/\sigma_{LO}$. 
}
\vspace{-0.5cm}
\label{Tab:cs}
\end{table}

\begin{figure}[t]
\begin{center}
\centerline{
\includegraphics[width=10.0cm,height=7.5cm]{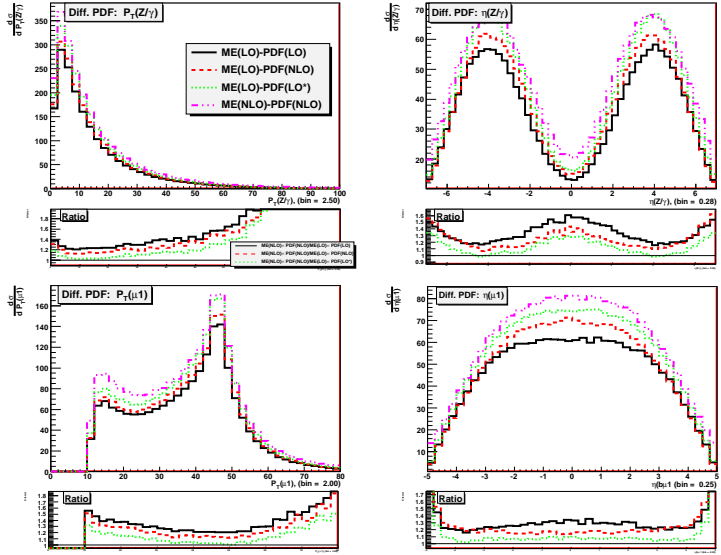}
}
\caption{ 
The comparison between the competing predictions for the differential 
cross-section for $Z/\gamma$-boson production at the LHC (upper plots) and for 
the resulting highest $p_t$ muon (bottom plots). 
}
\vspace{-1.2cm}
\label{Pic:z_phys}
\end{center}
\end{figure}
The first example is the production of $Z/\gamma$ bosons, decaying to muons. 
In order to exclude the dangerous region $m_{\mu\mu}\to 0$, where the ME 
at LO has a singularity, we apply some experimentally reasonable cuts cuts 
$p_T > 10$ GeV and $|\eta|<5.0$. These cuts are more or less appropriate for 
most analyses in CMS/ATLAS. The process is dominated by the $Z$ peak. The 
mechanism is rather similar to that for $W$ production, but now the initial 
quarks are the same flavour and the $x$ at zero rapidity is slightly higher, 
i.e. $x_0=0.0065$. The similarity is confirmed in the results. Again all the 
total cross-sections using the LO generators are lower than the {\it truth}, 
as seen in Table~\ref{Tab:cs}, but that using the LO* partons is easily closest. 
The distributions in terms of the final state boson or the highest-$p_T$ muon 
are shown in the upper and bottom plots of Fig.~\ref{Pic:z_phys} respectively. 
For the boson the LO* partons gives comparable, perhaps marginally 
better, quality of shapes as the NLO partons, but better normalization. 
The LO partons have the worst suppression at central rapidity, and all partons 
give an underestimate of the high-$p_T$ tail. For the muon the LO* partons 
give an excellent result for the rapidity distribution until $|\eta|>4$, 
better in shape and normalization that the NLO partons whilst the LO partons 
struggle at central $\eta$. Again, as in $W$ production, the $p_T$ distribution 
of the muon is better than for the boson, and in normalization is best 
described by the LO* PDFs. 

\begin{figure}
\begin{center}
\centerline{
\includegraphics[width=10.5cm,height=9.5cm]{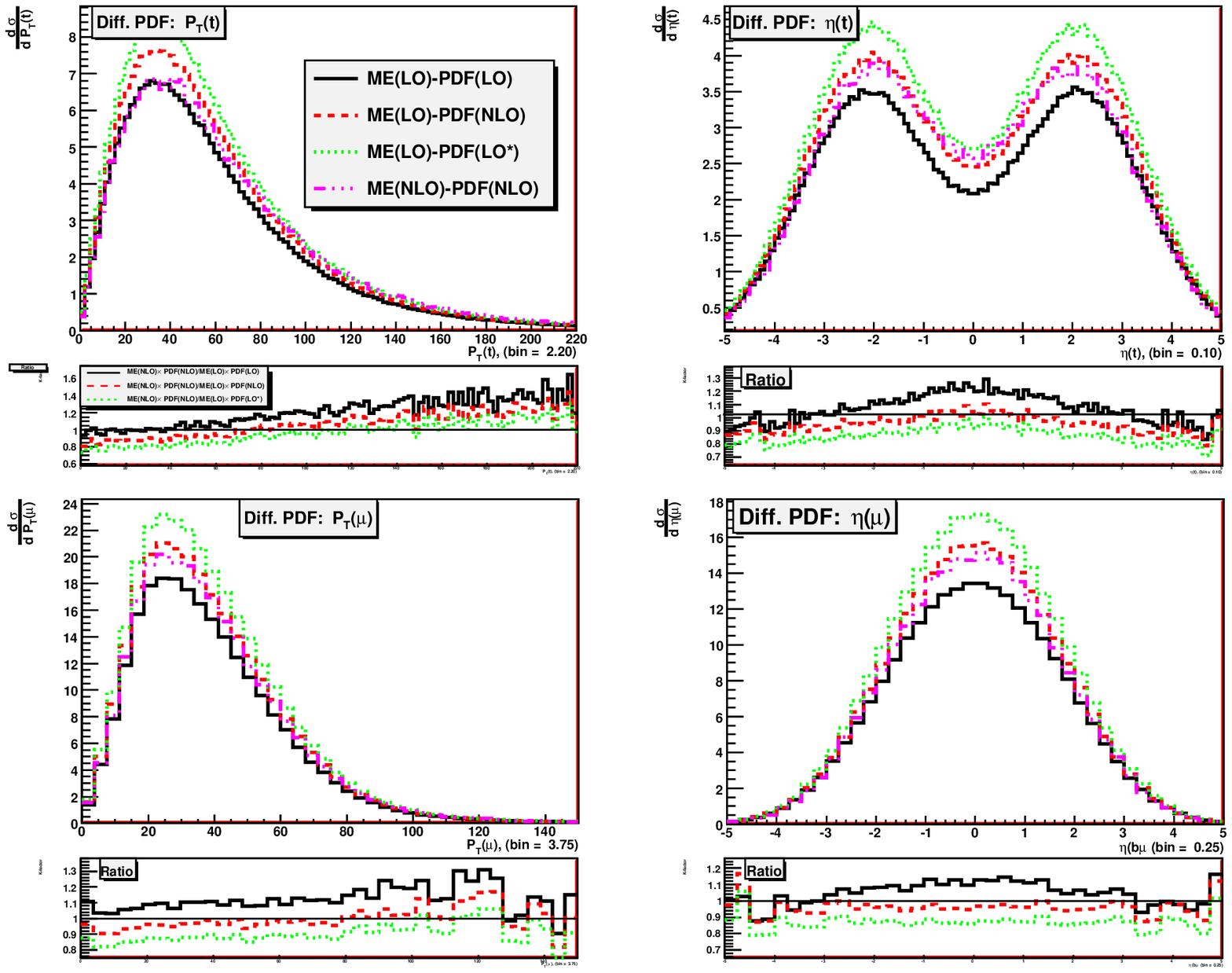}}
\centerline{
\includegraphics[width=10.5cm,height=9.5cm]{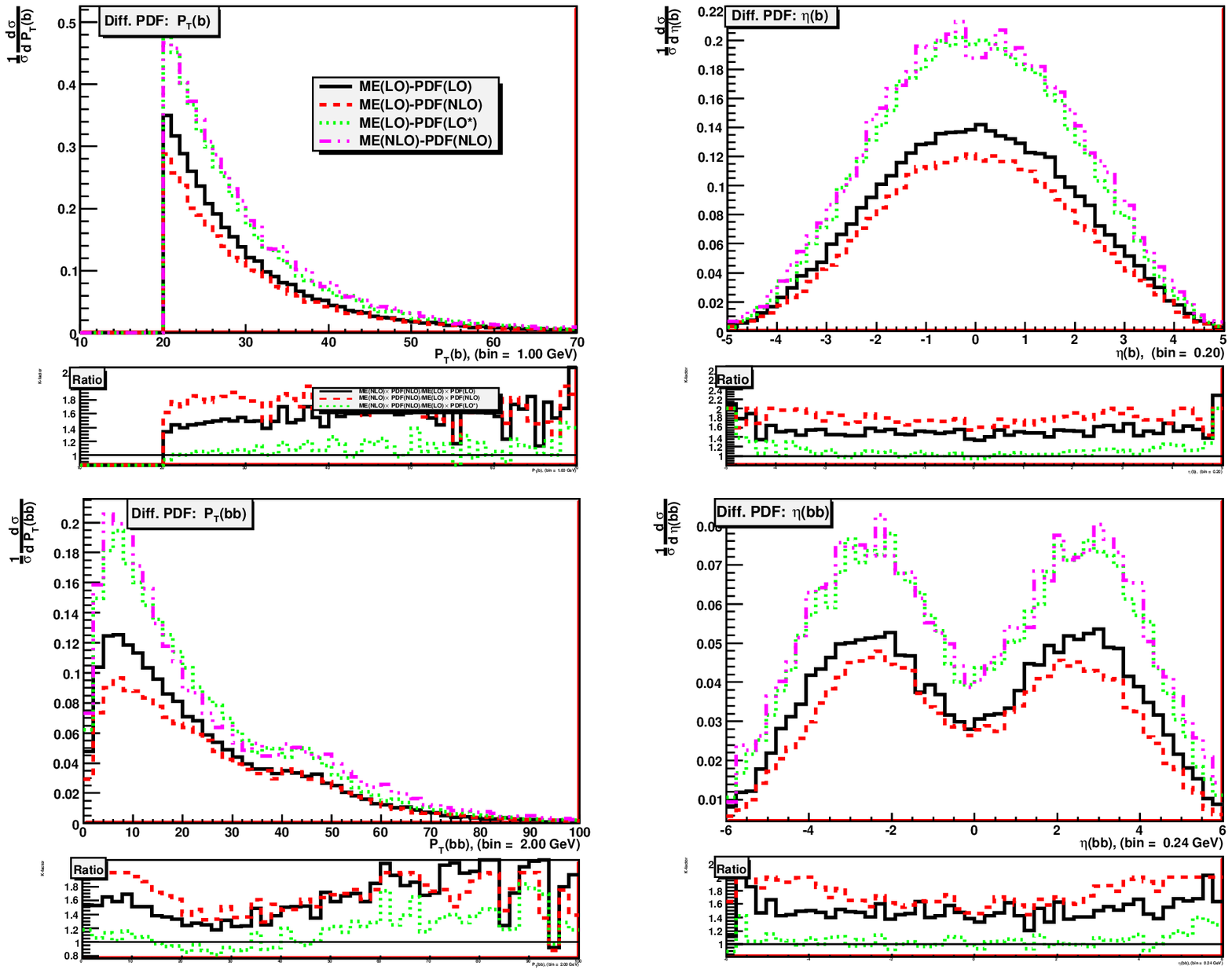}
}
\caption{ 
The comparison between the competing predictions for the differential 
cross-section for single top production at the LHC (left upper plots) and 
for the resulting $p_t$ muon (left bottom plots). Differential cross-sections 
for $b$ production at the LHC (right upper plots) and for a $b \bar b$ pair 
(right bottom plots). 
}
\label{Plot:st_bb_phys}
\vspace{-1.2cm}
\end{center}
\end{figure}

Now we consider a somewhat different process, i.e. the single top production in 
the $t$-channel. At the partonic level the dominant process is $qb(q\bar b)\to qt(q\bar t)$, 
where the $b$-quark has been emitted from gluon. Since the b-quark PDF is 
calculated based on gluon PDFs, this cross-section probes 
both the gluon distribution and the quark distributions for invariant masses 
of above about $200$ GeV, i.e. at central rapidity $x_0 \sim 0.05$. 
The $t$-channel nature of this process makes the invariant mass of the 
final state and the probed $x$ values less precise than the the $W-$boson 
production. The total cross-section for the various methods of calculation 
are seen in Table~\ref{Tab:cs}. In this case the result using the LO 
ME and the LO PDFs is suppressed, but that using the LO* PDFs is 
now larger than the {\it truth}. This is due to the large enhancement of 
the LO* gluon distribution. The NLO PDFs give the closest normalization. 
The distributions in terms of $p_T$ and $\eta$ of the final state top and 
$\mu$ originated from the top are shown in the left of Fig.~\ref{Plot:st_bb_phys}. 
For the top distribution the result using the LO generator and the LO* and 
NLO PDFs give a very similar result, being better than the LO PDF result both 
for normalization and for shape due to the suppression of the LO quarks at 
central rapidities. In the case of the $\mu$ (from the top) the distributions 
calculated with the LO generator look better then for the top, since the real 
NLO correction (irradiation if an extra parton ) plays lesser role for the top 
decay products. In this process there is a particular NLO enhancement at 
central rapidity, so it gives a total cross section larger than {\em the truth}. 

We now consider the $b\bar b$ production at the LHC. At LO the process 
consists of three contributions: $gg/q\bar q\to b\bar b$ 
(Flavour Creation, or FCR), $qb\to qb$, where the second b-quark is 
simulated by initial parton showers (Flavour Excitation, or FEX), and the 
QCD $2\to2$ process with massless partons, where the b-quarks arise from parton 
showers\footnote{For example, the total cross-section for the improved LO 
PDFs from Table~\ref{Tab:cs} has three terms: $\sigma_{tot} = 
\sigma_{FCR} + \sigma_{FEX} + \sigma_{GSL}$, where $\sigma_{FCR} = 1.6\;\mu$b, 
$\sigma_{FEX} = 0.57\;\mu$b, and $\sigma_{GSP} = 0.46\;\mu$b -- the total 
cross sections for the FCR, FEX, and GSP processes respectively.} (Gluon 
Splitting, or GSP). The 2nd and 3rd subprocesses have massless 
partons and, thus, soft and collinear singularities. In order to exclude 
the dangerous regions, we apply some reasonable cuts: $p_T(b)>20$ GeV, 
$|\eta(b)|<5.0$, $\Delta R(b,\bar b)>0.5$. At NLO we can not separate the 
subprocesses, so only the FCR process exists at NLO~\cite{Frixione:2003ei}. 
In $\bb$ we probe rather low $x \sim 10^{-3}-10^{-2}$ and the gluon-gluon 
initial state, so the process is sensitive to the small-$x$ divergence in 
the NLO MEs, and the NLO correction is very large. The total cross-sections 
are shown in Table~\ref{Tab:cs}. All the LO calculations are below the 
{\it truth}, but the reduced NLO gluon means that the NLO PDF gives by far 
the worst result. The best absolute prediction is obtained using the LO* 
partons. The differential distributions in terms of $p_T$ and $\eta$ of a 
single $b$ quark are shown on the upper plots and for the pseudo-rapidity and $p_T$ 
of a $b\bar b$ pair on the bottom plots in right of Fig.~\ref{Plot:st_bb_phys}. The LO* 
PDFs do well for the single $b$ rapidity distribution, but underestimate a 
little at high rapidity. The LO and NLO PDFs are similar in shape, but the 
normalisation is worse for NLO and it fails particularly at low $p_T$, i.e small $x$. 
All PDFs obtain roughly the right shape for the 
$\eta(b \bar b)$, except small underestimation at very high rapidity. 
However, for all partons there is a problem with the shape as a function of 
$p_T$. Obviously, all the ratio curves become higher as $p_T$ goes up. 
As for other processes this happens due to the different behaviour of the 
additional parton generated in the NLO matrix element compared to those 
generated by parton showers. In general, we conclude the LO* PDFs give the 
best results in the comparison. 

Another interesting heavy quark production process is the double top quark 
production. The total cross sections are reported in Table~\ref{Tab:cs}. 
At the LHC this process is dominated by the gluon contribution $gg\to t\bar t$. 
For example, $\sigma_{ME[LO]-PDF[LO]}=
\sigma_{gg\to t\bar t}+\sigma_{q\bar q\to t\bar t}=486.9\;pb+74.5\;pb$. 
The LO* PDFs appreciably enlarge the gluonic cross section, namely, 
$\sigma_{ME[LO]-PDF[LO*]}=
\sigma_{gg\to t\bar t}+\sigma_{q\bar q\to t\bar t}=622.1\;pb+77.3\;pb$. 
Again the LO* PDFs gives the best prediction. 

\subsection{Conclusions}
We have examined the effects of varying both the order of the MEs and the 
PDFs when calculating cross-section for hadron colliders. The intention is 
to find the best set of PDFs to use in current Monte Carlo generators. 
A fixed prescription of either LO or NLO PDFs with LO matrix elements is 
unsuccessful, with each significantly wrong in some cases. For LO PDFs this 
is mainly due to the depletion of quarks for $x \sim 0.1-0.001$ and the 
large LO gluon above $x\sim 0.01$, while for NLO partons the smallness in 
some regions compared to LO PDFs is a major problem if the large NLO matrix 
element is absent. To this end we have suggested an optimal set of partons 
for Monte Carlos, which is essentially LO but with modifications to make 
results more NLO-like, and are called LO* PDFs. The NLO coupling is used, 
which is larger at low scales, and helps give a good fit to the data used 
when extracting partons from a global fit. The momentum sum rule is also 
relaxed for the input parton distributions. This allows LO PDFs to be large 
where it is required for them to compensate for missing higher order 
corrections, but not correspondingly depleted elsewhere. 

We have compared the LO, NLO and LO* PDFs in LO calculations to the {\it truth}, 
i.e. full NLO, for a wide variety of processes which probe different types 
of PDF, ranges of $x$ and QCD scales (more examples are available 
in~\cite{Sherstnev:2007nd}). In general, the results are very positive. 
The LO* PDFs nearly always provide the best description compared to the 
{\it truth}, especially for the s-channel processes. This is particularly 
the case in terms of the normalization, but the shape is usually at least as 
good, and sometimes much better, than when using NLO PDFs. It should be 
stressed that no modification of the PDFs can hope to successfully 
reproduce all the features of genuine NLO corrections. In particular we 
noticed the repeating feature that the high-$p_T$ distributions are underestimated 
using the LO generators, and this can only be corrected by the inclusion of 
the emission of a relatively hard additional parton which occurs in the NLO 
matrix element correction. 
A preliminary version of the LO* PDFs, based on fitting the same data as 
in \cite{Martin:2004ir}, is available on request. A more up-to-date version, 
based on a fit to all recent data, and with uncertainty bands for the PDFs, 
will be provided in the MSTW08 PDF set. 

\subsection*{ACKNOWLEDGEMENTS}
We would like to thank J. Butterworth, S. Moch, C. Gwenlan, P. Bartalini, 
M. Cooper-Sarkar, J. Huston, A. Martin, S. Mrenna, T. Sj\"ostrand, 
J. Stirling, G. Watt and B. Webber for helpful discussions. RST would like to 
thank the Royal Society for the award of a University Research Fellowship. 
AS would like to thank the Science and Technology Facilities Council for the 
award of a Responsive RA position.

%% file: s_jet_intro/les-houches-jet-intro.tex
%

This introductory section is intended to help provide the reader with
some background to the current jet-related panorama at the LHC, in
particular as concerns the basic principles and properties of the main
jet algorithms currently in use within the Tevatron and LHC
experiments and in phenomenological and theoretical discussions.
Part of what is described here formed the basis of discussions during
the course of the workshop and subsequent work, but for completeness
additional material is also included.

Several other jet-related sections are present in these proceedings.
Section~\ref{sec:hadronic-accords} outlines two proposals for accords
reached during the workshop, one concerning general nomenclature for
jet finding, the other about the definition of the hadronic
final-state that should be adopted when quoting experimental
measurements. 
Section~\ref{sec:lhprocs_gavin} examines how to measure the
performance of jet algorithms at hadron level and determine optimal
choices in two physics cases, a fictional narrow $Z'$ over a range of
$Z'$ masses, and in top production, providing examples of simple and
complex quark-jet samples.
Section~\ref{sec:rabbertz} examines the performance of jet algorithms
at hadron level in inclusive jet and $Z$+jet production, and in $H\to
gg$ decays for a range of Higgs masses, which provides examples of
gluon-jet samples.
Section~\ref{sec:campanelli} instead examines the performance of jet
algorithms at detector level, using calibrated calorimetric clusters
as input four-vectors, also examining the influence on jet
reconstruction of the presence of a moderate pileup, as expected in
the first years of LHC running.
Other jet-related work that was discussed in part during the workshop,
but was not the focus of workshop-specific investigation includes
studies of non-perturbative effects in jets \cite{Dasgupta:2007wa} and
the use of jet substructure in the discovery of new particles
\cite{Butterworth:2008iy},
as well as methods for dealing with the problem of soft contamination
of jets in the presence of pileup or in heavy-ion
collisions~\cite{Blyth:2006vb,Kodolova:2007hd,D'Enterria:2007xr,Cacciari:2007fd}.
We note also related discussion of jet-finding in the context of the
Tev4LHC workshop~\cite{Albrow:2006rt}, as well as the recent
review~\cite{Ellis:2007ib}. For a review of jet algorithms for $ep$
and $e^+e^-$ colliders, see~\cite{Chekanov:2002rq}.

\subsection{Jet algorithms}
\label{jetalgs_jetalgorithms}

As per the accord in section~\ref{sec:jet-nomenc}, by \emph{jet
  algorithm} we refer to a generic ``recipe'' for taking a set of
particles (or other objects with four-vector like properties) and
obtaining jets. That recipe will usually involve a set 
of parameters (a common example being the jet-radius $R$). The
recipe plus specific values for the parameters provides a fully
specified \emph{jet definition}.

Many hadron-collider jet algorithms are currently being discussed and
used in the literature. This section provides an overview of the basic
principles underlying the jet algorithms for which we are aware of
experimental or theoretical use in the past couple of years.
There are two broad groups of jet algorithms, those based in one form
or another on cones and those that involve repeated
recombination of particles that are nearby in some distance measure.
The nomenclature used to distinguish the flavours of jet algorithm is
currently not always uniform across the field --- that
used here follows the lines set out in \cite{Cacciari:2008gp}.

\subsubsection{Cone algorithms}
\label{sec:cone-algs}

There are many different cone algorithms in use. Most are ``iterative
cones'' (IC). In such algorithms, a seed particle $i$ sets some initial
direction, and one sums the momenta of all particles $j$ within a cone
of radius $R$ around $i$ in azimuthal angle $\phi$ and rapidity $y$ (or
pseudorapidity $\eta$), i.e.\
taking all $j$ such that
\begin{equation}
  \label{eq:deltaij}
  \Delta_{ij}^2 = (y_i - y_j)^2 + (\phi_i-\phi_j)^2 < R^2\,,
\end{equation}
where $y_i$ and $\phi_i$ are respectively the rapidity and azimuth of
particle $i$.
The direction of the resulting sum is then used as a new seed
direction, and one iterates the procedure until the direction of the
resulting cone is stable.

Such a procedure, if applied to an ensemble of many particles can lead
to multiple stable cones that have particles in common (overlapping
cones). Cone algorithms fall into two groups, depending on how they
resolve this issue.

One approach is to start iterating from the particle (or calorimeter
tower) with the largest transverse momentum. Once one has found the
corresponding stable cone, one calls it a jet and removes from the
event all particles contained in that jet. One then takes as a new
seed the hardest particle/tower among those that remain, and uses that
to find the next jet, repeating the procedure until no particles are
left (above some optional threshold). 
A possible name for such algorithms is iterative cones with
progressive removal (IC-PR) of particles. Their use of the hardest
particle in an event gives them the drawback that they are collinear
unsafe:
the splitting of the hardest particle (say $p_1$) into a nearly
collinear pair ($p_{1a}$, $p_{1b}$) can have the consequence that
another, less hard particle, $p_2$ with 
$p_{t,1a},\,p_{t,1b}<p_{t,2}<p_{t,1}$, pointing in a different
direction suddenly becomes the hardest particle in the event, thus
leading to a different final set of jets.

A widespread, simpler variant of IC-PR cone algorithms is one that
does not iterate the cone direction, but rather identifies a fixed
cone (FC) around the seed direction and calls that a jet, starting
from the hardest seed and
progressively removing particles as the jets are identified (thus
FC-PR). It suffers from the same collinear unsafety issue as the IC-PR
algorithms.
Note that IC-PR and FC-PR algorithms are sometimes referred to as
UA1-type cone algorithms, though the algorithm described in the original UA1
reference~\cite{Arnison:1983gw} is somewhat different.

Another approach to the issue of the same particle appearing in many
cones applies if one chooses, as a first stage, to find the stable
cones obtained by iterating from all particles or towers (or those for
example above some threshold $\sim 1-2$GeV).\footnote{In one variant,
  ``ratcheting'' is included, which means that during iteration of a
  cone, all particles included in previous iterations are retained
  even if they are no longer within the geometrical cone.}
One may then run a split--merge (SM) procedure, which merges a pair
of cones if more than a fraction $f$ of the softer cone's transverse
momentum is in common with the harder cone; otherwise the shared
particles are assigned to the cone to which they are closer.%
\footnote{Commonly used values for the overlap threshold parameter are
  $f$ = 0.5, 0.75 (see also recommendations below).}
A possible generic name for such algorithms is IC-SM.
An alternative is to have a ``split-drop'' (SD) procedure where the
non-shared particles that belong to the softer of two overlapping
cones are simply dropped, i.e.\ are left out of jets altogether. 
The exact behaviour of SM and SD procedures depend on the precise
ordering of split and merge steps and a now standard procedure is
described in detail in~\cite{Blazey:2000qt} with the resolution of
some small ambiguities given in \cite{Salam:2007xv}.

IC-SM type algorithms have the drawback that the addition of an extra
soft particle, acting as a new seed, can cause the iterative process
to find a new stable cone. Once passed through the split--merge step
this can lead to the modification of the final jets, thus making
the algorithm infrared unsafe. 
A solution, widely used at Run II of the Tevatron, as recommended
in~\cite{Blazey:2000qt}, was to additionally search for new stable
cones by iterating from midpoints between each pair of stable cones
found in the initial seeded iterations (IC$_{mp}$-SM). While this
reduces the set of configurations for which a soft particle modifies
the final jets, it does not eliminate the problem entirely.
One full solution instead avoids the use of seeds and iterations, and
finds \emph{all} stable cones through some exact procedure. This type
of algorithm is often called a seedless cone (SC, thus SC-SM with a
split--merge procedure). Historically, the computational complexity of
seedless-cone algorithms had made their use impractical for use on
events with realistic numbers of particles, however, recently a
geometrically-based solution was found to this
problem~\cite{Salam:2007xv}.

Cone algorithms with split--merge or split--drop steps are subject
to a phenomenon of ``dark towers''~\cite{Ellis:2001aa}, regions
of hard energy flow that are not clustered into any jet. A solution to
this proposed in~\cite{Ellis:2001aa} 
--- referred to as the ``searchcone'' --- works around
the problem by using a smaller radius to find stable cones and then
expands the cones to their full radius without further iteration
before passing them to the SM procedure. 
It was subsequently discovered that this reintroduces IR safety
issues~\cite{Albrow:2006rt}, and an alternative solution is a
multi-pass algorithm, one that runs the cone algorithm again on the
set of all particles that do not make it into any of the ``first-pass''
jets (this can be repeated over and over until no particles are left
unclustered).

  \begin{table}
    \centering
    \begin{tabular}{|l|l|c|l|p{4cm}|}
      \hline
      Algorithm & Type  & IRC status  & Ref. & Notes \\
      \hline
      \hline
      inclusive $k_t$     & 
      SR$_{p=1}$ &
      OK        &
      \cite{Catani:1991hj,Catani:1993hr,Ellis:1993tq} &
      also has exclusive variant
      \\ \hline
      flavour $k_t$     & 
      SR$_{p=1}$ &
      OK        &
      \cite{Banfi:2006hf} &
      $d_{ij}$ and $d_{iB}$ modified when $i$ or $j$ is ``flavoured''
      \\ \hline
      Cambridge/Aachen &
      SR$_{p=0}$ &
      OK        &
      \cite{Dokshitzer:1997in,Wobisch:1998wt} &
      \\ \hline
      anti-$k_t$ %
      &
      SR$_{p=-1}$ &
      OK         &
      \cite{Cacciari:2008gp} &
      \\ \hline
      SISCone     &
      SC-SM  &
      OK     &
      \cite{Salam:2007xv}&
      multipass, with optional cut on stable cone $p_t$
      \\ \hline
      CDF JetClu      &
      IC$_r$-SM   &
      IR$_{2+1}$  &
      \cite{Abe:1991ui}&
      \\ \hline
      CDF MidPoint cone &
      IC$_{mp}$-SM  &
      IR$_{3+1}$    &
      \cite{Blazey:2000qt} &
      \\ \hline
      CDF MidPoint searchcone &
      IC$_{se,mp}$-SM  &
      IR$_{2+1}$    &
      \cite{Ellis:2001aa} &
      \\ \hline
      D0 Run II cone     &
      IC$_{mp}$-SM  &
      IR$_{3+1}$    &
      \cite{Blazey:2000qt} &
      no seed threshold, but cut on cone $p_t$
      \\ \hline
      ATLAS Cone  &
      IC-SM       &
      IR$_{2+1}$  &
      &
      \\ \hline
      PxCone     &
      IC$_{mp}$-SD  &
      IR$_{3+1}$    &
      &
      no seed threshold, but cut on cone $p_t$, 
      \\ \hline
      CMS Iterative Cone     &
      IC-PR  &
      Coll$_{3+1}$    &
      \cite{Bayatian:2006zz,CMS-Jet-Algs-Prelim} &
      \\ \hline
      PyCell/CellJet (from Pythia)     &
      FC-PR  &
      Coll$_{3+1}$    &
      \cite{Sjostrand:2006za}
      &
      \\ \hline
      GetJet (from ISAJET) &
      FC-PR  &
      Coll$_{3+1}$    &
      &
      \\ \hline
    \end{tabular}
    \caption{\small Overview of some jet algorithms used in experimental
      or theoretical work in hadronic collisions in the past couple of
      years. SR$_{p=x}=$
      sequential recombination (with $p=-1,0,1$ characterising the exponent
      of the transverse momentum scale, 
      eq.~(\ref{eq:genkt})); SC = seedless cone (finds all cones); IC = 
      iterative cone (with midpoints 
      $mp$, ratcheting $r$, searchcone $se$), using either
      split--merge (SM), split--drop (SD) or progressive removal (PR)
      in order to address issues with overlapping stable cones;
      FC = fixed-cone.
      In the characterisation of infrared and collinear (IRC) safety
      properties (for the algorithm as applied to particles), 
      IR$_{n+1}$ indicates that given $n$ hard particles in a common
      neighbourhood, the addition of 1 extra soft particle can modify
      the number of final hard jets; Coll$_{n+1}$ indicates that given
      $n$ hard 
      particles in a common 
      neighbourhood, the collinear splitting of one of the particles
      can modify the number of final hard jets.
      Where an algorithm is labelled with the name of an experiment,
      this does not imply that it is the only or favoured one of the
      above algorithms used within that experiment.
      Note that certain computer codes for jet-finding first project
      particles onto modelled calorimeters.
      \label{jetalgs_conefeatures}
    }
  \end{table}

\subsubsection{2$\to$1 Sequential recombination}

Sequential recombination (SR) algorithms introduce distances $d_{ij}$
between entities (particles, pseudojets) $i$ and $j$ and $d_{iB}$
between entity $i$ and the beam (B). The (inclusive) clustering
proceeds by identifying the smallest of the distances and if it is a
$d_{ij}$ recombining entities $i$ and $j$, while if it is $d_{iB}$
calling $i$ a jet and removing it from the list of entities.  The
distances are recalculated and the procedure repeated until no
entities are left.

The distance measures for several algorithms are of the form
\begin{subequations}
  \label{eq:genkt}
  \begin{align}
    d_{ij} &= \min(k_{ti}^{2p}, k_{tj}^{2p}) \frac{\Delta_{ij}^2}{R^2}\,,\\
    d_{iB} &= k_{ti}^{2p}\,,
  \end{align}
\end{subequations}
where $\Delta_{ij}^2$ was defined in (\ref{eq:deltaij}) and
$k_{ti}$  is the transverse momentum of particle $i$. Here $R$ is the
jet-radius parameter, while $p$ parametrises the type of algorithm. 
For $p=1$ one has the inclusive $k_t$ algorithm as defined
in~\cite{Ellis:1993tq}, while with $p=0$ one obtains the
Cambridge/Aachen algorithm as defined in \cite{Wobisch:1998wt}. Both
are related to corresponding ``exclusive'' algorithms ($k_t$
\cite{Catani:1991hj,Catani:1993hr}, Cambridge~\cite{Dokshitzer:1997in},
and also \cite{Pierce:1998sm}) with similar or identical distance measures
but additional stopping conditions.
A recent addition to the SR class is the anti-$k_t$
algorithm, with $p=-1$~\cite{Cacciari:2008gp}. Together with the PR cones, it has
the property that soft radiation does not affect the boundary of the
jet, leading to a high proportion of circular jets with actual radius
$R$. This property does not hold for SM and SD cones, nor SR
algorithms with $p\ge0$.

Other sequential recombination algorithms, used mainly in $e^+e^-$ and
DIS collisions, include the JADE
algorithm~\cite{Bethke:1988zc,Bartel:1986ua} which simply has a
different distance measure, and the ARCLUS
algorithm~\cite{Lonnblad:1992qd} which performs $3\to 2$
recombinations (the inverse of a dipole shower).

\subsubsection{General remarks}

A list of algorithms used in experimental or theoretical studies in
the past couple of years is given in table~\ref{jetalgs_conefeatures}.
Where possible references are provided, but some algorithms have not been
the subject of specific publications, while for others the description
in the literature may only be partial. Thus in some cases, to obtain
the full definition of the algorithm it may be advisable to consult
the corresponding computer code.

A point to be noted is that as well as differing in the underlying
recipe for choosing which particles to combine, jet algorithms can also
differ in the scheme used to recombine particles, for example direct
4-momentum addition (known as the $E$-scheme), or $E_T$ weighted
averaging of $\eta$ and $\phi$. In the past decade recommendations
have converged on the $E$-scheme (see especially the Tevatron Run-II
workshop recommendations~\cite{Blazey:2000qt}), though this is not
used by default in all algorithms of table~\ref{jetalgs_conefeatures}.

As discussed in section~\ref{sec:cone-algs} many of the algorithms
currently in used are either infrared or collinear unsafe. For an
algorithm labeled IR$_{n+1}$ or Coll$_{n+1}$, jet observables that
are non-zero starting with $m$ partons in the final state (or $m-1$
partons and one $W/Z$ boson) will be divergent in perturbation theory
starting from N$^{n-m+2}$LO. Given that these are usually
single-logarithmic divergences, the physics impact is that N$^{n-m}$LO
is then the last order that can be reliably calculated in perturbation
theory (as discussed for example in detail in \cite{Salam:2007xv}).

Because of the perturbative divergences and other non-perturbative
issues that arise with non infrared and collinear safe algorithms,
there have been repeated recommendations and accords, dating back to
the Snowmass accord~\cite{Huth:1990mi}, to use just infrared and
collinear safe jet algorithms.
This recommendation takes on particular importance at the LHC, because
multi-jet configurations, which will be far more widespread than at
previous colliders, are particularly sensitive to infrared and
collinear safety issues. 
Furthermore there is very significant investment by the theoretical
community in multi-leg NLO computations (see for example the
proceedings of the NLO Multi-leg working group of this workshop), and
the benefit to be had from such calculations will largely be
squandered if infrared or collinear unsafe jet algorithms are used for
analyses.
The set of IRC-safe algorithms that have been the subject of some
degree of recent study
includes $k_t$, Cambridge/Aachen,
SISCone (which can be used as a replacement for IC-SM type algorithms)
and anti-$k_t$ (which is a candidate for replacing IC-PR type
algorithms).


\subsubsection{Jet algorithm packages}

Given the many jet algorithms that are in use, and the interest in
being able to easily compare them, two packages have emerged that
provide uniform access to multiple jet algorithms. 
{\tt FastJet}~\cite{Cacciari:2005hq,FastJetWeb}, originally written to
implement fast strategies for sequential recombination, also has a
``plugin'' mechanism to wrap external algorithms and it provides a
number of cone algorithms in this manner, including
SISCone~\cite{Salam:2007xv}.
{\tt SpartyJet} \cite{spartyjet} provides a wrapper to the {\tt
  FastJet} algorithm implementations (and through it to SISCone) as well
as to a number of cone algorithms, together with specific
interfaces for the ATLAS and CDF environments.
Both packages are under active development and include various
features beyond what is described here, and so for up to date details of
what they contain, readers are referred to the corresponding web
pages.

\subsection{Validation of jet-finding}


During the Les Houches workshop, a validation protocol was defined in
order to ensure that all participants were using identical jet
algorithms and in the same way. For this purpose, a sample of 1000
events was simulated with Pythia 6.4~\cite{Sjostrand:2006za}, for the
production and subsequent hadronic decay of a $Z'$, $Z' \to q\bar{q}$
with $M_{Z'}=1000$~GeV. 
This was run through the different participants' jet software for each
of the relevant jet definitions, and it was checked that they obtained
identical sets of jets.\footnote{This statement holds for comparisons
  carried out with double-precision inputs; where, for data-storage
  efficiency reasons, inputs were converted to single precision,
  slight differences occasionally arose.}

The following jet algorithms were used in the jet validation
\begin{itemize}
\item $k_t$
\item Cambridge/Aachen
\item Anti-$k_t$ (added subsequent to the workshop)
\item SISCone
\item CDF Midpoint cone
\end{itemize}
For each, one uses values of $R$ from $R_{\rm min}=0.3$ to $R_{\rm
  max}=1.0$ in steps of $\Delta R =0.1$. In the two SM-type cone
algorithms, the SM overlap threshold $f$ was set to $0.75$. This
choice is recommended more generally because smaller values (including
the quite common $f=0.50$) have been seen to lead to successive
merging of cones, leading to ``monster-jets'' (see e.g.\
\cite{Cacciari:2008gn}). 

Readers who wish to carry out the validation themselves may obtain
the event sample and further details from
\begin{quote}\small
  \url{http://www.lpthe.jussieu.fr/~salam/les-houches-07/validation.php }
\end{quote}
together with reference results files and related tools.


%% file: s_jet_accords/les-houches-jet-accords.tex
\subsection{Jet nomenclature}
\label{sec:jet-nomenc}

\begin{figure}[th]
        \centering
       	\includegraphics[width=10cm]{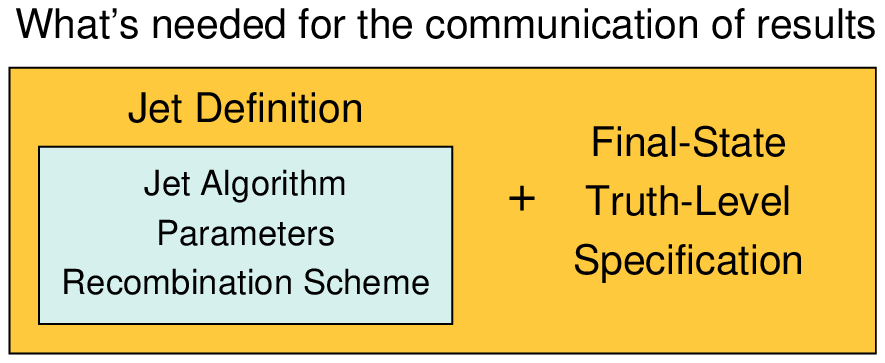}
        \caption{\small \label{jetalgs_nomenclature}
A summary of the elements needed to communicate 
jet observables in a non-ambiguous way.
}
\end{figure}

In this section we aim to establish a common and non-ambiguous 
nomenclature to be used when discussing jet physics. 
Such a basis is needed for the communication of experimental
results, in order to ensure that they can be reproduced exactly,
or that matching theory predictions can be made.
We propose that the following elements should always be specified
in experimental publications:
\begin{itemize}
\item {\bf The jet definition} which specifies all details of the procedure
  by which an arbitrary set of four-momenta from physical objects 
  is mapped into a set of jets.
  The jet definition is composed of a {\bf jet algorithm} 
  (e.g.\ the inclusive longitudinally boost-invariant $k_T$ algorithm), 
  together with {\em all} its {\bf parameters} 
  (e.g.\ the jet-radius parameter $R$,
  the split--merge overlap threshold $f$, the seed-threshold $p_T$ cut, etc.) 
  and the {\bf recombination scheme} (e.g.\ the four-vector 
  recombination scheme or ``E-scheme'') according to which the 
  four-momenta are recombined during the clustering procedure.
  We recommend that a reference to a {\em full} specification of the
  jet algorithm is given. 
  If this is not available, the jet algorithm should be described in detail.

\item {\bf The final state (``truth-level'') specification}.
  Consistent comparisons between experimental results, or between experimental
  results and Monte Carlo simulations, are only possible if the jet 
  definition is supplemented with an exact specification of the set of the 
  physical objects to which it was applied, or to which a quoted 
  jet measurement has been corrected.
  This could e.g.\  be the set of momenta of all hadrons with a 
  lifetime above some threshold.
  Discussions and recommendations of possible final state choices 
  are given below in section~\ref{jetalgs_finalstate}.

\end{itemize}
This nomenclature proposal
is summarised graphically 
in Fig.~\ref{jetalgs_nomenclature}.

\subsection{Final state truth level}
\label{jetalgs_finalstate}

Whenever experiments present ``corrected'' results for given
jet observables, the question arises 
{\em ``What exactly have these results been corrected for?''},
or in other words {\em ``On which set of four-vectors are the 
quoted results of this jet measurement defined?''}.
These questions address the ``truth-level'' to which experimental
results correspond to.
A detailed answer to this question is relevant since supposedly 
minor differences can be significant, and they certainly are for 
precision jet measurements\footnote{Note that the ambiguity addressed 
here does not include the jet definition, which is supposed to have 
already been agreed upon and fully specified.}.
In the history of jet physics at particle colliders, many different 
choices have been made on how jet results were presented.
Experiments have corrected their jet results
\begin{itemize}
\item back to the leading order matrix-elements in a Monte Carlo.
    The jets are supposed to correspond to the partons from the 
    $2\rightarrow 2$ scattering process. 
\item back to the level after the parton shower in a Monte Carlo.
    The jets are supposed to correspond to the result of the
    purely perturbative phase of the hadronic reaction.
\item back to the level of stable particles in a Monte Carlo,
      but excluding the particles from the ``underlying event''.  
\item for all detector effects and, in addition, also for the energies 
     observed in interactions triggered by ``minimum bias'' triggers.
     The latter contribution is supposed to correspond to the 
     ``underlying event''.
\item for all detector effects and nothing else. The corrected jet
     results correspond to jets defined on all (stable) particles
     from the hadronic interaction.
\end{itemize}
It would be useful for the LHC and the Tevatron experiments to have a 
common definition of what they call the ``truth'' final-state particle level 
(specifically for jets). 
While we cannot enforce any agreement, we can provide a set of 
recommendations, and make the following proposals:
\begin{itemize}
    \item The truth input to the jet clustering 
should always be physical (i.e.\ observable) final-state particles, 
not any kind of model-dependent partons (neither from a matrix-element
nor from a parton-shower). 

    \item For similar reasons, the final-state particles 
should include everything from the main hadronic scatter. 
Therefore the underlying event (defined as additional partonic 
interactions from the same hadron-hadron interaction plus interactions
of the hadron remnants) is included.
This is part of the hadronic interaction and cannot be unambiguously 
separated from the hard subprocess (see, however, next subsection).

   \item The contributions from pile-up due to additional
   hadronic collisions in the same bunch crossing, recorded in the same 
   event, should not be included.
   In other words, the jet observable should be corrected for
   contributions from multiple hadron interactions.

    \item 
A  standard lifetime cut on what is considered to be ``final state'' 
should be agreed upon. 
A lifetime of 10\,ps is used elsewhere, and we also recommend this value: 
only hadrons with a shorter lifetime will be allowed to decay in the 
Monte Carlo generators. 
All other particles will be considered to be stable.

    \item Neutrinos, muons and electrons 
     from hadronic decays should be included as part of the final state.

    \item However, prompt muons, electrons (and radiated photons), 
    neutrinos and photons 
    are excluded from the definition of the final state. 
    The same applies to the decay products of prompt taus. 

    \item The jet algorithm should be given as input the full physical four-vectors. 
    How it treats them is part of the jet definition and the recombination 
    scheme.

\end{itemize}
We acknowledge that these recommendations may
not be useful in all circumstances. During the process of 
understanding and calibrating detectors, other definitions
(e.g.\ including only visible energy in the calorimeter)
may be needed. 
But whenever a jet measurement is presented or a jet observable is 
quoted, we suggest that the jets it refers to are based on a specific 
(and clearly stated) jet definition and the final-state truth particle 
definition recommended above.

\subsection{A level behind the truth: Partons}
\label{sec:jets_presentation}

It should be noted that the above definitions about the final state 
truth level also apply to theoretical calculations.  
Some theoretical calculations are implemented in
Monte Carlo event generators, including the modelling of 
non-perturbative processes (hadronization and underlying event).
These can directly be compared to experimental results
that are obtained according to the recommendations from the
previous section. 

Other calculations provide purely perturbative results
(typically at next-to-leading order in the strong coupling constant,
sometimes accompanied by resummations of leading logarithms).
These results correspond to the ``parton level'' of the jet observable.
When trying to compare a perturbative calculation to an experimental 
result, one needs to at least estimate the size of the non-perturbative
corrections (consisting of underlying event and hadronization corrections).
Typically, these are obtained using Monte Carlo event generators.
We strongly recommend that each experiment should determine and publish
its best estimate of non-perturbative corrections together
with the data.
It should be kept in mind that these corrections should always be quoted
separately and not be applied to the data, but only to the perturbative
calculations.
Experiment and theory should meet at the level of an observable.
This seems to be an established procedure, which is used in most 
jet analyses at LEP, HERA, and also in Run~II of the Tevatron.


%% file: s_cacciari/les-houches-jet-algorithms.tex
\newcommand{\bea}{\begin{eqnarray}}
\newcommand{\eea}{\end{eqnarray}}
\newcommand{\be}{\begin{equation}}
\newcommand{\ee}{\end{equation}}

\newcommand{\comment}[1]{{\bf #1}}

\newcommand{\lp}{\left(}
\newcommand{\rp}{\right)}
\newcommand{\lc}{\left[}
\newcommand{\rc}{\right]}
\newcommand{\ttbar}{$t\bar{t}~$}
\newcommand{\Qa}[1]{$Q_{f=#1}^{w}(R)~$}
\newcommand{\Qb}{$Q_{w=1.25\sqrt{M}}^{f}(R)~$}

\title{Quantifying the performance of jet algorithms at the LHC}

\author{M. Cacciari$^1$, J. Rojo$^1$, G. Salam$^1$ and
G. Soyez$^2$}
\institute{$^1$LPTHE, CNRS UMR 7589; UPMC Univ. Paris 6;
Universit\'e Paris Diderot (Paris VII), 75252 Paris Cedex 05, France
\\$^2$ Brookhaven National Laboratory, Upton, NY, USA}

\maketitle

\subsection{General strategy}
\label{sec:zprime-top-performance}

The performance of a given jet algorithm depends on its parameters,
like the radius $R$, but it also depends on the specific 
process under consideration. For
example, a jet algorithm that gives good results in a simple
dijet environment
might perform less well in a more complex multi-jet situation.
In this contribution we wish to quantify the extent to which this is
the case in the context of a couple of illustrative reconstruction
tasks.
This is intended to help cast light on the following question: should
the LHC experiments devote the majority of their effort to calibrating
as best as possible just one or two jet definitions? Or should they
instead devote effort towards flexibility in their choice of jet
definition, so as to be able to adapt it to each specific analysis?

One of the main issues addressed in examining this question is that of
how, simply but generally, to quantify the relative performance of
different jet algorithms.
This physics analyses used as examples will be the reconstruction of
massive particles, because such tasks are central both to Standard
Model and to discovery physics at the LHC.
As quality measures we shall use the mass resolution, and the signal
size for fixed detector mass resolution, both defined in such a way as
to be insensitive to the exact signal shape (which depends
significantly on the jet definition).
As test cases we will take a hypothetical $Z'$ for different values of
its mass, and the $W$ boson and top quark in fully hadronic decays of
$t\bar{t}$ events.

A point that we wish to emphasise is that we have purposefully avoided
quality measures, used in the past, that consider the relation between
jets and the hard partons produced at matrix-element level in a
parton-shower Monte Carlo. This is because the relation between those
two concepts depends as much on approximations used in the parton
showering, as on the jet definition. Indeed in modern tools such as
MC@NLO~\cite{Frixione:2002ik} or POWHEG~\cite{Nason:2004rx} it becomes
impossible, even programmatically, to identify the single parton to
which one would want to relate the jet.
Note however that addressing the issue of the performance of jet
algorithms in contexts other than kinematic reconstructions (e.g.\ for
the inclusive jet spectrum) would require rather different strategies
than those we use here (see for example \cite{Dasgupta:2007wa} and
section~\ref{sec:rabbertz}). 
A strategy related to ours, to assess the performance of jet
algorithms based on the Higgs mass reconstruction from the invariant
mass of gluon jets in $H\to gg$ can be found in
Sect.~\ref{sec:rabbertz}.

We note that we do not address issues of experimental relevance
like the reconstruction efficiency of different jet algorithms
after detector simulation, which however are discussed in the contribution
of section~\ref{sec:campanelli}. 

\subsection{Figures of merit}\label{sec:jetalgs_figs}

We start by defining the figures of merit that quantify the quality of
the heavy object mass reconstruction through jet clustering
algorithms.

\begin{figure}[t]
  \centering
  \includegraphics[width=0.48\textwidth]{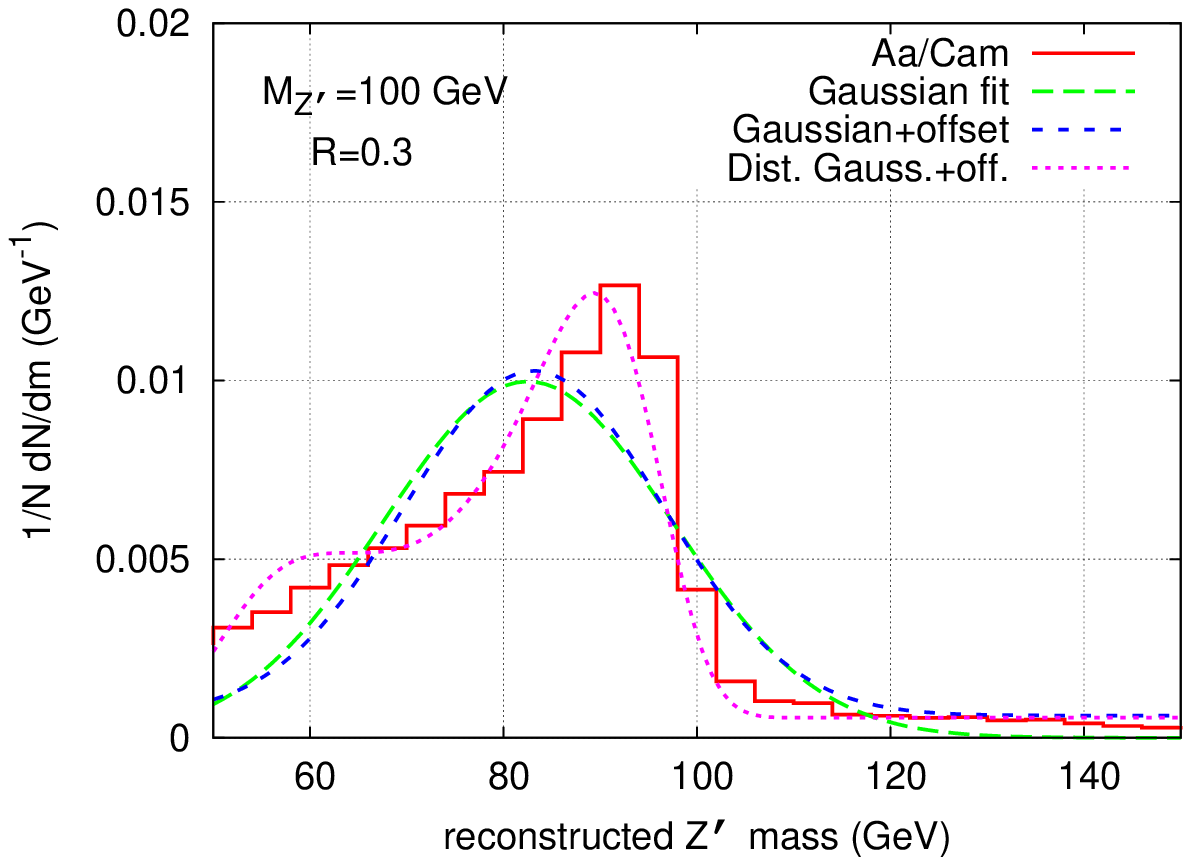}
  \includegraphics[width=0.48\textwidth]{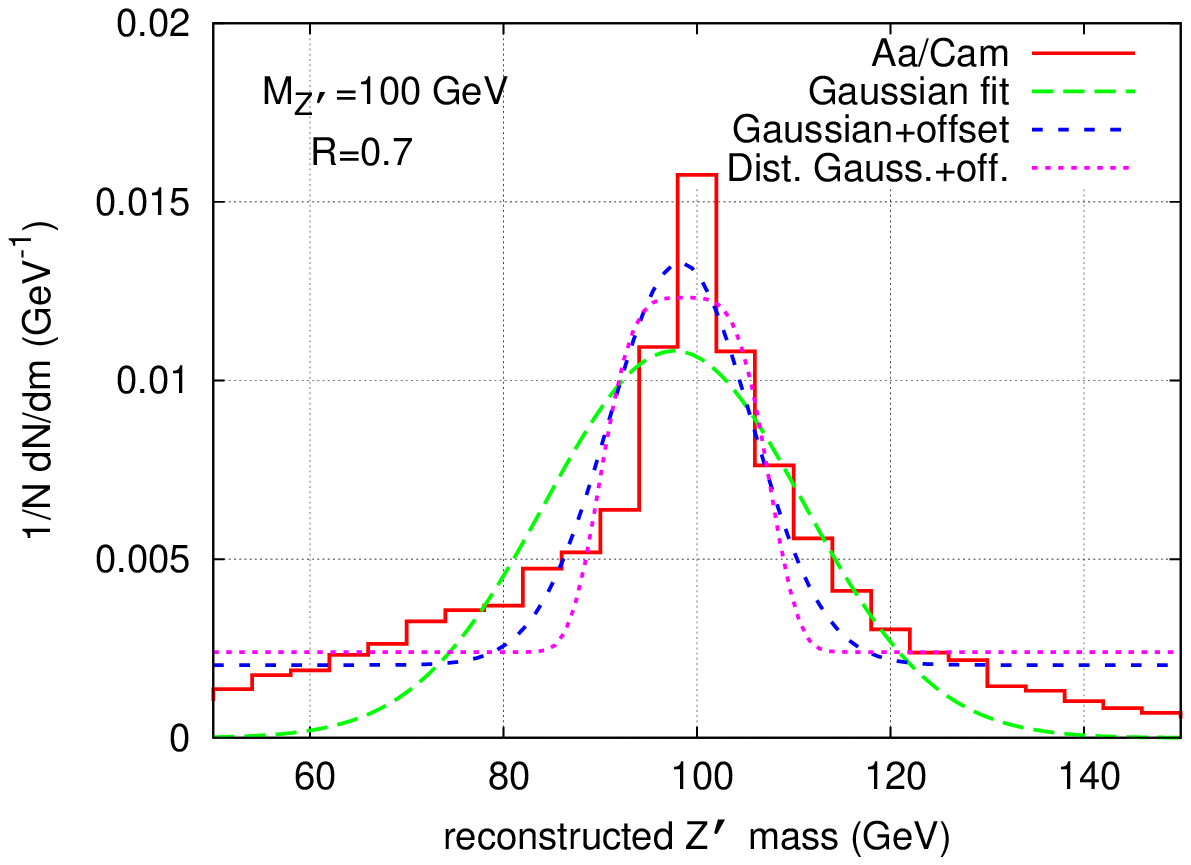}
  \caption{\small The mass of the reconstructed $Z'$ boson in the
$M_{Z'}=100$ case with the Cambridge/Aachen algorithm for
$R=0.7$ (left) and $R=0.3$ (right), together with various fits
of simple probability distributions.}
  \label{fig:jetalgs_fits} 
\end{figure}

We wish to avoid assumptions on the underlying shape of the
invariant mass distribution that we are 
reconstructing, such as whether it is Gaussian, asymmetric or has a pedestal,
since in general the reconstructed mass distributions cannot be described 
by simple functional forms. This is illustrated in
Fig.~\ref{fig:jetalgs_fits}, where different functions are
fitted to two reconstructed mass spectra from the $Z'\to q\bar{q}$ samples
for two different values of R.
One sees that even in the more symmetric situation, it is
difficult to reproduce it properly with any simple functional form.

Instead we shall use figures of merit that relate to
the maximisation of the signal over background ratio (more precisely,
$S/\sqrt{B}$), for the simplifying assumption that the background is
flat and is not affected by the jet clustering procedure.
Specifically, we propose the following two measures:
\begin{enumerate}
\item {\bf $Q_{f=z}^{w}(R)$}: The width of the smallest (reconstructed) mass window that
  contains a fraction $f=z$ of the generated massive objects,\footnote{Note that
in general the number of generated massive objects
differs from the total number of events, for example
if in the \ttbar samples we have $N_{\rm ev}=10^5$, the
number of generated W bosons (and top quarks) is $N_{\rm W}=2\cdot 10^5$.}
that is
\be
\label{eq:q1}
f = 
 \lp \frac{{\rm \#~reconstructed~massive~objects~in~window~of~width~}w}{
\rm Total ~\#~generated~massive~objects}\rp \ .
\ee
A jet definition that is more effective in reconstructing the majority
of massive objects within a narrow mass peak gives a lower value for
$Q_{f=z}^{w}(R)$, and is therefore a ``better'' definition.
The
  value that we will use for the fraction $f$ 
 will be adjusted in
  order to have around 25\% of the reconstructed objects inside the
  window.\footnote{The approximate fraction of events that pass the
event selection
  cuts for each physical process can be seen in Table
  \ref{tab:jetalgs_frac}, together with the value for the fraction $z$
  ensuring that approximately one quarter of the successfully
  reconstructed heavy objects are inside the window.}
  
\item {\bf $Q_{w=x\sqrt{M}}^{f}(R)$}: To compute this quality measure, 
first we displace over the mass distribution a window  of fixed
width given by $w=x\sqrt{M}$, where $M$ is the nominal heavy object mass
that is being reconstructed\footnote{Note that
we avoid using the reconstructed mass $M_{\rm reco}$, obtained from
the mean of the distribution for example, since in general
it depends strongly on the jet definition.}  until we find 
the maximum number of events of the mass distribution contained in it.
Then the figure of merit is given in terms of the ratio of this number
of events with respect to the
total number of generated events, 
\be
Q_{w=x\sqrt{M}}^{f}(R) \equiv
 \lp \frac{{\rm Max ~\#~reconstructed~massive~objects~in~window~of~width~} w=x\sqrt{M} }{
\rm Total ~\#~generated~massive~objects}\rp^{-1} \ ,
\ee
where we take the inverse so that the optimal result is a minimum of
$Q_{w=x\sqrt{M}}^{f}(R)$, as in the previous case.

 The default choice that will be used is $x=1.25$, that is
$w=1.25\sqrt{M}$ (for compactness we omit the dimensions on $x$, which
are to be understood as (GeV)$^{1/2}$).
This particular choice is motivated by experimental
considerations of the CMS and ATLAS experiments, in particular
the default value corresponds to the
jet resolution in CMS. This means that the default values
that will be used through this contribution will be 
$w=1.25\sqrt{M_{Z'}}$ for the
$Z'$ samples, $w=1.25\sqrt{M_W}\sim 10$~GeV for the W mass distributions and
$w =1.25\sqrt{M_W}\sim 15$~GeV for the top quark mass distributions.
\end{enumerate}

\begin{figure}[t]
  \centering
  \includegraphics[width=10cm]{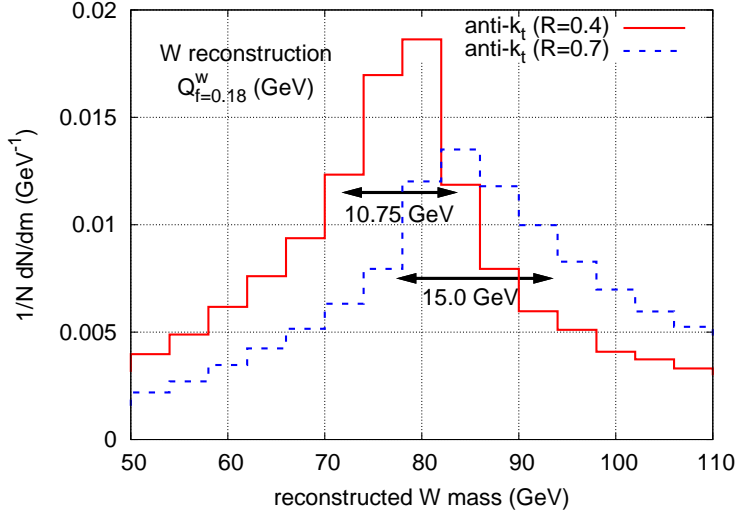}
  \caption{\small \label{jetalgs_measure_example} The quality measure
    $Q_{f=0.18}^{w}(R)$ in the case of $W$ mass
    reconstruction for hadronic \ttbar~ production.}
\end{figure}

In tests of a range of possible quality measures for mass
reconstructions (including Gaussian fits, and the width at half peak
height), the above two choices have been found to be the least
sensitive to the precise shape of the reconstructed mass distribution,
as well as to any kind of binning. Another encouraging feature,
which will be seen below, is that the two measures both lead to similar
conclusions on the optimal algorithms and $R$ values.

As an example of the behaviour of
these quality measures in an actual
mass distribution, we show in Fig.~\ref{jetalgs_measure_example} the
quality measure $Q_{f=0.18}^{w}(R)$ 
in the case of $W$ mass reconstruction for hadronic \ttbar~ production.
We observe that indeed in the case where the mass reconstruction is
clearly poorer (blue dashed histogram), the value of 
$Q_{f=0.18}^{w}(R)$ is sizably larger.

With the aim of better comparing the
performances of different jet definitions,
we can establish a mapping between variations 
of these quality measures and variations in effective
luminosity needed to achieve constant signal-over-background ratio
for the mass peak reconstruction, working with the
assumption that the background is flat and constant,
and not affected by the
jet clustering. We define the effective power to discriminate
the signal with respect to the background $\Sigma^{\rm eff}$ 
for a given jet definition (JA,$R$) as
\be
\Sigma^{\rm eff}\lp {\rm JA},R\rp \equiv \frac{N_{\rm signal}}{
\sqrt{N_{\rm back}}} \ ,
\ee 
where $N_{\rm signal}$ and $N_{\rm back}$ are respectively the number
of signal and background events. 
We can establish the following matching between variations
in quality measures and in the effective luminosity ratios 
$\rho_{\mathcal{L}}$ as follows. Suppose that a quality measure calculated
with (JA$_2$,$R_2$) gives a worse (i.e. larger) result than with 
(JA$_1$,$R_1$).

\begin{itemize}
\item In the case of \Qa{z} a larger value of this quality measure 
(i.e.\ a larger window width) will correspond to a larger 
number of background events for a given, fixed number of signal events.
The jet definition $(\mathrm{JA}_1,R_1)$ will then need a lower
luminosity to deliver the same effective discriminating power as
$(\mathrm{JA}_2,R_2)$, since it deals with a smaller number of
background events.
So if we define 
\be
r_w \equiv \frac{Q_{f=z}^{w}\lp{\rm JA_2}, R_2\rp}
     {Q_{f=z}^{w}\lp{\rm JA_1}, R_1 \rp}
=   \frac{N_{\rm back}\lp {\rm JA_2}, R_2 \rp}
        {N_{\rm back}\lp {\rm JA_1}, R_1 \rp}
> 1\,,   
\ee
then at equal luminosity the discriminating power for $(\mathrm{JA}_1,R_1)$ will  
be better by a factor
\be
\label{eq:jet-discr-power}
   \frac{\Sigma^{\rm eff}\lp {\rm JA_1}, R_1 \rp}
        {\Sigma^{\rm eff}\lp {\rm JA_2}, R_2 \rp}
= \sqrt{r_w} \,, 
\ee
or equivalently the same discriminating power as $(\mathrm{JA}_2,R_2)$
can be obtained with a smaller luminosity ${\cal L}_1 = \rho_{\cal L} {\cal
  L}_2$, where
$\rho_{\cal L}$ is given by the inverse square of the ratio
eq.~(\ref{eq:jet-discr-power}).
\be
\rho_{\cal L} =  \frac{1}{r_w} \; .
\ee

\item In the case of $Q_{w=x\sqrt{M}}^{f}(R)~$ it is instead the number
of signal events that varies when the quality measure changes. Defining
\be
r_f \equiv  \frac{Q_{w=x\sqrt{M}}^{f}\lp {\rm JA_2}, R_2\rp}
     {Q_{w=x\sqrt{M}}^{f}\lp {\rm JA_1}, R_1\rp} 
=
  \frac{N_{\rm signal}\lp {\rm JA_1},R_1 \rp}
    {N_{\rm signal}\lp {\rm JA_2},R_2 \rp}
> 1\,,
\ee
then at equal luminosity 
the discriminating power for $(\mathrm{JA}_1,R_1)$ will be better by a factor
\be
\label{eq:jet-discr-power-2}
   \frac{\Sigma^{\rm eff}\lp {\rm JA_1}, R_1 \rp}
        {\Sigma^{\rm eff}\lp {\rm JA_2}, R_2 \rp}
= r_f \, ,
\ee
or equivalently the same discriminating power as $(\mathrm{JA}_2,R_2)$
can be obtained with a smaller luminosity ${\cal L}_1 = \rho_{\cal L} {\cal
  L}_2$, where $\rho_{\cal L}$ is now given by the inverse square of
the ratio eq.~(\ref{eq:jet-discr-power-2})
\be
\rho_{\cal L}  = \frac{1}{r_f^2} \,.
\ee
\end{itemize}
In the remainder of this study we shall see that for the
processes under consideration, the two quality measures 
indicate similar effective luminosity improvements to be gained by going
from $(\mathrm{JA}_2,R_2)$ to $(\mathrm{JA}_1,R_1)$, once one takes into
account the different functional dependence indicated above (e.g.\ a
gain (i.e. {\sl smaller}) by a factor of 2 in $Q_{w=x\sqrt{M}}^{f}(R)~$ 
should correspond with good approximation to a gain of a factor of 
$ 2^2= 4$ in \Qa{z}).

\subsection{Jet algorithms}\label{ref:jetalgs_jetalgs}
With the help of the quality measures defined in the
previous section,
we will study the performance of the following jet algorithms:
\begin{enumerate}
\item longitudinally invariant inclusive $k_t$ algorithm \cite{Catani:1991hj,Catani:1993hr,Ellis:1993tq}.
\item Cambridge/Aachen algorithm \cite{Dokshitzer:1997in,Wobisch:1998wt}.
\item Anti-$k_t$ algorithm \cite{Cacciari:2008gp}.
\item SISCone \cite{Salam:2007xv} 
with split--merge overlap threshold $f=0.75$, an
  infinite number of passes and no $p_T$ cut on stable cones.
\item The Midpoint cone algorithm in CDF's implementation 
\cite{Blazey:2000qt} with an area
    fraction of 1 and a maximum number of iterations of 100,
  split--merge overlap threshold $f=0.75$ and seed threshold of 1 GeV.
\end{enumerate}
In every case, we will add four-momenta using a $E$-scheme (4-vector)
recombination. 
Each jet algorithm will be run with several values of 
$R$ varying by steps of 0.1 within a range  $\lc R_{\rm min},
R_{\rm max}\rc$ adapted to observe a
well defined preferred $R_{\rm best}$ value. 
Practically, we will have  $R_{\rm min}=0.3$ and $R_{\rm max}=1.3$
for the $Z'$ analysis and
 $R_{\rm min}=0.1$ and $R_{\rm max}=1.0$ for the \ttbar samples.

Note that we have fixed the value of the overlap parameter of the
cone algorithms to $f=0.75$. This rather large value is motivated
(see {\em e.g.} \cite{Cacciari:2008gn}) by the fact that ``monster jets''
can appear for smaller values of $f$.
For sequential recombination
clustering algorithms we use their inclusive longitudinally-invariant versions,
suited for hadronic collisions. The jet algorithms have been
obtained via the implementations and/or plugins in the {\tt FastJet}
package \cite{Cacciari:2005hq}.

The infrared-unsafe CDF midpoint algorithm is only included here for
legacy comparison purposes.

\subsection{Physical processes}\label{ref:jetalgs_process}
We consider the following physical processes:
$Z' \to q\bar{q}$ for various values of $M_{Z'}$ and fully
hadronic  \ttbar~ production, and we reconstruct the
mass of the $Z'$ boson and that of the $W$ boson and the top quark
to assess the performance of the jet algorithms
described in Sect. \ref{ref:jetalgs_jetalgs}. 
We should emphasise again that the performance of a given
jet definition depends on the process under consideration,
thus it is  important to study different
jet algorithms for diverse processes with different
mass scales, kinematics and jet structure.

All the samples have been generated with {\tt Pythia} 6.410
\cite{Sjostrand:2006za} with
the DWT tune~\cite{Albrow:2006rt}. For the $t\bar{t}$ samples the B mesons have been
kept stable to avoid the need of B decay reconstruction for B
tagging\footnote{The effects of imperfect B tagging should
be addressed in the context of detector simulation studies.}.
The top
quark mass used in the generation is $M_t=175$ GeV while the 
$W$ mass is $M_W = 80.4$ GeV.

Now we describe for
each process the main motivations to examine it and the mass
reconstruction techniques employed, while results
are discussed in the next section. The fraction
of events that pass the selection cuts discussed above is
to a good approximation independent of the particular
jet definition, and their values can be seen in Table
\ref{tab:jetalgs_frac}.

\begin{table}
\begin{center}
\begin{tabular}{|c|c|c|c|c|}
\hline
Process &  \# Gen. events  & \# Acc. events    &  Fraction acc.
vs. gen.  & Fraction $f$ in Eq. \ref{eq:q1}\\
\hline
$Z'\to q\bar{q}$    &  50$\,$000 & $\sim$ 23$\,$000 & $\sim 0.46$ & 0.12 \\
Hadronic \ttbar     & 100$\,$000 & $\sim$ 75$\,$000 & $\sim 0.75$ &  0.18 \\
\hline
\end{tabular}
\end{center}
\caption{\small \label{tab:jetalgs_frac}  Number of
generated and accepted events for each process, the corresponding approximate
fraction of accepted events and the fraction $f$ of the
total number of generated events which correspond to a 25\% of
the selected events.}
\end{table}

\begin{itemize}
\item $Z' \to q\bar{q}$ for various values of $M_{Z'}$.

This process serves as a physically well-defined source 
of monochromatic quarks. By reconstructing the dijet invariant
 mass one effectively obtains a measure of the $p_T$ 
resolution and offset for each jet definition.
The range of $Z'$ masses is: 100, 150, 200, 300, 500, 700, 1000, 2000 and 
4000 GeV. Many of these values are already excluded, 
but are useful to study as measures of resolution at different energies. 
Note also that the generated $Z'$ particles have a narrow width
 ($\Gamma_{Z'}\le 1$ GeV). This is not very physical but
 useful from the point of view of providing monochromatic jet sources.

For each event, the reconstruction procedure is the following:\vspace{0.5em}
\begin{enumerate}
\item Carry out the jet clustering based on the list of all
  final-state particles
\item Keep the two hardest jets with $p_T \ge 10$ GeV. If no such
  two jets exist, reject the event.
\item Check that the two hard jets have rapidities
  $|y| \le 5$, and that
 the rapidity difference between them satisfies $|\Delta y| \le 1$. If not, reject the event.
\item The $Z'$ is reconstructed by summing the two jets' 4-momenta.\vspace{0.5em}
\end{enumerate}

\item Fully hadronic $t\bar{t}$ decay. 

This process provides a 
complex environment involving many jets in which one can test a 
jet definition's balance between quality of energy reconstruction
 and ability to separate multiple jets. The reconstruction of
$M_W$ and $M_t$ is obtained as follows:\vspace{0.5em}
\begin{enumerate}
\item Carry out the jet clustering based on the list of all final-state particles
\item Keep the 6 hardest jets with $p_T \ge 10$ GeV and $|y|\le 5$. If
  fewer than 6 jets pass these cuts, reject the event.
\item Among those 6 jets, identify the $b$ and the $\bar{b}$ jets. If the
  number of $b/\bar b$ jets is not two, then reject the event.
\item Using the four remaining jets, form two pairs to reconstruct the
  two $W$ bosons. Among the 3 possible pairings, choose the one that
  gives masses as close as possible to the nominal $W$ mass.
\item Reconstruct the two top quarks by pairing the $b$ and $W$ jets.
  Pairing is done by minimising the mass difference between the two
  candidate $t$ jets.
\end{enumerate}

%
\end{itemize}

\subsection{Results}\label{sec:results}

Now we discuss the results for the mass reconstruction of the
processes described in section~\ref{ref:jetalgs_process} 
with the jet algorithms of section~\ref{ref:jetalgs_jetalgs}.
We quantify the comparison between different jet definitions
using the quality measures defined in section~\ref{sec:jetalgs_figs}.
We note that in the various histograms of this section, the lines
corresponding to different jet algorithms have been
slightly shifted in order to improve legibility.

\subsection{Analysis of the $Z'$ samples}

\begin{figure}[ht]
  \centering
  \includegraphics[width=0.48\textwidth]{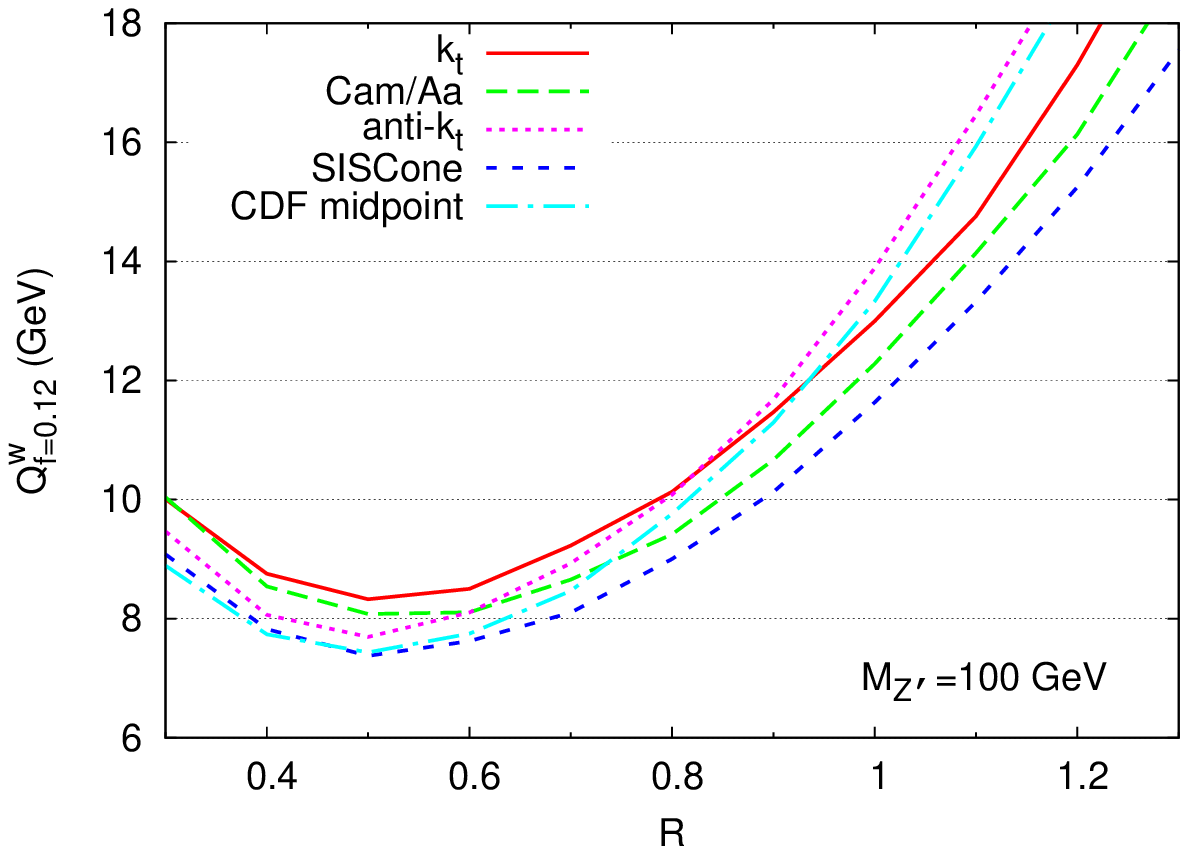}
  \includegraphics[width=0.48\textwidth]{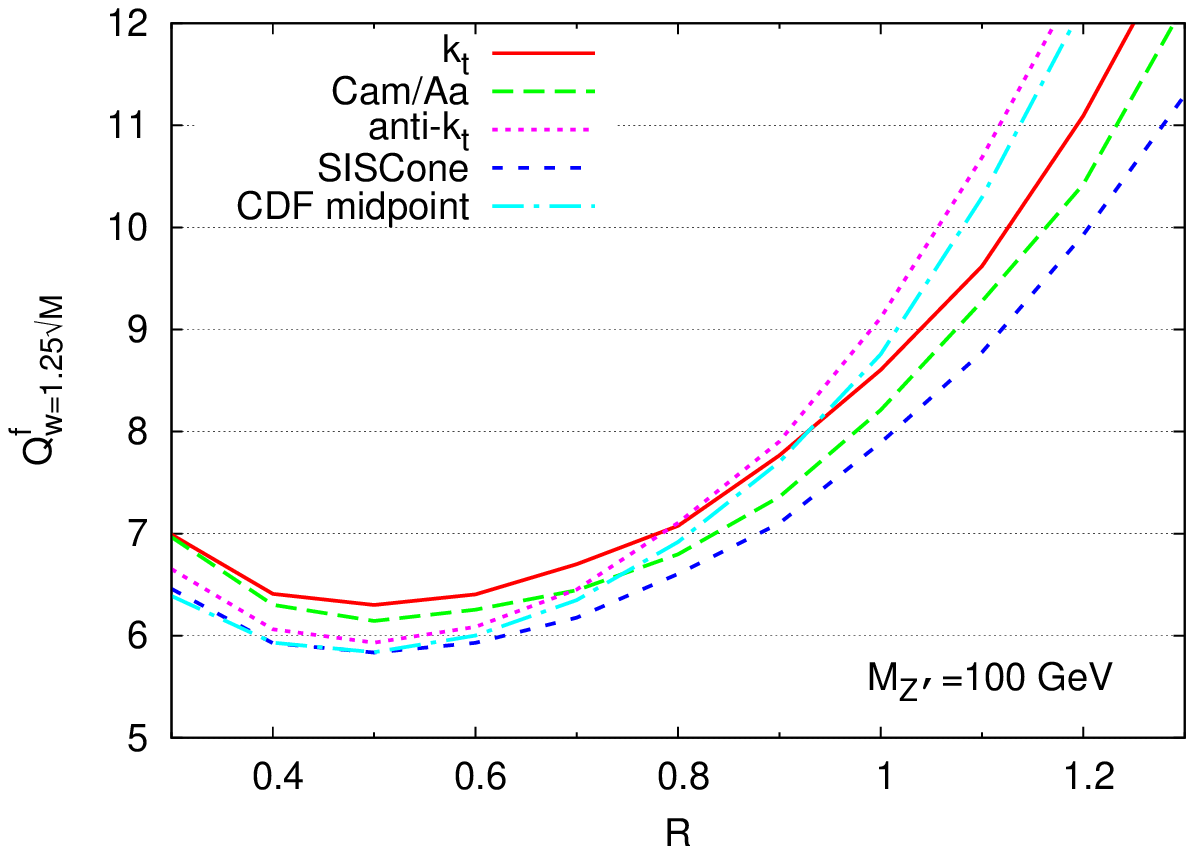}
  \includegraphics[width=0.48\textwidth]{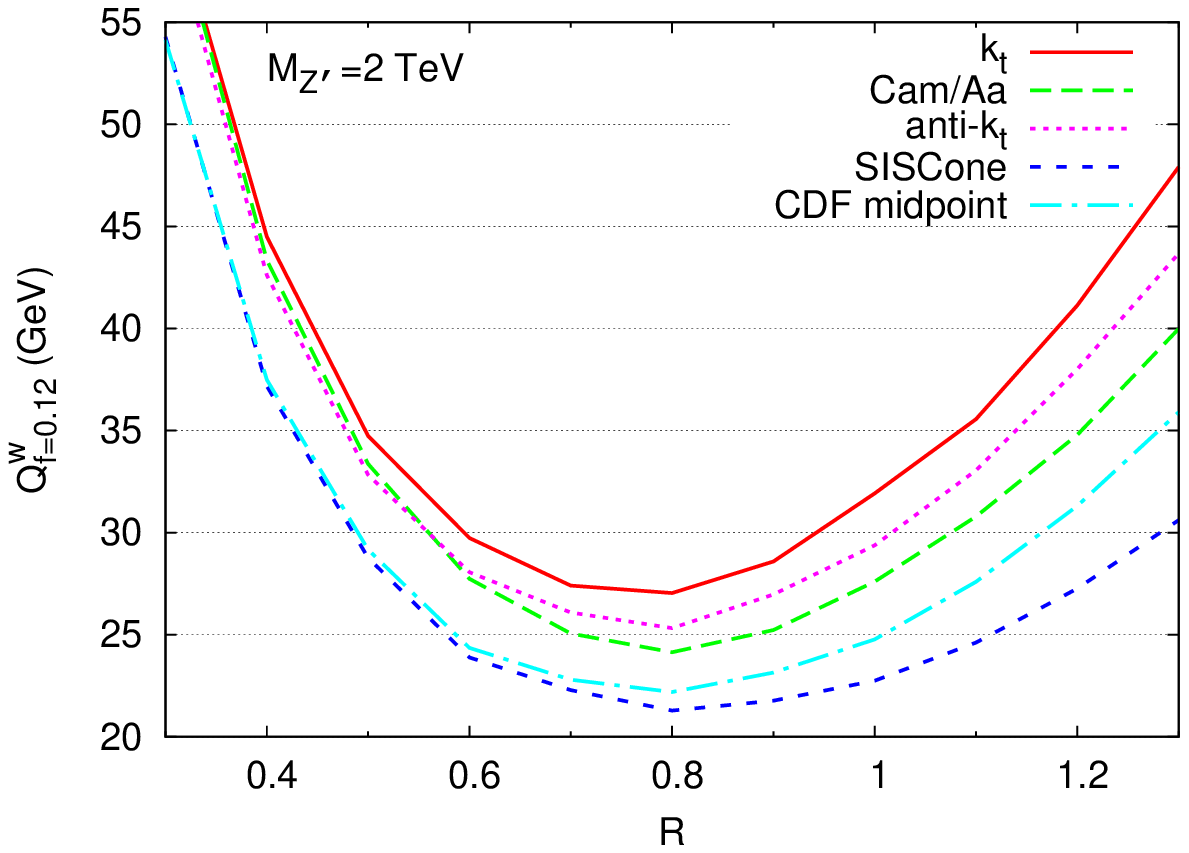}
  \includegraphics[width=0.48\textwidth]{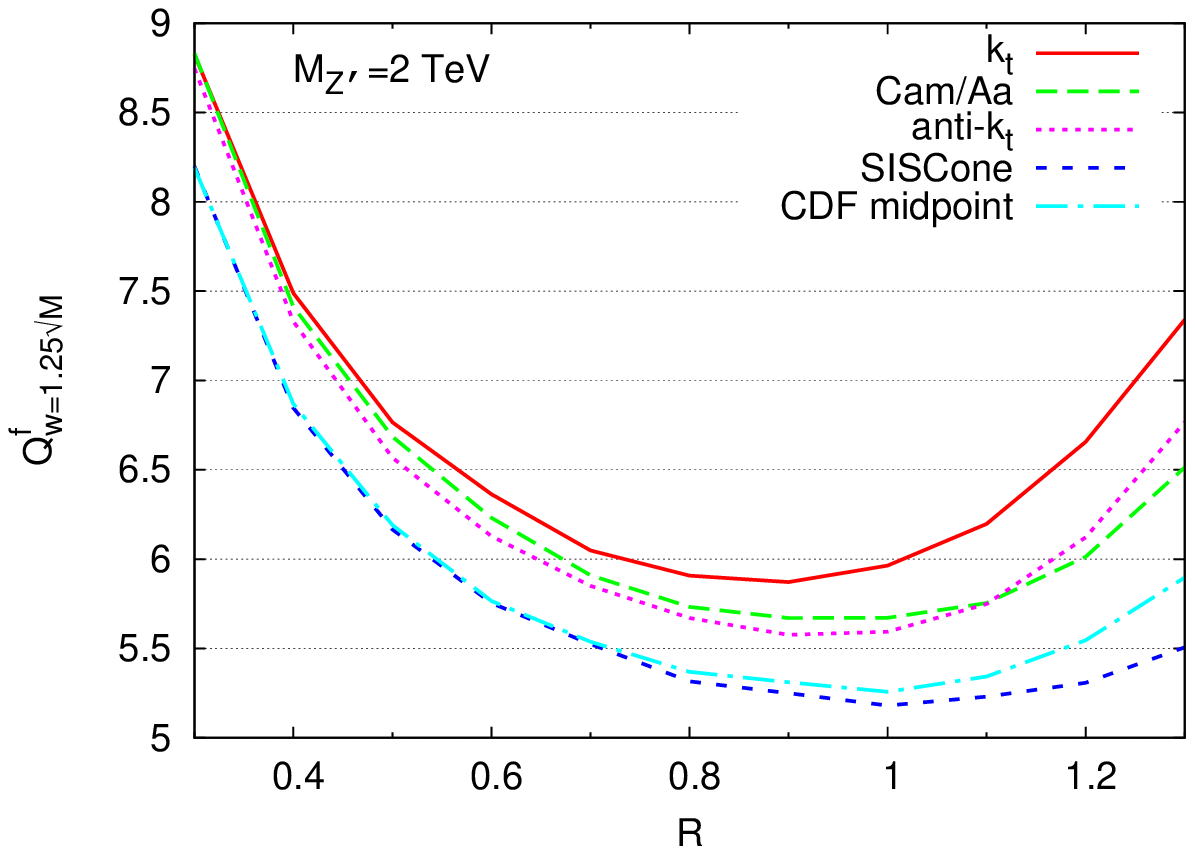}
  \caption{\small The figures of merit \Qa{0.12} and \Qb
 for the $Z'$ samples
    corresponding to $M_{Z'}=100$~GeV (upper plots) 
and $M_{Z'}=2$ TeV (lower plots).}
  \label{fig:jetalgs_zprime_merit} 
\end{figure}

\begin{figure}[ht]
  \centering
  \includegraphics[width=0.48\textwidth]{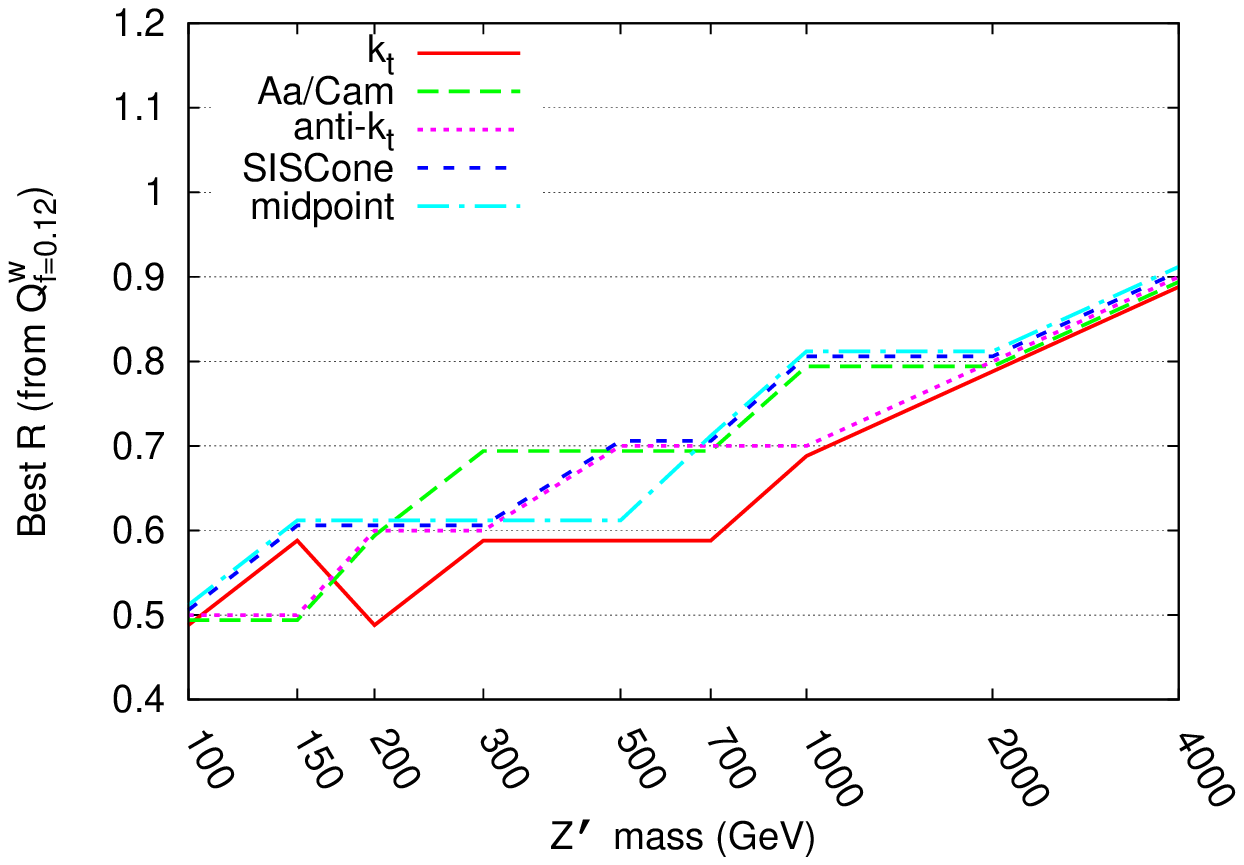}
  \includegraphics[width=0.48\textwidth]{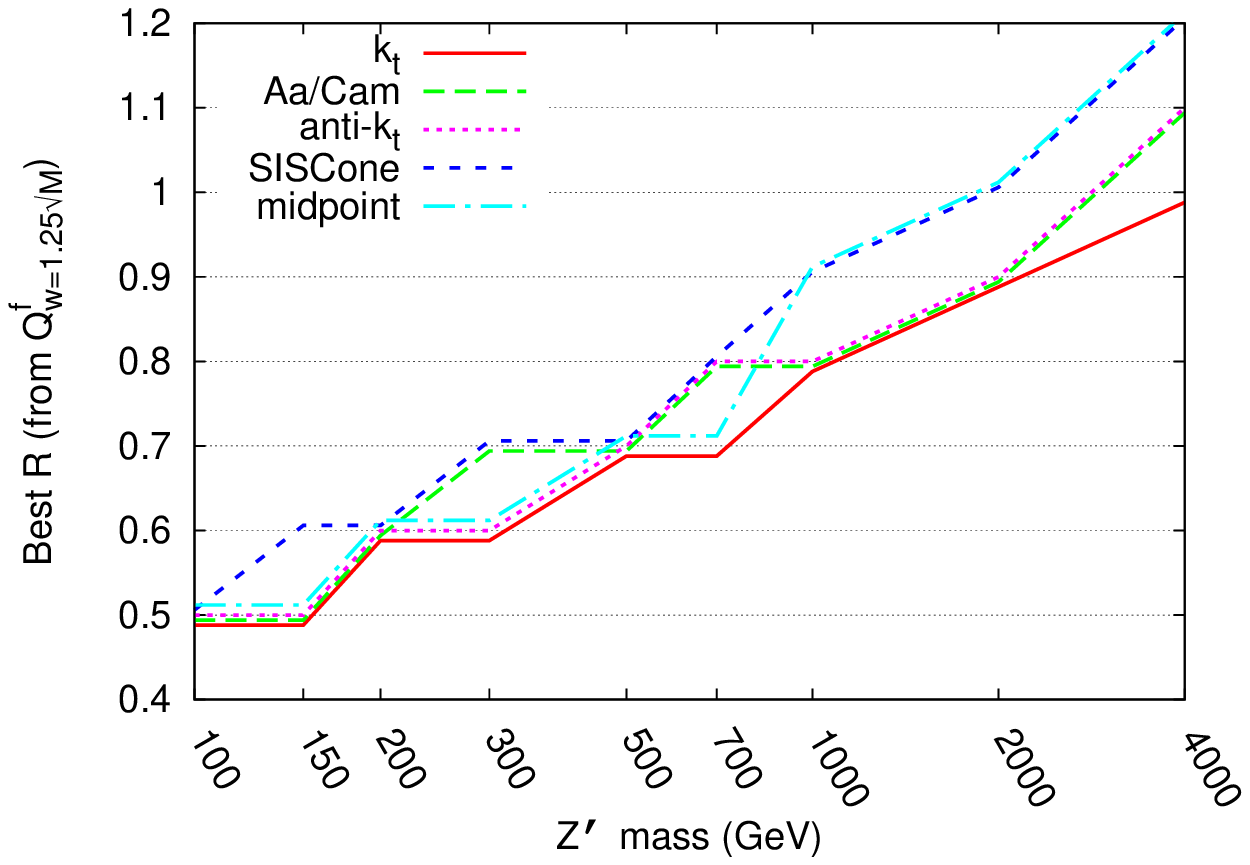}
  \caption{\small The best value of the jet radius $R_{\rm best}$ 
(defined as the
    minimum of the corresponding figure of merit) as determined from
    \Qa{0.12} (left plot) and \Qb (right plot) as a function of
    $M_{Z'}$.}
  \label{fig:jetalgs_zprime_bestR} 
\end{figure}

\begin{figure}[ht]
  \centering
  \includegraphics[width=0.48\textwidth]{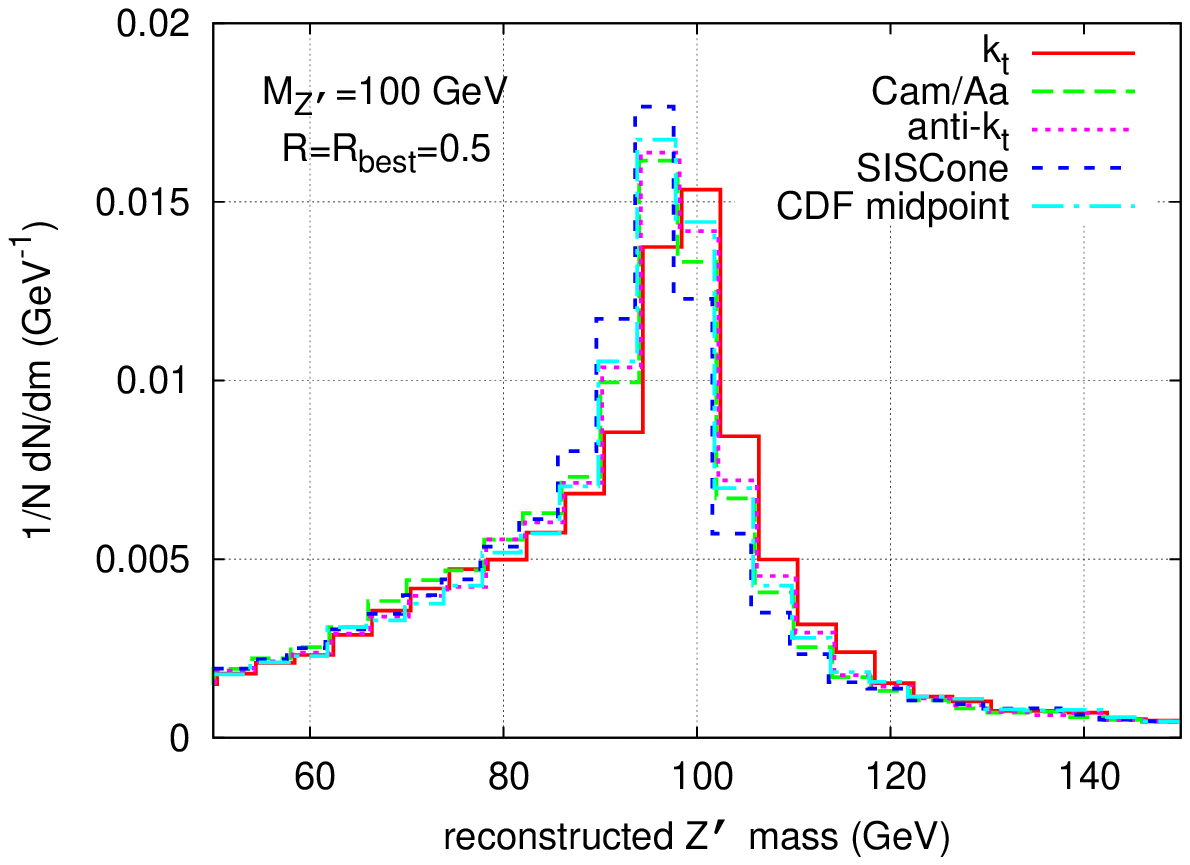}
  \includegraphics[width=0.48\textwidth]{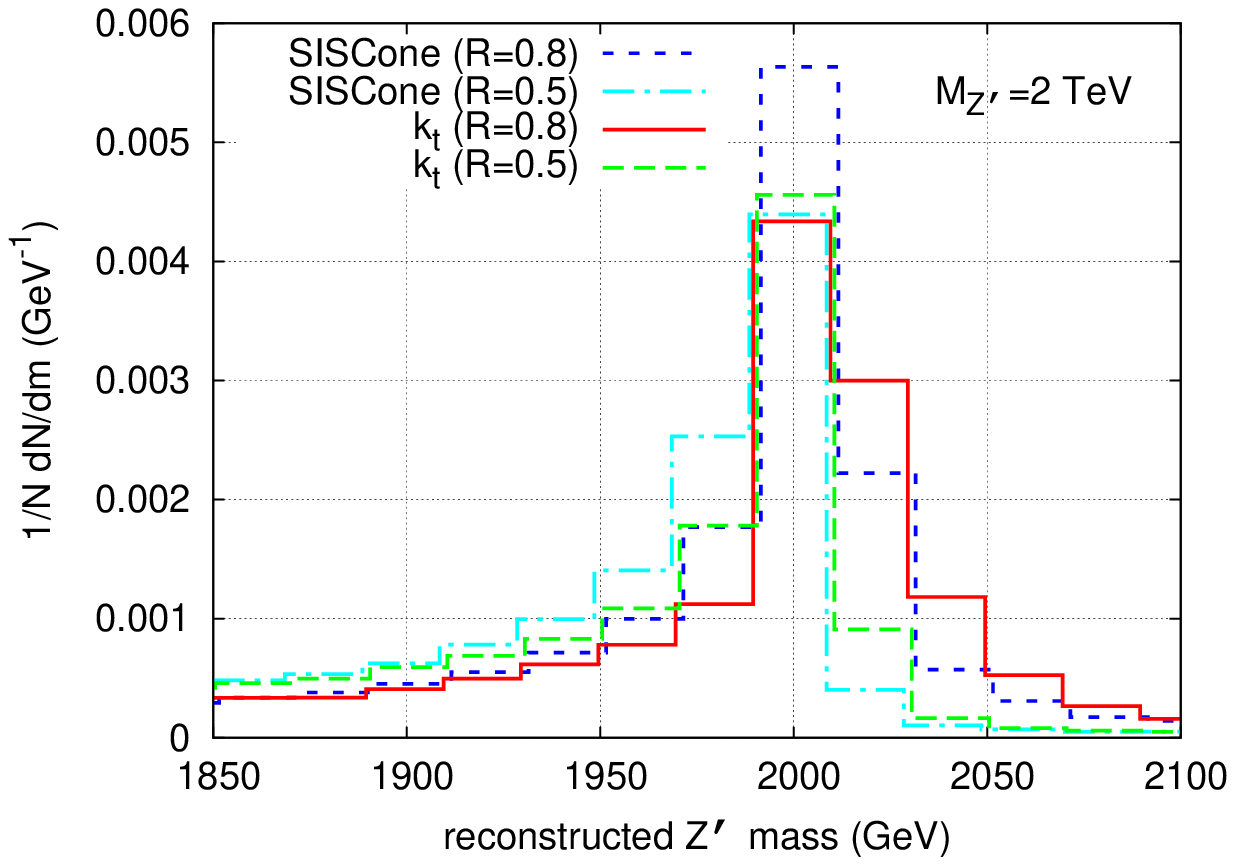}
  \caption{\small The invariant mass distribution in the $Z'$ samples
    for two different values of $M_{Z'}$.}
  \label{fig:jetalgs_zprime_histograms} 
\end{figure}

\begin{figure}[ht]
  \centering
  \includegraphics[width=0.48\textwidth]{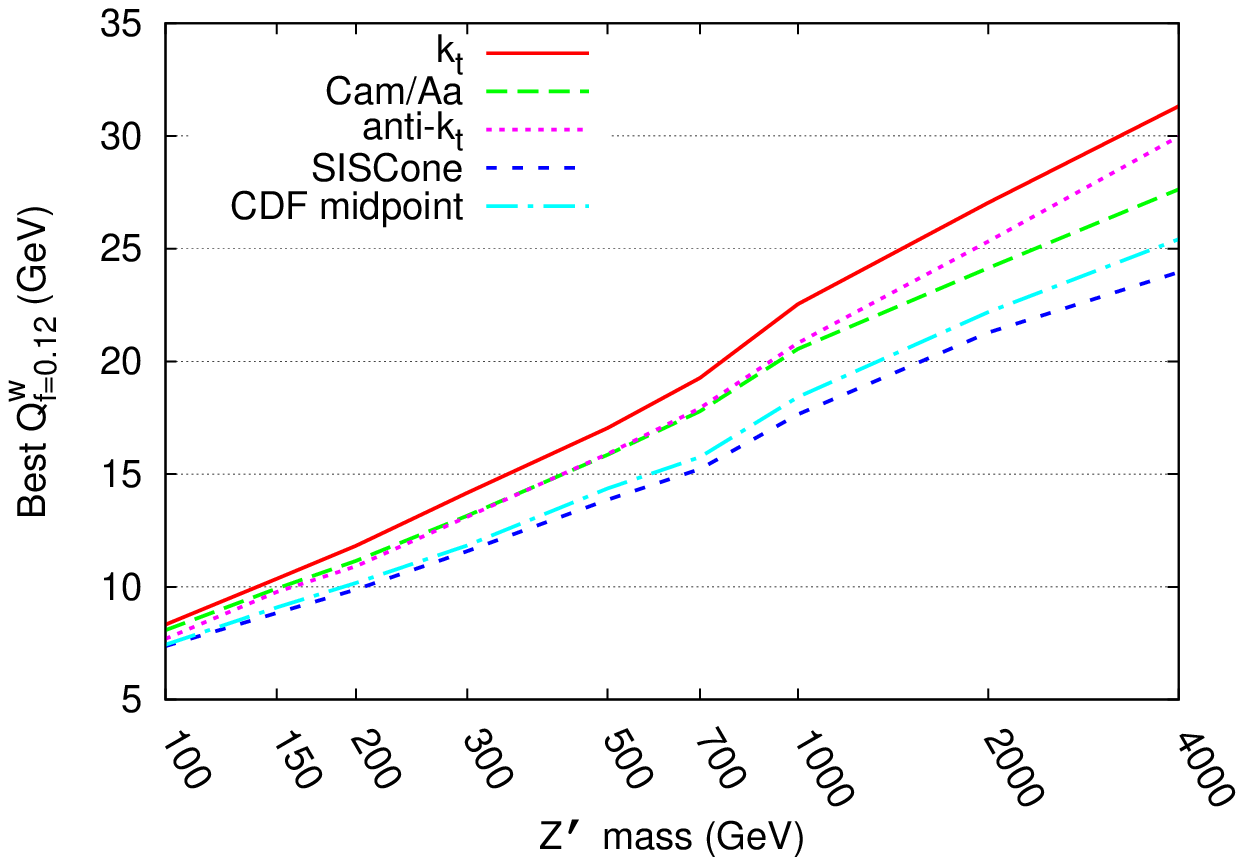}
  \includegraphics[width=0.48\textwidth]{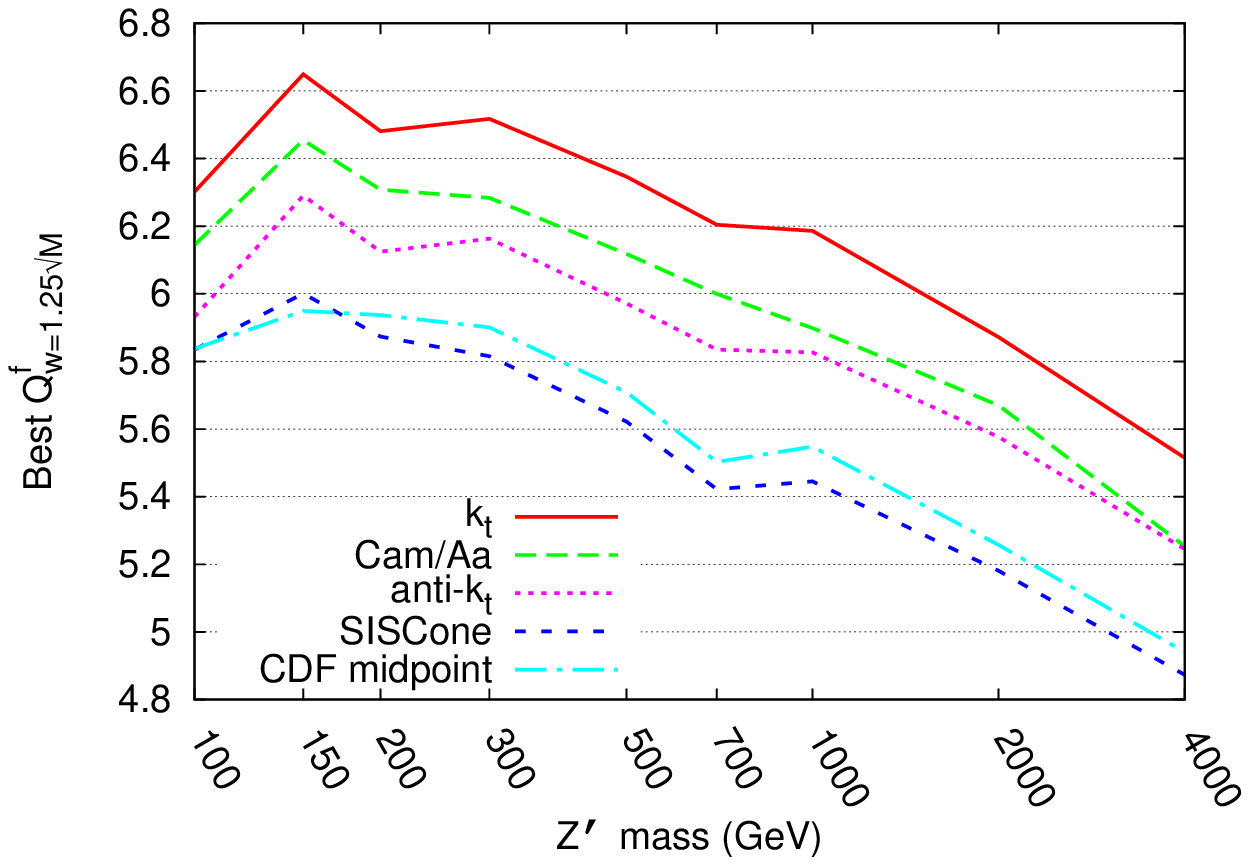}
  \caption{\small The best value of the figures of merit \Qa{0.12} and
    \Qb within all possible values of $R$ as a function of
    $M_{Z'}$.}
  \label{fig:jetalgs_zprime_bestQ} 
\end{figure}

The figures of merit for $Q_{f=0.12}^{w}(R)$
 and
$Q_{w=1.25\sqrt{M}}^{f}(R)$ are plotted 
in Fig.~\ref{fig:jetalgs_zprime_merit}, as a
function of the radius $R$ for a $Z'$ of 100 GeV and 2 TeV. Each plot
includes the results for the five jet algorithms under consideration.
There are two lessons we can learn from this figure. Firstly, even
though some algorithms give better quality results than others (we
will come back on this later), the main source of quality differences
does not come from the choice of algorithm but rather from the adopted
value for $R$. Secondly, the minimum of the quality
measures gives, for each jet
algorithm, a preferred value $R_{\text{best}}^{M_Z'}$ for $R$.

That preferred value over the whole range of $Z'$ masses is
shown\footnote{Varying $R$ continuously between $0.3$ and $1.3$ would
  probably result in a smoother curve for $R_{\text{best}}$ as a
  function of $M_{Z'}$. However, there is no real interest in
  determining an $R$ parameter with more than one decimal figure.} in
Fig.~\ref{fig:jetalgs_zprime_bestR}. We observe that the two quality
measures roughly agree on the extracted preferred value, with the
possible exception of the largest values of $M_{Z'}$ for which we
observe small differences. Furthermore, when the mass of the $Z'$
becomes larger, the best quality is also achieved using larger values
for $R$: $R_{\text{best}}$ goes from 0.5 for low $Z'$ masses, to
$R_{\text{best}}\approx$ 0.9 for high $Z'$ masses.

This behaviour can be explained by the fact the as $M_{Z'}$ increases,
perturbative radiation (which favours larger $R$) grows larger
(roughly as $M$) while the underlying event contribution (which
favours smaller $R$) stays fixed, thus resulting in an overall larger
value for the optimal $R$~\cite{Dasgupta:2007wa}.
Another relevant point
is that $Z'$ decays are mostly dijet events,
so the invariant mass reconstruction is in general not affected
by the accidental merging of hard partons that takes
place for larger
values of $R$ in multi-jet environments like hadronic $t\bar{t}$ decays.

Given our method to quantitatively analyse the performance of jet
algorithms and to extract a preferred value for $R$, there are a few
more interesting figures we want to look at. The first one,
Fig.~\ref{fig:jetalgs_zprime_histograms}, is simply the histogram of
the reconstructed $Z'$ mass. The left plot shows the reconstructed
$Z'$ peaks for the five algorithms at $R=R_{\text{best}}$ and though
some slight differences exist all algorithms give quite similar
results. In the right plot we show the reconstructed $Z'$ histogram
for the $k_t$ and the SISCone algorithms using either
$R=R_{\text{best}}^{2 \text{ TeV}}=0.8$, as extracted from the quality
measures at 2 TeV, or $R=R_{\text{best}}^{100\text{ GeV}}=0.5$,
extracted at 100 GeV. The behaviour is again what one expects from
Fig.~\ref{fig:jetalgs_zprime_merit}, namely that SISCone with $R=0.8$
performs a bit better than SISCone with $R=0.5$ and $k_t$ with
$R=0.8$, which themselves give a better peak than the $k_t$ algorithm with
$R=0.5$.

Let us now consider again the whole range of $Z'$ masses and discuss
the initial point of interest which is finding the best algorithm to
be used in jet analysis, at least from the point of view of $Z'$
reconstruction. To that aim, we look at the quality measure at
$R_{\text{best}}$ as a function of the $Z'$ mass and for each jet
algorithm.  The results are presented in 
Fig.~\ref{fig:jetalgs_zprime_bestQ} for \Qa{0.12} (left plot) and \Qb
(right plot). Note that \Qa{0.12} has an approximately linear increase
with $\ln M_{Z'}$, while \Qb has a similar behaviour but in the
opposite direction.

The generic conclusion is that cone algorithms  split--merge
(SM) steps perform better
than the recombination-type algorithms, though we again emphasise that
the difference is rather small and, in particular, smaller than the
dependence on the parameter $R$. This conclusion is valid for all $Z'$
masses and for both quality measures. In general, among the cone
algorithms, SISCone produces results slightly better than CDF-Midpoint
while, among the recombination-type algorithms, $k_t$ is a bit worse
than Cambridge/Aachen and anti-$k_t$, the ordering between those two
depending on the mass and quality measure under consideration. 

This can
be understood due to the fact that SISCone has
a reduced sensitivity to the underlying event 
(smaller effective area \cite{Cacciari:2008gn})
 while stretching out up
to larger distances\footnote{
In the limiting case it can merge two equally hard partons
separated by a angular distance $2R$.}, 
thus is able to merge emitted partons
even at relatively large angles. Note that this feature, which
is advantageous in a clean environment like $Z'\to q\bar{q}$, 
essentially a dijet event, is on the other hand
something that degrades jet clustering with SISCone on
denser environments like \ttbar.

We can quantify the differences between jet algorithms at $R_{\rm
  best}$ using the mapping between quality measures and effective
luminosity ratios introduced in Sect.~\ref{sec:jetalgs_figs}.  For
$M_{Z'}=100$ GeV, both quality measures coincide in that when comparing the
best jet algorithm (SISCone) with the worst ($k_t$) one finds
$\rho_{\mathcal{L}} \approx 0.85$, while for the $M_{Z'}=2$ TeV case,
one finds that the effective luminosity ratio is $\rho_{\mathcal{L}}
\approx 0.8$.

An important consequence that can be drawn for this analysis is
that optimising the value of $R$ for a given jet algorithm
is crucial to optimise the potential of a physics analysis.
For example, in the $M_{Z'}~=2$~TeV case, if one chooses $R=0.5$
(based e.g. on considerations for the $M_{Z'}~=100$~GeV process)
instead of the optimal value $R_{\rm best}\simeq 0.8$, it is equivalent to 
losing a factor $\rho_{\mathcal{L}}~\approx~0.75$ in luminosity (for all
algorithms and both quality measures).
We note that the optimal value of $R$ at high masses is somewhat
larger than what is being considered currently in many studies by the
LHC experiments.

\subsection{Analysis of the hadronically decaying $t\bar t$ sample}

\begin{figure}[ht]
  \centering
  \includegraphics[width=0.48\textwidth]{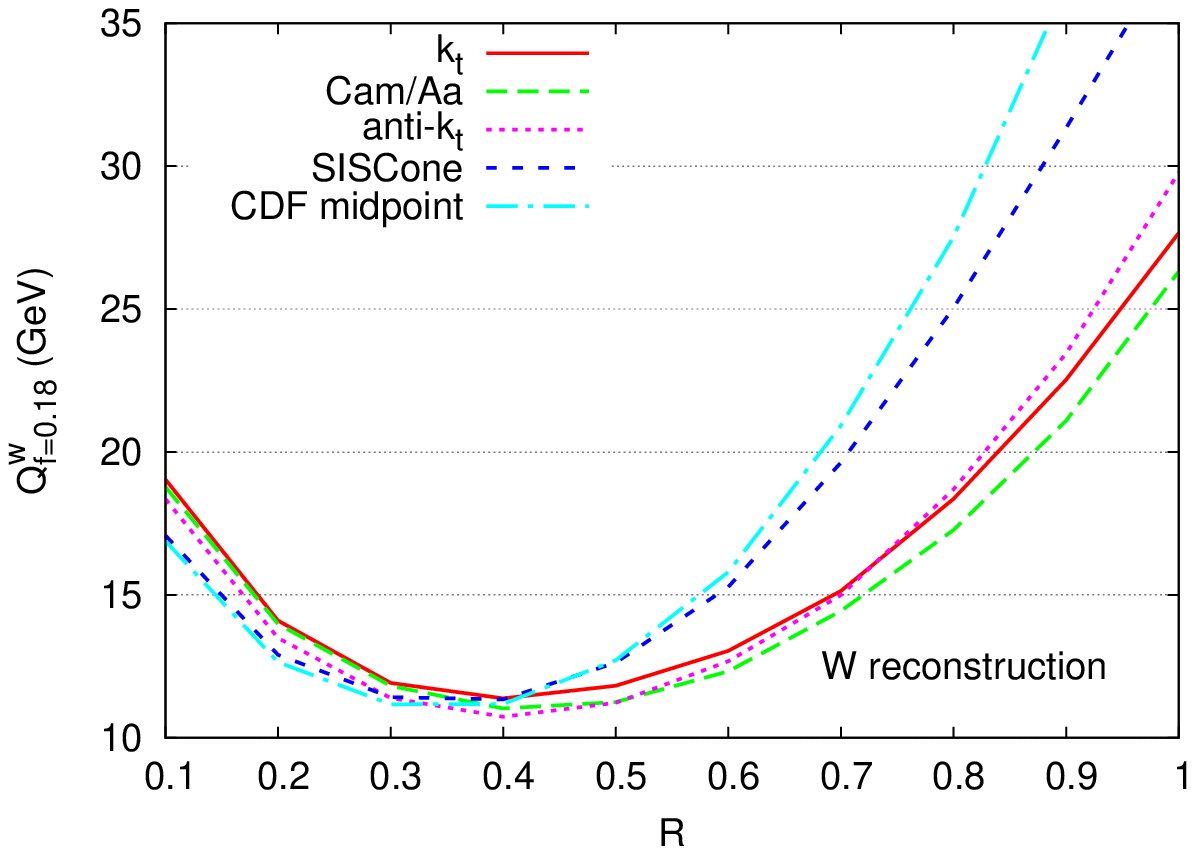}
  \includegraphics[width=0.48\textwidth]{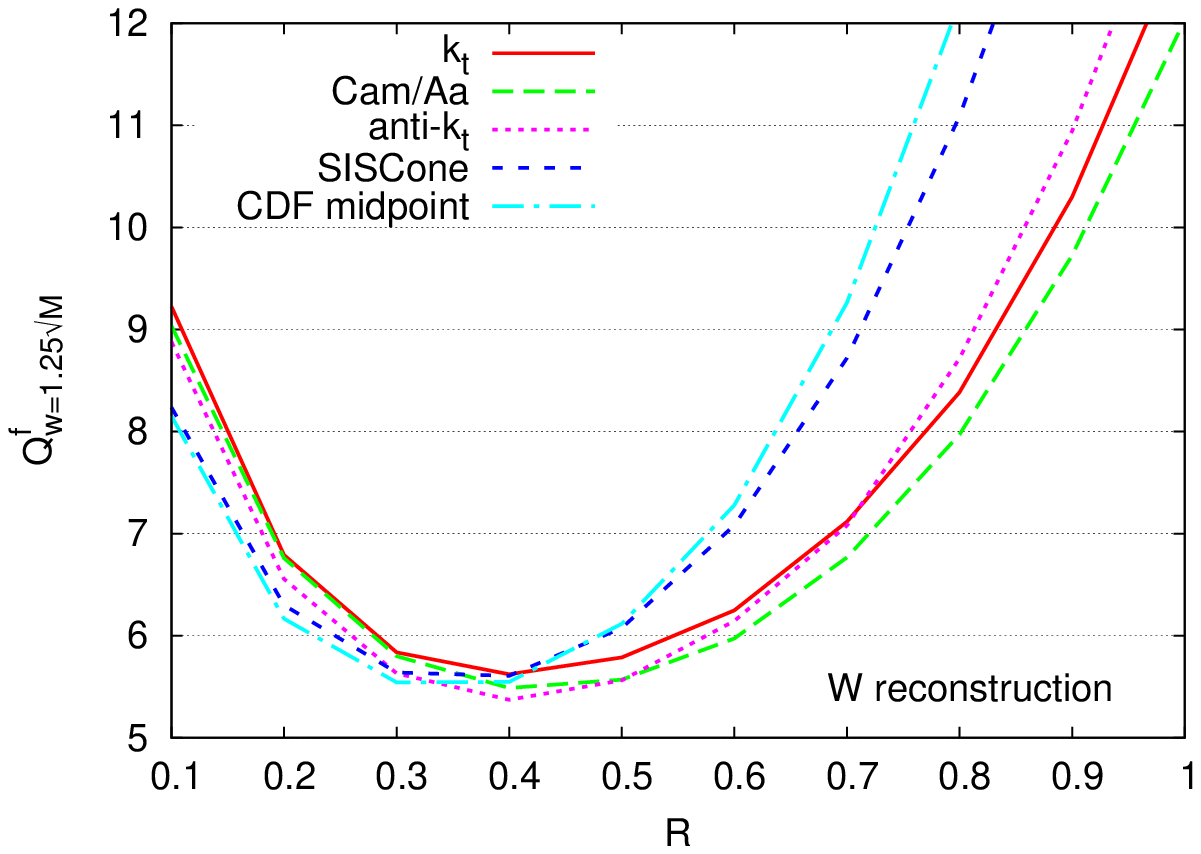}
  \includegraphics[width=0.48\textwidth]{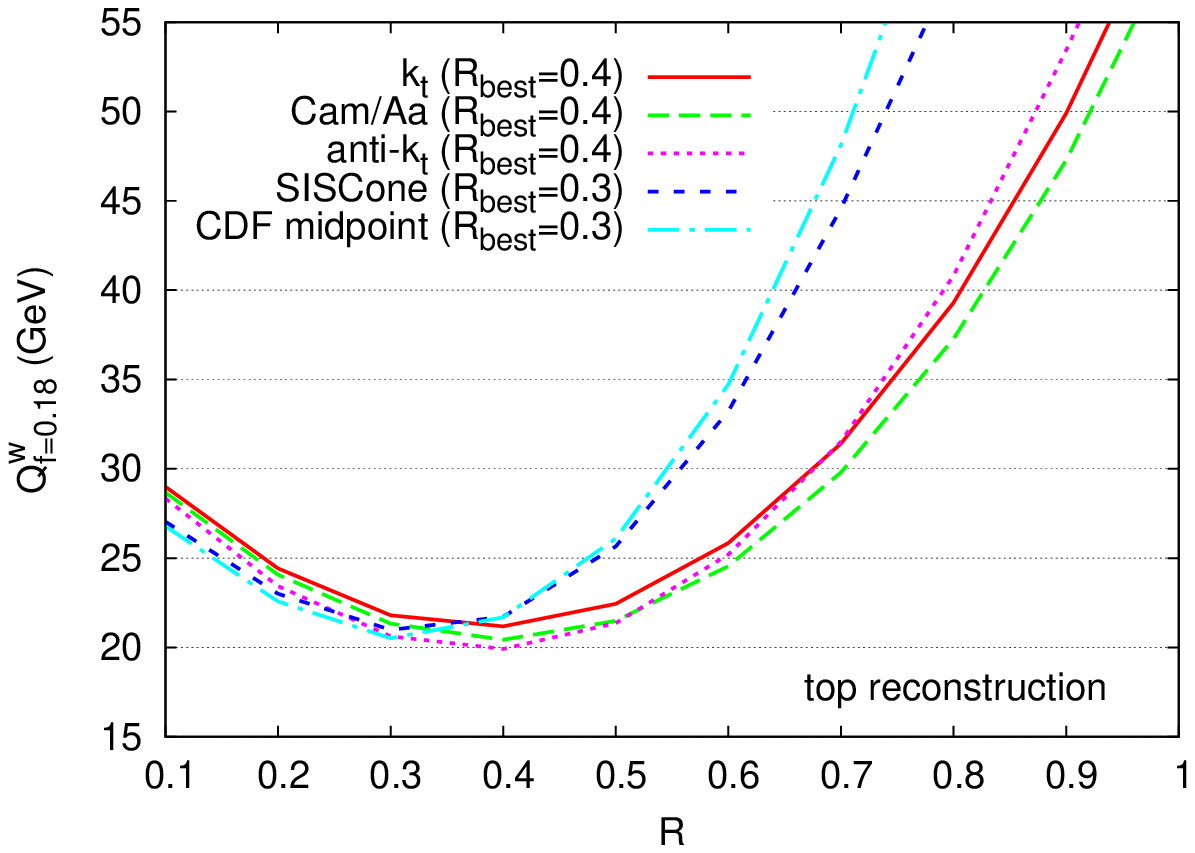}
  \includegraphics[width=0.48\textwidth]{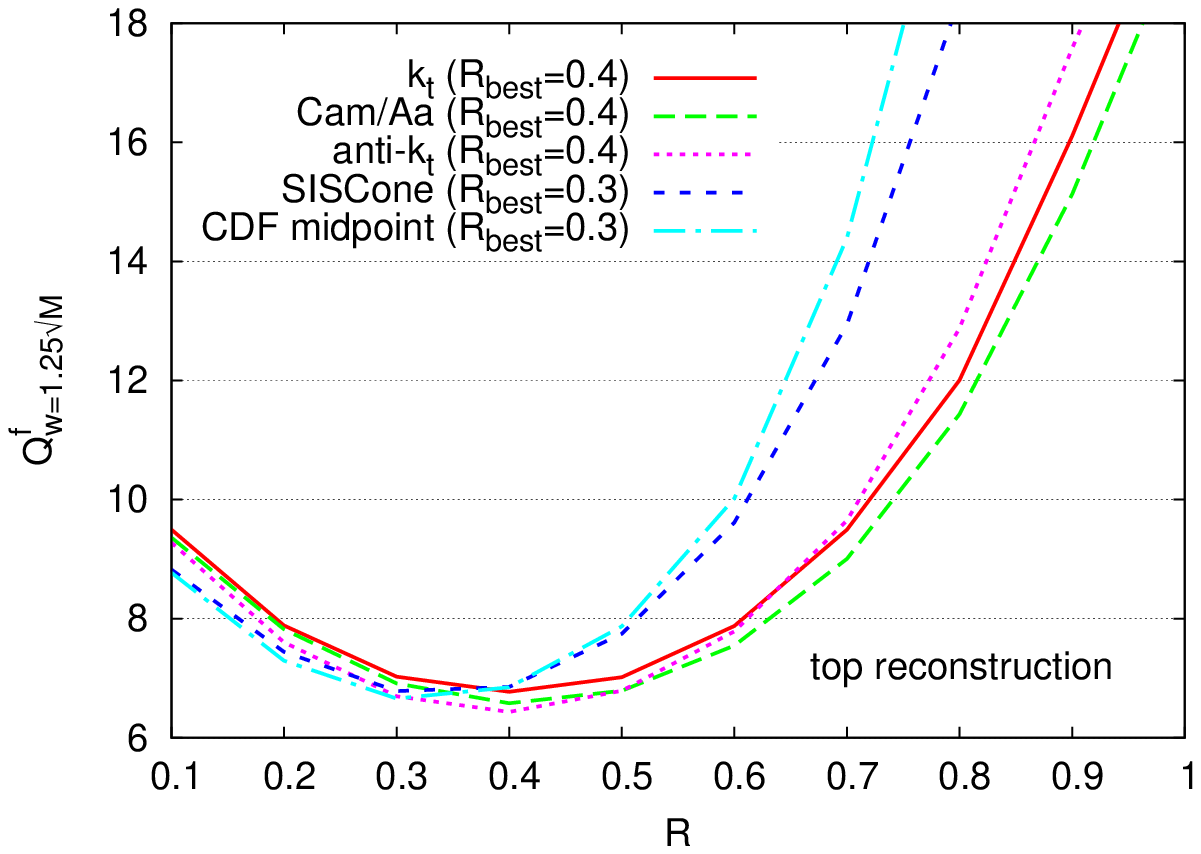}
  \caption{\small The figures of merit \Qa{0.18} and \Qb for the
    invariant mass distributions of the hadronic $t\bar{t}$ samples.}
  \label{jetalgs_ttbar1} 
\end{figure}

\begin{figure}[ht]
  \centering
  \includegraphics[width=0.48\textwidth]{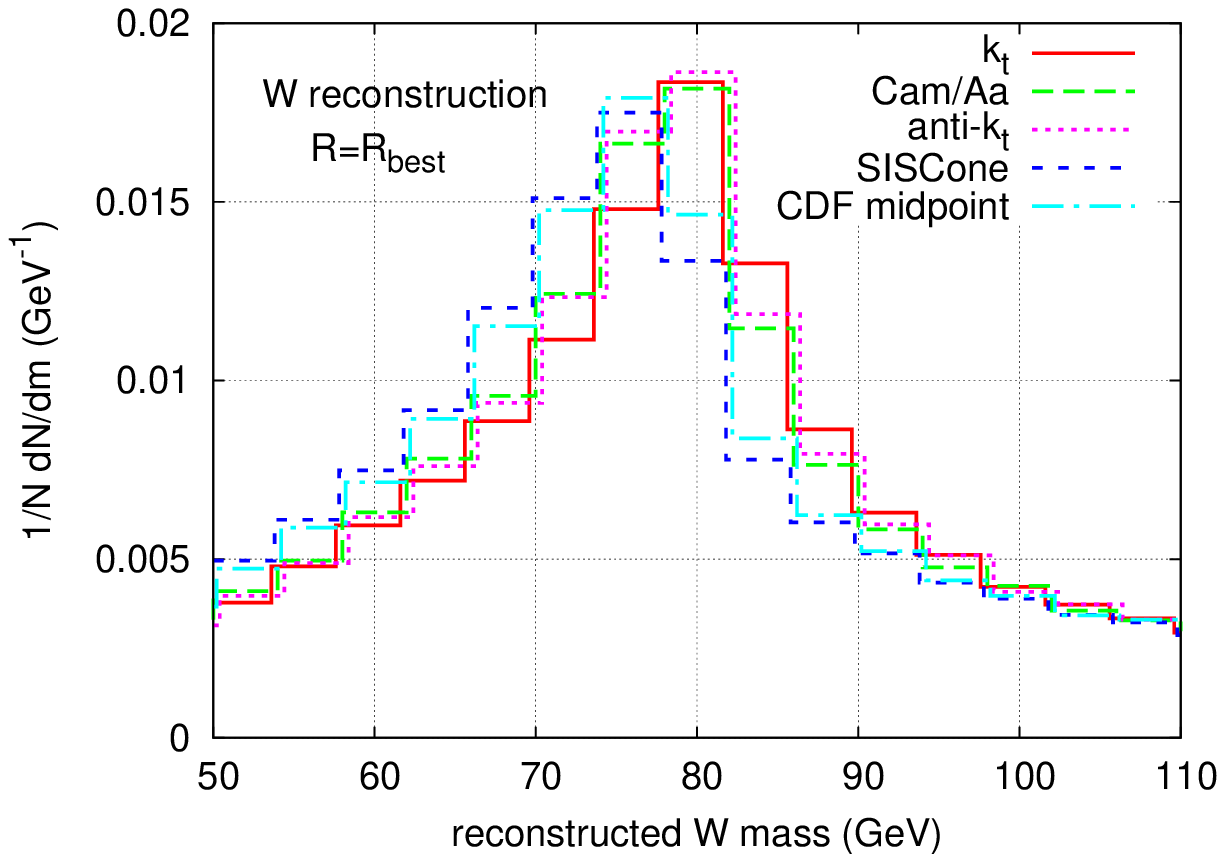}
  \includegraphics[width=0.48\textwidth]{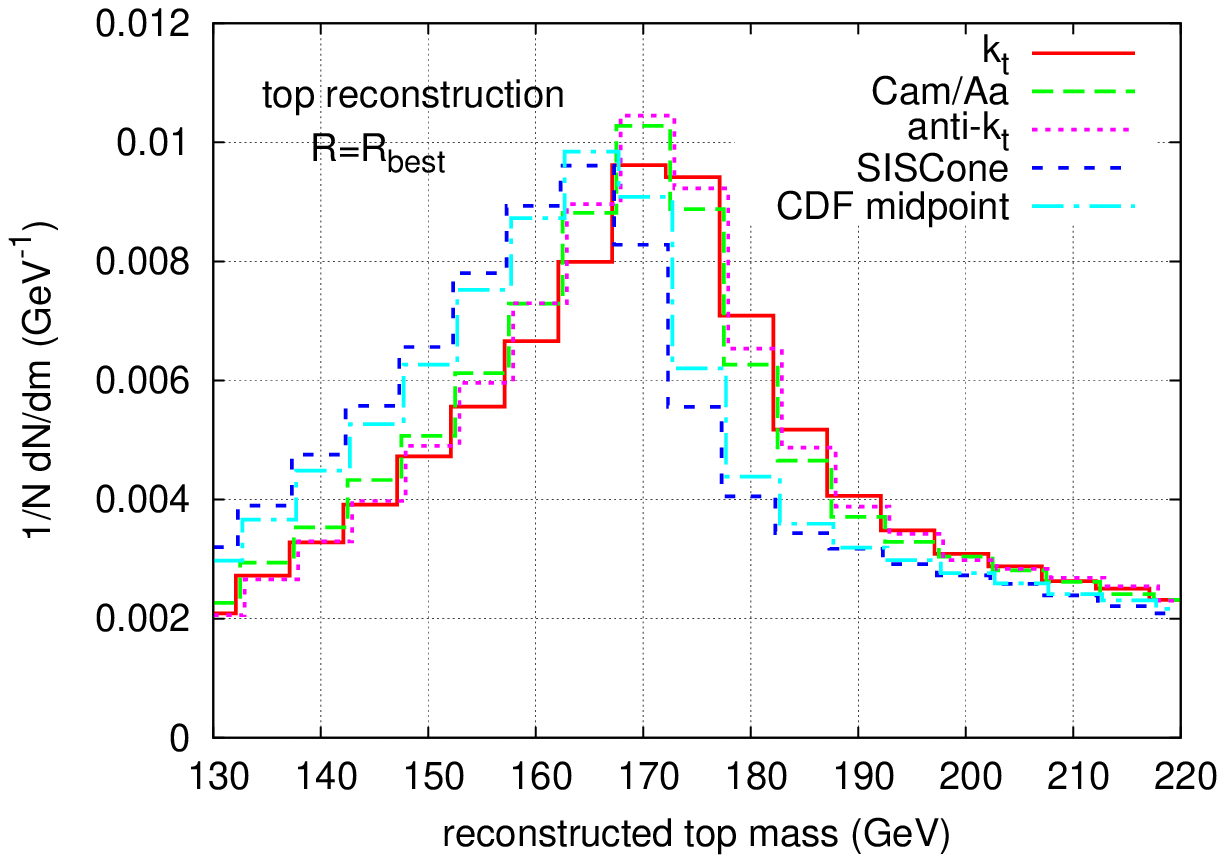}
  \caption{\small The $W$ and $t$ invariant mass distributions for the
    hadronic $t\bar{t}$ samples for $R_{\rm best}=0.4$.}
  \label{jetalgs_ttbar2} 
\end{figure}

\begin{figure}[ht]
  \centering
  \includegraphics[width=0.48\textwidth]{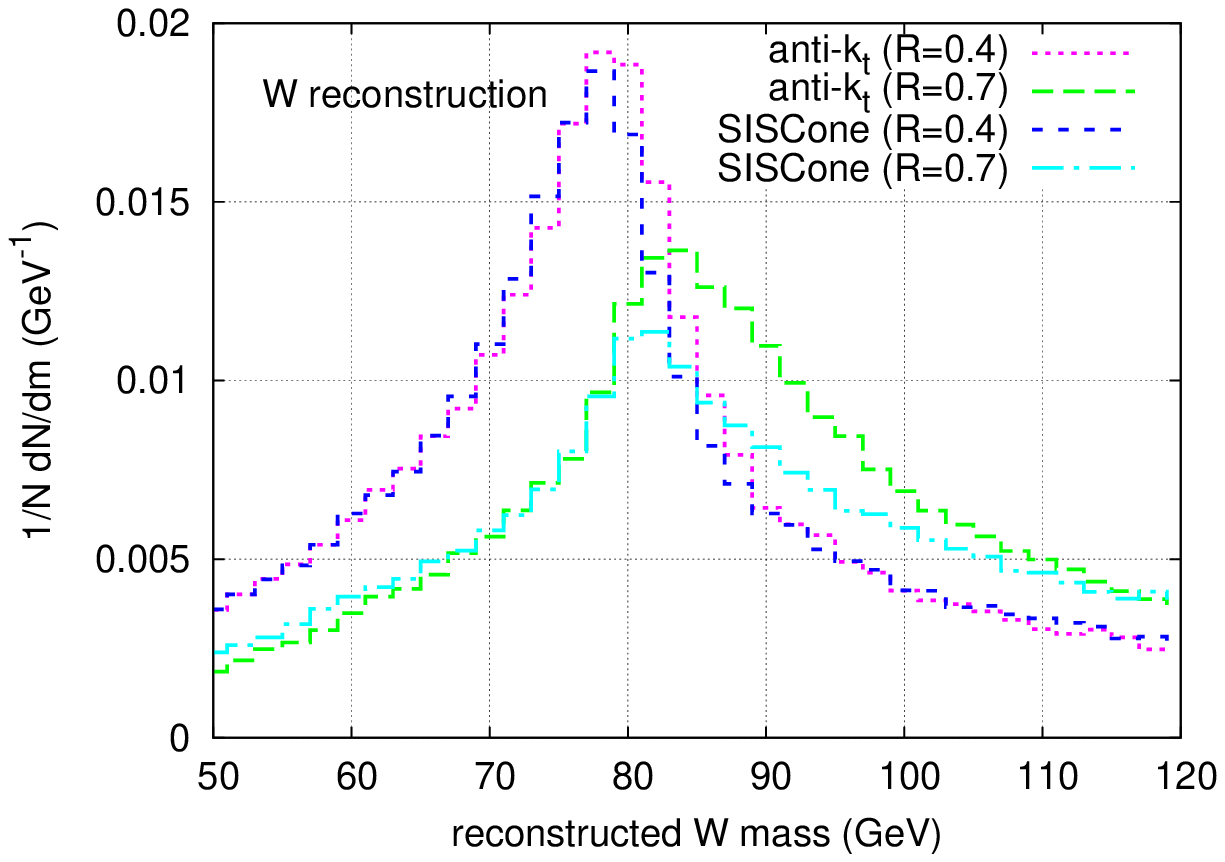}
  \includegraphics[width=0.48\textwidth]{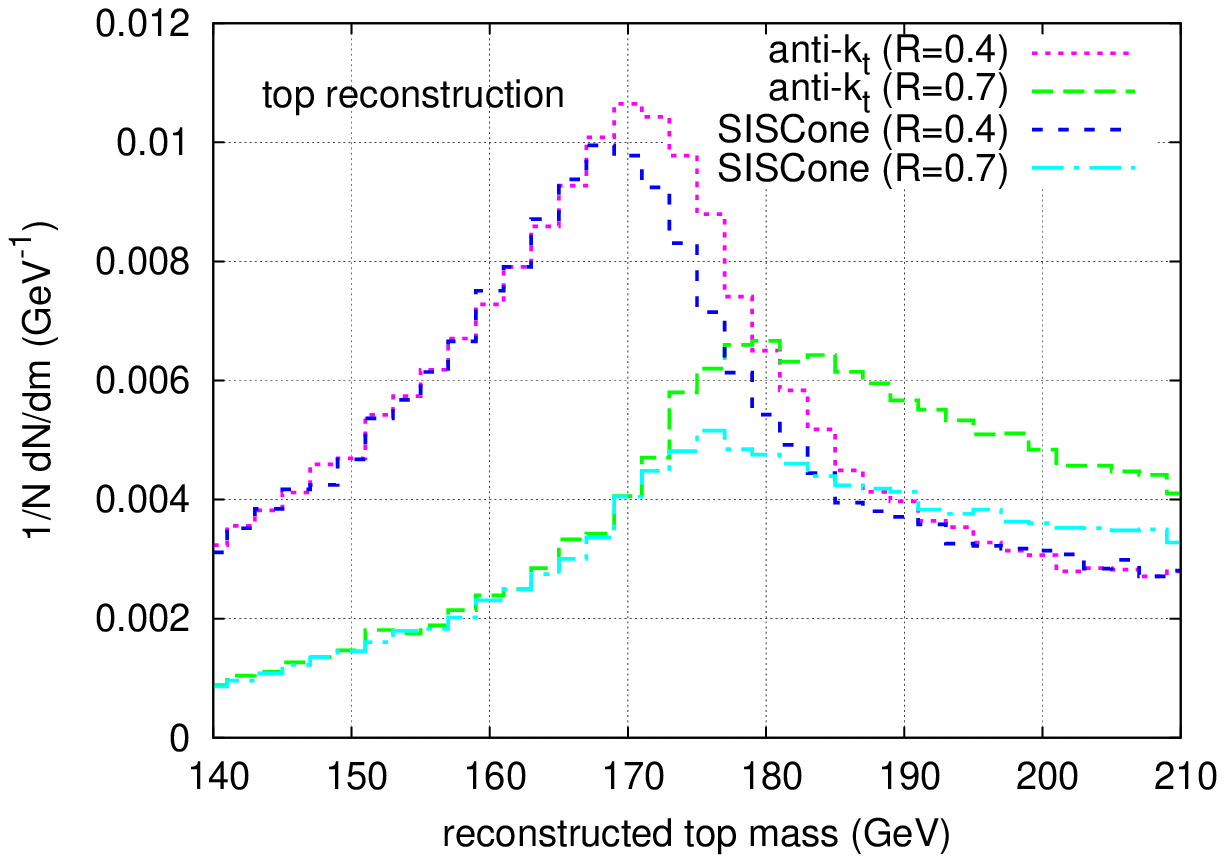}
  \caption{\small The $W$ and $t$ invariant mass distributions for the
    hadronic $t\bar{t}$ samples for $R_{\rm best}=0.4$ for the 
best overall algorithm anti-$k_t$ and SISCone compared to
their counterparts for $R=0.7$.}
  \label{jetalgs_ttbar_comp} 
\end{figure}

Hadronic  $t\bar t$ production is a challenging environment since
the jet algorithm has to reconstruct at least $6$ hard jets.
In this process one can test a 
jet definition's balance between quality of energy reconstruction
and ability to separate multiple jets.

For each of the mass distributions that we reconstruct in this case
(that of the $W$ boson and that of the top quark), we show the plots of
the corresponding figures of merit \Qa{0.18} and \Qb in
Fig.~\ref{jetalgs_ttbar1}.
Although all jet algorithms perform rather similarly at $R_{\rm
  best}$, there is a slight preference for the anti-$k_t$ algorithm.
The resulting effective luminosity ratio computed for the top
reconstruction between the two limiting algorithms is
$\rho_{\mathcal{L}}\approx 0.9$.  Note that at larger values of $R$
the cone algorithms perform visibly worse than the sequential
recombination ones, probably because they tend to accidentally cluster
hard partons which should belong to different jets. In the same
spirit, the preferred radius is $R_{\rm best}=0.4$ for sequential
recombination algorithms, while cone algorithms tend to prefer a
somewhat smaller optimal value $R_{\rm best}= 0.3$.

For the hadronic $t\bar{t}$ samples, we show the invariant mass
distributions at $R_{\rm best}$ in each case for $M_W$ and $M_t$ in
Fig.~\ref{jetalgs_ttbar2}. We observe that all algorithms lead to
rather similar results at the optimal value of the jet radius.

Then, in Fig.~\ref{jetalgs_ttbar_comp}, we compare the $W$ and $t$
invariant mass distributions for the hadronic $t\bar{t}$ samples for
the best overall algorithm anti-$k_t$ and for SISCone, both with
$R=R_{\rm best}$, compared to their counterparts for $R=0.7$. We
observe that, as indicated by the figures of merit, the choice
$R=R_{\rm best}$ for the anti-$k_t$ algorithm leads to a somewhat
larger number of events in the peak than for SISCone, but in any case this
difference is small compared with the difference between $R=R_{\rm
  best}$ and $R=0.7$. The degradation of the mass peak at large $R$ is
both due to contamination from the UE and to the fact that hard
partons are sometimes accidentally merged (more often in cone
algorithms with SM steps).

As in the $Z'$ case, one of the main results of this
study is that choosing a non-optimal value of $R$ can result in a
severe degradation of the quality of the reconstructed mass peaks. For
example, comparing in Fig.~\ref{jetalgs_ttbar_comp} the results for
$R=R_{\rm best}$ and $R=0.7$, we observe that the degradation of the
mass peak can be of the order of $\sim 40-50 \%$, confirmed by the
quality measures, for which we obtain $\rho_{\mathcal{L}}\sim
0.3-0.6$. 
Thus our analysis confirms that the relatively small values of $R$
currently being used by the LHC experiments in top reconstruction are
appropriate.  Specific care is needed with cone algorithms with
split-merge stages, for which one should make sure that $R$ is not
larger than $0.4$.

As a final remark we note that we have also examined semi-leptonic
$t\bar t$ decays. Though there are fewer jets there, the results are
rather similar (with slightly larger differences between algorithms),
mainly because the semileptonic case resembles a single hemisphere of
the fully hadronic case.

\subsection{Summary}

We have presented in this contribution a general technique to quantify
the performance of jet algorithms at the LHC in the case of the mass
reconstruction of heavy objects.  

One result is that for simple events, as modelled by a fake $Z'$ decay
at a range of mass scales, SISCone and the midpoint algorithm behave
slightly better than others, presumably because they reach furthest
for hard perturbative radiation, but without taking in extra of
underlying event contamination.
Quantitatively, our performance measures suggests that one can obtain
equivalent signal$/\sqrt{\mathrm{background}}$ with a factor
$\rho_L\simeq 0.8-0.9$ less luminosity than for (say) the $k_t$
algorithm. The Cambridge/Aachen and anti-$k_t$ algorithms are
intermediate.

An effect of sometimes greater significance is the dependence of the results on
the choice of $R$ parameter. In particular we find that the optimal
$R$ increases significantly with mass scale $M_{Z'}$, most probably
for the reasons outlined in~\cite{Dasgupta:2007wa}, namely an
interplay between perturbative effects (which scale $M_{Z'}$ and
prefer a larger $R$) and non-perturbative effects (independent of
$M_{Z'}$ and favouring smaller $R$). If one takes $R=0.5$, which is
optimal at $M_{Z'}=100$~GeV, and uses it at $M_{Z'}=2$~TeV, it's
equivalent to a loss of luminosity of a factor of $\rho_L\simeq 0.75$
compared to the optimal $R \simeq 0.9$. The need for large $R$ is
likely to be even more significant for resonances that decay to
gluons, as suggested by the study in section~\ref{sec:rabbertz}. 

We have also examined more complex events, hadronic decays of $t\bar
t$ events. Here the need to resolve many different jets modifies the
hierarchy between algorithms, with anti-$k_t$ performing best.
Overall the differences between algorithms are however fairly small,
with an effective luminosity reduction from best to worst of $\rho_L
\simeq 0.9$. The choice of the correct $R$ is even more important here
than in the $Z'$ case, with small values $R\simeq 0.4$ being optimal.

Let us emphasise that our results should be taken with some care,
since in general the jet clustering procedure will affect the
background as well as the signal, and our measures ignore this
effect. Nevertheless, while our analysis cannot replace a proper
experimental  $S/\sqrt{B}$ study, it does provides  an
indication of the typical variations that might be found in different
jet definition choices at the LHC, and points towards the need for
flexibility in jet finding at the LHC.

The strategy presented in this contribution can be readily applied to
quantify the performance of different ideas and strategies for
improving jet finding at the LHC. One possibility is the use of subjet
capabilities of sequential clustering algorithms, similar to what was
done in \cite{Butterworth:2008iy}, but extended beyond that context. This
potential for future progress in jet-finding methods is yet another
reason for encouraging flexibility in LHC jet-finding.

Finally, all the MC data samples used in this contribution, together
with the results of mass reconstruction using different jet algorithms
can be found at the following webpage:
\begin{center}
\begin{verbatim}
   http://www.lpthe.jussieu.fr/~salam/les-houches-07/
\end{verbatim}.
\end{center}

\subsection*{Acknowledgements}
This work has been supported in part by the grant ANR-05-JCJC-0046-01
from the French Research Agency and under Contract
No. DE-AC02-98CH10886 with the U.S. Department of Energy.


%% file: s_rabbertz/lhkajets.tex
\subsection{Introduction}

With the advent of the LHC, a new regime in center-of-mass energy for
hadron-hadron collisions will be accessed and the by far dominant
feature of the events to be measured is the abundant production of
jets, i.e.\ collimated streams of hadrons that are supposed to
originate from a common initiator. In theory, these initiators are
usually the outgoing partons of a hard interaction calculable in
perturbative QCD (pQCD). Limitations of QCD perturbation theory,
however, make it impossible to unambiguously assign a bunch of
observed hadrons to such a hard parton. To achieve nevertheless the
comparability of our best theoretical knowledge with experimental
results, jet algorithms are employed that define a distance measure
between objects and uniquely determine which of them are sufficiently
close to each other to be considered to come from the same origin and
hence to combine them into a jet. This same procedure is applied
equally to the partons of theory calculations, the final state
particles of Monte-Carlo generators, that serve as input to experiment
simulations, as well as measured deposits of energy in calorimeters or
tracks of charged particles. Provided the jet algorithms are well
behaved, i.e.\ they are especially collinear- and infrared-safe (CIS),
the measured jets can now be related to jets constructed of the theory
objects.

However, a number of residual effects of either experimental origin or
of theoretical nature, the latter comprising perturbative radiation,
hadronization and the underlying event (UE), still have to be taken
into account.  Recent overviews showing how these have been dealt with
in the past, especially at Tevatron, can be found in e.g.\
Refs.~\cite{Albrow:2006rt, Ellis:2007ib}.  Since energies reachable at
the LHC are much larger though than everything investigated so far,
the best choices of jet algorithms and parameters to delimit and/or
control these residual effects have to be reevaluated. In this work we
contribute to this effort by examining the influence of different jet
algorithms and jet sizes on the reconstruction of characteristics of a
hard process. More precisely, we have varied the respective jet size
parameters, usually labelled as $R$ or $D$ and generically denoted as
$R$ further on, from $0.3$ to $1.0$ in steps of $0.1$ for the
following four algorithms:
\begin{itemize}
\item The Midpoint cone algorithm, Ref.~\cite{Blazey:2000qt} (with
  split-merge overlap threshold $f$ of $0.75$ and a seed threshold of
  $1\,{\rm GeV}$)
\item The SISCone algorithm, Ref.~\cite{Salam:2007xv} (with
  split-merge overlap threshold $f$ of $0.75$, an infinite number of
  passes and no transverse momentum cut on stable cones)
\item The $k_T$ algorithm, Refs.~\cite{Ellis:1993tq, Catani:1992zp,
    Catani:1993hr}, in the implementation of Ref.~\cite{Cacciari:2005hq}
\item The Cambridge/Aachen algorithm, Refs.~\cite{Dokshitzer:1997in,
    Wobisch:1998wt}
\end{itemize}
In all cases the four-vector recombination scheme or E scheme was used.  
We note that Midpoint cone is not collinear and infrared-safe and is
included primarily for comparison.

In this first step, we restrict the analysis to examine the transition
from leading-order (LO) pQCD events to fully hadronized ones using
Pythia version 6.4, Ref.~\cite{Sjostrand:2006za}, as event generator.
The parameter set of tune~DWT, Ref.~\cite{Acosta:2006bp}, has been
chosen to represent a possible extrapolation of the underlying event
to LHC energies. On occasion we have employed the S0 tune,
Refs.~\cite{Skands:2007zg, priv:Skands2008},\footnote{In addition to
  the settings given in table~I of Ref.~\cite{Skands:2007zg}, the
  parameters MSTP(88) and PARP(80) have been set to the non-default
  values of $0$ and $0.01$ resp.\ as they would be set by a call to
  the corresponding PYTUNE routine.} as an alternative. A more
complete study is foreseen including further models as given by Herwig
plus JIMMY, Refs.~\cite{Corcella:2002jc, Butterworth:1996zw}, or
Herwig++, Ref.~\cite{Gieseke:2006ga, Bahr:2007ni}.

With this set-up, three primary types of reactions have been
considered representing typical analysis goals:
\begin{itemize}
\item Inclusive jet production for comparison with higher-order
  perturbative calculations and fits of parton density functions,
\item $Z$ boson production in association with a balancing jet for a
  similar usage but in addition for jet calibration purposes and
\item production of heavy resonances with the aim of finding new
  particles and measuring their masses.
\end{itemize}
The choice of resonance produced, $H \rightarrow gg$, has been made so
as to serve as well-defined source of monochromatic gluons and less as
a realistic analysis scenario.  Finally, we adopt a final state truth
definition for the jet finding taking all
stable\footnote{Particles with lifetimes $\tau$ such that $c \tau \geq
  10\,{\rm mm}$.} particles as input apart from prompt leptons or
leptons from decays of heavy resonances.
 
Additional requirements imposed by the experimental set-up and e.g.\
the jet energy calibration or pile-up have to be investigated in
further studies.

\subsection{Inclusive jets}

For inclusive jet transverse momentum spectra one emphasis is on the
comparison of measured data with QCD perturbation theory to higher
order, see for example Refs.~\cite{Abulencia:2005jw, Abulencia:2005yg,
  Abulencia:2007ez, Abazov:2008hu}.  Currently, calculations up to NLO
are at disposal in the form of JETRAD, Ref.~\cite{Giele:1994gf}, or
NLOJET++, Refs.~\cite{Nagy:2001fj, Nagy:2003tz}, which, like most
programs of the cross section integrator type, remain at the parton
level and do not allow to attach perturbative parton showers with
subsequent hadronization so that a full simulation of these events is
excluded.\footnote{Additionally, it would be necessary to perform an
  unweighting step in order to avoid simulating huge amounts of events
  with positive and negative weights.}  As a consequence, when
referring calibrated experimental data unfolded for detector effects
to the NLO calculation, the required corrections cannot be determined
in a completely consistent way. The theoretical "truth", i.e.\ NLO
in this case, lies inbetween the LO matrix element (ME) cross section
and the LO cross section with attached parton showers. Therefore we
present in the following ratios of the inclusive jet $p_T$ spectra of
fully hadronized events with respect to a LO matrix element
calculation. To focus on the hadronization step alone, the same was
performed with respect to the spectrum derived from events including
parton showers but without fragmentation. In the latter case one
should note that the parton radiation has been performed for the hard
interaction as well as for the underlying event so that this
corresponds only to one part of the desired correction.  Most
interesting would be a comparison to the correction achievable with a
NLO program with matched parton showers like MC@NLO,
Refs.~\cite{Frixione:2002ik, Frixione:2006gn}, for which unfortunately
the inclusive jets have not yet been implemented.  A theoretical study
going into more detail on the subject of the composition of
perturbative (parton showers) and non-perturbative (underlying event,
hadronization) corrections to hard interactions can be found in
Ref.~\cite{Dasgupta:2007wa}.

\begin{figure}[tbh]
  \begin{center}
    \includegraphics[width=0.50\textwidth]{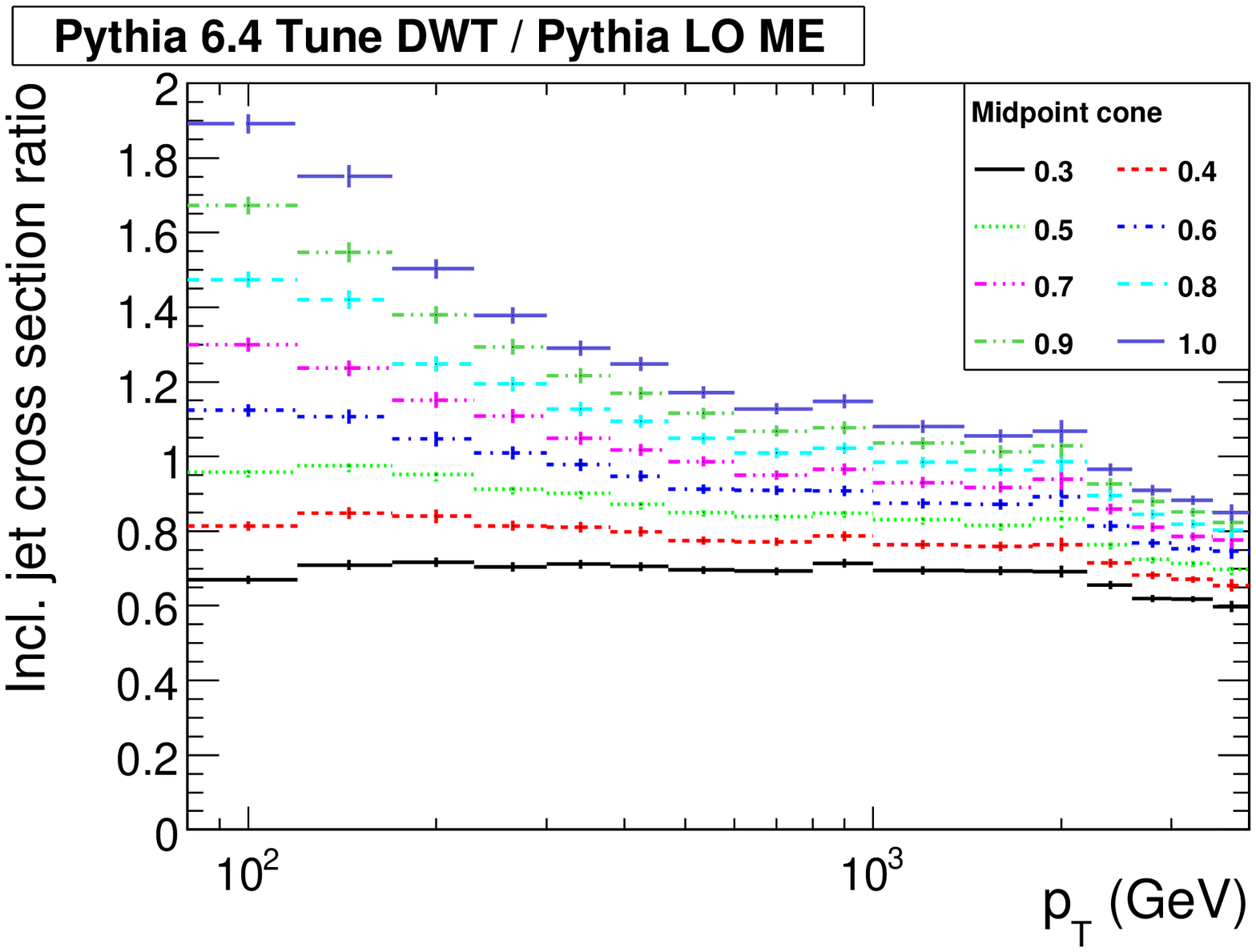}%
    \includegraphics[width=0.50\textwidth]{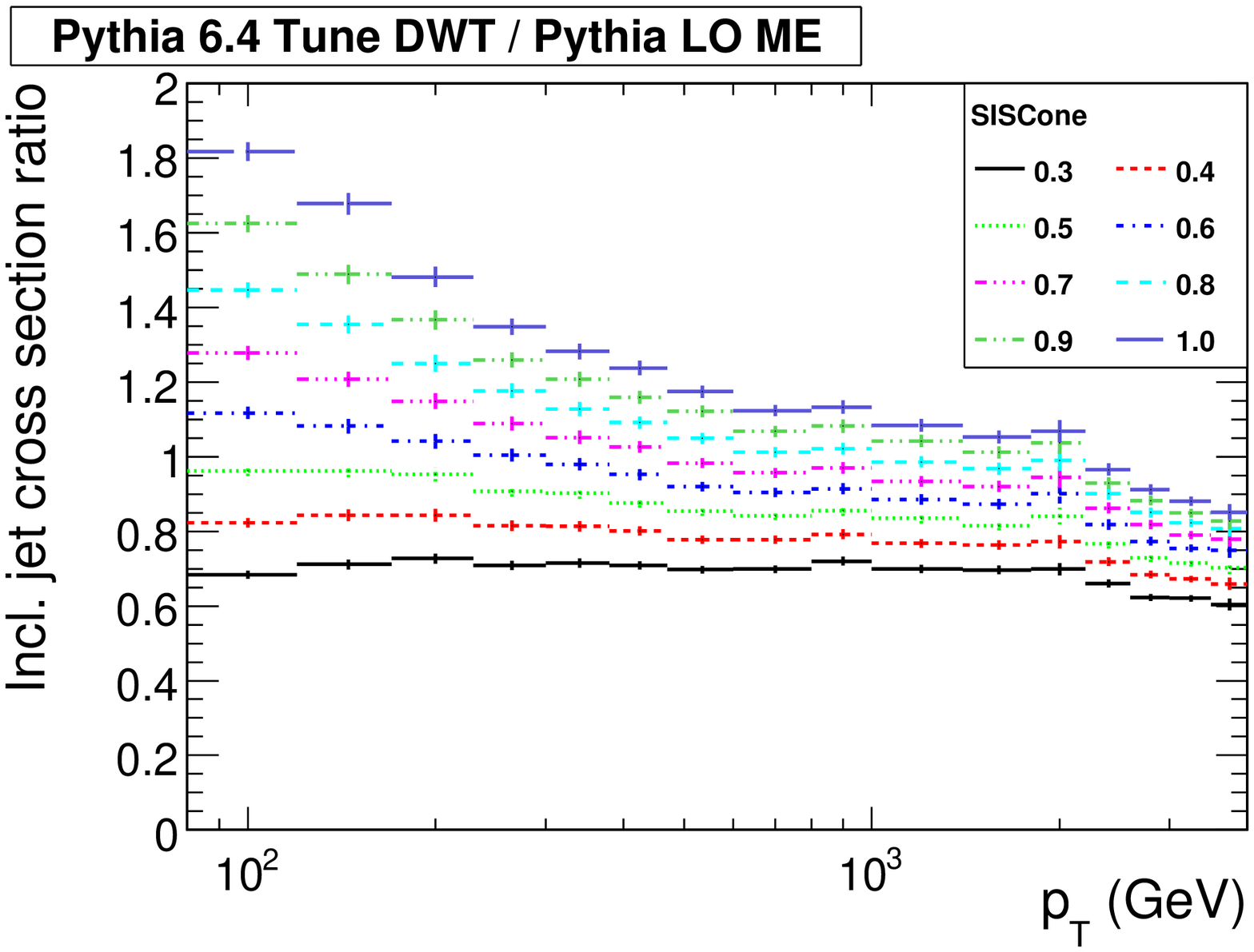}\\
    \includegraphics[width=0.50\textwidth]{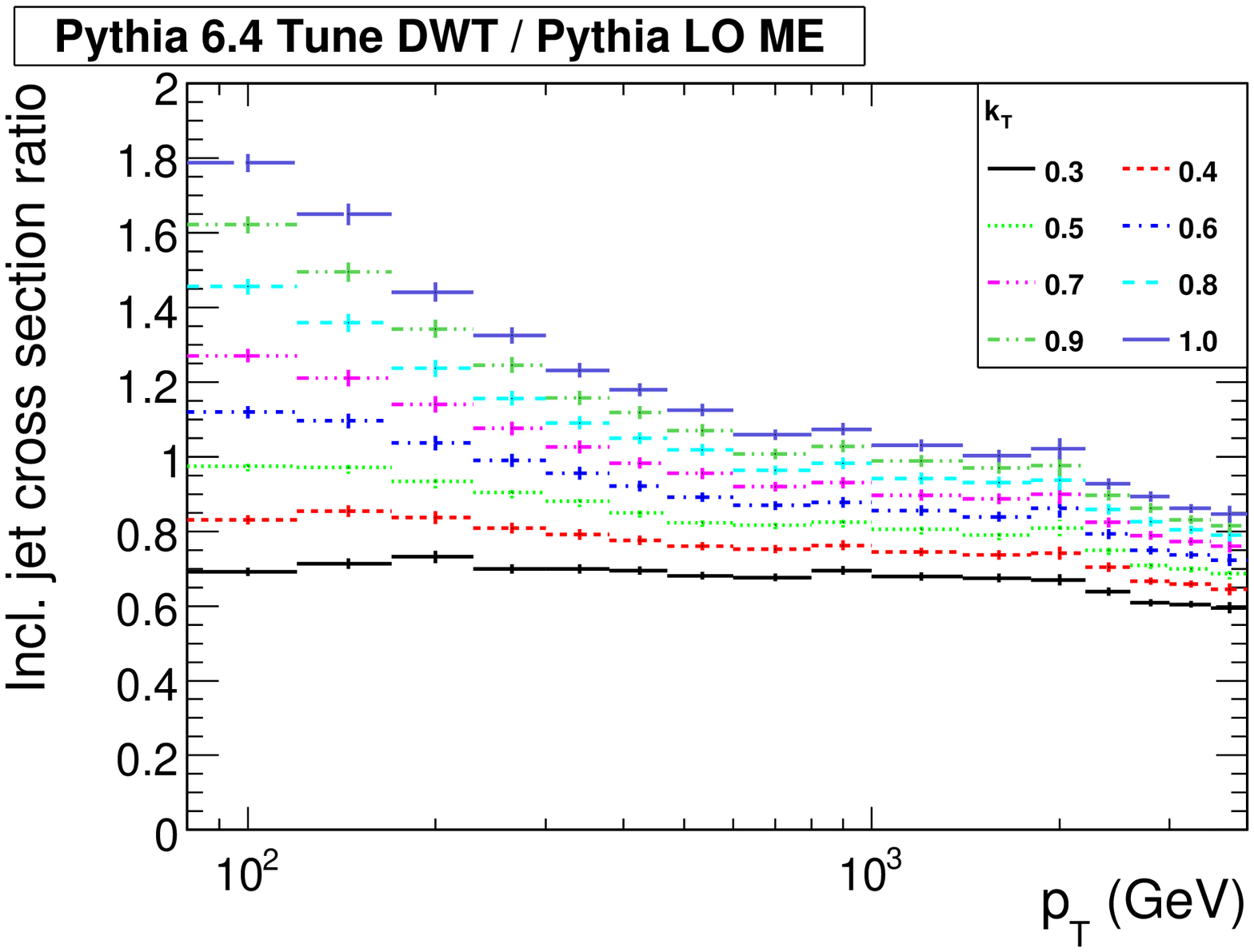}%
    \includegraphics[width=0.50\textwidth]{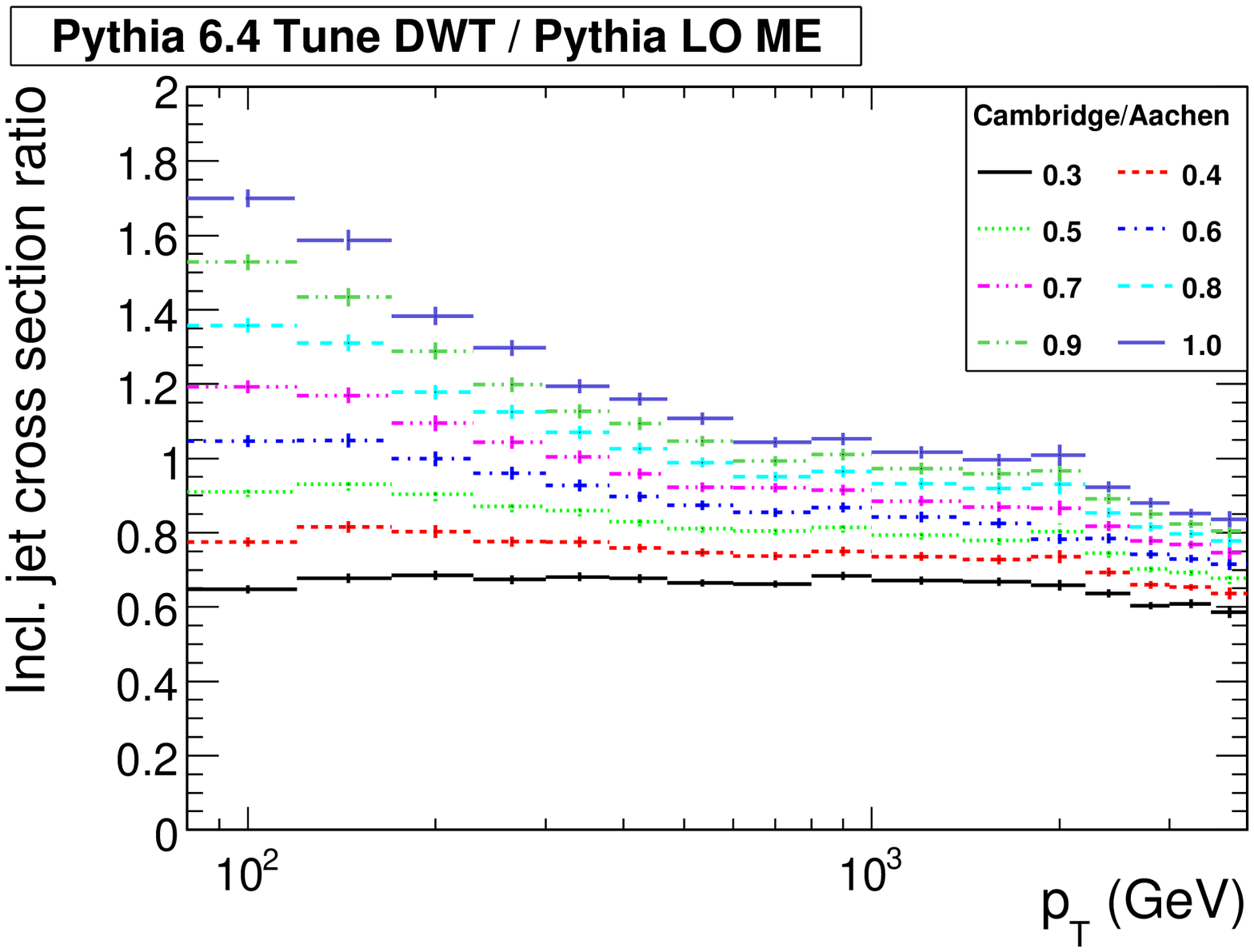}
    \caption{Ratio of inclusive jet cross sections of fully hadronized
      events by Pythia DWT tune over Pythia LO ME for jet sizes $R$ of
      $0.3$ up to $1.0$ for the Midpoint cone (upper left), SISCone
      (upper right), $k_T$ (lower left) and Cambridge/Aachen algorithm
      (lower right).}
    \label{fig:incljets_hadbyme}
  \end{center}
\end{figure}

In this section, the jets have been required to have a minimal
transverse momentum $p_T$ larger than $50\,{\rm GeV}$. No cut on the
jet rapidity or polar angle was imposed.
Figure~\ref{fig:incljets_hadbyme} shows the ratio of inclusive jet
cross sections of fully hadronized events by Pythia DWT tune over
Pythia LO ME for jet sizes $R$ of $0.3$ up to $1.0$ for the
investigated jet algorithms. For the latter, the respective parameters
of the Pythia program controlling the parton shower, initial and final
state radiation, multiple parton interactions (MPI) and the
fragmentation have been switched off. It becomes obvious, that the
effects increasing the jet $p_T$, initial state radiation and multiple
parton interactions, and the effects reducing the jet $p_T$ are
relatively well balanced for $R$ around $0.5$ to $0.6$ for Midpoint
cone and SISCone as well as for $k_{T}$ and Cambridge-Aachen. For
smaller $R$, the jets tend to lose $p_{T}$ due to out-of-cone effects
during the evolution from LO ME to hadronized events, while larger $R$
result in an increase of $p_{T}$ due to the jets collecting particles
from other sources. Corrections to derive the LO ME jet cross section
from the hadronized final state will have to take these effects into
account.

\begin{figure}[tbh]
  \begin{center}
    \includegraphics[width=0.50\textwidth]{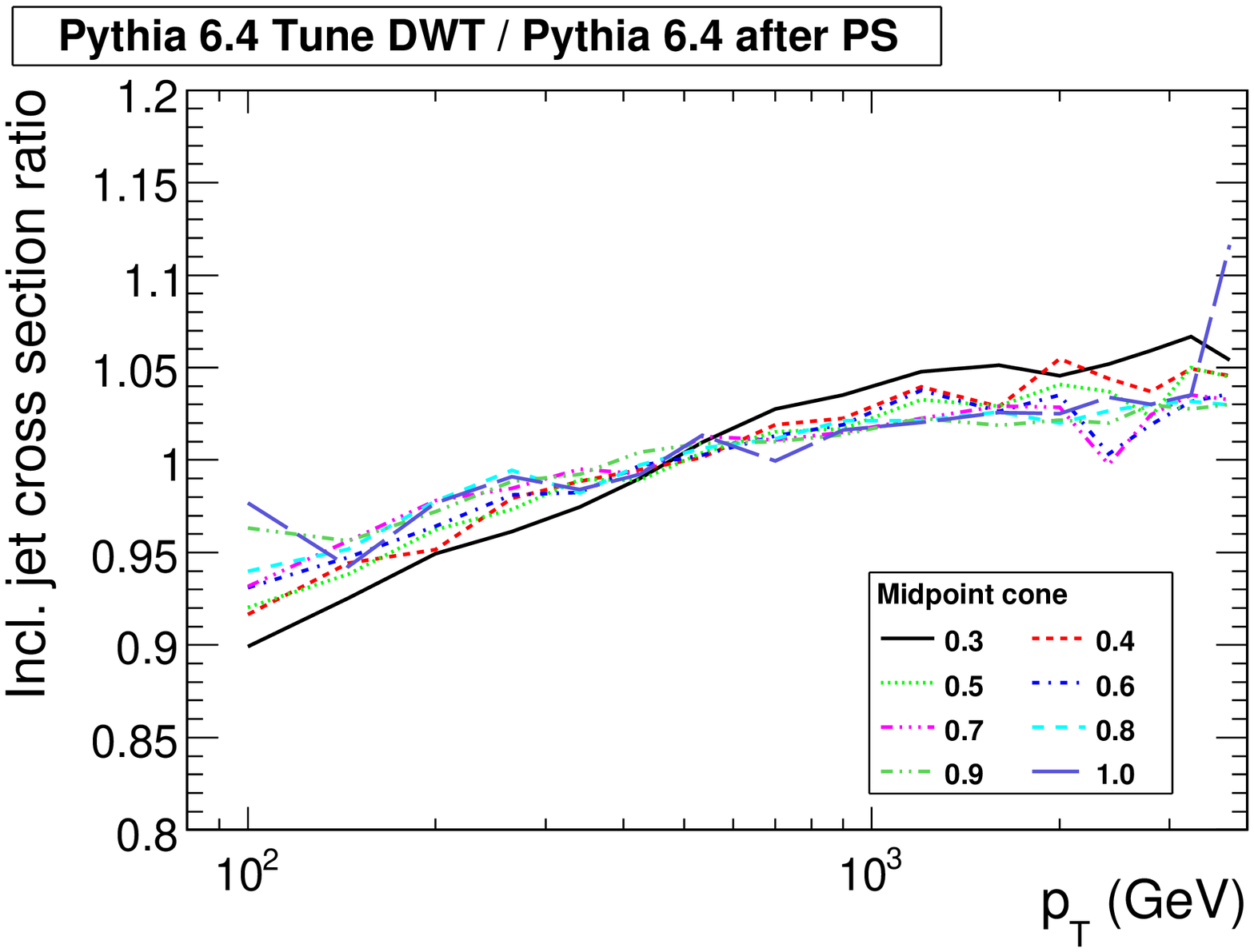}%
    \includegraphics[width=0.50\textwidth]{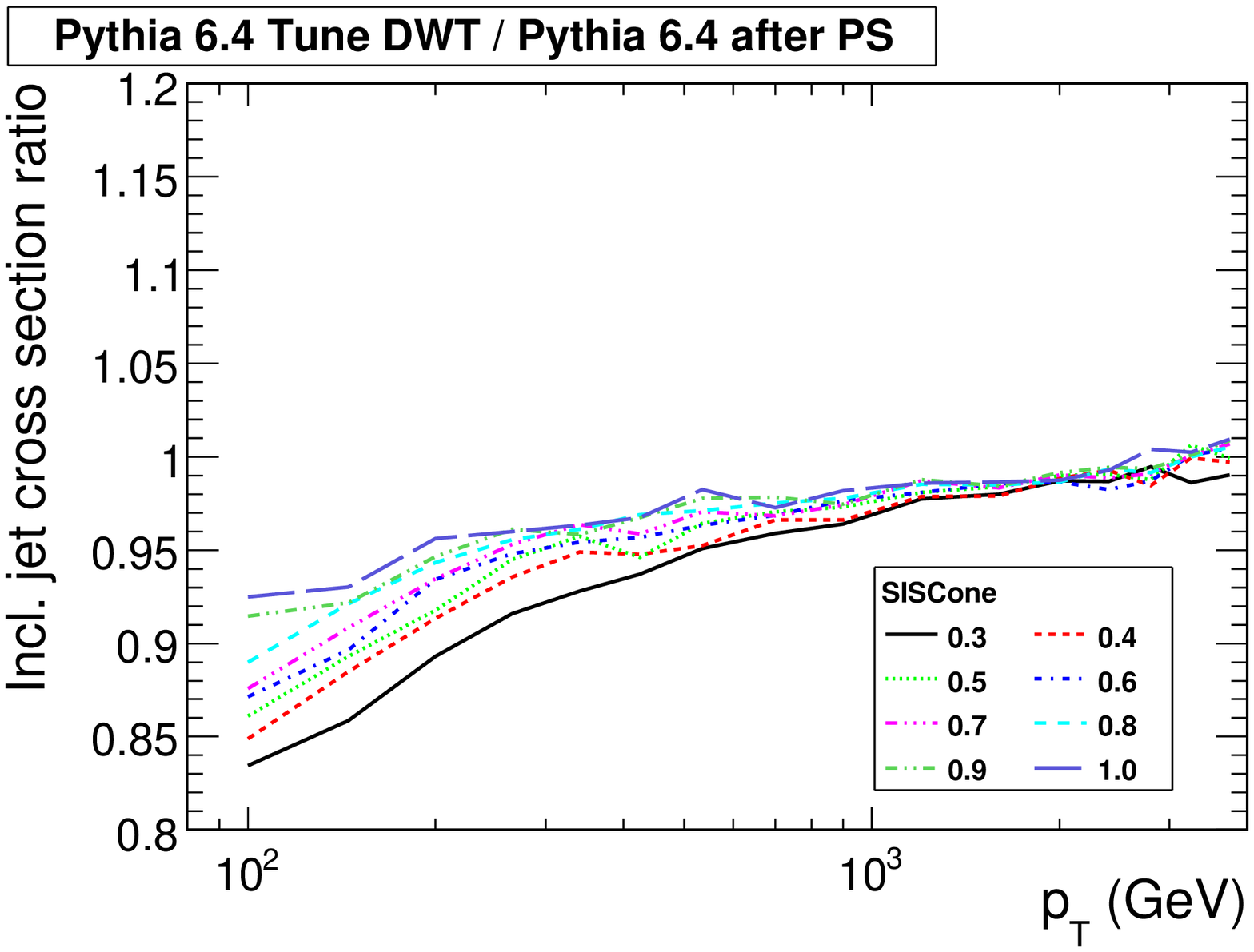}\\
    \includegraphics[width=0.50\textwidth]{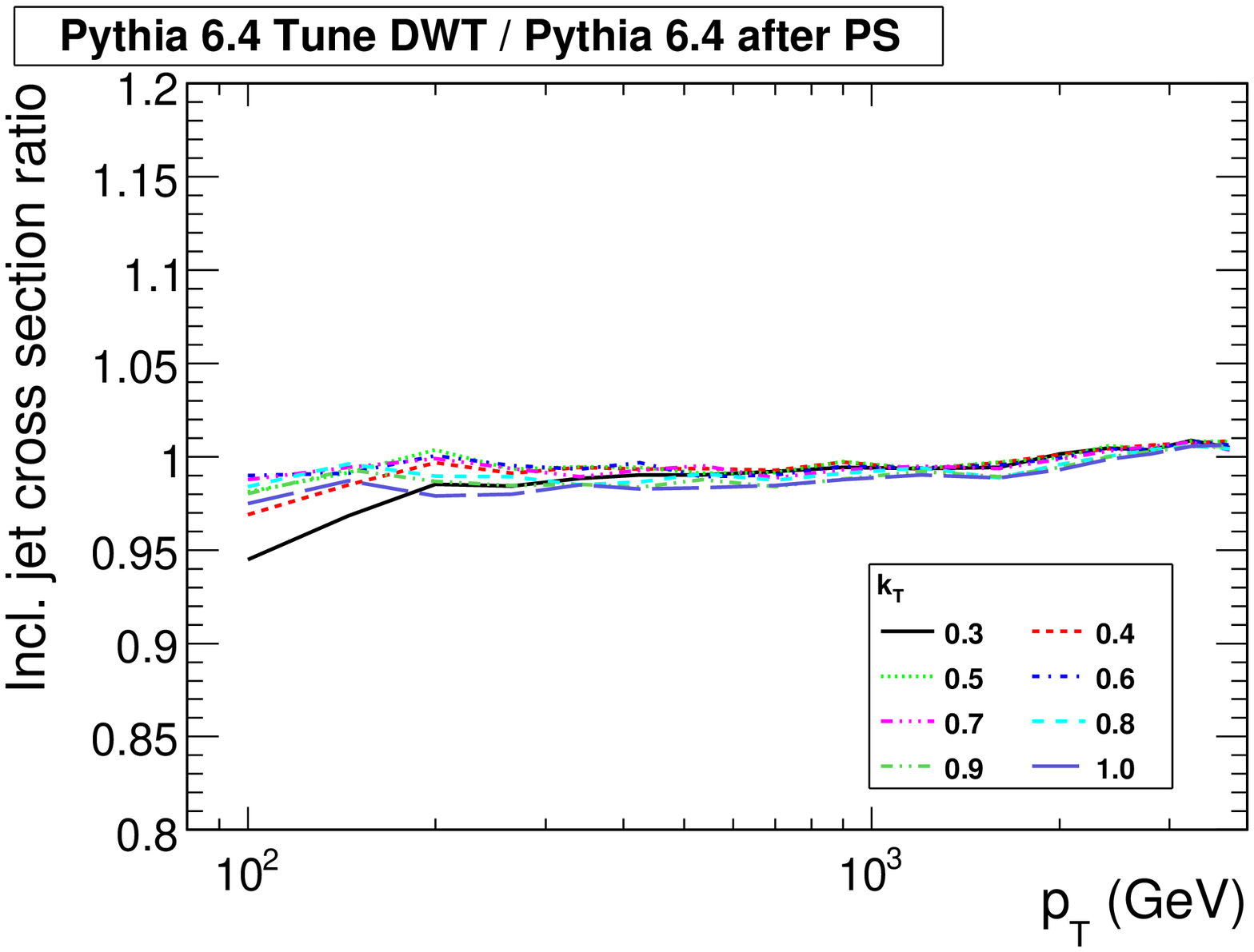}%
    \includegraphics[width=0.50\textwidth]{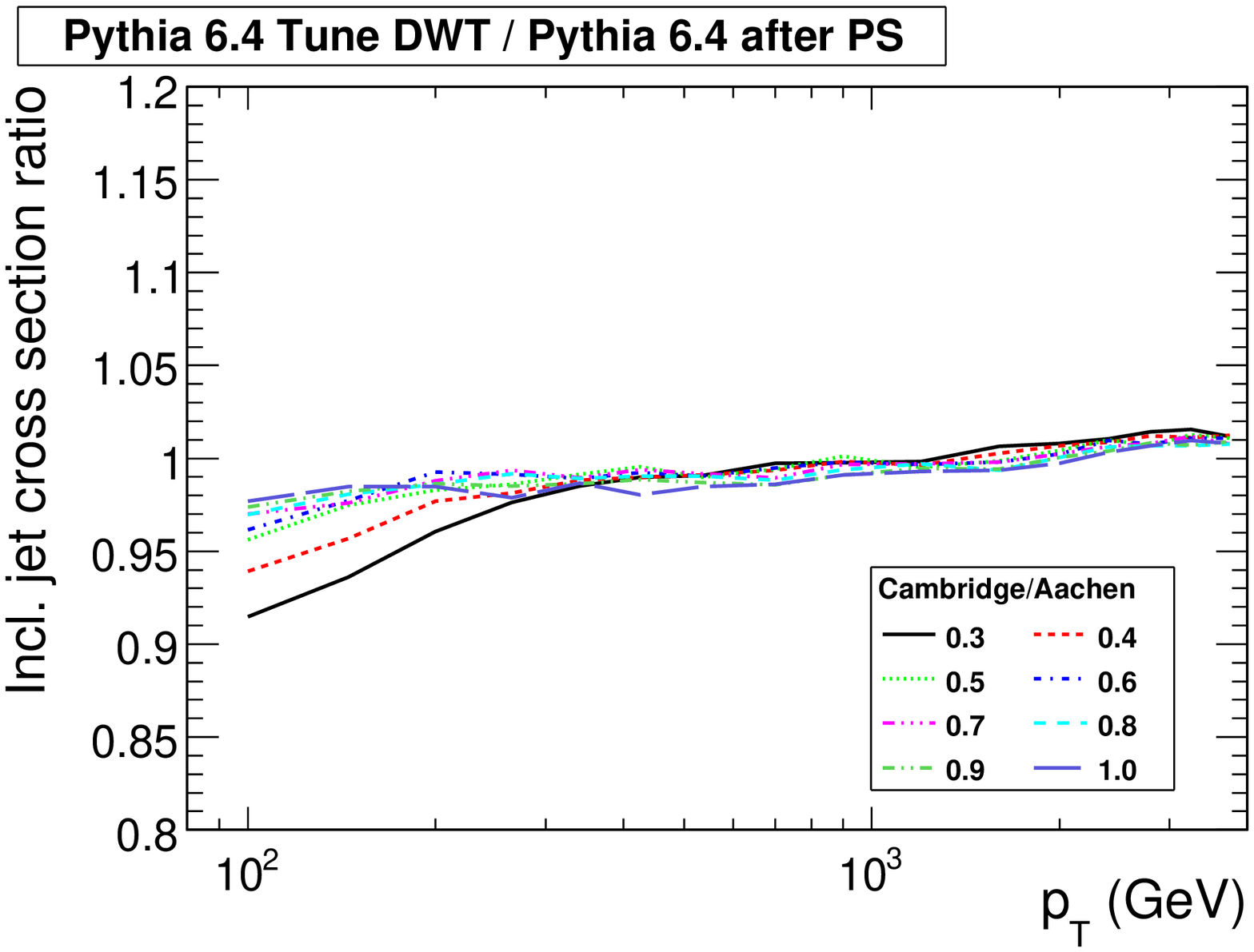}
    \caption{Ratio of inclusive jet cross sections as in
      Fig.~\ref{fig:incljets_hadbyme} but divided by Pythia tune~DWT
      after parton showers (including the underlying event). This
      shows predominantly the influence of the hadronization model.}
    \label{fig:incljets_hadbypfs}
  \end{center}
\end{figure}

In Figure~\ref{fig:incljets_hadbypfs} the jet $p_T$ distribution of
fully hadronized events has been divided by the spectrum after parton
showers (including the underlying event) for the same range of jet
sizes $R$ as above. This shows predominantly the influence of the
hadronization model, Lund string fragmentation in the case of Pythia,
on the jets, usually leading to a loss in $p_T$ especially for
cone-type algorithms and more pronounced for smaller cone sizes due to
out-of-cone effects. The sequential recombination type algorithms like
$k_T$ and Cambridge/Aachen are almost unaffected for all choices of
$R$.

\begin{figure}[tbh]
  \begin{center}
    \includegraphics[width=0.5\textwidth]{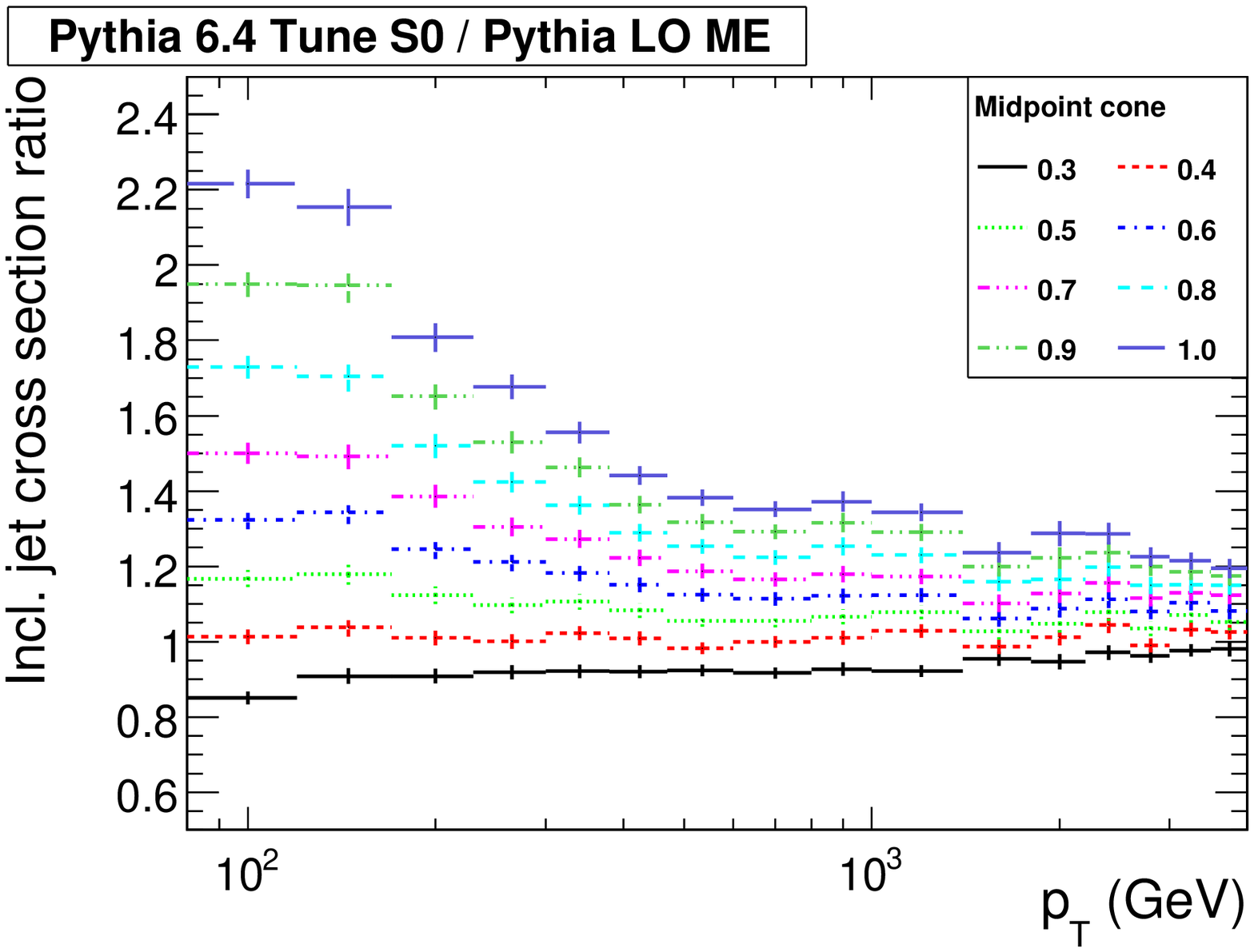}%
    \includegraphics[width=0.5\textwidth]{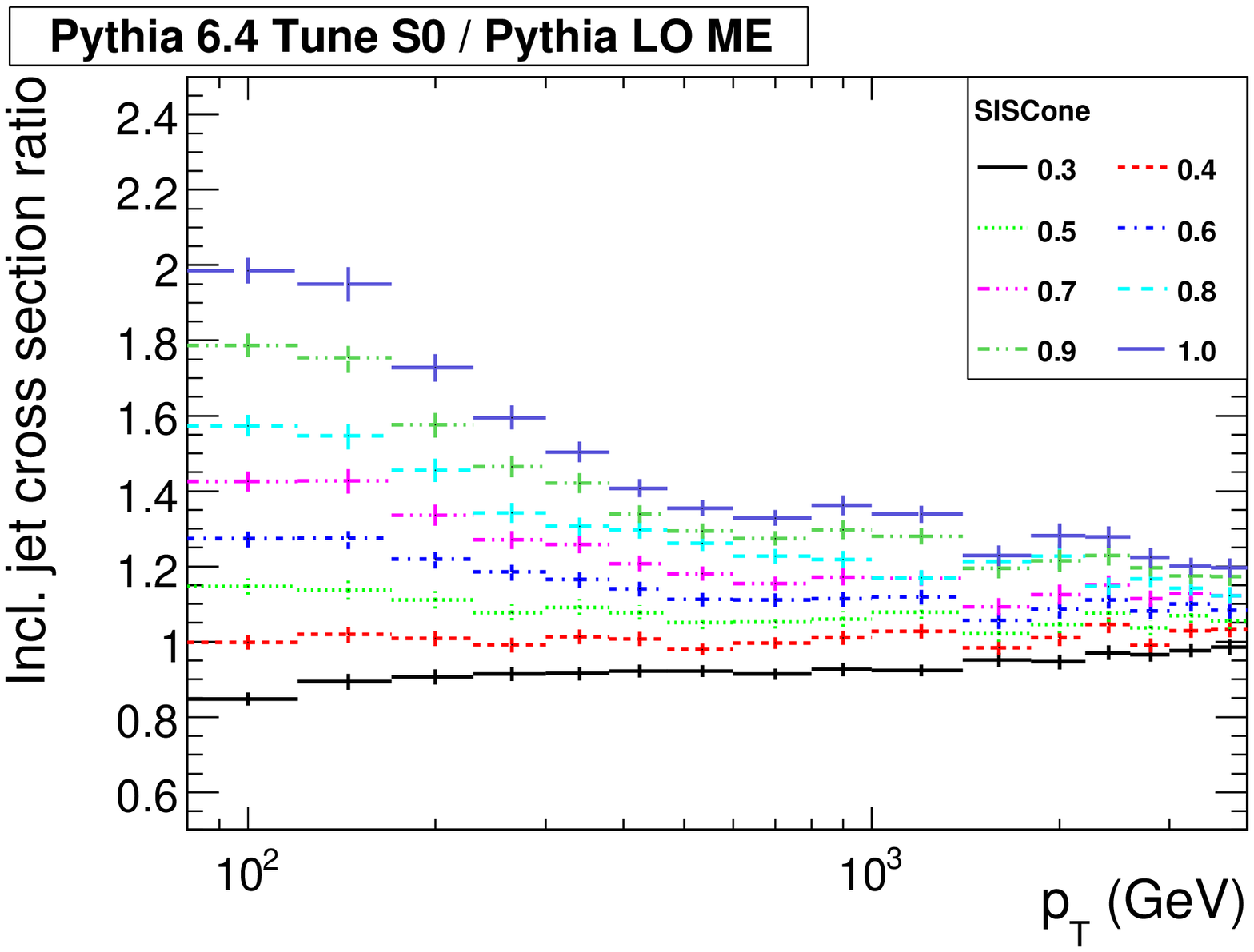}\\
    \includegraphics[width=0.5\textwidth]{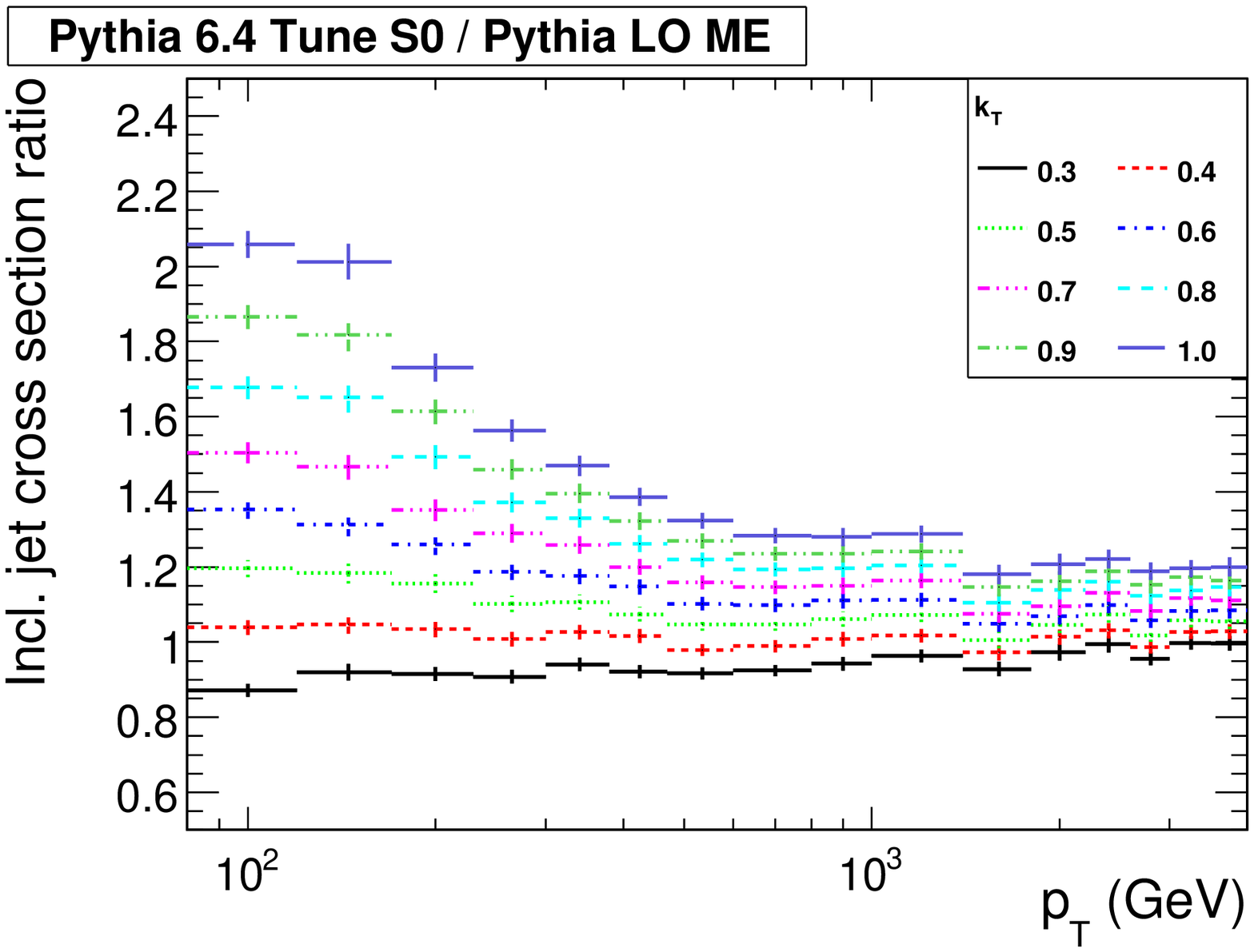}%
    \includegraphics[width=0.5\textwidth]{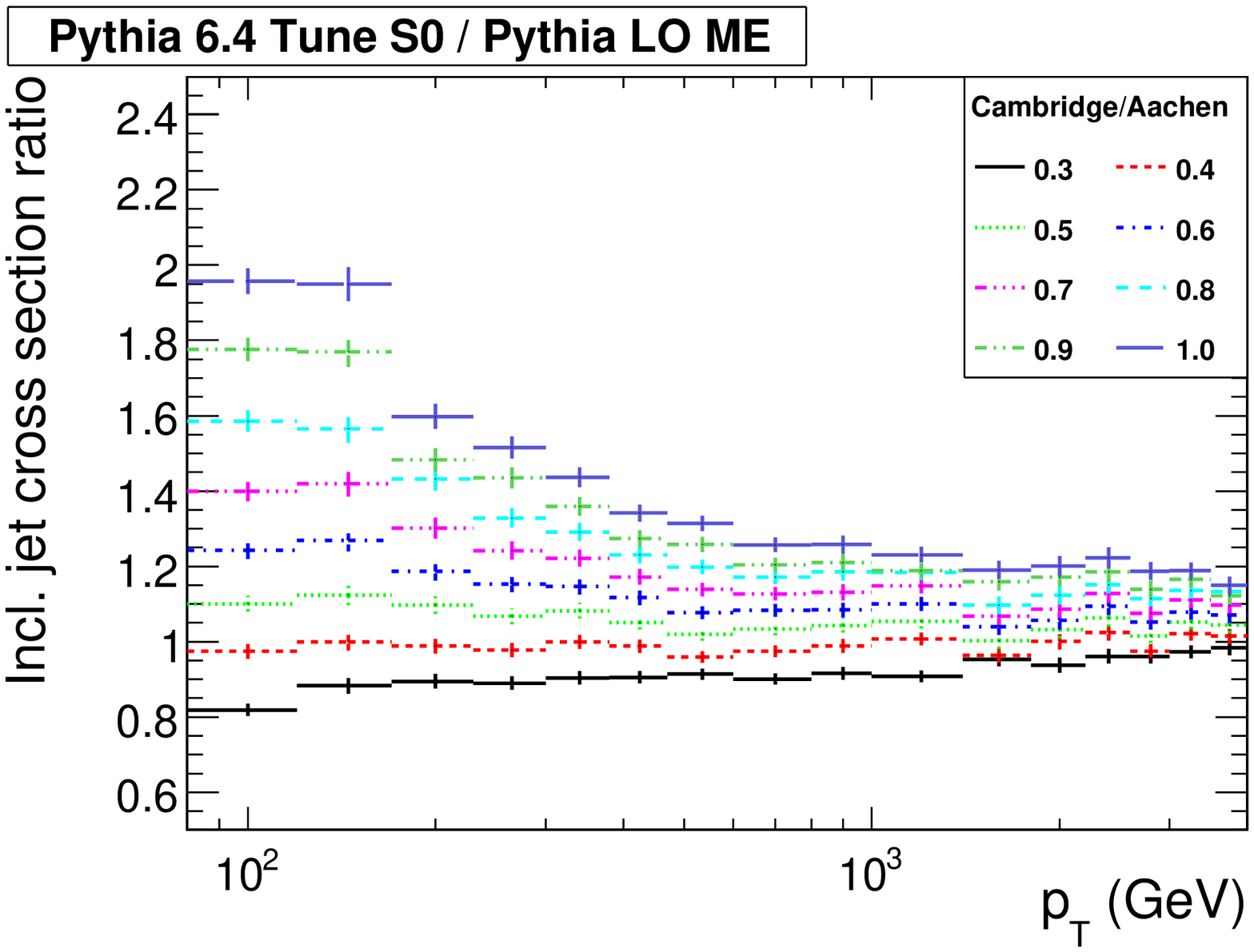}
    \caption{Ratio of inclusive jet cross sections as in
      Fig.~\ref{fig:incljets_hadbyme} but for events with Pythia
      tune~S0.}
    \label{fig:incljets_hadbyme_s0}
  \end{center}
\end{figure}

Finally, to emphasize the importance of the underlying event we
present in Fig.~\ref{fig:incljets_hadbyme_s0} the same ratios as in
Fig.~\ref{fig:incljets_hadbyme} but for the alternative tune~S0
employing a completely new model for both, parton shower and multiple
parton interactions. Events produced with this tune contain a small
fraction of jets with $p_{T}$ significantly higher than it would be
expected from the imposed phase space restrictions on the event
generation. These events had to be removed manually to avoid artefacts
in the inclusive jet cross sections due to their high weights and the
procedure to combine event samples generated separately in bins of the
hard momentum scale.  The number of discarded events is well below one
percent for all algorithms and jet sizes $R$.

As can be seen, the fully hadronized tune~S0 events generally contain
jets with higher $p_{T}$ than the events produced with tune~DWT, which
is mainly due to an increased amount of energy spread into the event by
the new MPI model. This yields the somewhat surprising consequence
that an $R$ of $0.4$ delivers a ratio that is very close to unity for
all applied jet algorithms over the whole $p_{T}$ range.

\subsection{\boldmath $Z$ plus jets\unboldmath}

At LHC energies, events with $Z$ bosons and jets will be much more
abundant than at the Tevatron. Therefore the aspect of calibrating jet
energies using the balancing transverse momentum of a reconstructed
$Z$ boson will become more important. In addition, $Z$ plus jet
reconstruction suffers much less from backgrounds than the similarly
useful photon plus jets process, where the huge cross section for
di-jet production requires, due to misidentified jets, to impose
strong isolation criteria on the photons.  Restricting the analysis to
decays of the $Z$ boson into two muons, as done here, has the further
advantage to decouple completely the jet energy scale from
calorimetric measurements and to relate it to the muon track
reconstruction instead.\footnote{Nevertheless, $Z$ decays into
  electron-positron pairs are very useful, since already the
  electromagnetic energy scale is known more precisely than the
  hadronic one and also here track information can be exploited.}

\begin{figure}[tbh]
  \begin{center}
    \includegraphics[width=0.5\textwidth]{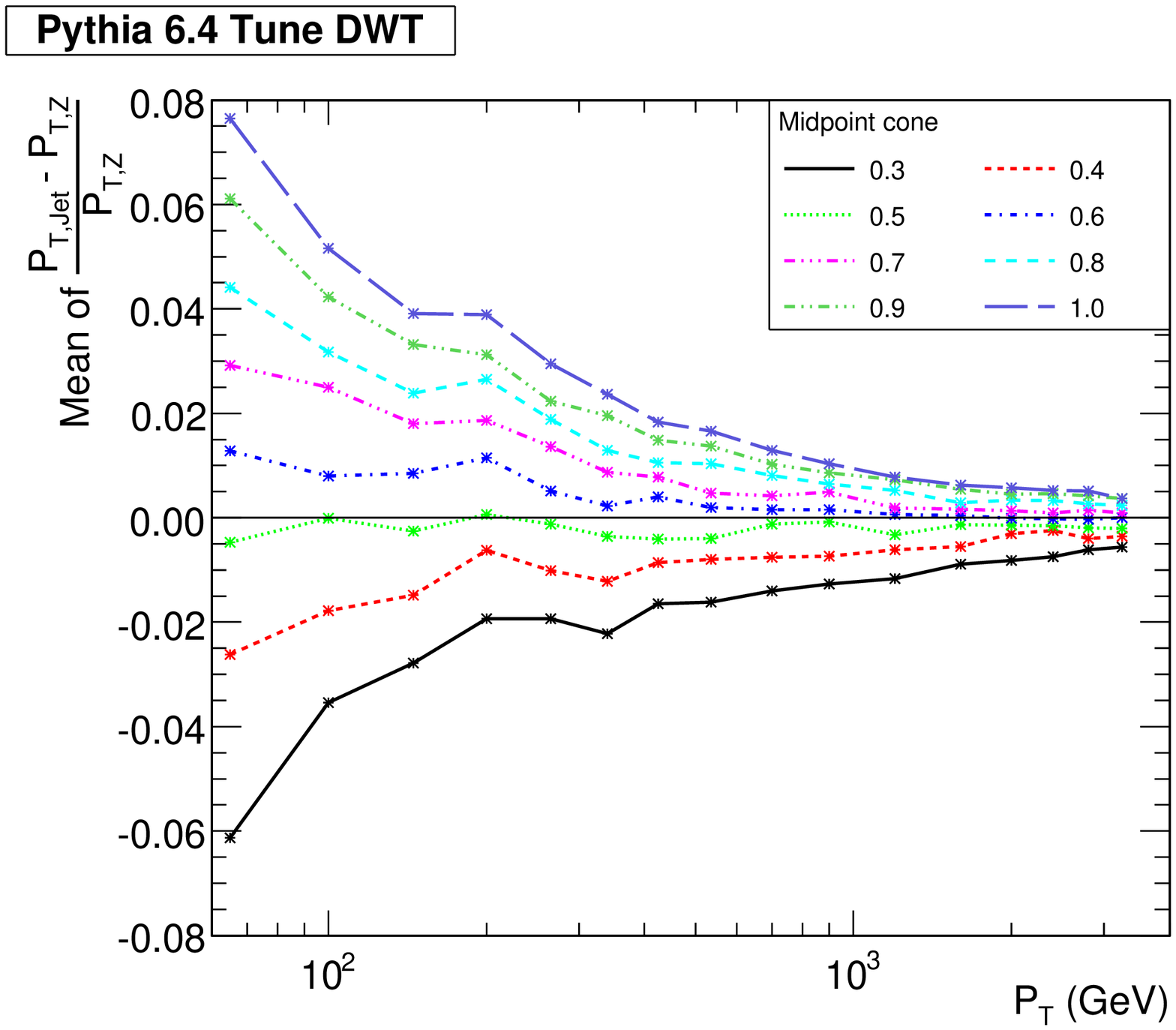}%
    \includegraphics[width=0.5\textwidth]{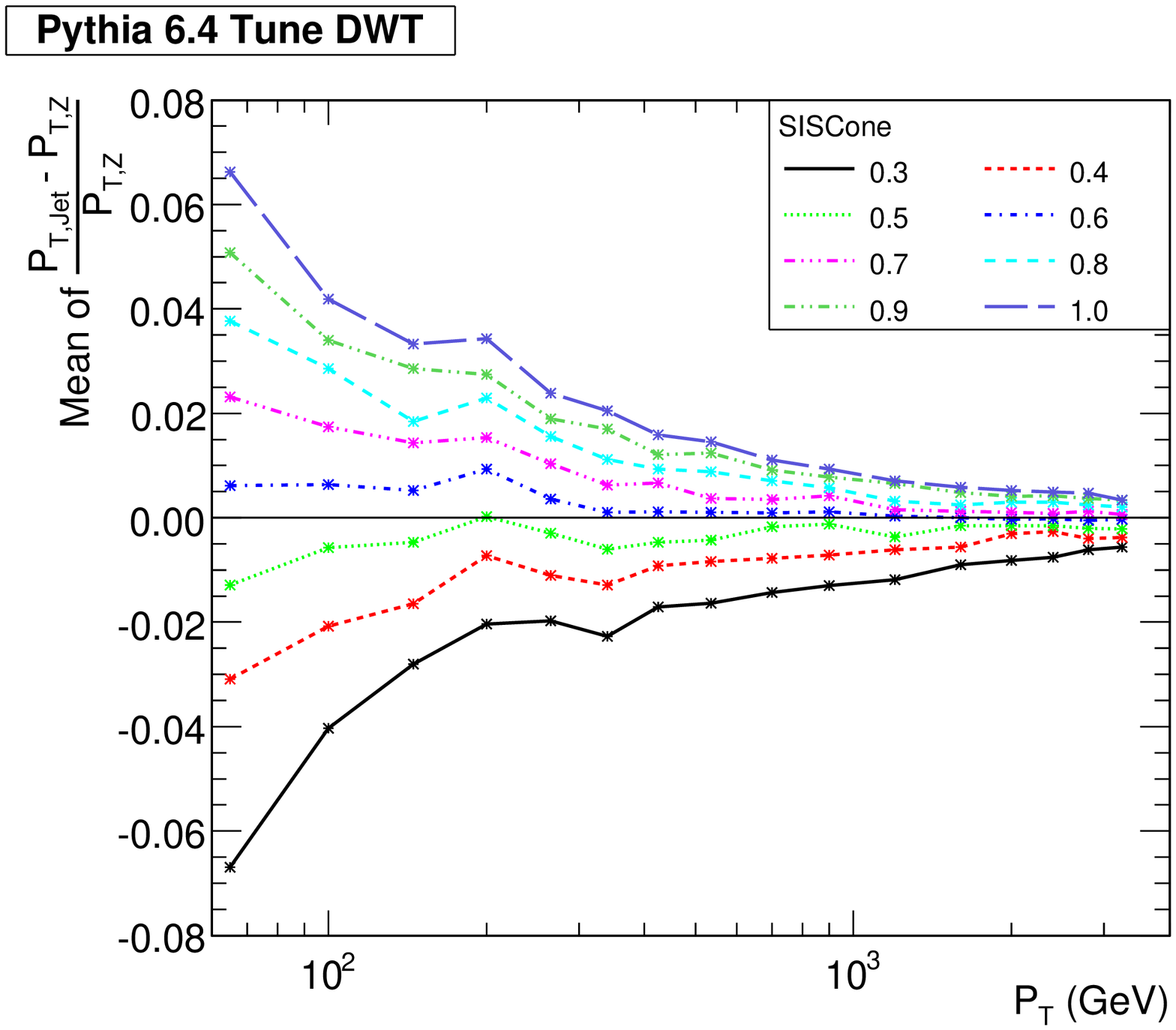}\\
    \includegraphics[width=0.5\textwidth]{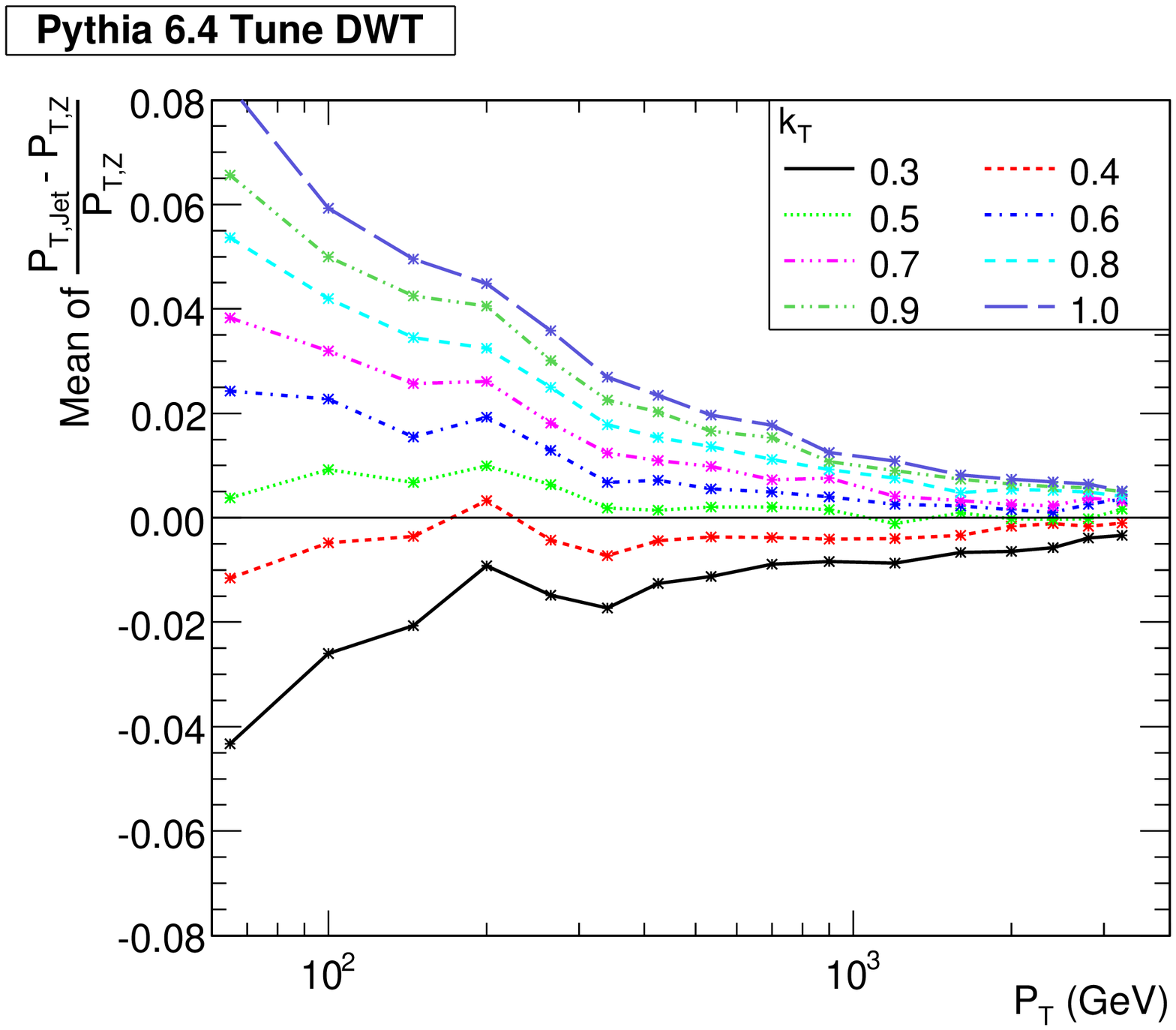}%
    \includegraphics[width=0.5\textwidth]{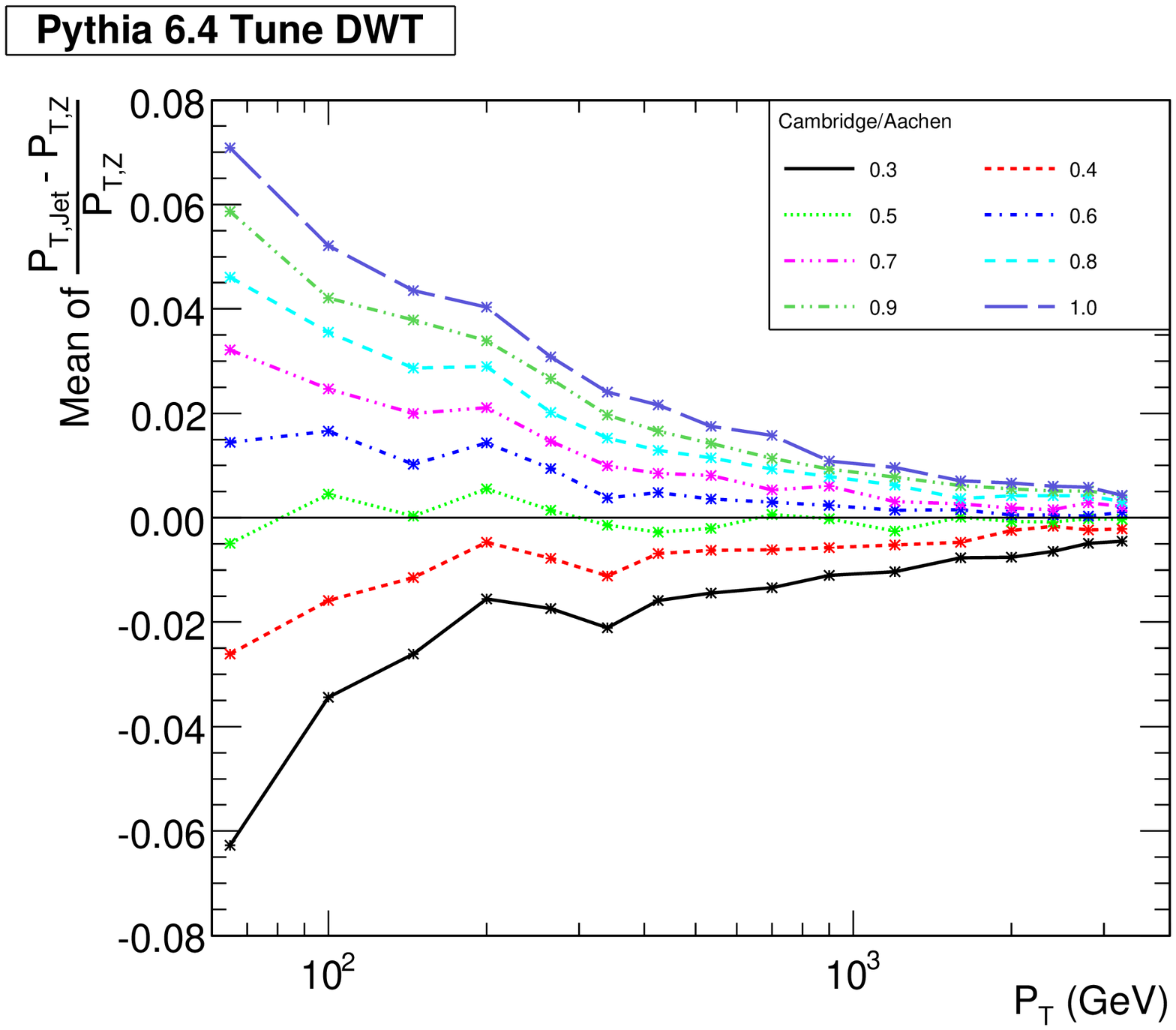}
    \caption{Relative deviation between the transverse momentum of the
      jet and the balancing $Z$ boson from a Gaussian fit of fully
      hadronized Pythia tune~DWT events versus $p_T$ for jet sizes $R$
      of $0.3$ up to $1.0$ for the Midpoint cone (upper left), SISCone
      (upper right), $k_T$ (lower left) and Cambridge/Aachen algorithm
      (lower right).}
    \label{fig:Zjets_had_DWT}
  \end{center}
\end{figure}

In the following, events will be selected with respect to the best
possible jet calibration.  The quantity we will be looking at is the
average relative deviation of the reconstructed jet $p_T$ from the
transverse momentum of the balancing $Z$~boson $(p_{T,{\rm jet}} -
p_{T,Z})/p_{T,Z}$. As this is only valid for events, in which the
$Z$~boson is exactly balanced by one jet of the hard process, one has
to extract a clean sample of $Z$ plus one jet events. Additional
selection criteria are imposed due to geometrical and triggering
limitations of a typical LHC detector.

\begin{figure}[tbh]
  \begin{center}
    \includegraphics[width=0.5\textwidth]{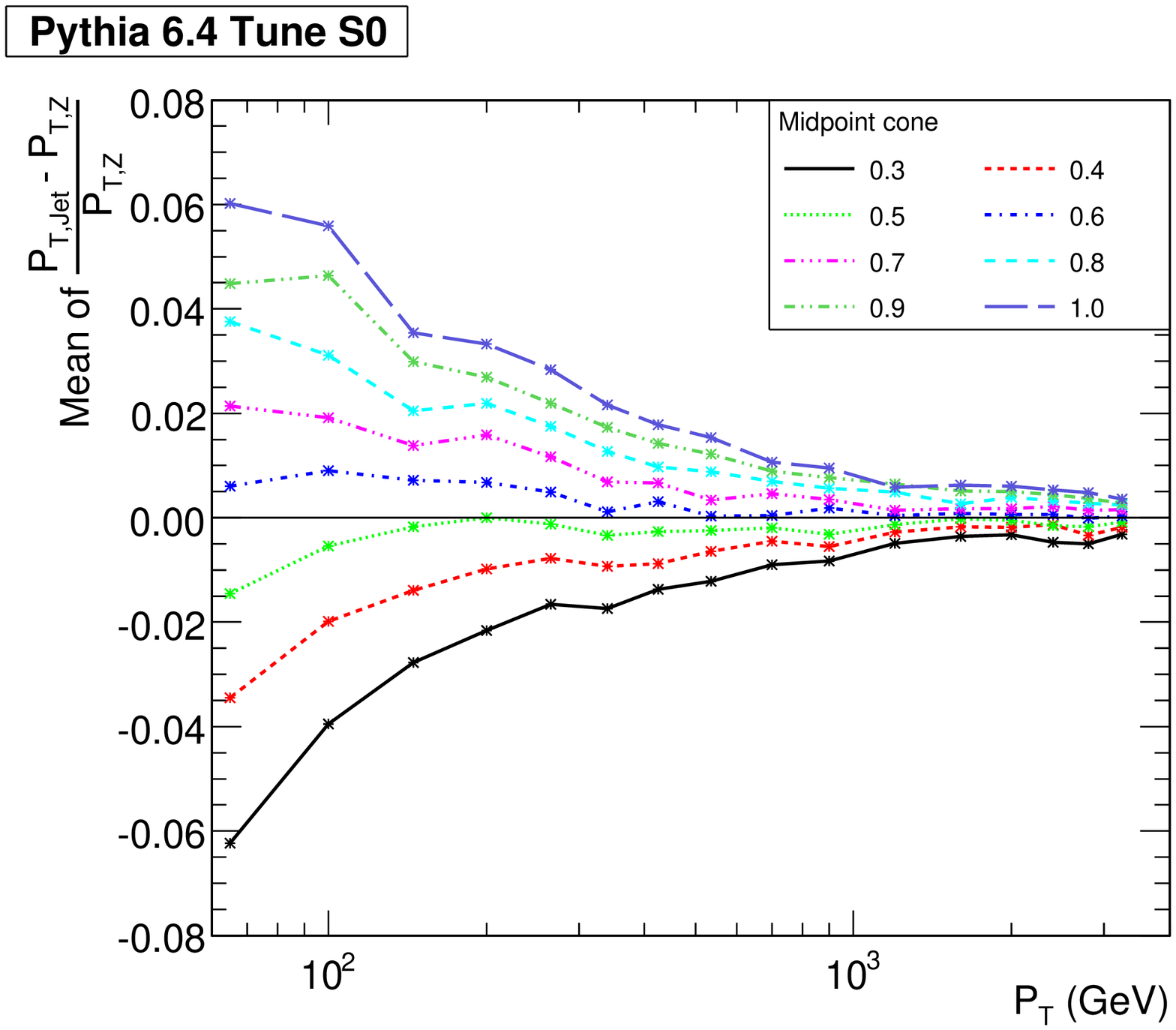}%
    \includegraphics[width=0.5\textwidth]{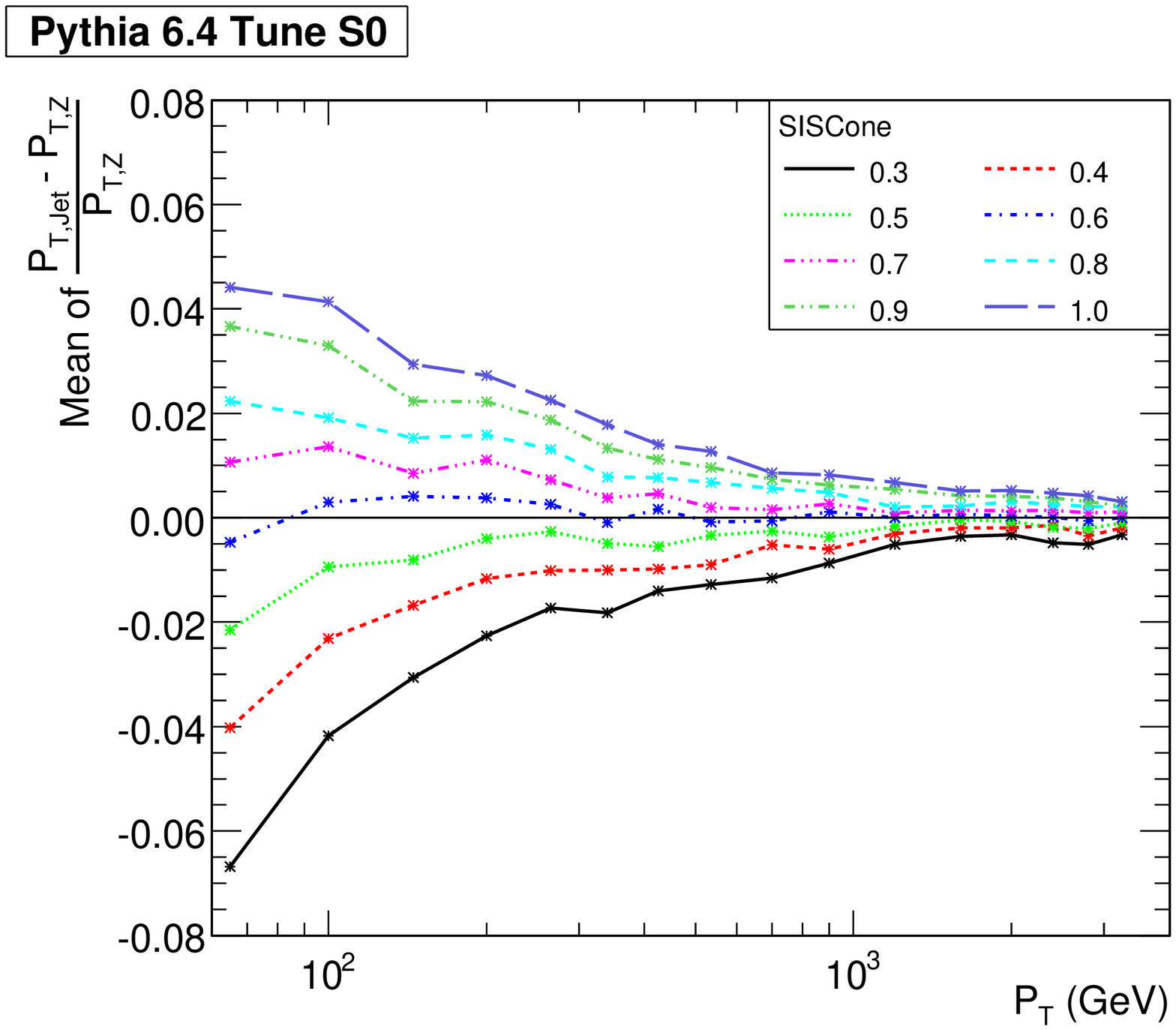}\\
    \includegraphics[width=0.5\textwidth]{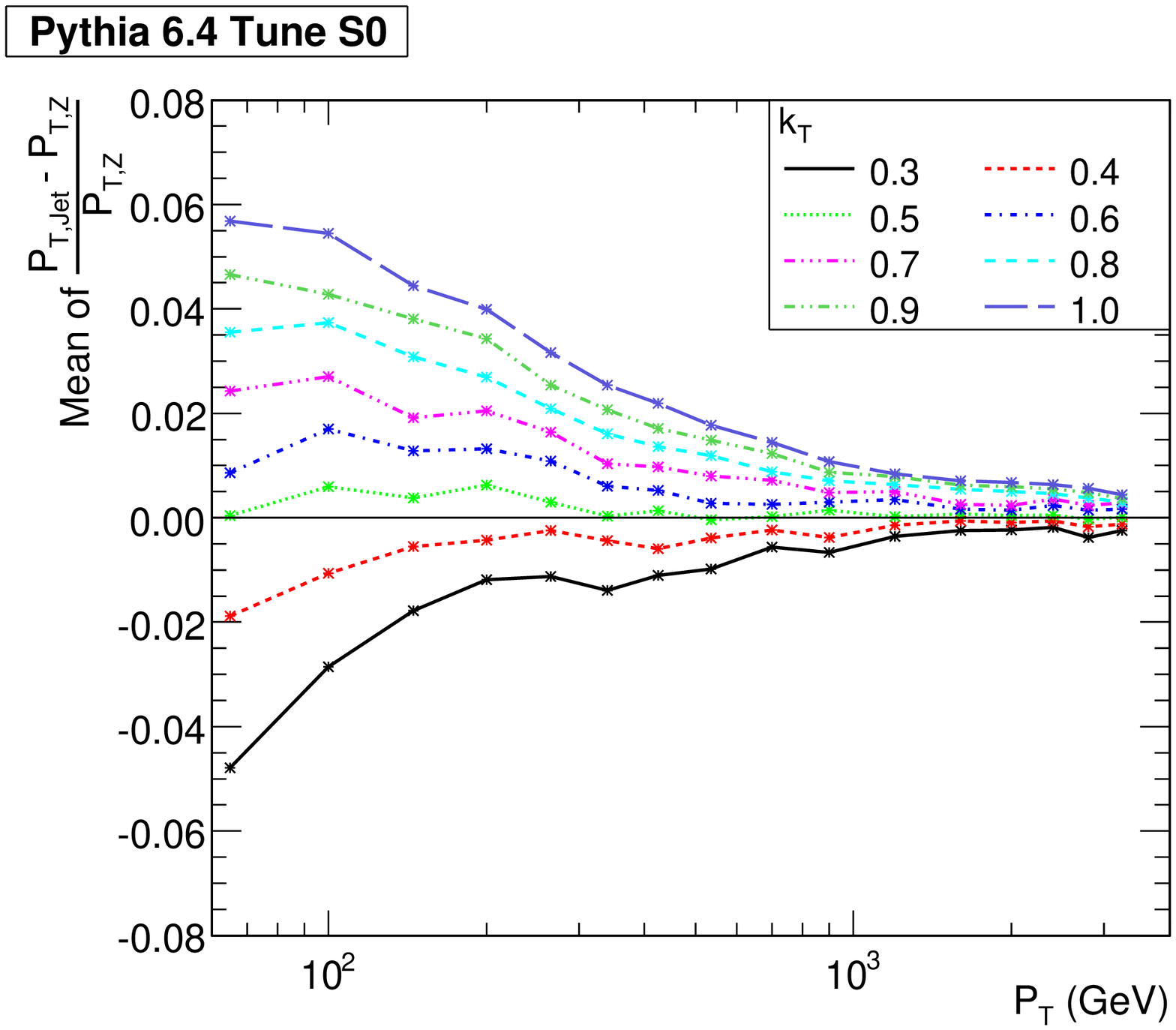}%
    \includegraphics[width=0.5\textwidth]{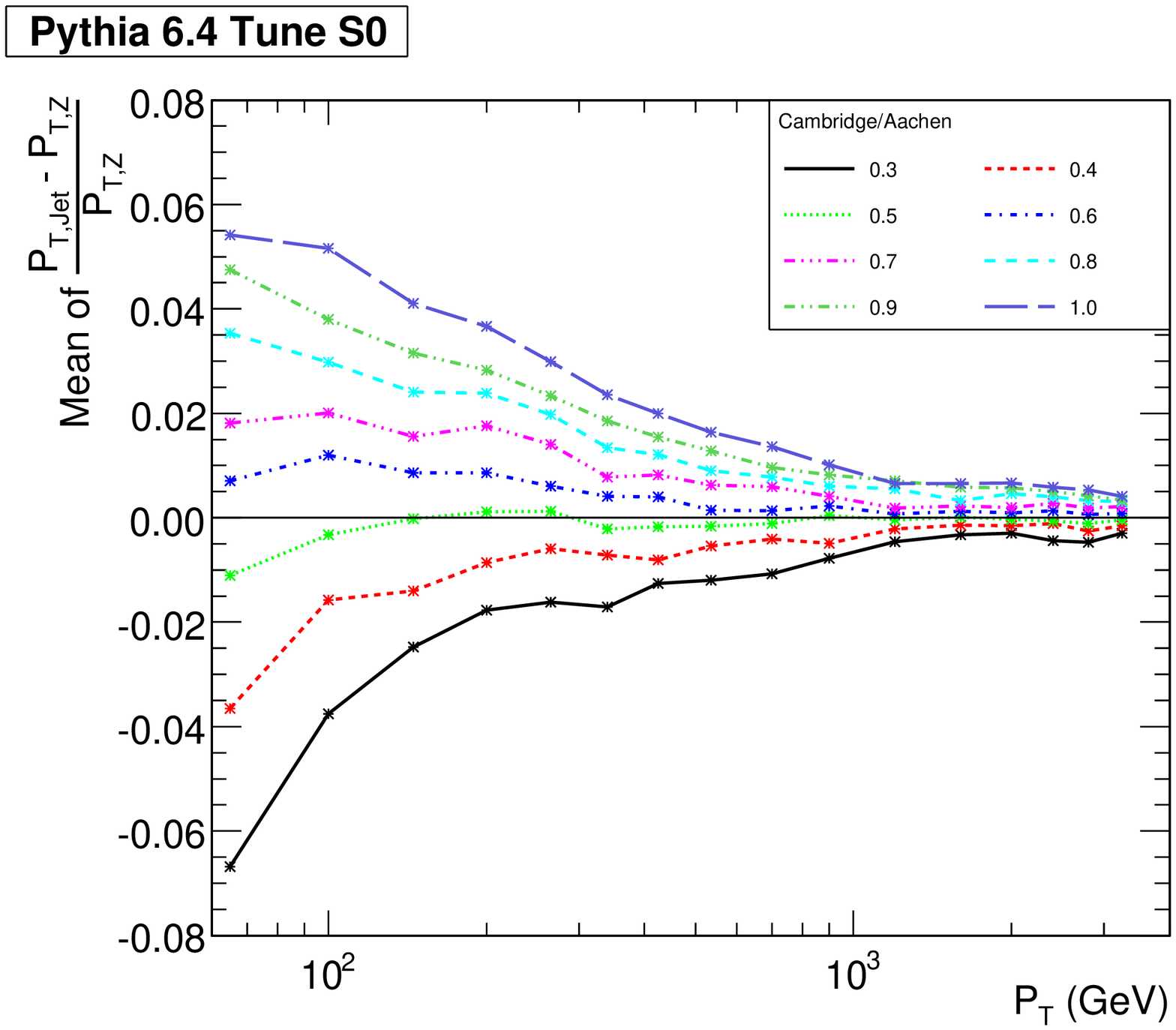}
    \caption{Same as Fig.~\ref{fig:Zjets_had_DWT} but for Pythia
      tune~S0.}
    \label{fig:Zjets_had_S0}
  \end{center}
\end{figure}

A precise measurement of the muon kinematics with a tracking system is
assumed to be feasible in the region in pseudo-rapidity\footnote{$\eta
  = -\ln(\tan\frac{\Theta}{2})$}~$|\eta|$ of up to~$2.4$. Due to
possible trigger constraints, only events are considered where both
muons have transverse momenta larger than $15\,{\rm GeV}$. Having
identified two or more muons in an event, the pair of muons with
opposite charge and an invariant mass closest to the $Z$~mass is
chosen. The event is accepted if the invariant mass of this di-muon
system is closer to the $Z$~mass than $20\,{\rm GeV}$.  Likewise, from
the jet collection only jets in the central region with $|\eta| < 1.3$
are selected where uncalibrated but otherwise reliable jet energy
measurements are expected. In addition, the jets are required to have
a minimal transverse momentum of $20\,{\rm GeV}$.

\begin{figure}[tbh]
  \begin{center}
    \includegraphics[width=0.5\textwidth]{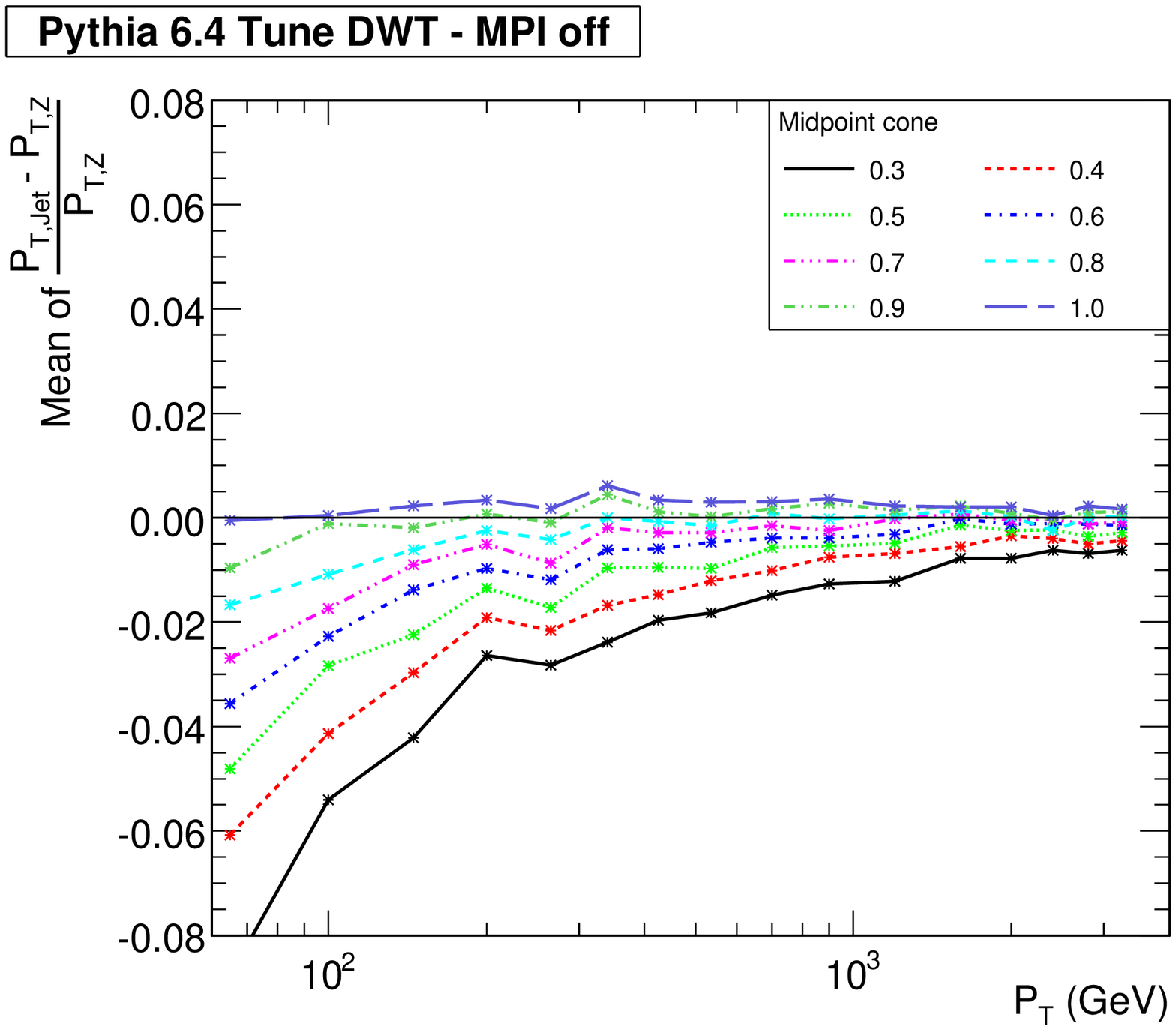}%
    \includegraphics[width=0.5\textwidth]{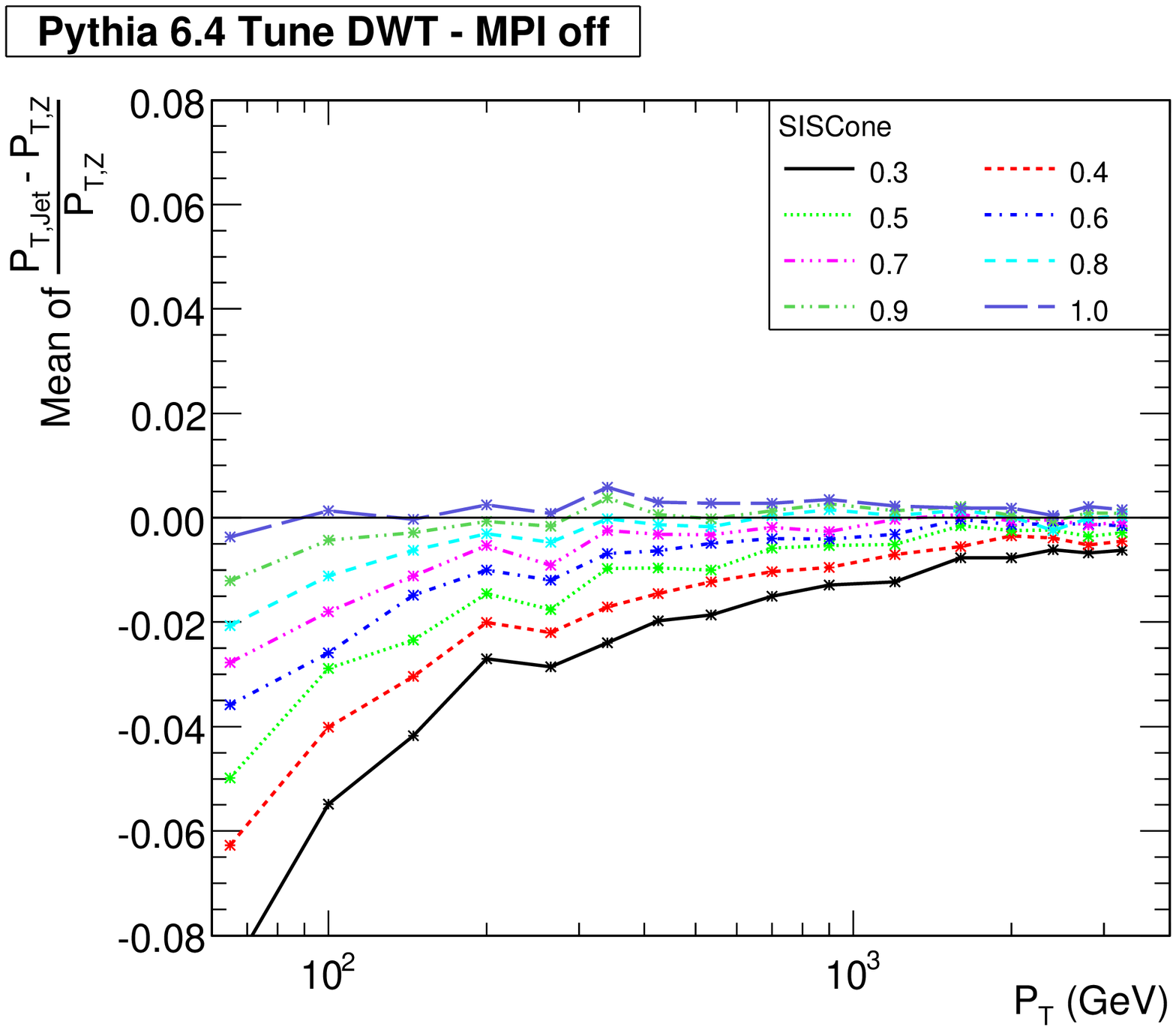}\\
    \includegraphics[width=0.5\textwidth]{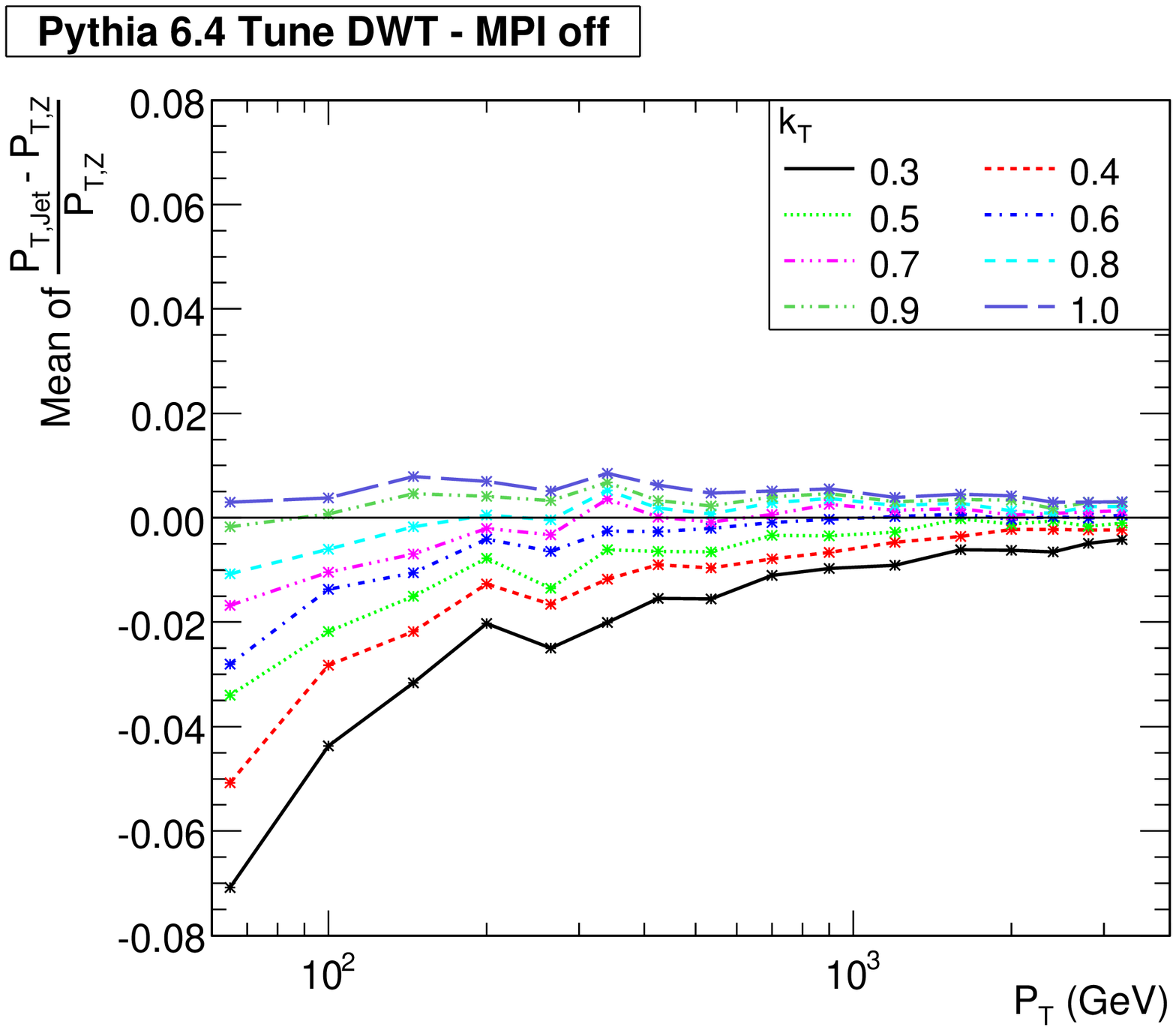}%
    \includegraphics[width=0.5\textwidth]{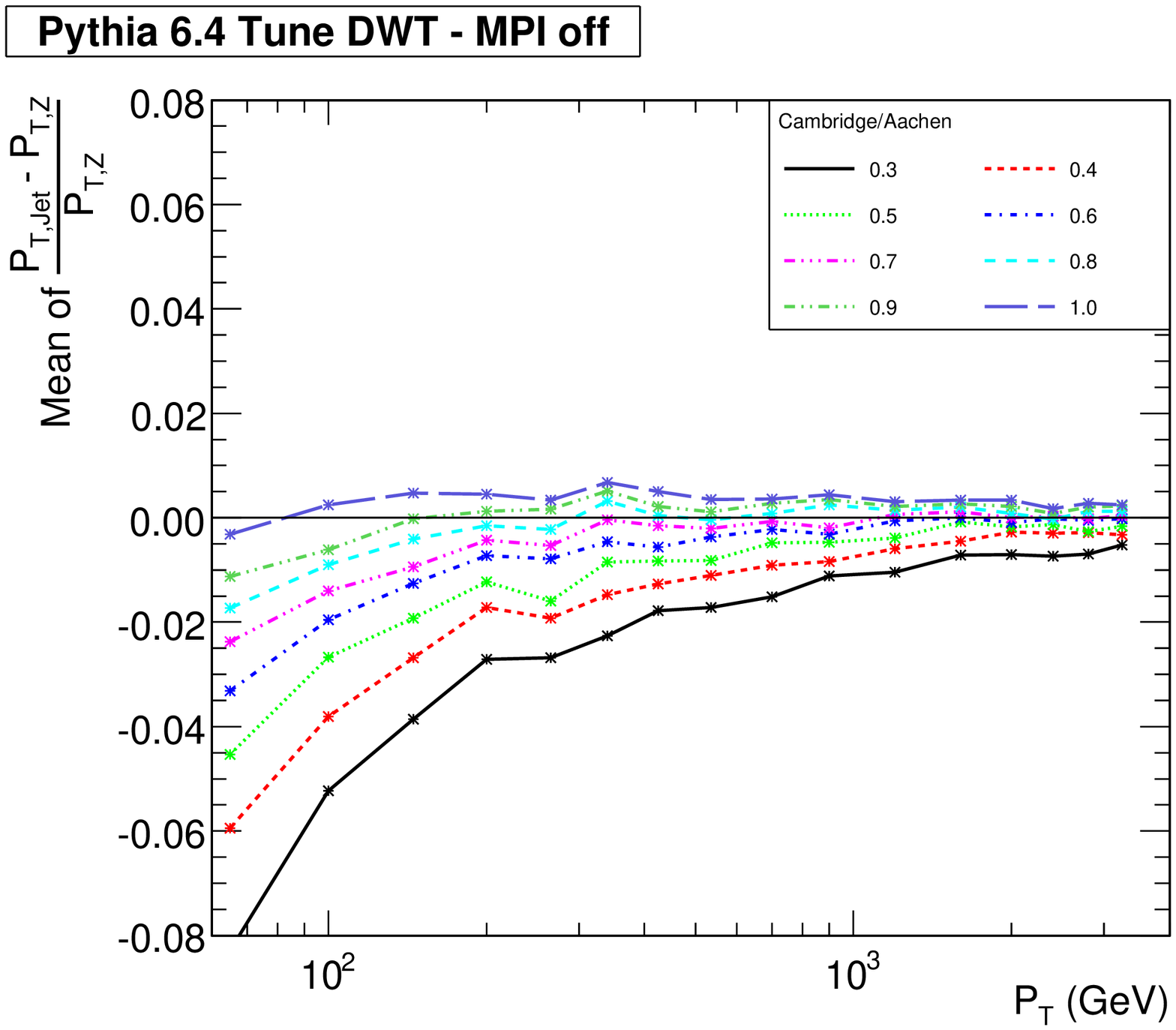} \\
    \caption{Same as Fig.~\ref{fig:Zjets_had_DWT} but with multiple
      parton interactions switched off.}
    \label{fig:Zjets_UUOnOff}
  \end{center}
\end{figure}

In the current implementation of the analysis, all stable particles
are selected as input objects to the jet algorithm, including the two
muons from the decay of the $Z$~boson. This leads to two fake jets in
the event which have to be removed manually from the jet collection.
This is done by discarding jets which lie inside a cone of $\Delta R <
0.5$ around the directions of the two muons.\footnote{$\Delta R =
  \sqrt{(\Delta\eta)^2 + (\Delta\Phi)^2}$} As the $Z$-jet system is
balanced in azimuth $\Phi$, the muon fake jets are in the opposite
hemisphere and therefore do not interfere with the determination of
the properties of the jet balancing the $Z$~boson so that the final
state truth definition given in the introduction still holds.

In order to ensure a clean sample of events in which the Z boson is
exactly balanced against one jet of the hard process, the second
leading jet in transverse momentum is required to have less than
$20\%$ of the transverse momentum of the Z boson. In addition, the
leading jet in $p_T$ is required to be opposite in azimuthal angle
$\Phi$ by complying with $|\Delta\Phi({\rm jet},Z) - \pi|<0.15$.

The relative deviation of the reconstructed jet $p_T$ from the
transverse momentum of the balancing $Z$~boson $(p_{T,{\rm jet}} -
p_{T,Z})/p_{T,Z}$ is determined independently for each range in the
hard transverse momentum scale set for the event generation.  The mean
and width of the relative difference of jet and boson $p_T$ is
performed in a two step procedure employing Gaussian fits where the
first one is seeded with the mean and root mean squared~(RMS) of the
corresponding histogram. The second fit then uses the result of the
first step as input.

Figure~\ref{fig:Zjets_had_DWT} presents this observable for fully
hadronized Pythia tune~DWT events versus $p_T$ for jet sizes $R$ of
$0.3$ up to $1.0$ of the investigated algorithms. All four exhibit a
very similar behaviour that small jet sizes on average under- and
large jet sizes overbalance the transverse momentum of the $Z$. Above
$\approx 500\,{\rm GeV}$ this difference remains well below $2\%$. To
smaller transverse momenta the balance gets increasingly worse. No
particular advantage can be identified for any of the four algorithms
and it is always possible to choose a suitable jet size to minimize
the deviations. But any such choice depends, of course, heavily on the
interplay of jet energy loss due to parton showers and hadronization
and energy gain because of the underlying event.

To give an estimate of the influence of the underlying event, the same
quantity is shown for comparison in Fig.~\ref{fig:Zjets_had_S0} for
the alternative Pythia tune~S0 for the four algorithms.  The smaller
jet sizes show nearly the same behaviour for both tunes.  For the
larger jet sizes a slight loss in energy for the tune~S0 compared to
DWT is exhibited. The effect decreases for larger transverse momenta.

In order to examine the influence of the underlying event on the jet
energy in dependence of the jet size, the mean of the relative
deviation between the transverse momentum of the jet and the balancing
$Z$~boson is shown in Fig.~\ref{fig:Zjets_UUOnOff} for fully
hadronized Pythia tune~DWT events with and without multiple
interactions. Having disabled multiple interactions, the transverse
momentum of the jet is systematically underestimated compared to the
$Z$~boson. This effect decreases for larger $R$ parameters but remains
visible which indicates that the jet algorithms do not accumulate the
whole energy of the parton into the jet. So without the MPI even the
largest employed jet size hardly suffices to collect all energy to
balance the boson~$p_T$.  This feature is compensated by acquiring
additional energy from the underlying event into the jet. Enabling
multiple interactions, the larger jet sizes now overestimate the
transverse momentum as shown in Fig.~\ref{fig:Zjets_had_DWT}.

Concluding, no particular advantage of any jet algorithm can be
derived with respect to the jet and $Z$ boson momentum balance.
Preferred jet sizes depend heavily on the multiple parton interactions
and can only be selected once the underlying event has been determined
more precisely at the LHC\@.

\subsection{\boldmath$H\,\rightarrow\,gg\,\rightarrow\,\rm{jets}$
  \unboldmath}

In the last section, we evaluate the impact of the jet algorithms and
jet sizes on the mass reconstruction of a heavy resonance.  More
specifically, we look at the process
$H\,\rightarrow\,gg\,\rightarrow\,\rm{jets}$ as a "monochromatic"
gluon source. In order to reduce to a large degree the effect of the
finite Higgs width, on the one hand side we allow the actual Higgs
mass in an event to deviate from the nominal one only by $\pm 50\,{\rm
  GeV}$, on the other hand, when comparing the mass reconstructed from
the two gluon jets, the remaining difference to the nominal mass is
compensated for. The two jets are required to be the leading jets in
transverse momentum with a separation in rapidity\footnote{$y =
  \frac{1}{2} \ln\frac{E+p_z}{E-p_z}$}~$y$ of $|y_{\rm jet 1} - y_{\rm
  jet 2}|$ smaller than $1$.  To avoid potential problems with the
$gg$ production channel for large Higgs masses we decided to enable
only the weak boson fusion, process numbers $123$ and $124$ in Pythia,
Ref.~\cite{Sjostrand:2006za}.

\begin{figure}[tbh]
  \begin{center}
    \includegraphics[width=0.5\textwidth]{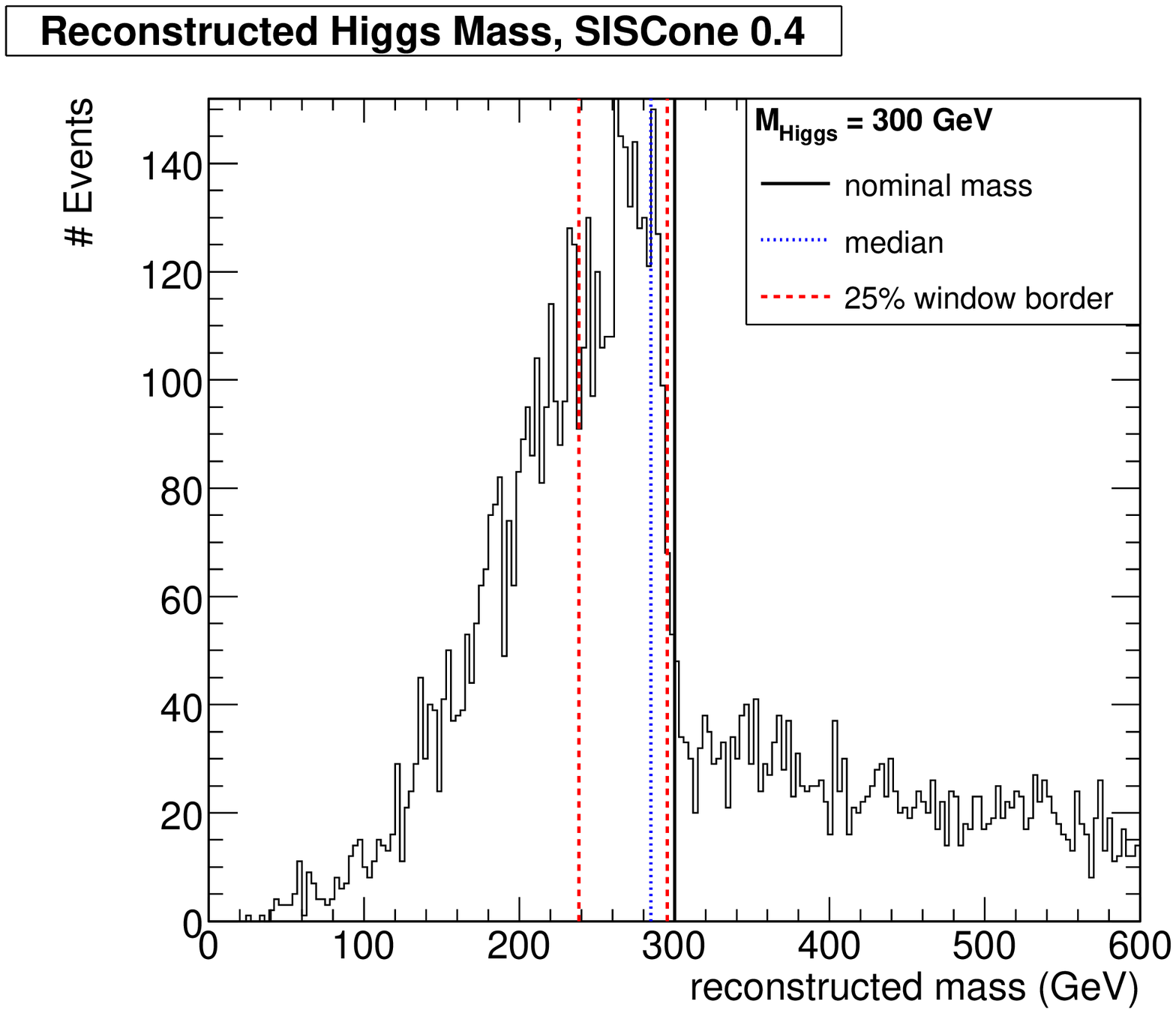}%
    \includegraphics[width=0.5\textwidth]{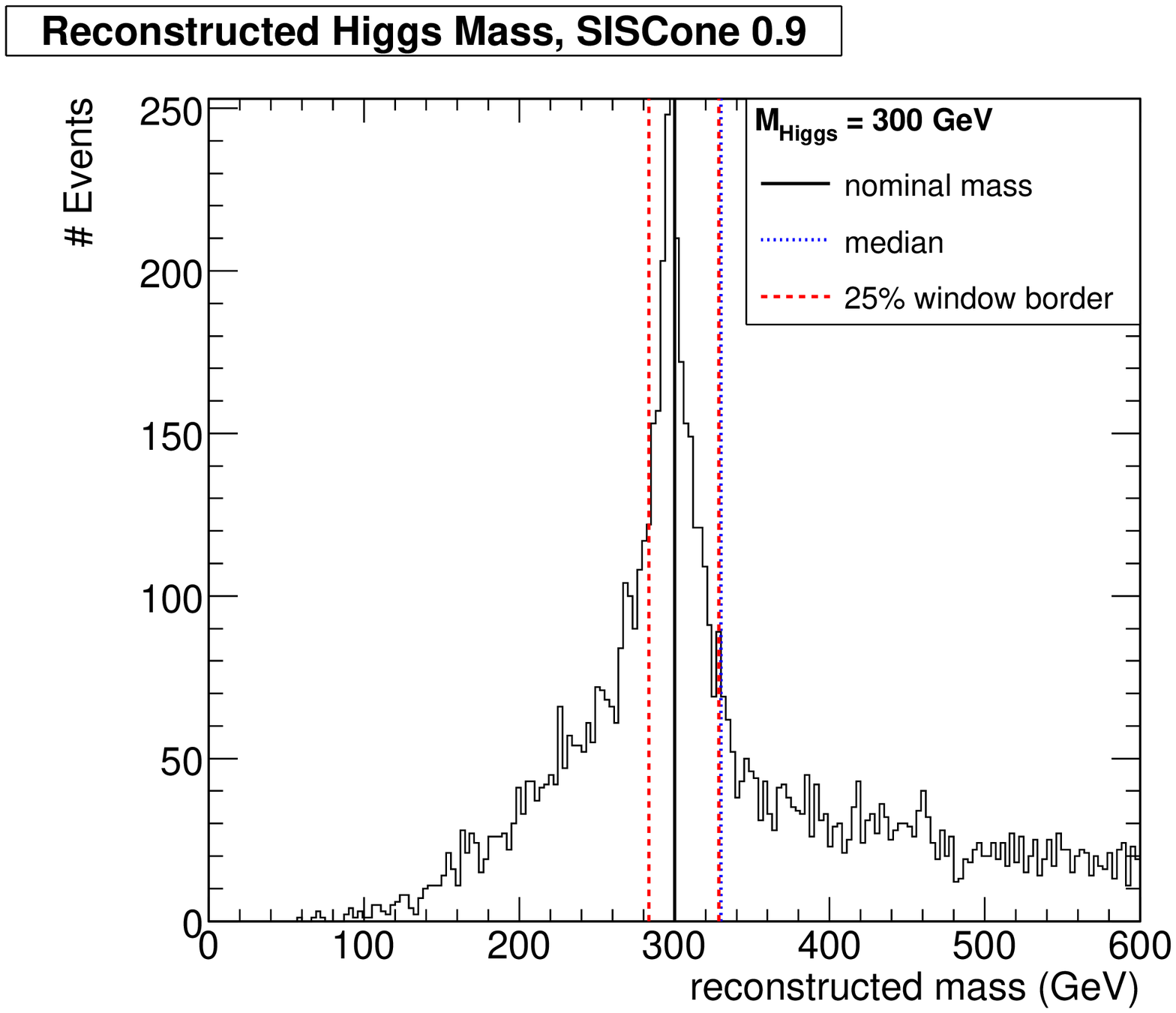}\\
    \includegraphics[width=0.5\textwidth]{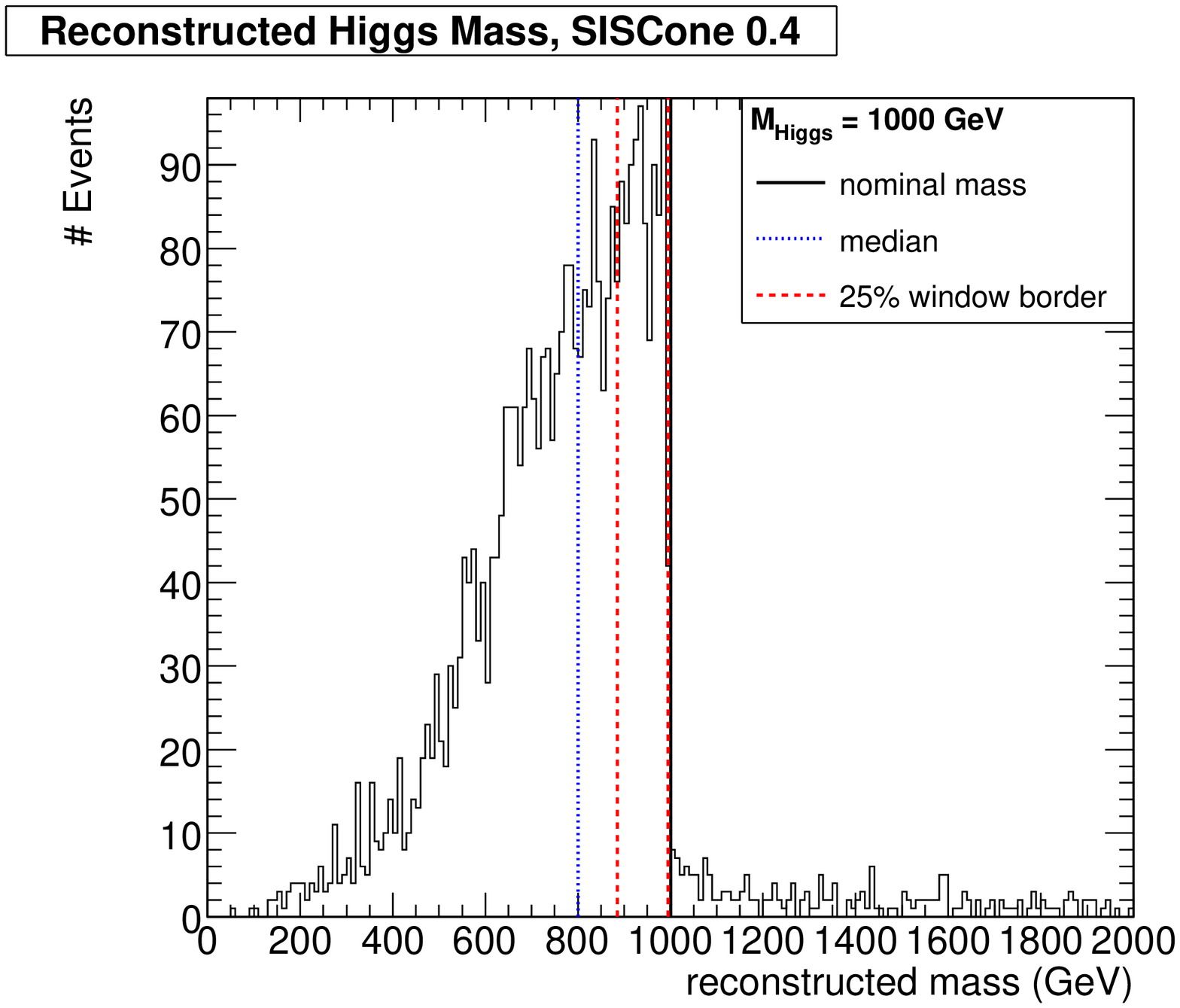}%
    \includegraphics[width=0.5\textwidth]{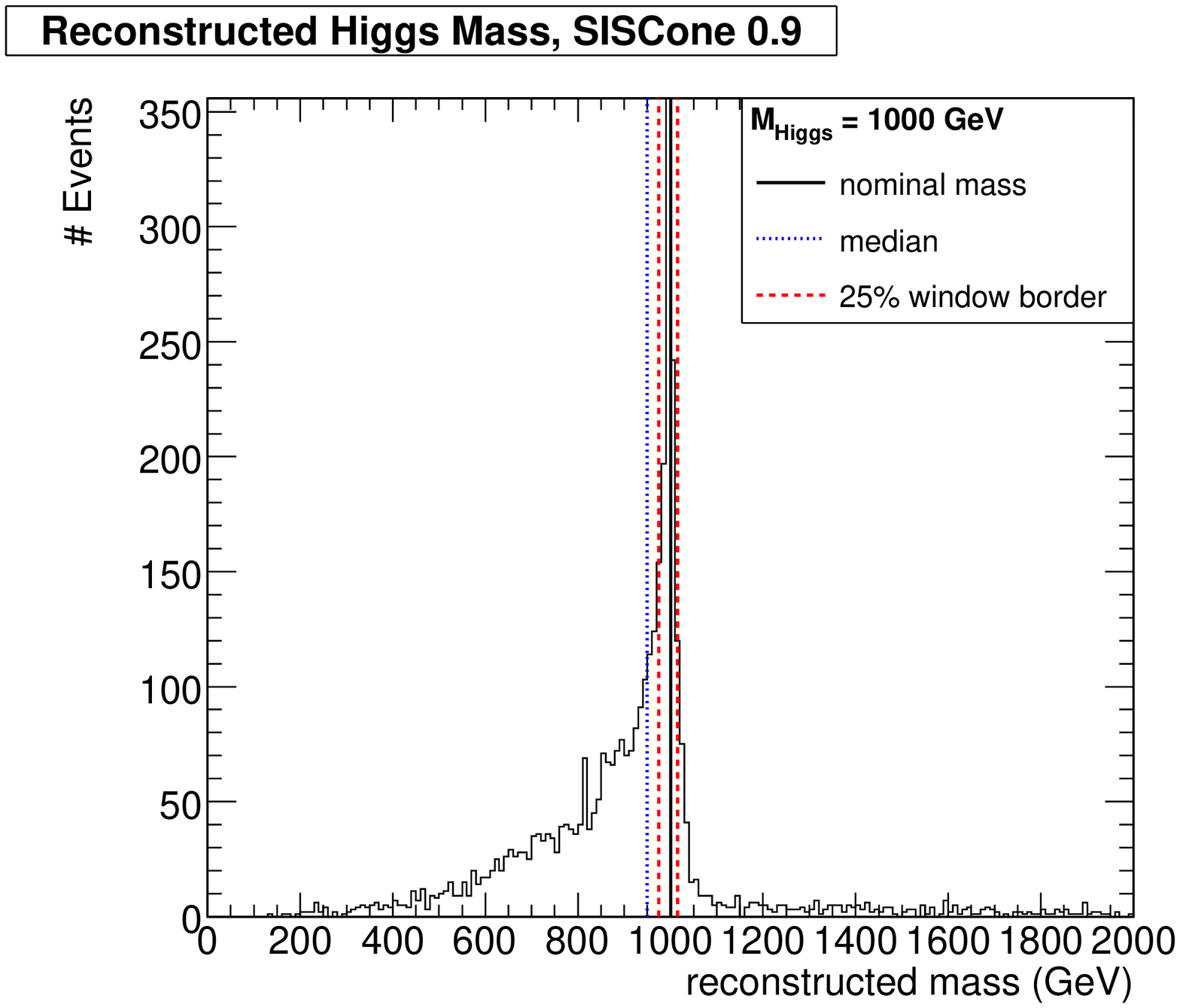}\\
    \caption{Reconstructed Higgs mass distributions for the SISCone
      algorithm with cone sizes $0.4$ (left) and $0.9$ (right) for the
      two nominal Higgs masses of $300$ (upper) and $1000\,{\rm GeV}$
      (lower). The full black line indicates the nominal mass, the
      dashed red lines show the location of the determined minimal
      mass window and the dotted blue line corresponds to the median.}
    \label{fig:Hgg_smw}
  \end{center}
\end{figure}

Nevertheless it proved to be difficult to define quality observables,
since Breit-Wigner as well as Gaussian fits or combinations thereof do
not in general well describe the mass distributions for all jet sizes.
At small $R$ up to $0.4$ the substructure of gluon jets is resolved
instead of features of the hard process. At intermediate resolutions a
small mass peak starts to reappear leading to asymmetric distributions
which are especially awkward to deal with.  The same problems arise in
the reconstruction of a $Z'$ mass which is investigated in more detail
in chapter~\ref{sec:lhprocs_gavin} of these proceedings.  For
comparison we use a similar approach here and look for the smallest
mass window containing $25\%$ of all events. As reconstructed mass
value we simply chose the median, which may lie outside the location
of the smallest window, since we primarily consider the width as
quality measure and not the obtained mass. Figure~\ref{fig:Hgg_smw}
displays as example the determined mass windows and medians for the
SISCone algorithm with cone sizes $0.4$ and $0.9$ for the two nominal
Higgs masses of $300$ and $1000\,{\rm GeV}$.

\begin{figure}[tbh]
  \begin{center}
    \includegraphics[width=0.5\textwidth]{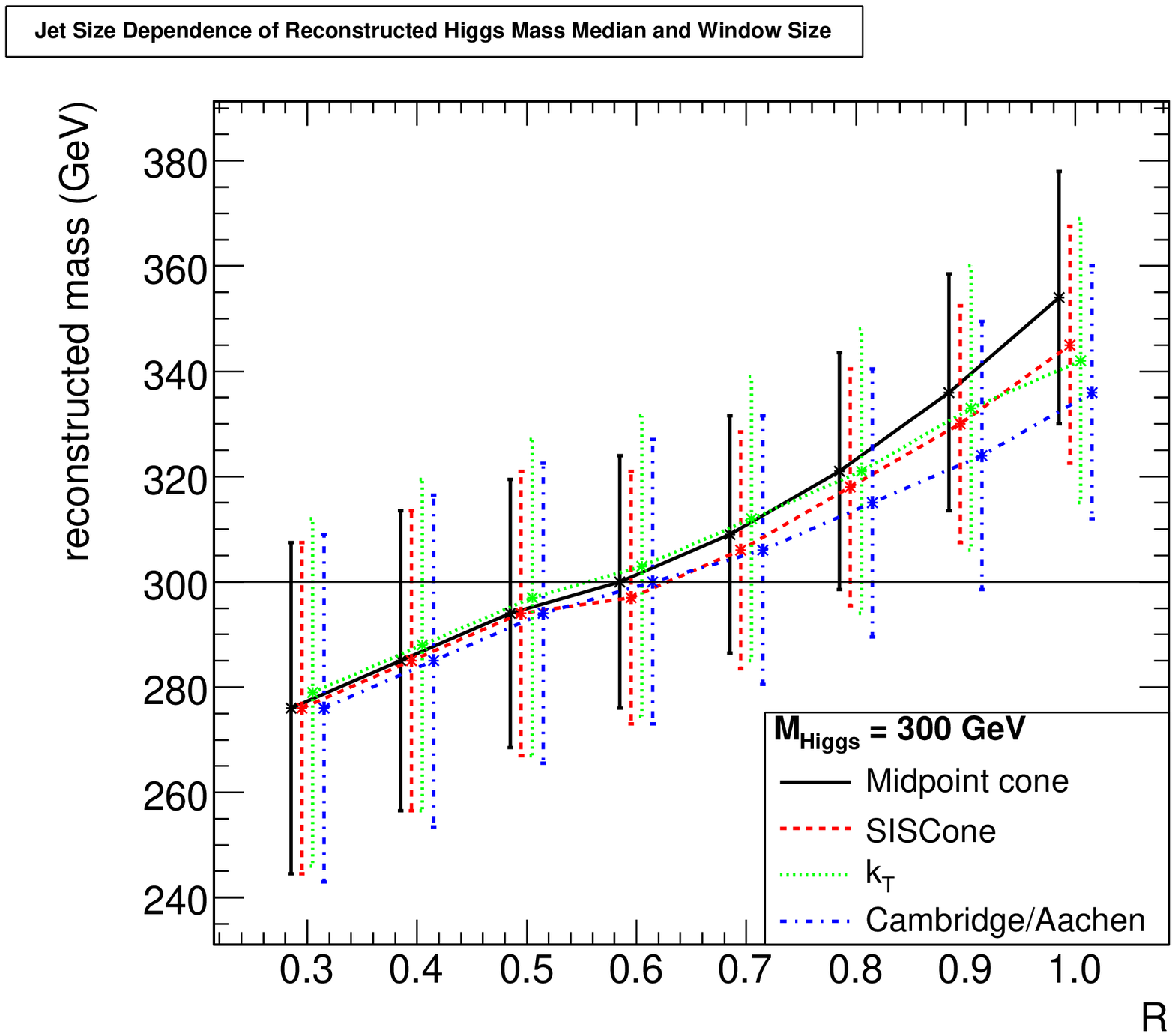}%
    \includegraphics[width=0.5\textwidth]{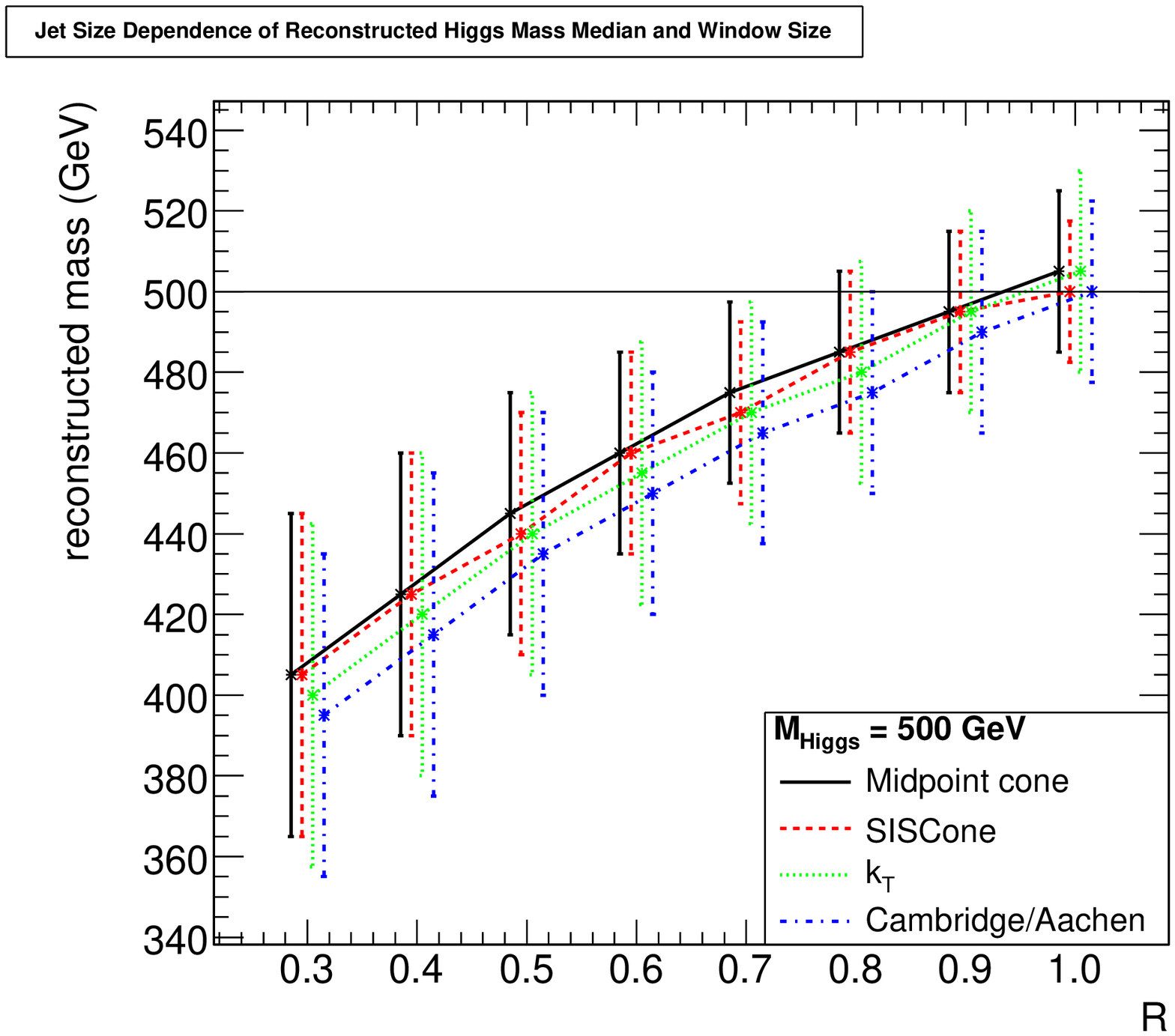} \\
    \includegraphics[width=0.5\textwidth]{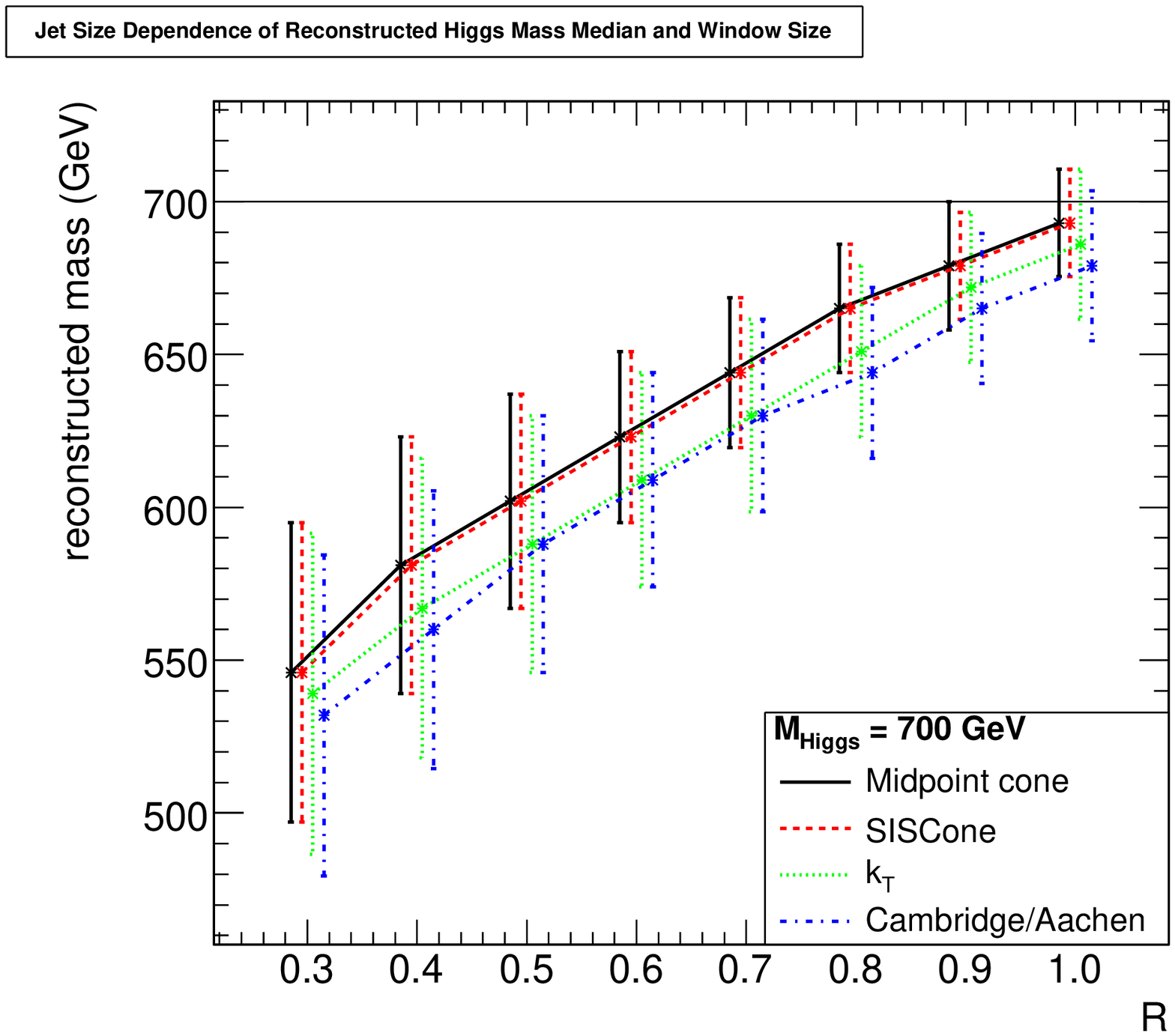}%
    \includegraphics[width=0.5\textwidth]{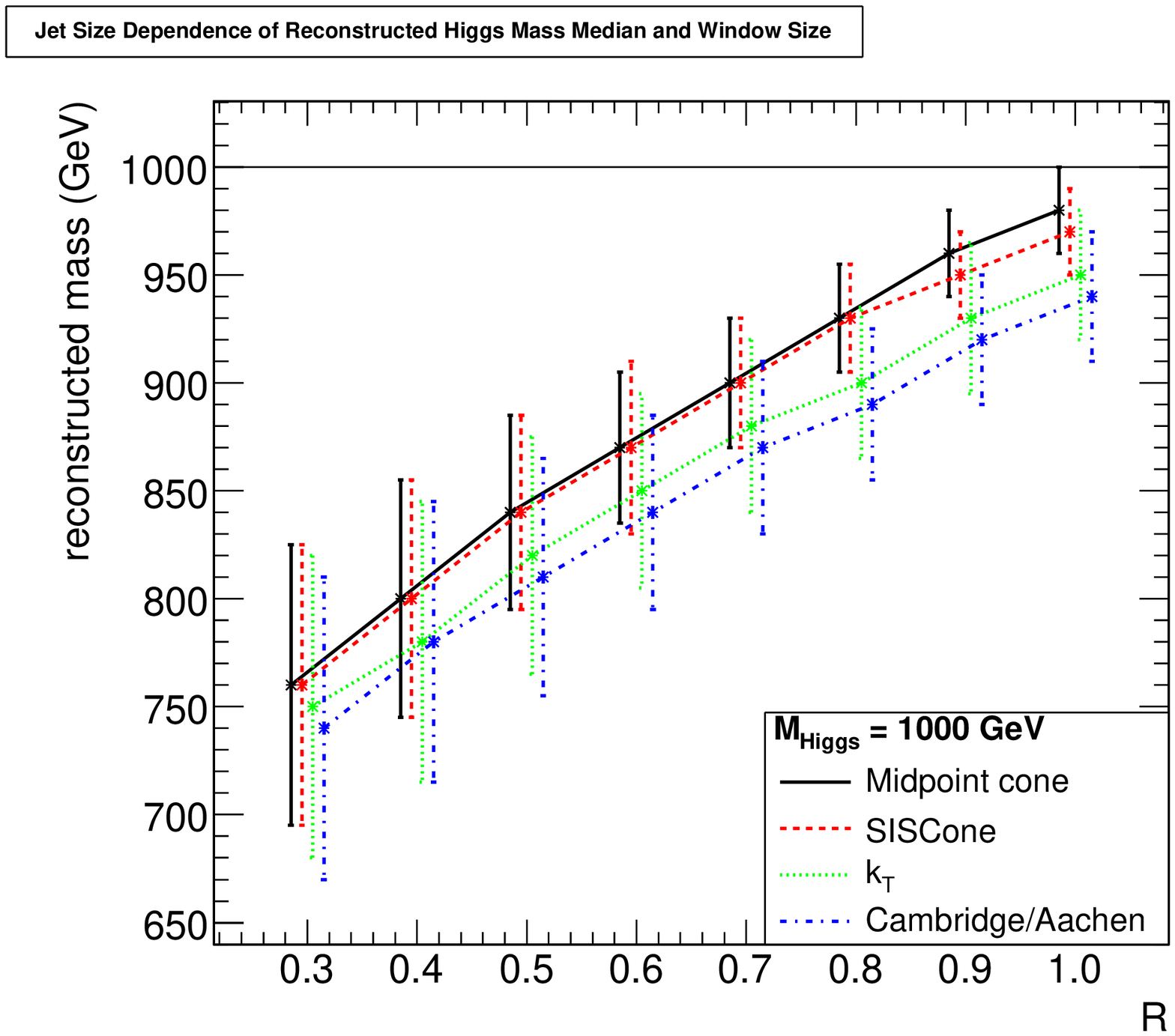}%
    \caption{Reconstructed Higgs mass and width when defined as median
      and minimal mass window containing $25\%$ of all events versus
      the jet size. For better visibility the points have been
      slightly displaced in $R$ for the different jet algorithms.}
    \label{fig:Hgg_Median25}
  \end{center}
\end{figure}

In Figure~\ref{fig:Hgg_Median25} the reconstructed Higgs mass and
width, defined as median and the minimal mass window, is shown for all
four jet algorithms versus the jet size for the four nominal resonance
masses of $300$, $500$, $700$ and $1000\,{\rm GeV}$. Obviously, the
median systematically underestimates the nominal mass for larger Higgs
masses.

\begin{figure}[tbh]
  \begin{center}
    \includegraphics[width=0.5\textwidth]{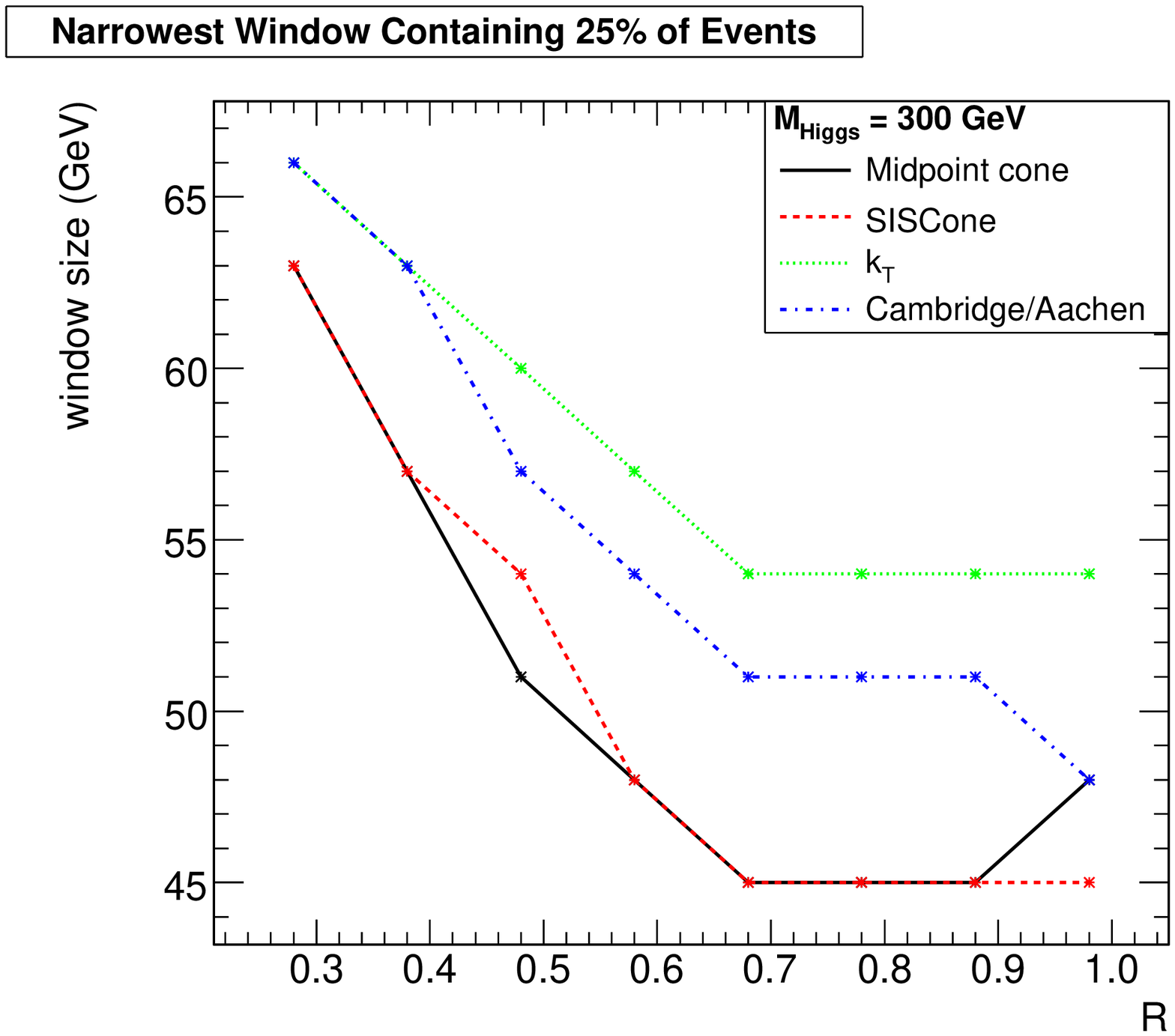}%
    \includegraphics[width=0.5\textwidth]{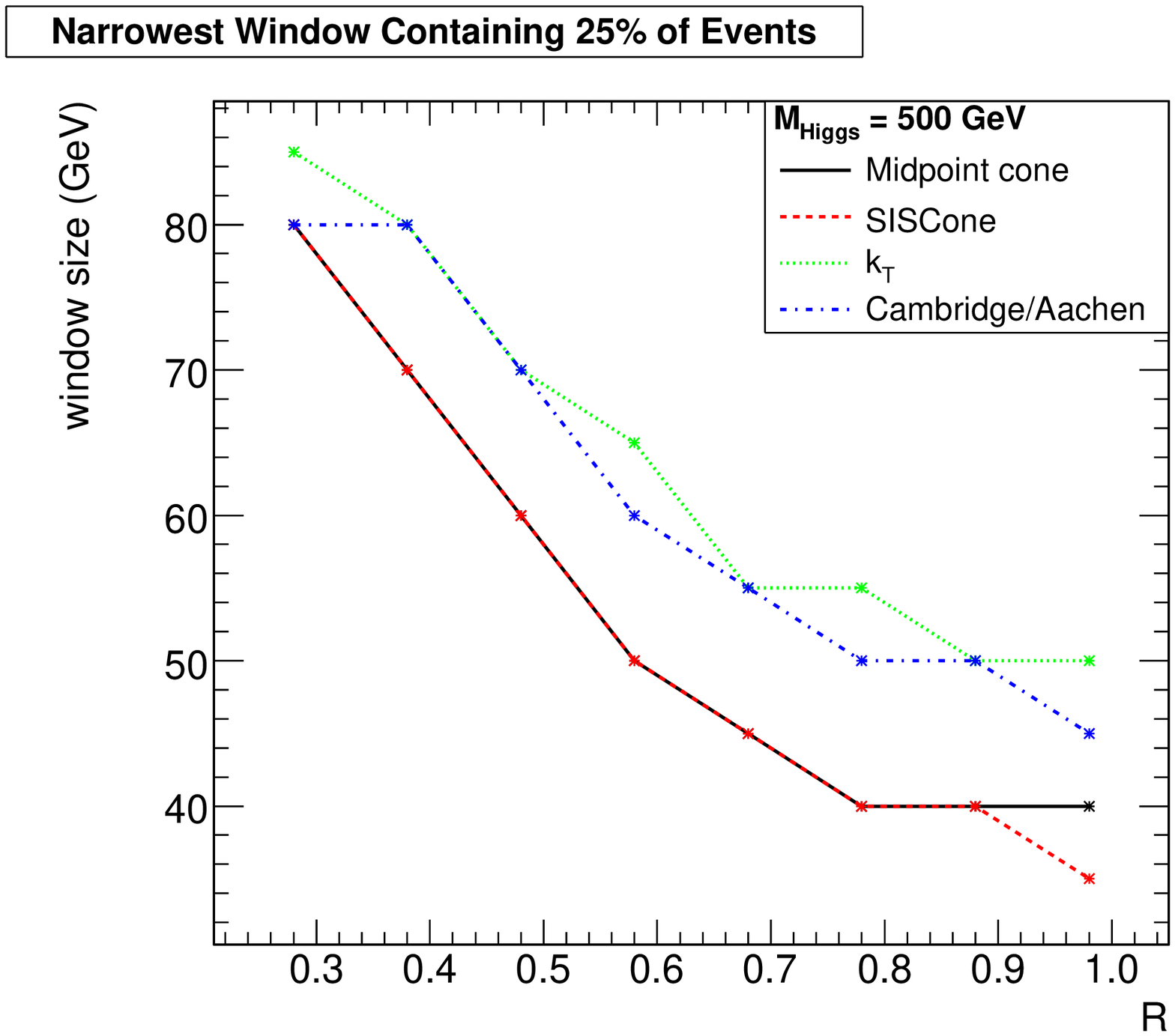}\\
    \includegraphics[width=0.5\textwidth]{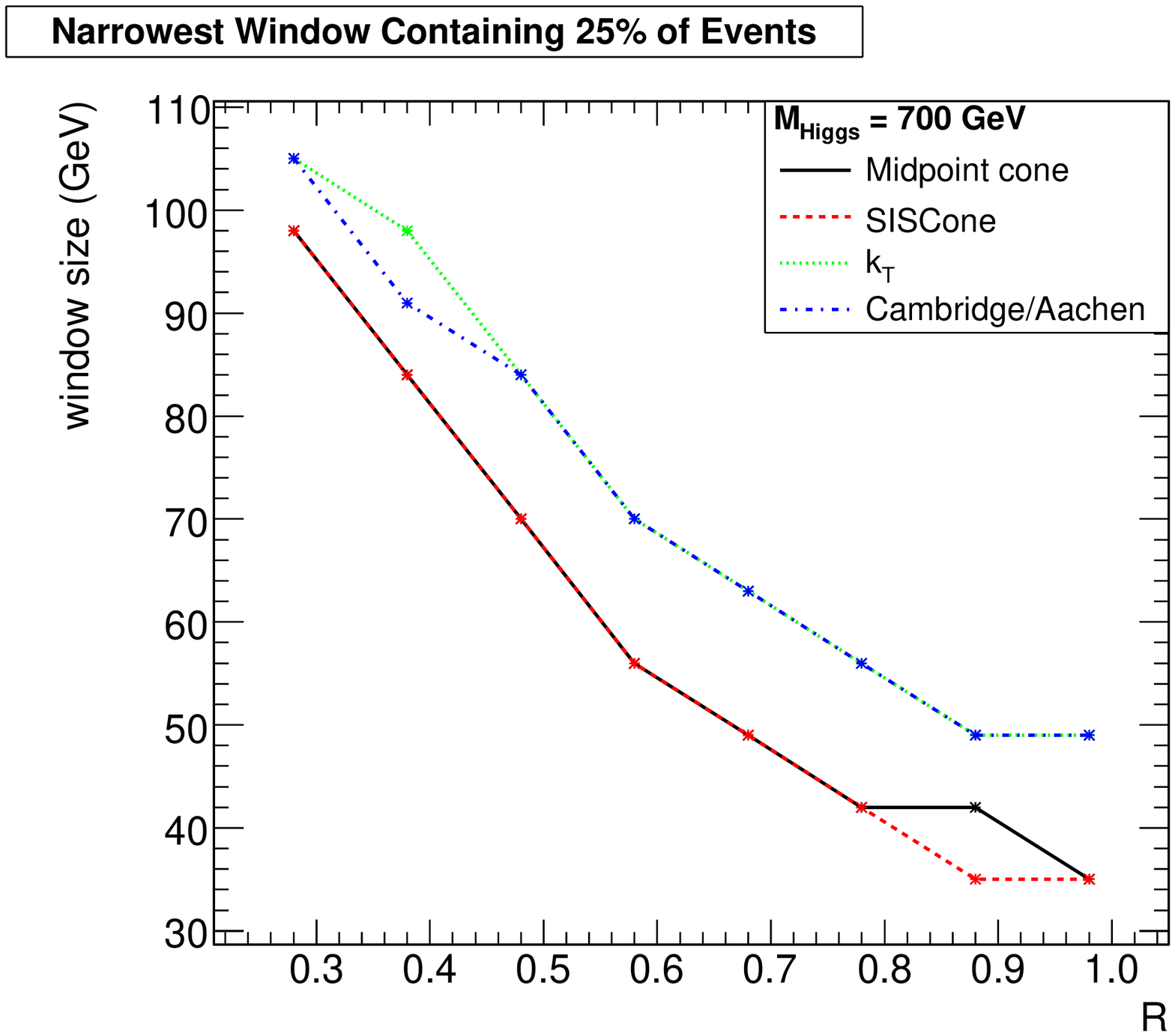}%
    \includegraphics[width=0.5\textwidth]{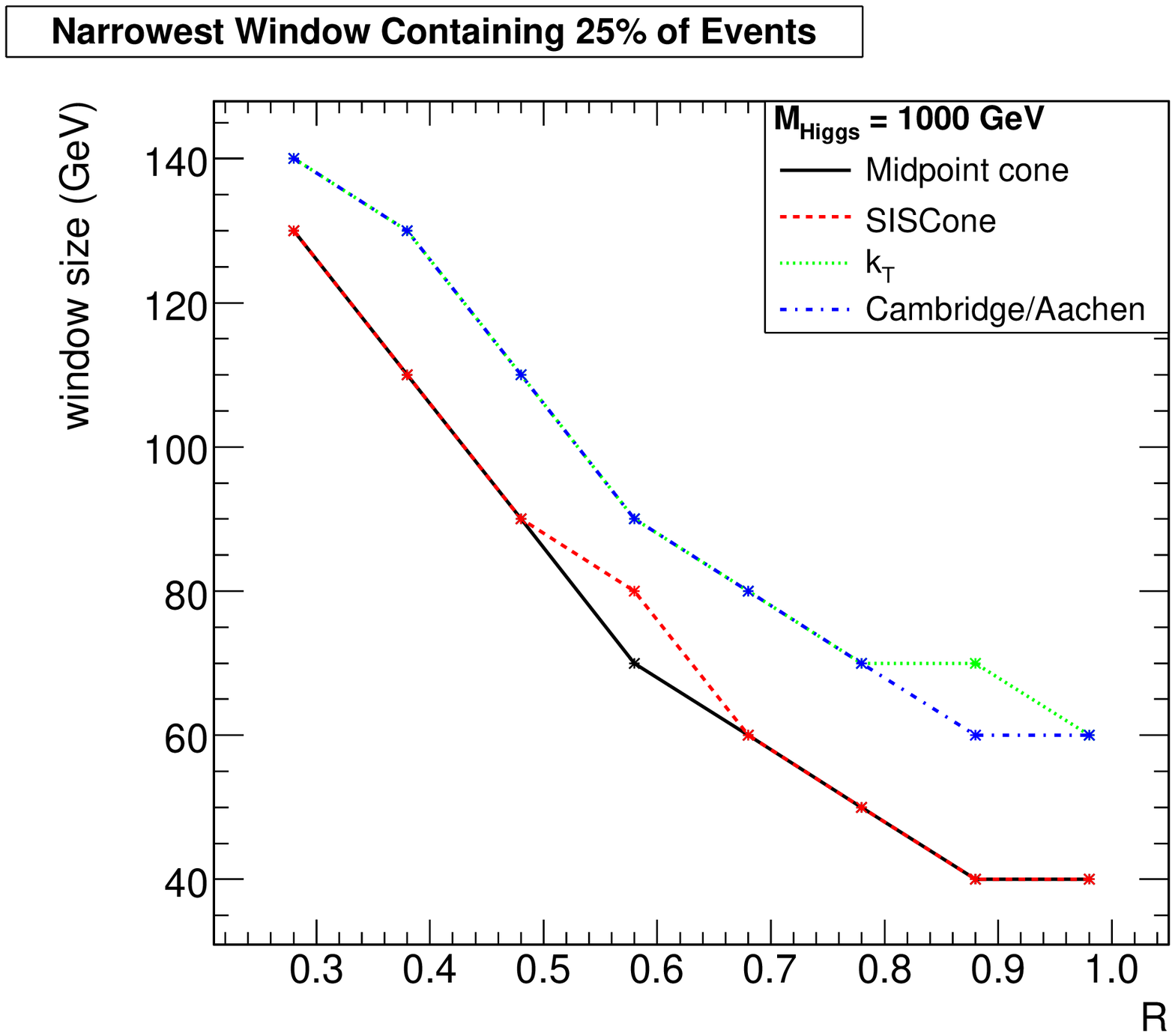}\\
    \caption{Minimal mass window sizes in dependence of the jet size
      $R$ for all jet algorithms and four nominal masses of the Higgs
      boson.}
    \label{fig:Hgg_wq}
  \end{center}
\end{figure}

Finally, in Fig.~\ref{fig:Hgg_wq} the derived minimal mass window
sizes are presented in dependence of the jet size $R$ for all jet
algorithms and four nominal masses of the Higgs boson.
Systematically, the cone type algorithms perform somewhat better than
the sequential recombination ones in the sense that they lead to 
smaller reconstructed widths.

\subsection*{Conclusions}

As already observed previously, hadronization corrections for
inclusive jets, especially at low transverse momenta, are smaller for
jet algorithms of the sequential recombination type ($k_T$,
Cambridge/Aachen). For the purpose of inclusive jet spectra, however,
one is predominantly interested in the newly accessible regime of
transverse momenta above $\approx 600\,{\rm GeV}$ or just below.  In
addition, in the complete correction a partial cancellation occurs of
hadronization effects and contributions from the underlying event
where no algorithm showed a distinctly better performance than the
others. So provided the current extrapolations of the underlying
event, one of the largest unknowns, are roughly comparable to what
will be measured, all algorithms are equally well suited.  For the
analysis of the $Z$ plus jets momentum balance no particular advantage
of any jet algorithm was observed neither. In the case of the
characterization of the reconstructed Higgs resonance via the median
and the minimal mass window containing $25\%$ of the events as
proposed in chapter~\ref{sec:lhprocs_gavin}, the cone type algorithms
(Midpoint cone, SISCone) exhibit smaller widths.

Concerning jet sizes, the inclusive jets analysis and the $Z$ plus jet
balance prefer medium jet sizes $R$ of $0.4$ to $0.8$, i.e.\ somewhat
smaller than the habitual value of $R \approx 1$ before. This is in
agreement with the expected higher jet multiplicities and larger
underlying event contributions at LHC energies which require a higher
jet resolution power. For the reconstruction of the Higgs resonance,
especially here from two gluon jets, larger jet sizes $R$ of $0.8$ or
$0.9$ are required. For jet sizes below $\approx 0.5$ one resolves the
substructure of the gluon jets instead of recombining all decay
products of the resonance.

Concluding, the suitability of the considered four jet algorithms was
investigated for three types of analyses and no decisive advantage for
a particular one was found within the scope of this study.  So apart
from the fact that Midpoint cone is not collinear- and infrared-safe
and was merely used for comparison, further investigations have to be
performed with respect to experimental aspects. We have shown that
especially the underlying event can be expected to have a significant
impact on the presented analyses.

\subsection*{Acknowledgements}

We thank Florian Bechtel, Ulrich Felzmann, Andreas Oehler, G\"unter
Quast, Gavin Salam, Peter Skands and Markus Wobisch for discussions.
We are grateful to our home institutions, the Graduiertenkolleg
Hochenergiephysik und Teilchenastrophysik of the Deutsche
Forschungsgemeinschaft and the Bundesministerium f\"ur Bildung und
Forschung for their support.


%% file: s_campanelli/campanelli.tex
%
%

\subsection{Introduction}

Almost all LHC physics channels will contain jets in the final state. For this reason, jet clustering algorithms deserve a great deal of  attention. Even though hadron collider experiments have reconstructed jets for over 30 years, until recently the
precision reached at hadron machines was not sensitive to the differences between
the different jet algorithms. In addition, the available computing power often limited the choice of
jet algorithms that were practical to use.

With the recent precision measurements from  the Tevatron, and in light
of the expectations for the LHC, it is worthwhile to re-examine the impact  jet algorithms do make at hadron colliders, especially as new algorithms and ideas are being developed.
Our aim in this contribution is to provide a systematic study of some characteristics
of representative jet clustering algorithms and parameters, using as an input
one of the closest analogues an experiment can provide to four-vectors,  the
ATLAS topological clusters. These are calorimeter clusters already calibrated for
detector measurement effects, to effectively the hadron level. These topoclusters are passed to the
clustering algorithms by the SpartyJet 
~\cite{spartyjet} tool, an interface to the major clustering algorithms that
allows easy change and control over relevant parameters.

\subsection{Algorithms considered}

Jet clustering algorithms can be divided into two main classes: cones and
iterative recombination (as for example the $k_T$ algorithm). Historically,
in hadron colliders, primarily cone algorithms have been used, being the only
algorithm fast enough for use at trigger level, and for fear of large systematic
effects in busy multi-jet environments from recombination algorithms. Fast implementations of the $k_T$
algorithm~\cite{Cacciari:2005hq}, as well as the first papers performing precision
measurements with it ~\cite{Abulencia:2007ez,Abulencia:2005jw}  call for a detailed comparison of
the $k_T$ algorithm with cone-based ones.\par
Many implementations of cone algorithms have been developed over the years
(and the experiments). Many of them have been shown to suffer from infrared
safety issues, i.e. the results of the algorithm can change if very soft
particles, that do not affect the overall topology of the event, are added
or subtracted. Unfortunately, algorithms that have long been the default
for large experiments, such as JetClu for CDF and the Atlas cone for Atlas, belong
to this category. Other algorithms, such as Midpoint ~\cite{Blazey:2000qt,Ellis:2001aa}  are stable
under infrared correction for most (but still not all) cases.  But, since they
start clustering jets around energy depositions larger than a given value (seed threshold), 
the outcome will depend in principle on the value of this threshold. The manner in which this will affect
clustering under real experimental conditions is one of the questions we
will attempt to address in this study. Finally, a seedless infrared-safe cone algorithm
has recently emerged~\cite{Salam:2007xv}, providing most of the desirable features
a cone algorithm needs from the theoretical point of view and a similar ease of use as previous cone algorithms. Its adoption
by the experimental community has been slow due to the lack of a comprehensive
comparison with more traditional approaches. Most of the studies presented
in the following sections will involve comparisons between the $k_T$ algorithm
(for the two different cone sizes of 0.4 and 0.6), the legacy Atlas cone and
the Midpoint cone algorithm (for a cone size of 0.4), the Cambridge/Aachen algorithm (similar to the $k_T$ algorithm, 
but only using the distance between clusters and not their energy) and
the seedless infrared cone algorithm (SISCone; cone size of 0.4). Throughout this contribution, 
these algorithms will be identified by the same color, i.e. black for Kt04,
red for Kt06, green for the Atlas cone(04), dark blue for SISCone(04), pink for 
MidPoint(04) and light blue for Cambridge/Aachen.

\subsection{Datasets}

To perform our studies, we have used the Monte Carlo datasets produced in the context of the Atlas
CSC notes exercise. In particular, we are interested in the behavior of
jet algorithms in a multi-jet environment and in the endcap region where
small changes in cluster position can result in large rapidity differences.
It was therefore natural to use samples from $W$+jets and VBF Higgs channels.
The former were generated with ALPGEN~\cite{Mangano:2002ea},  (interfaced to Herwig), for the case of a $W$
boson decaying into a muon and a neutrino, produced in association with a
number of partons ranging from 0 to 5; the latter are Herwig~\cite{Marchesini:1991ch} samples,
with a Higgs ($M_H$ = 120 GeV) decaying into  tau pairs, with each of the 
taus decaying into an electron or a muon and neutrinos. \par
Unless otherwise specified, the different algorithms are run on the same
datasets; therefore, the results obtained are not statistically
independent, and even small differences can be significant.
Jets reconstructed with the jet axis closer than $\Delta R = 0.4$ with respect to the 
closest lepton (either from W decay or a $\tau$ from $H\to \tau \tau$) are
discarded, in to avoid biasing the jet reconstruction performances either by
inclusion of those leptons in the jet, or  by calling jet a lepton or a tau decay product altogether.

\subsection{Jet Multiplicity}

The first variable we examined is the jet multiplicity for events with a
leptonically decaying W and a number of partons varying from 0 to 5. 

\begin{figure}[tbh]
\begin{center}
\epsfig{file=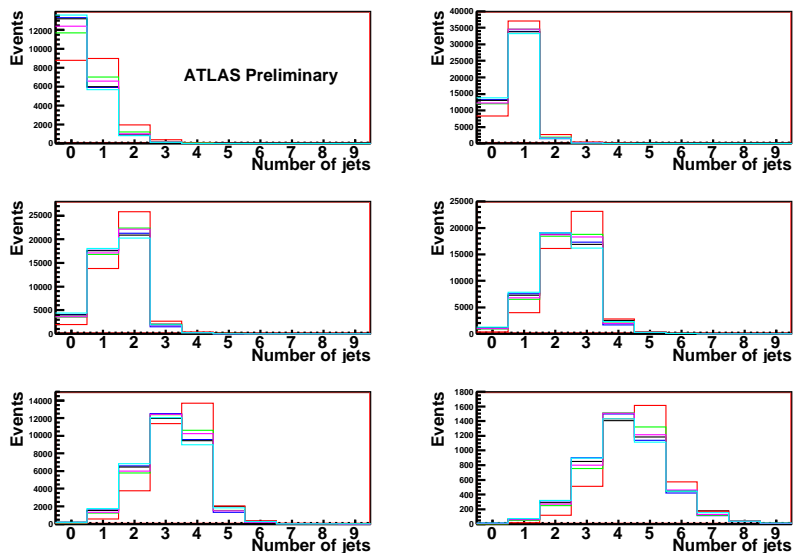,width=0.7\linewidth}
\end{center}
\caption{Number of reconstructed jets for W + n
partons Monte Carlo, with the number of partons increasing (from 0 to 5) as the
plot order.}
\label{fig:njets}
\end{figure}

\begin{figure}[tbh]
\begin{center}
\epsfig{file=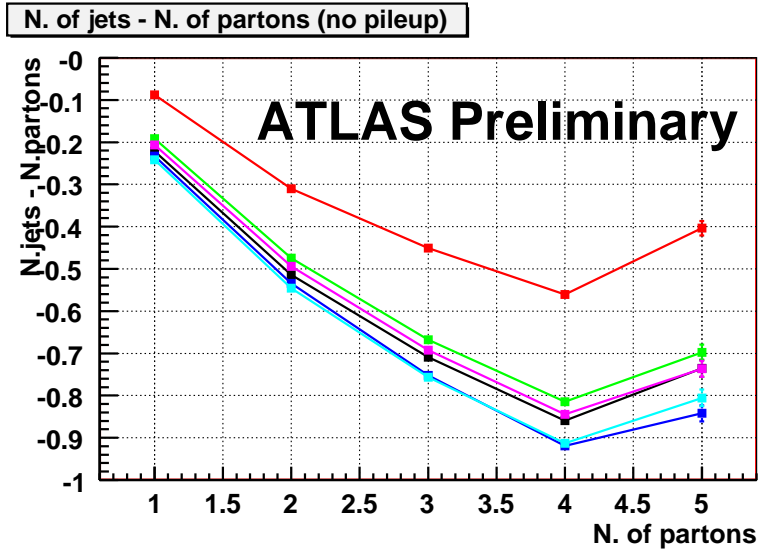,width=0.7\linewidth}
\end{center}
\caption{Difference between the number of reconstructed jets and the number of reconstructed partons vs this latter quantity, for W + n
partons Monte Carlo}
\label{fig:njetgraph}
\end{figure}

\begin{figure}[tbh]
\begin{center}
\epsfig{file=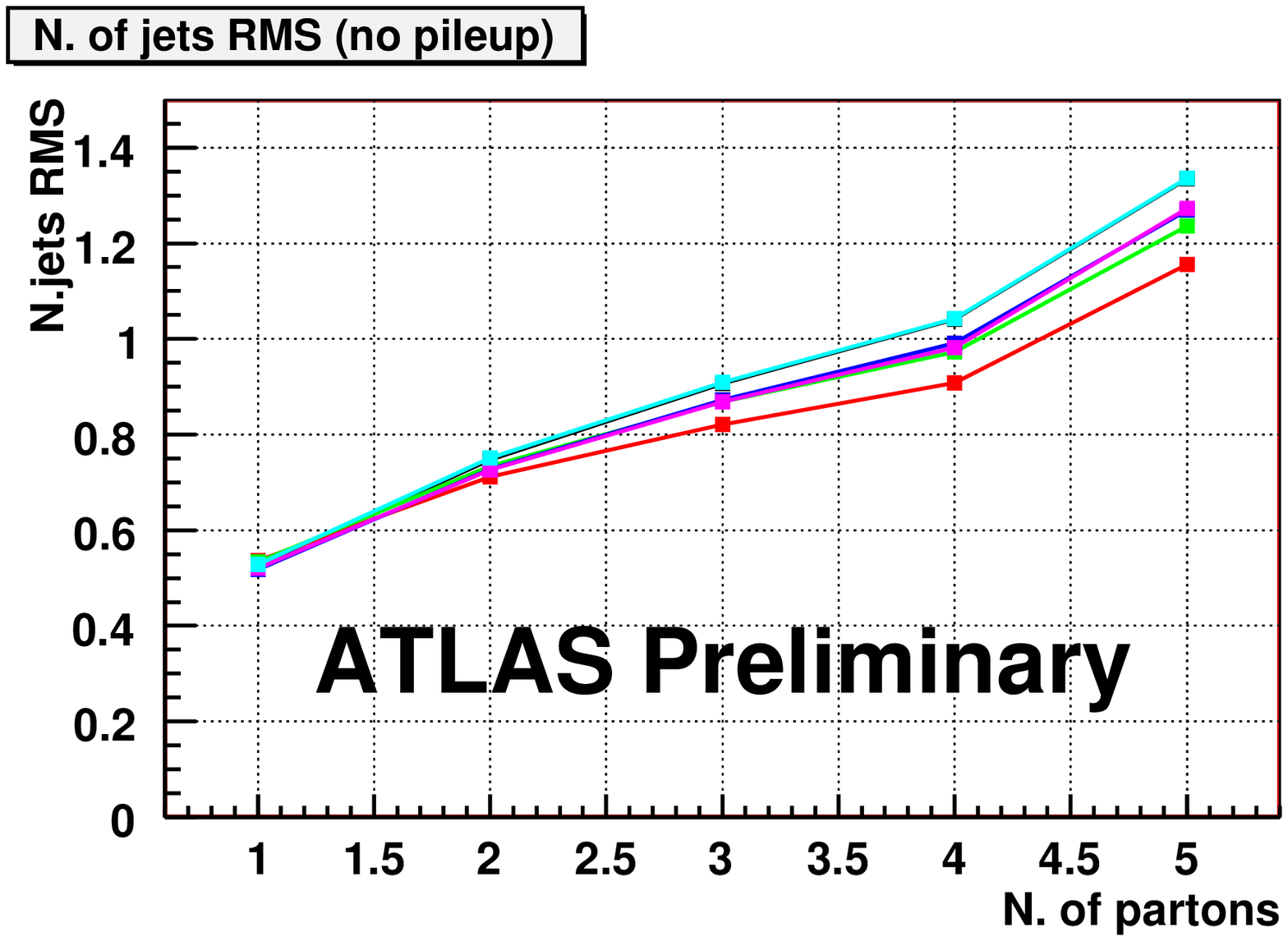,width=0.7\linewidth}
\end{center}
\caption{RMS of the distribution of the number of reconstructed jets,
as a function of the number of generated partons for W + n partons Monte Carlo}
\label{fig:rmsjetgraph}
\end{figure}

The reconstructed number of jets with $p_T > 20$ GeV for the various algorithms
(with color code as in the end of the ``Algorithms considered'' session)
is shown in Figure
~\ref{fig:njets}, where each plot represents a different number of generated
partons.  
To understand the trends somewhat better, Figure ~\ref{fig:njetgraph} shows the
difference between the mean number of reconstructed jets and the number of partons,
while Figure ~\ref{fig:rmsjetgraph} shows the RMS of this distribution.
As expected, the distribution of the 
number of reconstructed jets broadens as the number of partons increases, both at reconstructed
and generator level.
Since only jets passing the 20 GeV $p_T$ cut are included,
it is understandable that the multiplicity is higher for the Kt06 than for the Kt04
algorithm. This is true as well for large jet multiplicities, where the effect
of the smaller available phase space for the larger jet size is not relevant
for the multiplicities considered.
On the other hand, SISCone tends to reconstruct a smaller number of jets 
than the other algorithms.

\subsection{Matching efficiency}

One of the most important characteristics of a jet algorithm is the ability
to correctly find, after detector effects, jet directions as close as possible to the 
generated ones. Since a parton does not have a well-defined physical meaning,
we stress again here that all comparisons between generated and reconstructed
quantities are done with jets reconstructed from stable particles at the
hadron level, using the same algorithm as at detector level.
Matching efficiencies are defined as the number of hadron level jets in a
given $p_T$ or $\eta$ bin that have a reconstructed jet within a given $\Delta R$
cut.

\begin{figure}[tbh]
\begin{center}
\epsfig{file=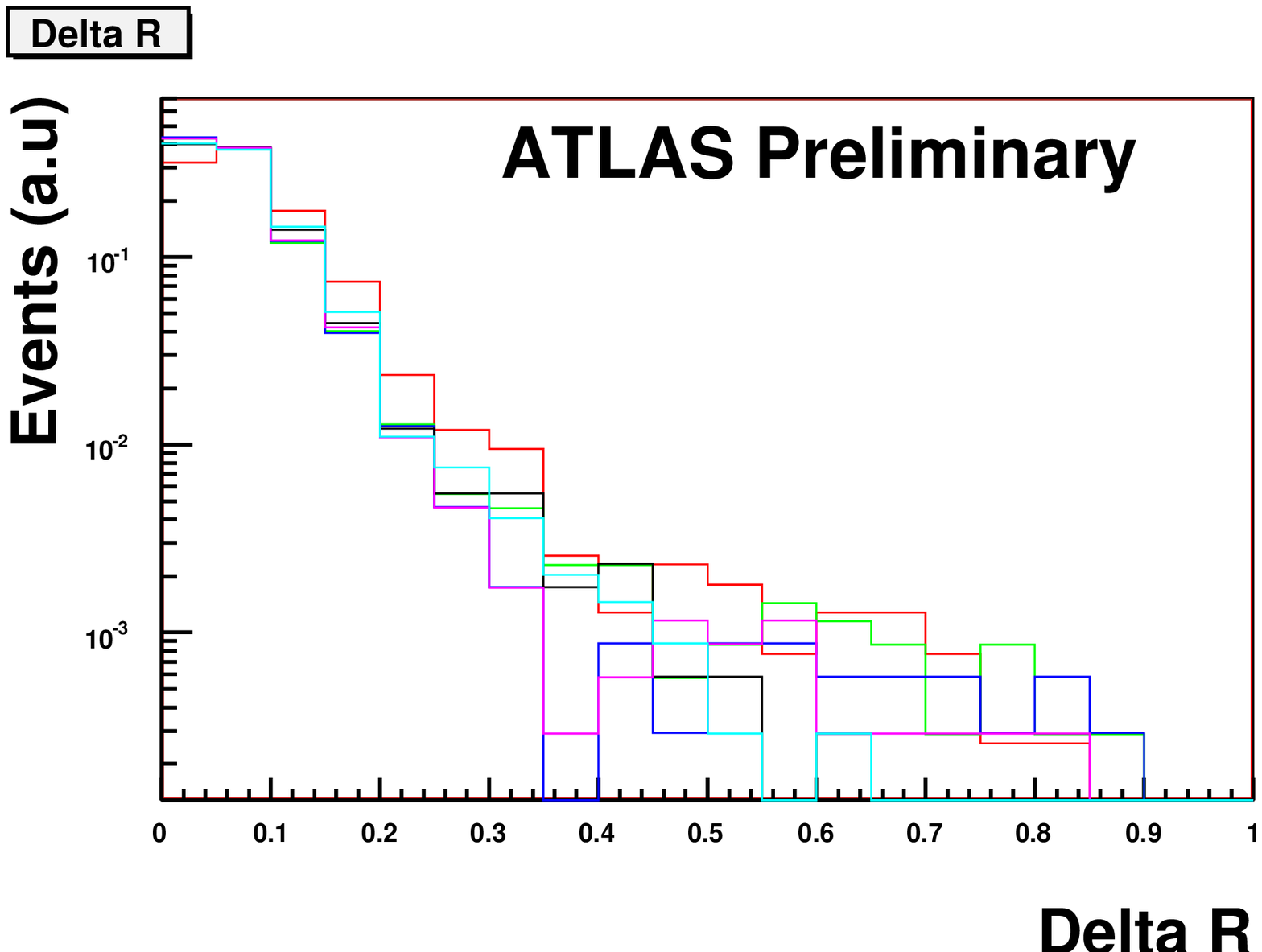,width=0.7\linewidth}
\end{center}
\caption{Distribution of $\Delta$R between generated and reconstructed jets,
 for W + 2 partons Monte Carlo.}
\label{fig:deltaR}
\end{figure}

\begin{figure}[tbh]
\begin{center}
\epsfig{file=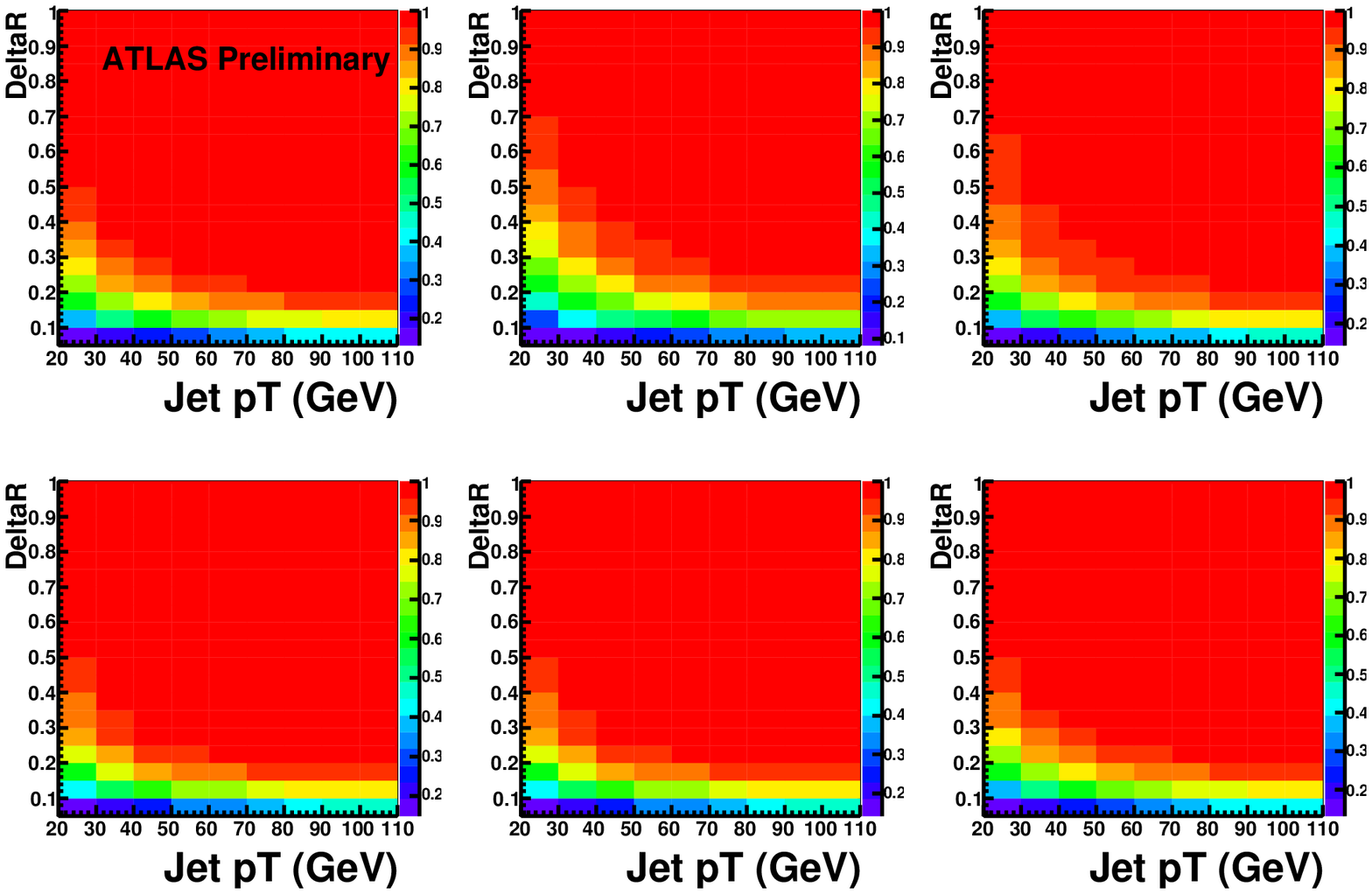,width=0.7\linewidth}
\end{center}
\caption{Matching efficiency as a function of jet $p_T$ and $\Delta$R cut for $W$ + 2 partons.}
\label{fig:eff}
\end{figure}

\begin{figure}[tbh]
\begin{center}
\epsfig{file=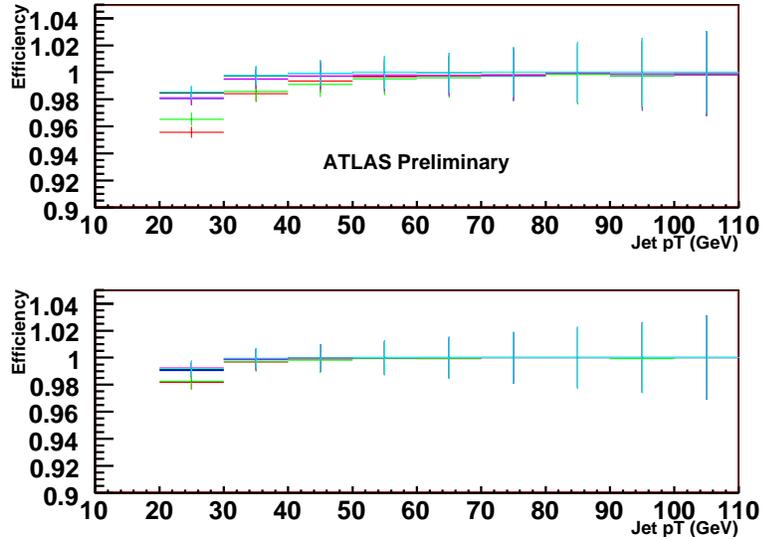,width=0.7\linewidth}
\end{center}
\caption{Matching efficiency as a function of jet $p_T$ for $W$ + 2 partons. The matching requirement is that $\Delta$ R $<$ 0.3 (above) and $\Delta$ R $<$ 0.4 (below).}
\label{fig:eff1d}
\end{figure}

The $\Delta R$ distribution between the generated and the closest reconstructed
jet is shown on the left side of Figure ~\ref{fig:deltaR} for the four algorithms studied in the
previous section, for a dataset of W + 2 partons Monte Carlo. 
We see that the Kt06 algorithm has the largest mean value of $\Delta$ R,
and therefore the worst matching, probably because of 
fluctuations far from the core of the jet. The same distribution
for jets in VBF Higgs events shows a smaller $\Delta$ R for all clustering
algorithms, showing that, in general, matching between generated and reconstructed
jets is better in VBF Higgs than in $W$ + parton events. 
To better understand the properties of matching, we will
study its behaviour as a function of jet kinematics. Figure ~\ref{fig:eff} 
shows the efficiency for various $p_T$ bins and for
a range of $\Delta R$ cuts for the algorithms considered in the previous
session, on a dataset of W + 2 partons Monte Carlo. For all algorithms, an
efficiency higher than 95\% (in red) is reached at high jet momenta even for
quite tight $\Delta R$ cuts, while small differences among algorithms emerge at
lower jet momenta. If we take the slices of this 2d plot corresponding to the
cuts $\Delta R < 0.3$  and $\Delta R < 0.4$, respectively, we obtain the 
results in Figure ~\ref{fig:eff1d}. 

\begin{figure}[tbh]
\begin{center}
\epsfig{file=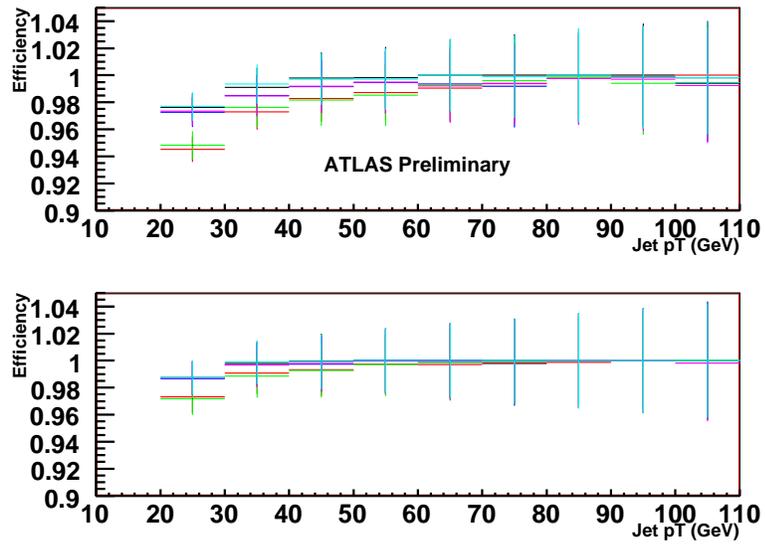,width=0.7\linewidth}
\end{center}
\caption{\label{fig:eff5p}Matching efficiency for $W$ + 5 parton events. 
The efficiency is smaller for all algorithms with respect to the $W$ + 1 parton case, 
but recombination-based algorithms show no worse behavior than the cone-based ones.}
\end{figure}

\begin{figure}[tbh]
\begin{center}
\epsfig{file=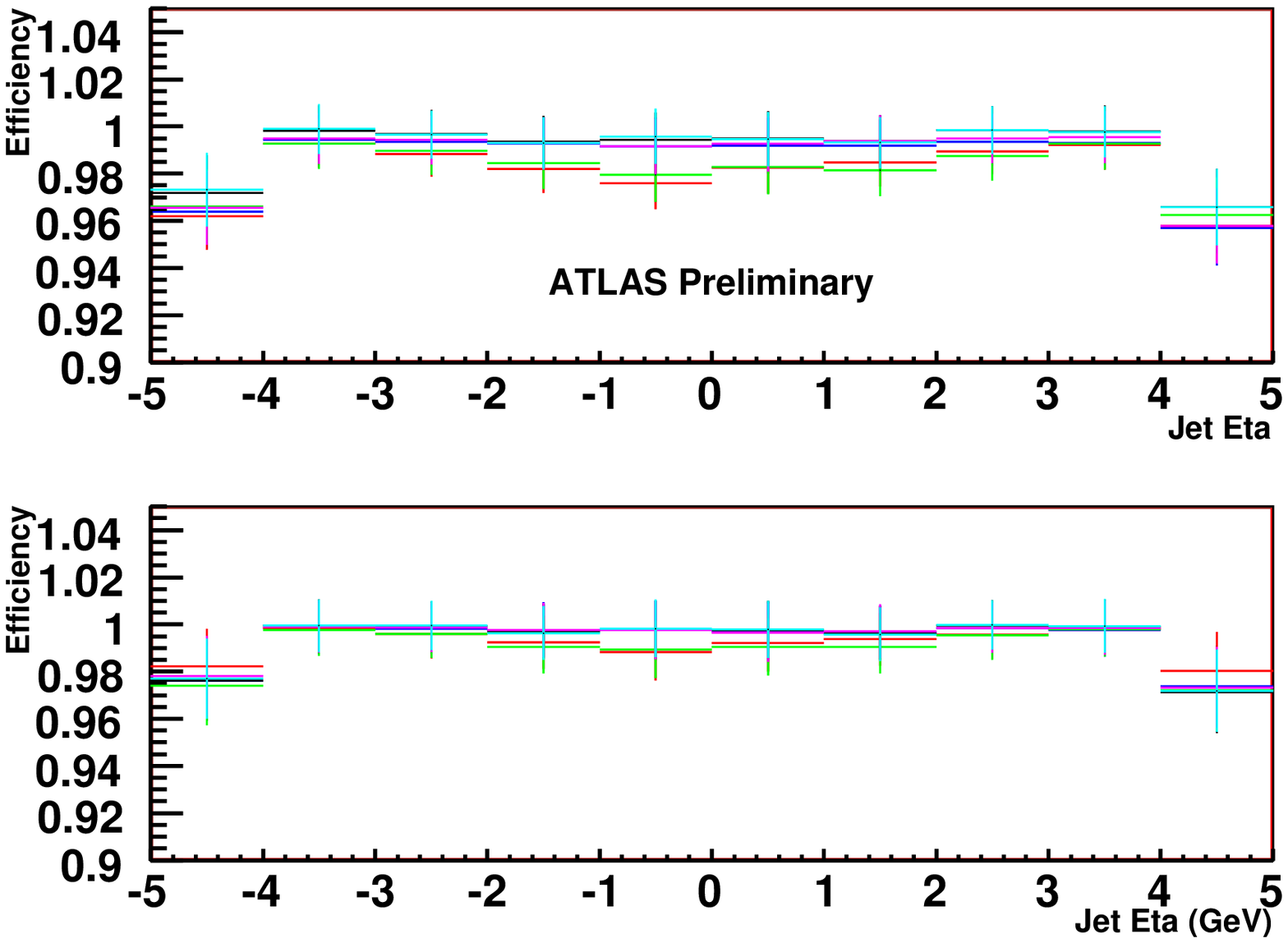,width=0.7\linewidth}
\end{center}
\caption{Matching efficiency for a fixed $\Delta$R cut as a function of jet $\eta$ 
for VBF Higgs Monte Carlo. The matching requirement is $\Delta$ R $<$ 0.3 (above) and
$\Delta$ R $<$ 0.4 (below).}
\label{fig:effvseta}
\end{figure}

These plots were produced from a $W$ + 2 partons dataset, but all other datasets
exhibit a similar behaviour, even for large parton multiplicities (see Figure 
~\ref{fig:eff5p} for W + 5 partons). SISCone does
a very good job under these difficult situations, and fears of the $k_T$ algorithm
picking up too much underlying event seem justified only in the case of large 
jet size. 
The matching efficiency as a function of the jet $\eta$ for VBF Higgs events
is shown in Figure~\ref{fig:effvseta}.
It is interesting to note how the endcap region, with $2<|\eta|<3$, equipped
with a Liquid Argon calorimeter with good pointing capabilities, is on average
more efficient than the barrel and the very forward endcap. The different
$\eta$ distribution, as well as the harder spectrum, may explain why jets 
from VBF Higgs events have a better matching efficiency than those from
$W$ + parton events.

\begin{figure}[tbh]
\begin{center}
\epsfig{file=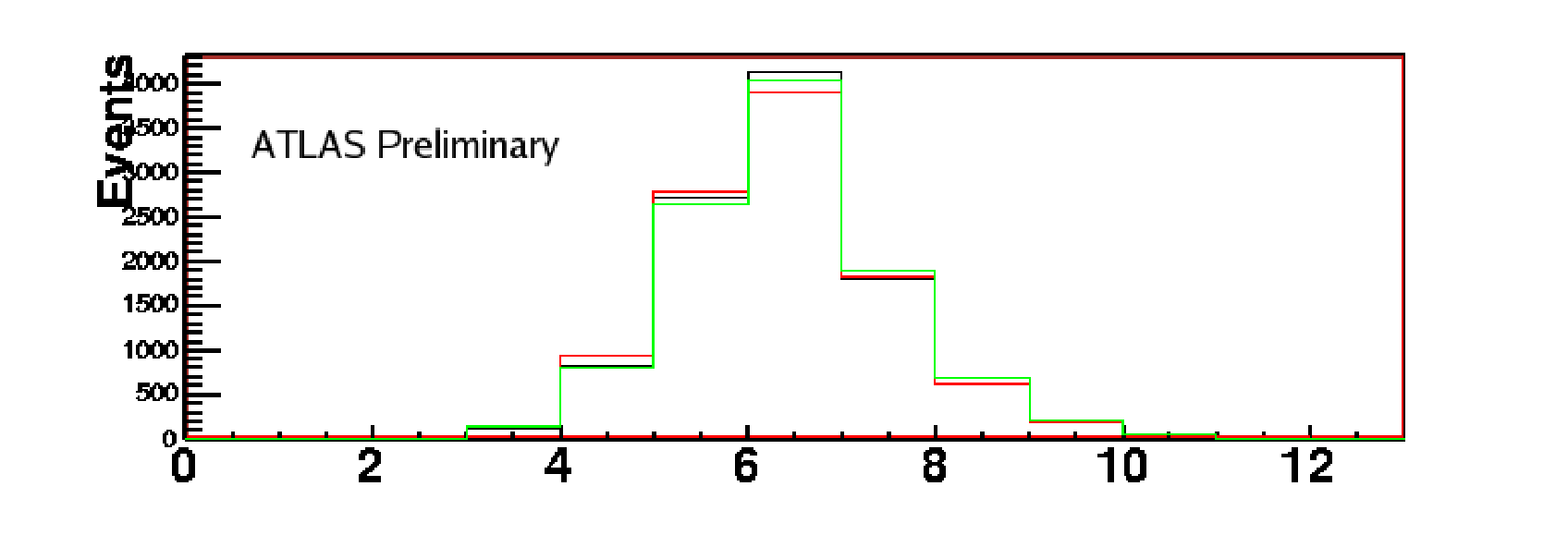,width=0.6\linewidth}
\end{center}
\caption{Number of jets 
using the Midpoint algorithm with seed threshold of 0.1, 1 and 2 GeV.}
\label{fig:njetsseed}
\end{figure}

\subsection{Seed threshold and split/merge parameter}

An obvious argument in favour of a seedless clustering algorithm is that the
seed threshold is in principle an arbitrary parameter, and the dependence of
jet reconstruction on arbitrary parameters should be avoided as much as possible.
On the other hand, from the experimental point of view, any seed below the
calorimeter noise-suppression cut should be equivalent, and no dependence on
seed threshold should be observed for reasonable values of this parameter.
To test this hypothesis, we looked at $W$ + 5 parton events, with very low jet
$p_T$ threshold (10 GeV).
The number of jets 
reconstructed with the MidPoint algorithm with seed 
thresholds of 0.1, 1 and 2 GeV is  shown in Figure 
~\ref{fig:njetsseed}. 
We see that
no significant difference is found for the different seed values, so the claim
that reasonable seed values lead to similar results seems justified, at least for inclusive distributions of the type examined here.
This fact does not reduce the merits of the seedless algorithm.\par

\begin{figure}[tbh]
\begin{center}
\epsfig{file=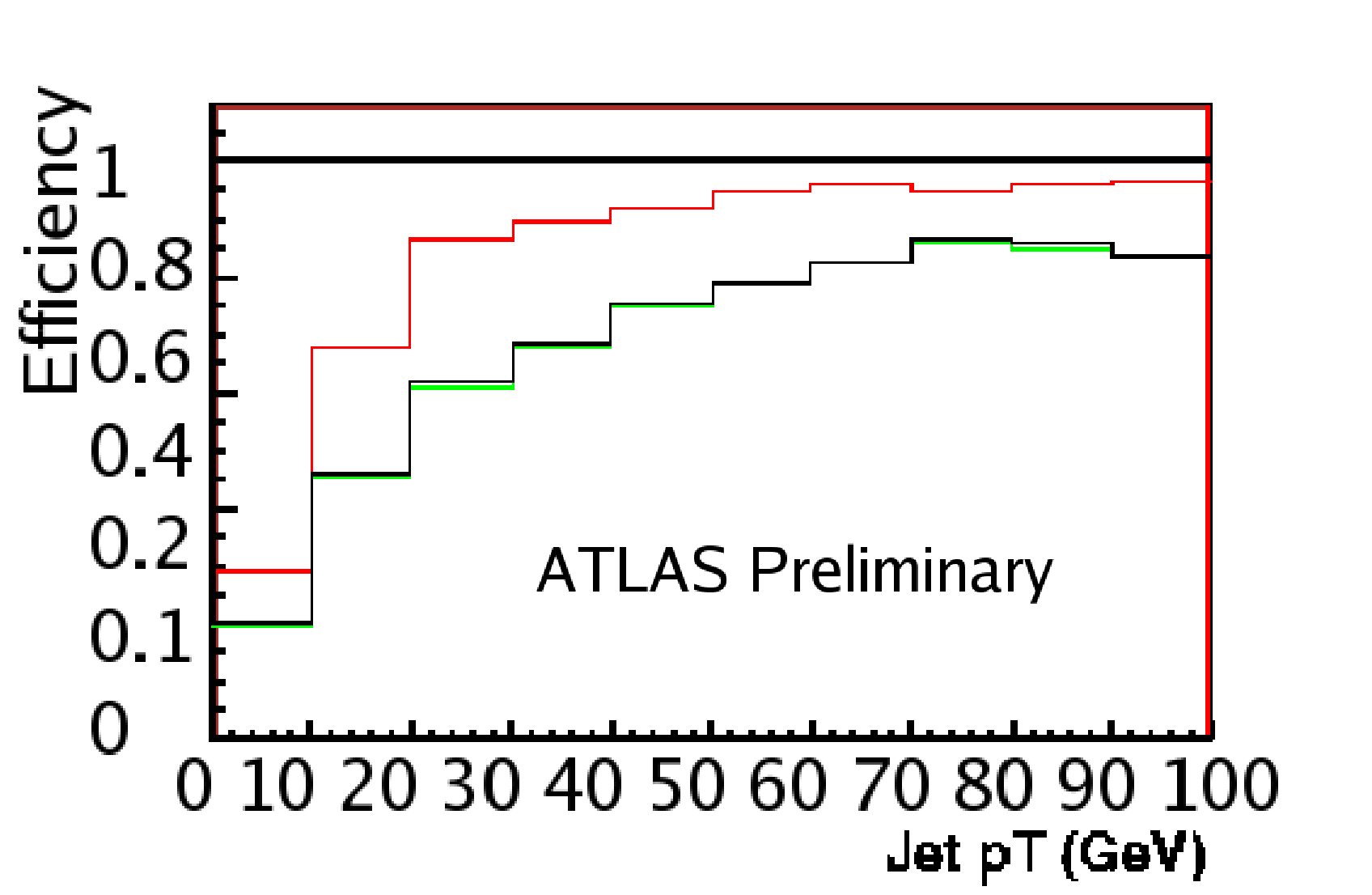,width=0.7\linewidth}
\end{center}
\caption{Matching efficiency for $\Delta$ R$<0.1$ for
SISCone, for values of the split/merge parameter of 0.5, 0.625 and
0.75.}
\label{fig:splitmergesc}
\end{figure}

To address the issue of the dependence of jet clustering on the split/merge 
parameter, we clustered $W$ + 2 parton events using the Atlas cone and SISCone algorithms
with this parameter set to 0.5, 0.625 and 0.75. Large differences are observed,
as seen for example for the SISCone case in Figure  ~\ref{fig:splitmergesc}.  
Perhaps a systematic study to fine tune this parameter could be useful.
We noticed that, out of the three options considered here, the best value
of this parameter is algorithm-dependent, and is in fact 0.5 for the Atlas cone
and 0.75 for SISCone, which are presently the default values for these 
algorithms.

\subsection{Energy reconstruction}

\begin{figure}[tbh]
\begin{center}
\epsfig{file=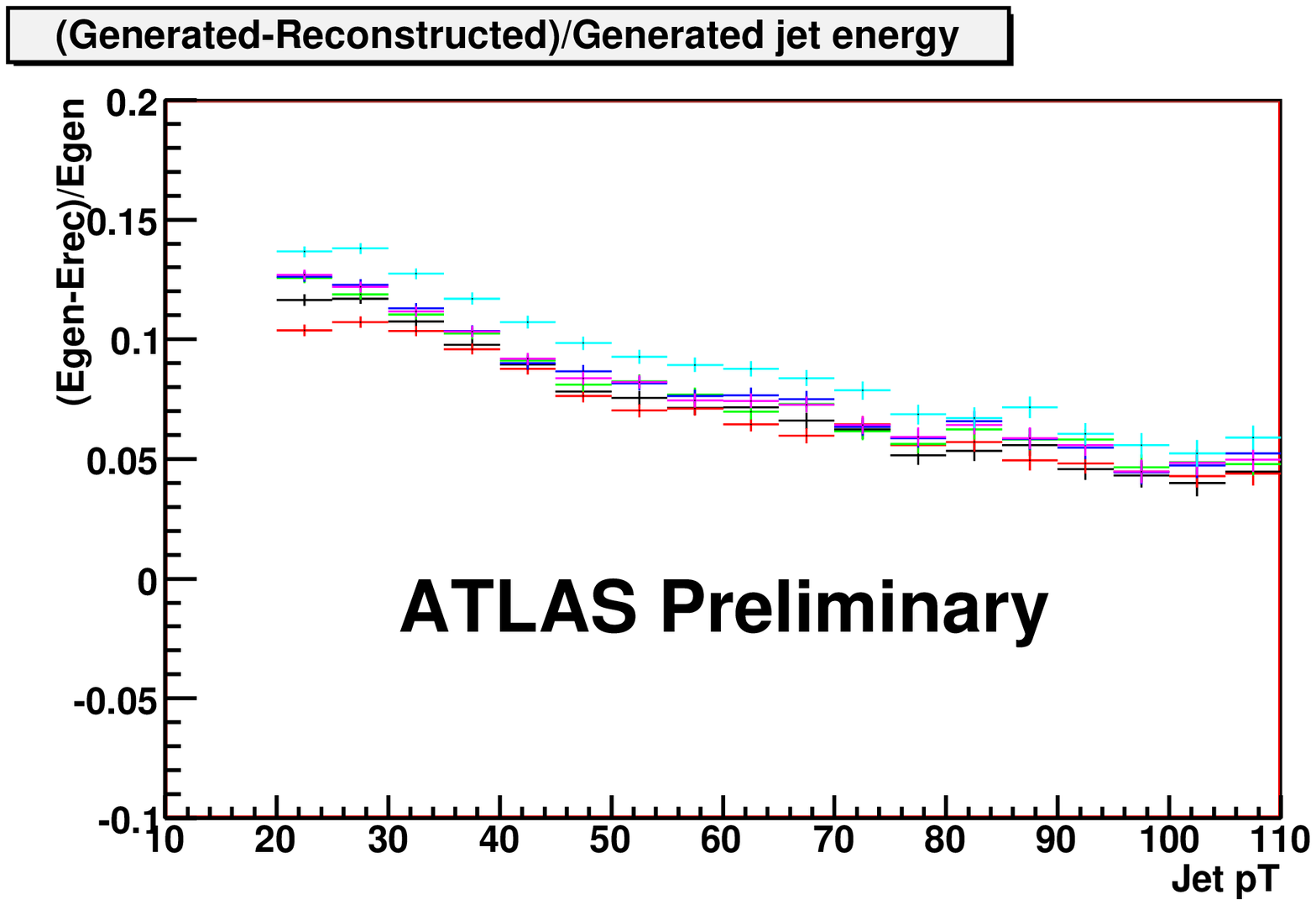,width=0.7\linewidth}
\end{center}
\caption{Difference between hadron and detector level jet $p_T$, divided by the
hadron level one, as a function of jet $p_T$. The observed bias is due 
to a small residual correction needed for topoclusters, especially at low energy.}
\label{fig:biasvspt}
\end{figure}

\begin{figure}[tbh]
\begin{center}
\epsfig{file=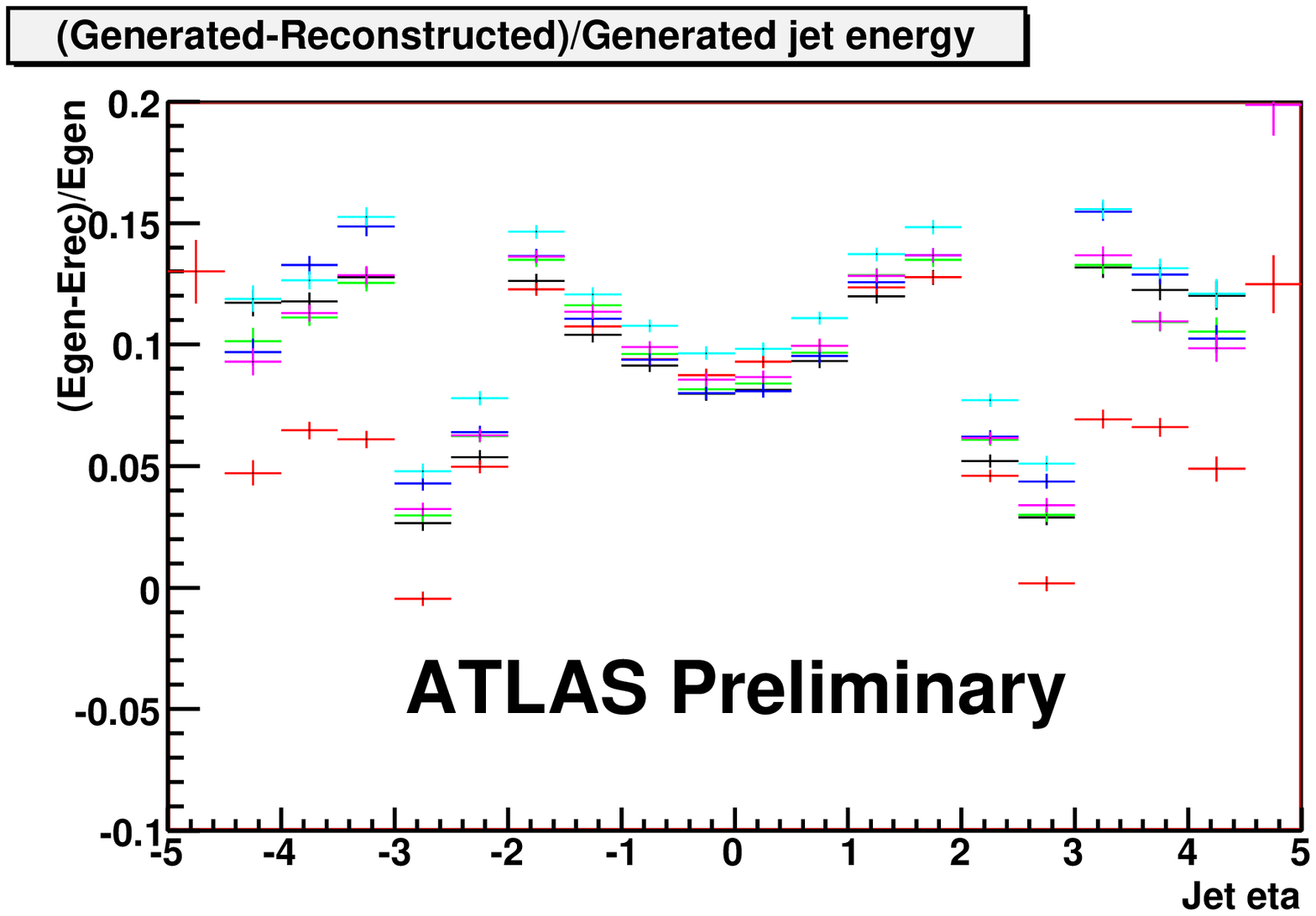,width=0.7\linewidth}
\end{center}
\caption{Same plot as before, as a function of jet eta.}
\label{fig:biasvseta}
\end{figure}

Even after compensation for the different calorimeter response to 
electromagnetic and hadronic showers, Atlas topological clusters currently underestimate 
the total visible energy by about 5\% due to noise-suppression thresholds, 
particle losses, inefficiencies etc.
This effect results in a systematically higher hadron-level energy with 
respect to the detector-level one, and is visible as a function of
jet $p_T$ and $\eta$ for $W$ + 2 parton events in Figures \ref{fig:biasvspt} and
\ref{fig:biasvseta}.
As expected, this bias is larger for low-energy jets where
the relative importance of low-energy clusters (more prone to losses etc.)
is higher. Also, the behavior in regions close to the badly-instrumented
parts of the detector differs considerably between the various algorithms.

\subsection{Cross sections}
\begin{figure}[tbh]
\begin{center}
\epsfig{file=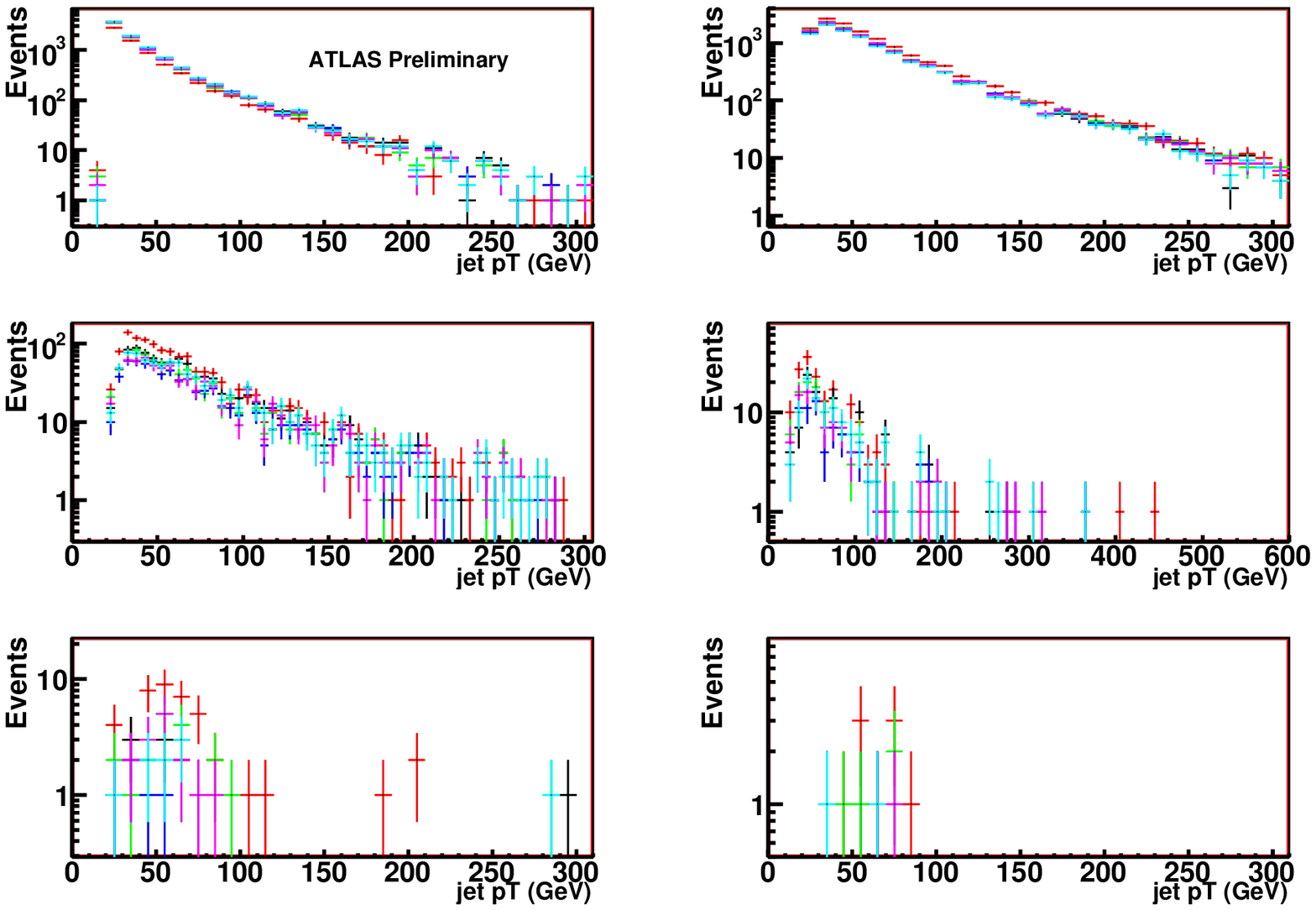,width=0.7\linewidth}
\end{center}
\caption{Reconstructed cross sections for the $W$ + 2 partons sample, as a function of the $p_T$ of the leading jet, for six jet multiplicities (A.U.)}
\label{fig:xsecrec}
\end{figure}

The study of $W$ + n jet cross sections, i.e. the $p_T$ distributions of the most
energetic jet for various jet multiplicities, allows a study of the effect of jet
clustering on energy distributions as well as on jet multiplicities. To select events
with W boson decays into a muon and a neutrino, we require the presence in the
event of a muon of at least 25 GeV in the acceptance region $|\eta|<2.4$ and 
missing transverse energy of at least 25 GeV. We accept jets if they have
transverse momentum larger than 15 GeV, $|\eta|<5$ and $\Delta R > 0.4$ with
respect to the muon. Events are classified according to the number of 
reconstructed jets. We studied the  distribution of the $p_T$ of the leading jet for
$W$ + n parton events. For space reasons, we show here only those obtained with
the $W$ + 2 parton sample, but all other distribution show similar 
characteristics. The reconstructed spectra of the leading jet are
shown in Figures ~\ref{fig:xsecrec}.
We see that the different behavior observed for the jets reconstructed with  the
KT06 algorithm is mainly due to the very soft region. Since, with this jet
size, there is the tendency of reconstructing a larger average number of jets, 
there are fewer events placed in the $W$ + 1 jet category (the red histogram
is always below the others for the first plot), and more in the cases where the
reconstructed multiplicity is higher than the generated one (all plots from
the third on). However, looking at the $p_T$ spectra, we realize that this effect
is mainly present for events with a soft leading jet, while for hard events
(i.e. for higher $p_T$ of the leading jets) all distributions tend to converge.

\subsection{Pileup}
We know that in the first phases of LHC operation, the proton density
in the bunches will be already high enough for the events to exhibit 
non-negligible pileup. No study of clustering algorithms would be complete
without an assessment on the behaviour under realistic running conditions.
Assuming that pileup can be added linearly to the event, we overlapped three
minimum-bias events to the $W$ + partons and Higgs VBF events considered in the
previous sections, and examined how the quantities considered above are
modified for the various algorithms.
\par
\begin{figure}[tbh]
\begin{center}
\epsfig{file=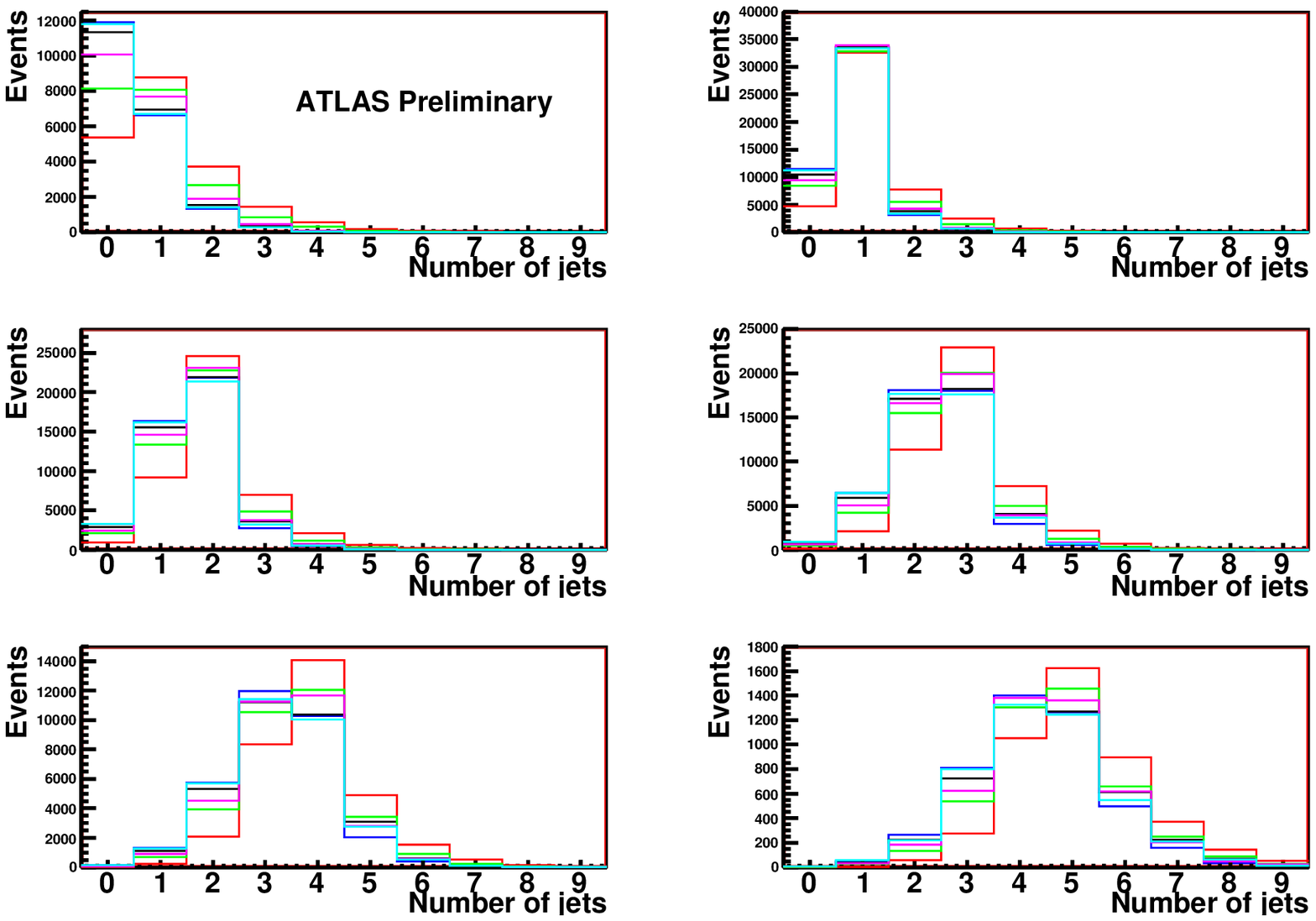,width=0.7\linewidth}
\end{center}
\caption{Number of reconstructed jets for the various $W$ + n parton samples, 
in the presence of three pileup events.}
\label{fig:njets_pu}
\end{figure}

\begin{figure}[tbh]
\begin{center}
\epsfig{file=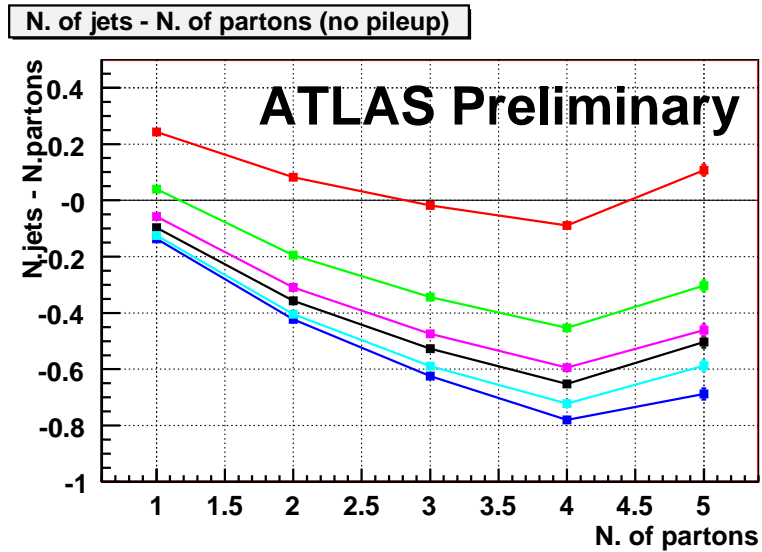,width=0.7\linewidth}
\end{center}
\caption{Difference between number of reconstructed jets and generated partons 
vs number of partons (with pileup)}
\label{fig:njetsgraphpu}
\end{figure}

\begin{figure}[tbh]
\begin{center}
\epsfig{file=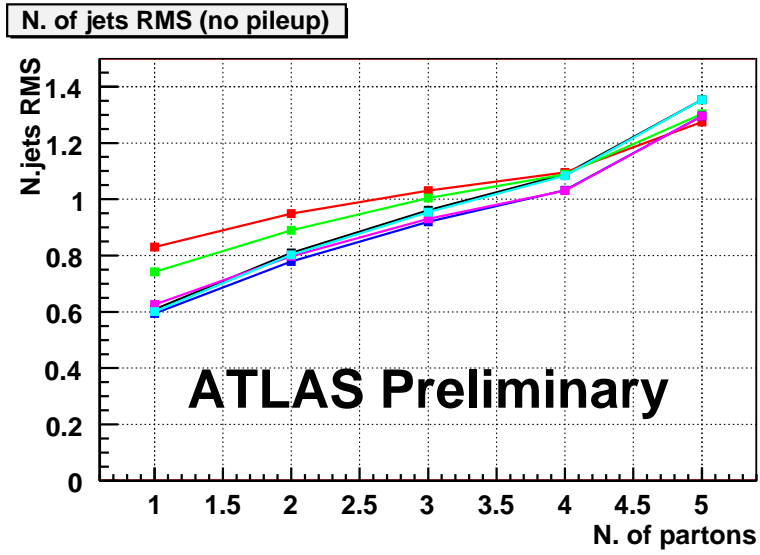,width=0.7\linewidth}
\end{center}
\caption{RMS of the distribution of reconstructed jets  
vs number of partons (with pileup)}
\label{fig:rmsjetsgraphpu}
\end{figure}

\begin{figure}[tbh]
\begin{center}
\epsfig{file=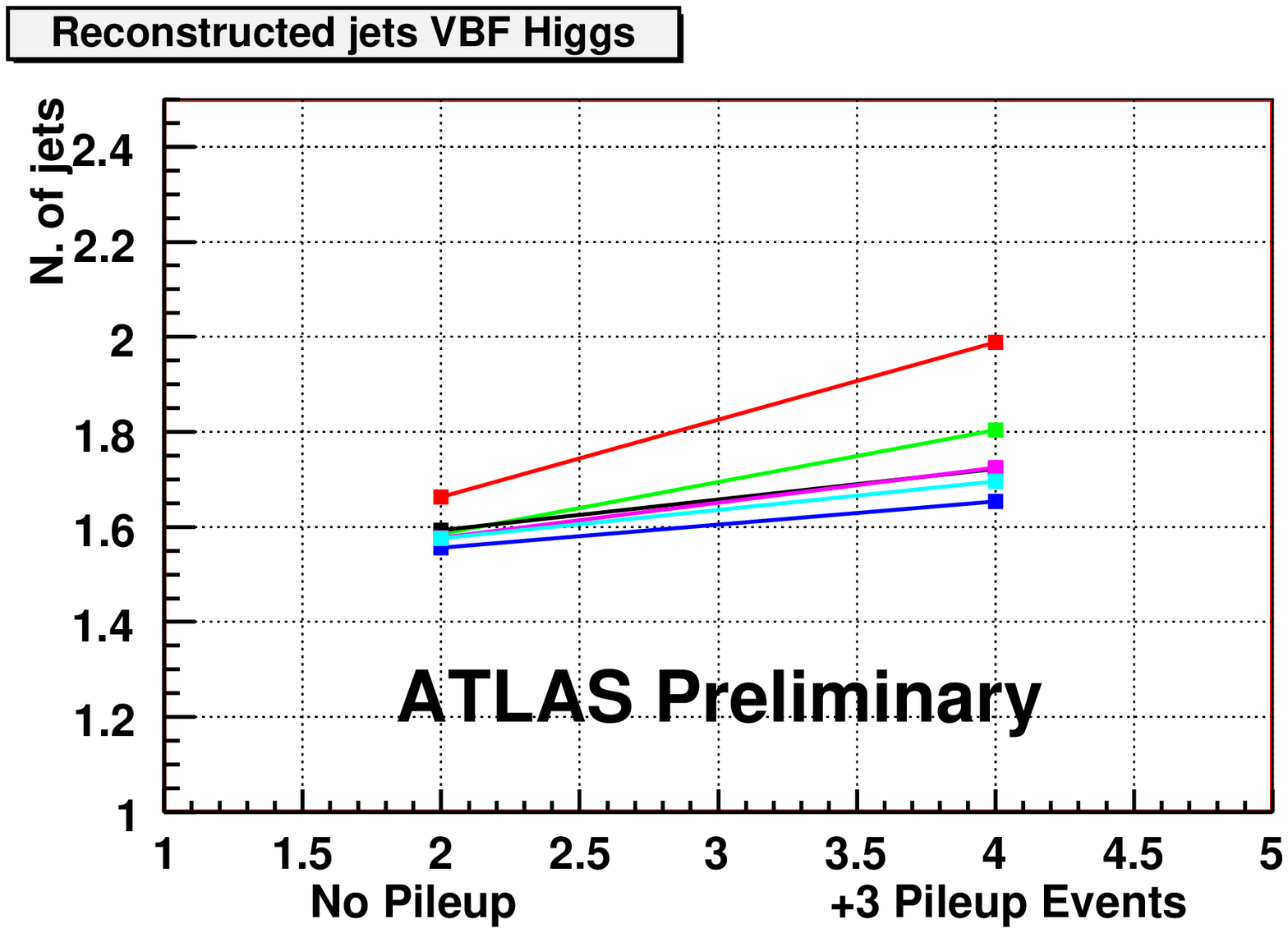,width=0.7\linewidth}
\end{center}
\caption{Number of reconstructed jets for VBF Higgs events with/out pileup}
\label{fig:njetgraphvbf_pu}
\end{figure}

The first property studied here is the jet multiplicity. We see that the
distribution of the number of jets for the $W$ + partons sample (Fig. 
\ref{fig:njets_pu}) is modified.  The behavior of the various algorithms
can be seen in the mean value and RMS of the reconstructed multiplicity as
a function of the number of partons (Figures ~\ref{fig:njetsgraphpu} and
\ref{fig:rmsjetsgraphpu}).
A direct comparison between the no-pileup and pileup case is made in 
Figure~\ref{fig:njetgraphvbf_pu}, where we show the average number of 
reconstructed jets for Higgs VBF events without and with pileup. Kt04 and
SISCone are the two algorithms that are less sensitive to the presence of pileup.
\par

\begin{figure}[tbh]
\begin{center}
\epsfig{file=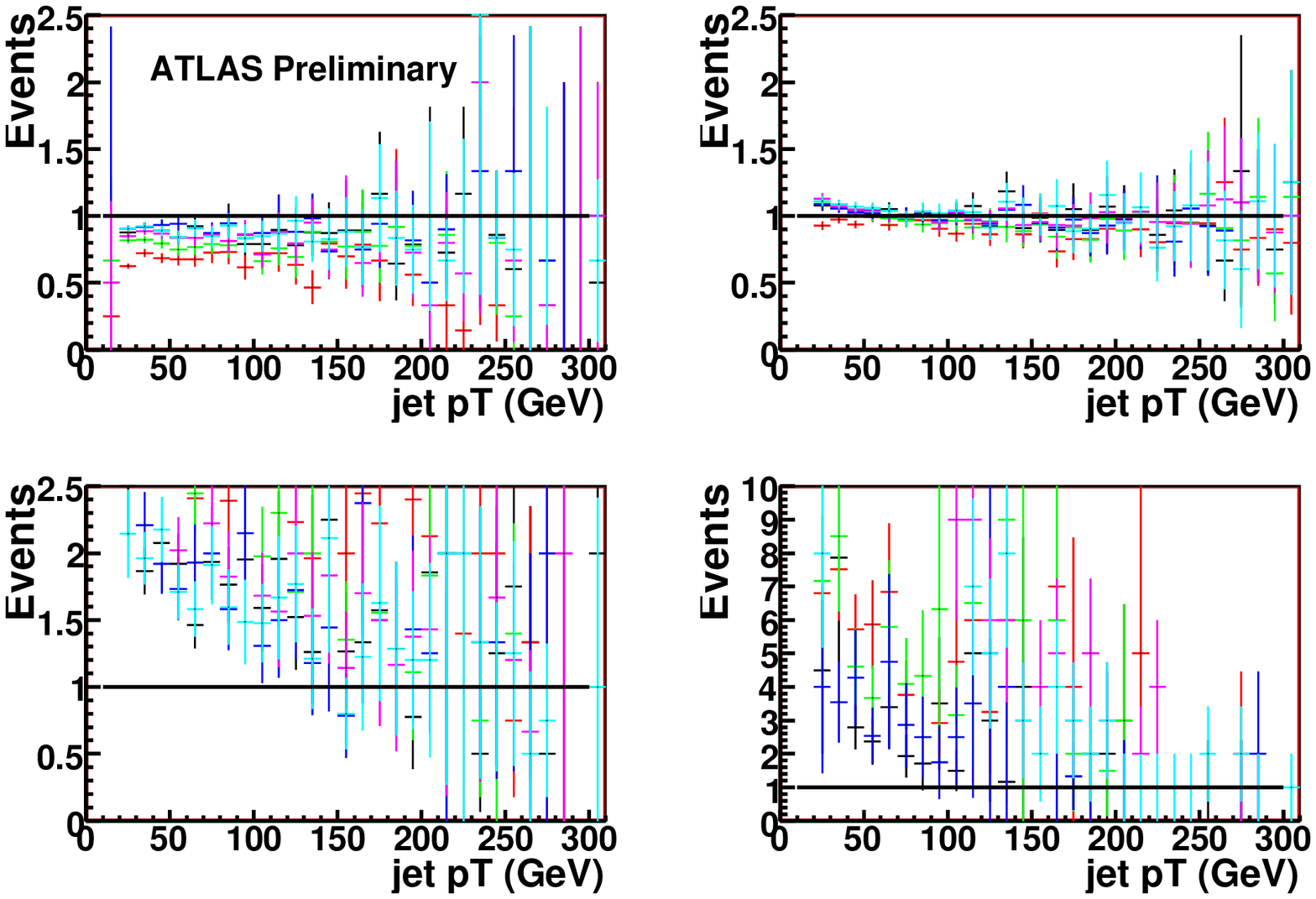,width=0.7\linewidth}
\end{center}
\caption{Ratio between cross section with and without pileup for all algorithms (W + 2 parton sample)}
\label{fig:xsecratio}
\end{figure}

In order to study the influence of pileup on the 
 kinematic distributions for the reconstructed jets, Figure ~\ref{fig:xsecratio} shows the ratio of the $p_T$ 
distributions with and without pileup for each reconstructed jet multiplicity,
for W + 2 parton events.
\par

\begin{figure}[tbh]
\begin{center}
\epsfig{file=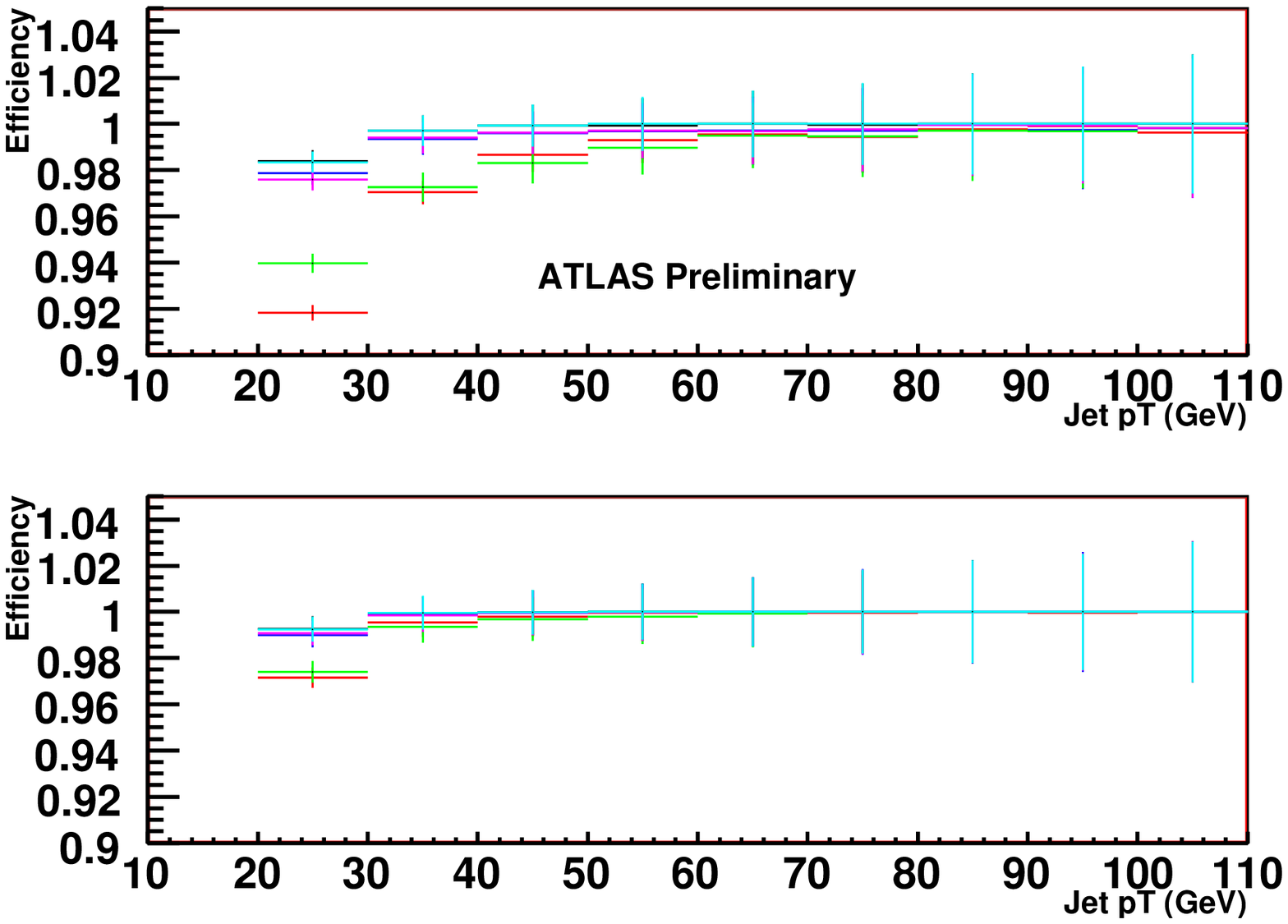,width=0.7\linewidth}
\end{center}
\caption{Efficiency vs jet $p_T$ with pileup ($\Delta$R$<$ 0.3 and 0.4)}
\label{fig:eff2pu}
\end{figure}

\begin{figure}[tbh]
\begin{center}
\epsfig{file=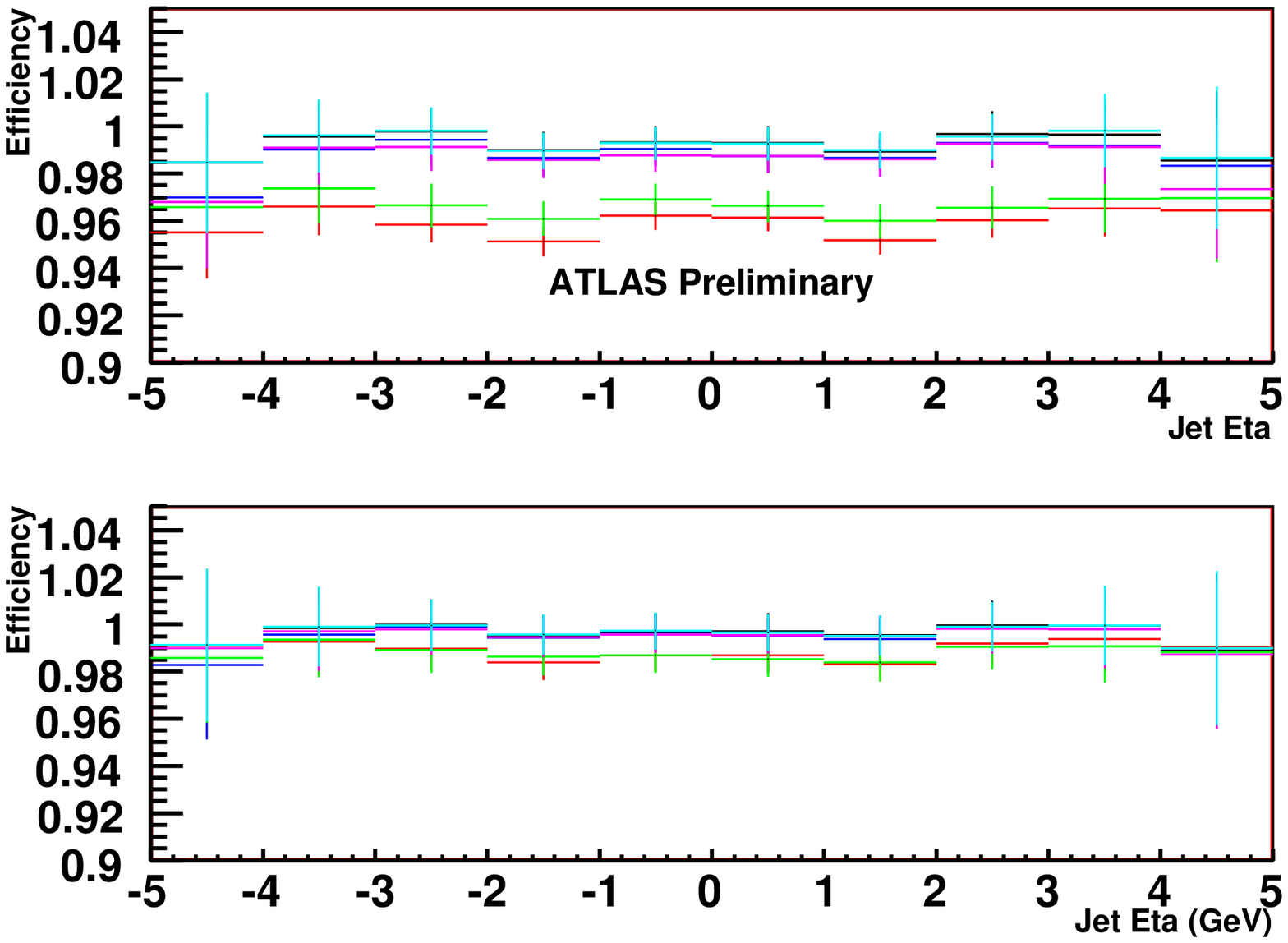,width=0.7\linewidth}
\end{center}
\caption{Efficiency vs jet $\eta$ with pileup ($\Delta$R$<$ 0.3 and 0.4)}
\label{fig:effetapu}
\end{figure}

The presence of pileup, leading to a modification of the jet axis direction,
also influences the matching efficiency between hadron level and detector
level jets. The efficiency as a function of jet $p_T$ and $\eta$, computed
using the same definition as in the previous sections, is shown in Figures
\ref{fig:eff2pu} and \ref{fig:effetapu}. Again, the scale of robustness
of the various algorithms to the presence of pileup obtained from the other
tests is confirmed.
\par

Finally, we tested the effect of using different algorithms on a simple
forward jet selection, aiming at a discrimination of  VBF Higgs events from the
background. The following cuts were applied to the VBF Higgs and to the $W$ + 2
partons and the $W$ + 3 partons Monte Carlo:

\begin{itemize}
\item Two jets with $P_T^1 > 40$ GeV and $P_T^2 > 20$ GeV
\item Both jets have $\Delta R > 0.4$ with respect to tau decay products
\item $\Delta \eta_{1,2} > 4.4$
\item Invariant mass between the two jets $>$ 700 GeV
\item No third jet with $|\eta|<3.2$ and $P_T > 30$
\end{itemize}
The efficiencies obtained in the three samples for three of the jet algorithms
under study here are summarized in Table
\ref{tab:seleff}.

\begin{table}[h]
\begin{center}
\begin{tabular}{|l|l|l|l|} \hline
Algorithm&VBF Higgs&W + 2p& W + 3p\\ \hline
Cone 04&15.9$\pm$0.4&0.37$\pm$0.03&1.17$\pm$0.05\\
KT 04&15.1$\pm$0.4&0.17$\pm$0.02&0.85$\pm$0.04\\
SISCOne 04&14.2$\pm$0.4&0.17$\pm$0.02&0.76$\pm$0.04\\ \hline
\end{tabular}
\label{tab:seleff}
\caption{Selection efficiency for the forward jet cuts described in the
text, for the various algorithms applied to the three Monte Carlo samples
of VBF Higgs, $W$ + 2 and $W$ + 3 partons}
\end{center}
\end{table}
\par
While the change in efficiency for the Higgs signal is quite marginal, the same
cannot be said for the difference in background rejection. Here the algorithms
that have proven to be more robust under the influence of pileup exhibit a
much better background rejection, and can improve the power of the analysis.

\subsection{Conclusions}
In this note we have systematically explored the behavior of several jet algorithms,
Kt (with different jet sizes, corresponding to the choice of D parameter of 0.4 and
0.6), the Atlas Cone, SISCone, MidPoint (all for cone size of 0.4) and Cambridge/Aachen, on several benchmarks with and without
the presence of pileup. The 
comparison of the smaller and larger jet sizes in the $k_T$ algorithm has shown that
the use of larger jets deteriorates the resolution in jet direction, and is 
more vulnerable to the presence of pileup, so  should be avoided for the 
purpose of jet finding, even if
it may be more accurate in determining the jet energy.
\par
The comparison of the different algorithms with approximately the same
jet size, corresponding to a radius of 0.4, indicates that the $k_T$ and SISCone 
algorithms have proven to be as good or better than algorithms more 
traditionally used in hadron colliders. 
